\documentclass[a4paper,10pt,twoside, titlepage,openright,colorlinks]{book}
\input epsf

\usepackage[usenames,dvipsnames]{color}
\definecolor{DarkRed}{RGB}{195,0,0} 
\usepackage[latin1]{inputenc}
\usepackage[english]{babel}
\usepackage{graphics}
\usepackage{amsfonts,amssymb,amsmath,latexsym,epsfig}
\usepackage{verbatim}
\usepackage{natbib}
\usepackage{makeidx}
\usepackage{sidecap} 
\usepackage{fancyhdr}
\usepackage[bf]{caption}
\usepackage{fix-cm}

\usepackage{lettrine}
\input GoudyIn.fd
\newcommand*\initfamily{\usefont{U}{GoudyIn}{xl}{n}}
\newcommand{\init}[2]{\lettrine[lines=6,slope=0pt,nindent=0pt]{\fontsize{85}{100}\selectfont {\initfamily {#1}}}{{\large {\sc #2}}}}

\makeindex

\defcitealias{lau09a}{Paper\,I}
\defcitealias{lau09b}{Paper\,II}
\defcitealias{lau10a}{Paper\,III}
\defcitealias{fyn10}{Paper\,IV}

\newcommand{\aap}[3]  {\href{http://adsabs.harvard.edu/cgi-bin/nph-abs_connect?version=1&warnings=YES&partial_bibcd=YES&sort=BIBCODE&db_key=ALL&bibstem=A\%26A&year=#1&volume=#2&page=#3&nr_to_return=100&start_nr=1}{#1, A\&A, #2, #3}}
\newcommand{\aipc}[3] {\href{http://adsabs.harvard.edu/cgi-bin/nph-abs_connect?version=1&warnings=YES&partial_bibcd=YES&sort=BIBCODE&db_key=ALL&bibstem=AIPC&year=#1&volume=#2&page=#3&nr_to_return=100&start_nr=1}{#1, AIP Conf.~Proc., #2, #3}}
\newcommand{\aj}[3]   {\href{http://adsabs.harvard.edu/cgi-bin/nph-abs_connect?version=1&warnings=YES&partial_bibcd=YES&sort=BIBCODE&db_key=ALL&bibstem=aj&year=#1&volume=#2&page=#3&nr_to_return=100&start_nr=1}{#1, AJ, #2, #3}}
\newcommand{\apjl}[3] {\href{http://adsabs.harvard.edu/cgi-bin/nph-abs_connect?version=1&warnings=YES&partial_bibcd=YES&sort=BIBCODE&db_key=ALL&bibstem=apjl&year=#1&volume=#2&page=L#3&nr_to_return=100&start_nr=1}{#1, ApJL, #2, L#3}}
\newcommand{\apjs}[3] {\href{http://adsabs.harvard.edu/cgi-bin/nph-abs_connect?version=1&warnings=YES&partial_bibcd=YES&sort=BIBCODE&db_key=ALL&bibstem=apjs&year=#1&volume=#2&page=#3&nr_to_return=100&start_nr=1}{#1, ApJS, #2, #3}}
\newcommand{\apj}[3]  {\href{http://adsabs.harvard.edu/cgi-bin/nph-abs_connect?version=1&warnings=YES&partial_bibcd=YES&sort=BIBCODE&db_key=ALL&bibstem=apj&year=#1&volume=#2&page=#3&nr_to_return=100&start_nr=1}{#1, ApJ, #2, #3}}
\newcommand{\apss}[3] {\href{http://adsabs.harvard.edu/cgi-bin/nph-abs_connect?version=1&warnings=YES&partial_bibcd=YES&sort=BIBCODE&db_key=ALL&bibstem=ap\%26SS&year=#1&volume=#2&page=#3&nr_to_return=100&start_nr=1}{#1, Ap\&AA, #2, #3}}
\newcommand{\araa}[3] {\href{http://adsabs.harvard.edu/cgi-bin/nph-abs_connect?version=1&warnings=YES&partial_bibcd=YES&sort=BIBCODE&db_key=ALL&bibstem=ARA\%26A&year=#1&volume=#2&page=#3&nr_to_return=100&start_nr=1}{#1, ARA\&A, #2, #3}}
\newcommand{\arxiv}[2]{\href{http://arxiv.org/abs/#2}{#1 (arXiv:#2)}}
\newcommand{\aspc}[3] {\href{http://adsabs.harvard.edu/cgi-bin/nph-abs_connect?version=1&warnings=YES&partial_bibcd=YES&sort=BIBCODE&db_key=ALL&bibstem=ASPC&year=#1&volume=#2&page=#3&nr_to_return=100&start_nr=1}{#1, ASP Conf.~Proc., #2, #3}}
\newcommand{\ban}[3]  {\href{http://adsabs.harvard.edu/cgi-bin/nph-abs_connect?version=1&warnings=YES&partial_bibcd=YES&sort=BIBCODE&db_key=ALL&bibstem=BAN&year=#1&volume=#2&page=#3&nr_to_return=100&start_nr=1}{#1, BAN, #2, #3}}

\newcommand{\eas}[3]  {\href{http://adsabs.harvard.edu/cgi-bin/nph-abs_connect?version=1&warnings=YES&partial_bibcd=YES&sort=BIBCODE&db_key=ALL&bibstem=EAS&year=#1&volume=#2&page=#3&nr_to_return=100&start_nr=1}{#1, EAS, #2, #3}}
\newcommand{\iaus}[3] {\href{http://adsabs.harvard.edu/cgi-bin/nph-abs_connect?version=1&warnings=YES&partial_bibcd=YES&sort=BIBCODE&db_key=ALL&bibstem=iaus&year=#1&volume=#2&page=#3&nr_to_return=100&start_nr=1}{#1, IAUS, #2, #3}}
\newcommand{\iauss}[3]{\href{http://adsabs.harvard.edu/cgi-bin/nph-abs_connect?version=1&warnings=YES&partial_bibcd=YES&sort=BIBCODE&db_key=ALL&bibstem=iauss&year=#1&volume=#2&page=#3&nr_to_return=100&start_nr=1}{#1, IAUSS, #2, #3}}
\newcommand{\jkas}[3] {\href{http://adsabs.harvard.edu/cgi-bin/nph-abs_connect?version=1&warnings=YES&partial_bibcd=YES&sort=BIBCODE&db_key=ALL&bibstem=JKAS&year=#1&volume=#2&page=#3&nr_to_return=100&start_nr=1}{#1, JKAS, #2, #3}}
\newcommand{\mnras}[3]{\href{http://adsabs.harvard.edu/cgi-bin/nph-abs_connect?version=1&warnings=YES&partial_bibcd=YES&sort=BIBCODE&db_key=ALL&bibstem=mnras&year=#1&volume=#2&page=#3&nr_to_return=100&start_nr=1}{#1, MNRAS, #2, #3}}
\newcommand{\msais}[3]{\href{http://adsabs.harvard.edu/cgi-bin/nph-abs_connect?version=1&warnings=YES&partial_bibcd=YES&sort=BIBCODE&db_key=ALL&bibstem=MSAIS&year=#1&volume=#2&page=#3&nr_to_return=100&start_nr=1}{#1, Mem.~S.~A.~It.~Suppl., #2, #3}}
\newcommand{\nat}[3]  {\href{http://adsabs.harvard.edu/cgi-bin/nph-abs_connect?version=1&warnings=YES&partial_bibcd=YES&sort=BIBCODE&db_key=ALL&bibstem=natur&year=#1&volume=#2&page=#3&nr_to_return=100&start_nr=1}{#1, Nature, #2, #3}}
\newcommand{\pasj}[3] {\href{http://adsabs.harvard.edu/cgi-bin/nph-abs_connect?version=1&warnings=YES&partial_bibcd=YES&sort=BIBCODE&db_key=ALL&bibstem=PASJ&year=#1&volume=#2&page=#3&nr_to_return=100&start_nr=1}{#1, PASJ, #2, #3}}
\newcommand{\pasp}[3] {\href{http://adsabs.harvard.edu/cgi-bin/nph-abs_connect?version=1&warnings=YES&partial_bibcd=YES&sort=BIBCODE&db_key=ALL&bibstem=PASP&year=#1&volume=#2&page=#3&nr_to_return=100&start_nr=1}{#1, PASP, #2, #3}}
\newcommand{\rvmp}[3] {\href{http://adsabs.harvard.edu/cgi-bin/nph-abs_connect?version=1&warnings=YES&partial_bibcd=YES&sort=BIBCODE&db_key=ALL&bibstem=RvMP&year=#1&volume=#2&page=#3&nr_to_return=100&start_nr=1}{#1, Rev.~Mod.~Phys., #2, #3}}
\newcommand{\spie}[3] {\href{http://adsabs.harvard.edu/cgi-bin/nph-abs_connect?version=1&warnings=YES&partial_bibcd=YES&sort=BIBCODE&db_key=ALL&bibstem=SPIE&year=#1&volume=#2&page=#3&nr_to_return=100&start_nr=1}{#1, SPIE, #2, #3}}
\newcommand{\zap}[3]  {\href{http://adsabs.harvard.edu/cgi-bin/nph-abs_connect?version=1&warnings=YES&partial_bibcd=YES&sort=BIBCODE&db_key=ALL&bibstem=ZA&year=#1&volume=#2&page=#3&nr_to_return=100&start_nr=1}{#1, ZA, #2, #3}}

\newcommand\ion[2]{#1{\sc #2}}
\newcommand\fion[2]{$[$#1{\sc #2}$]$}
\newcommand{\Nhi}{N_{\textrm{{\scriptsize H}{\tiny \hspace{.1mm}I}}}}
\newcommand{\nhi}{n_{\textrm{{\scriptsize H}{\tiny \hspace{.1mm}I}}}}

\newcommand{\nd}{n_{\textrm{d}}}
\newcommand{\F}{F}
\newcommand{\Flam}{\F(\lambda)}
\newcommand{\T}{\mathcal{T}}
\newcommand{\zre}{z_{\mathrm{re}}}

\newcommand{\Msun}{M_\odot}
\newcommand{\Mpyr}{\ensuremath{\Msun}~yr\ensuremath{^{-1}}}
\newcommand{\h}{h_{\mathrm{SPH}}}
\newcommand{\farcs}{\mbox{\ensuremath{.\!\!^{\prime\prime}}}}
\newcommand{\kms}{km~s\ensuremath{^{-1}}}
\newcommand{\ergs}{erg~s\ensuremath{^{-1}}}
\newcommand{\cmsq}{cm\ensuremath{^{-2}}}
\newcommand{\cmcb}{cm\ensuremath{^{-3}}}
\newcommand{\esca}{\ergs~\cmsq~{\AA}\ensuremath{^{-1}}}

\newcommand{\red}[1]{\textcolor{red}{#1}} 
\newcommand{\KUcol}[1]{\textcolor{OliveGreen}{{\sf #1}}} 
\newcommand{\black}[1]{\textcolor{black}{#1}} 

\renewcommand{\sec}[1]{Sec.~\ref{sec:#1}}

\newcommand{\cha}[1]{Chap.~\ref{cha:#1}}
\newcommand{\app}[1]{App.~\ref{app:#1}}
\newcommand{\Cha}[1]{Chapter \ref{cha:#1}}
\newcommand{\fig}[1]{Fig.~\ref{fig:#1}}
\newcommand{\Fig}[1]{Figure \ref{fig:#1}}
\newcommand{\tab}[1]{Tab.~\ref{tab:#1}}
\newcommand{\Tab}[1]{Table \ref{tab:#1}}
\newcommand{\eq}[1]{Eq.~\ref{eq:#1}}
\newcommand{\Eq}[1]{Equation \ref{eq:#1}}

\renewcommand{\cap}{\footnotesize}
\newcommand{\partL}[1]{\part[\Large{#1}]{#1}}

\fancypagestyle{plain}{%
  \fancyhead{} 
 }
\usepackage[colorlinks,bookmarks=true,citecolor=OliveGreen,filecolor=cyan,linkcolor=DarkRed,urlcolor=blue,pdftex]{hyperref}

\begin{document}

\thispagestyle{empty}
\begin{figure}[!h]
\centering
 \vspace*{-4cm}
  \hspace*{-3cm}
  \includegraphics [width=1.7\textwidth] {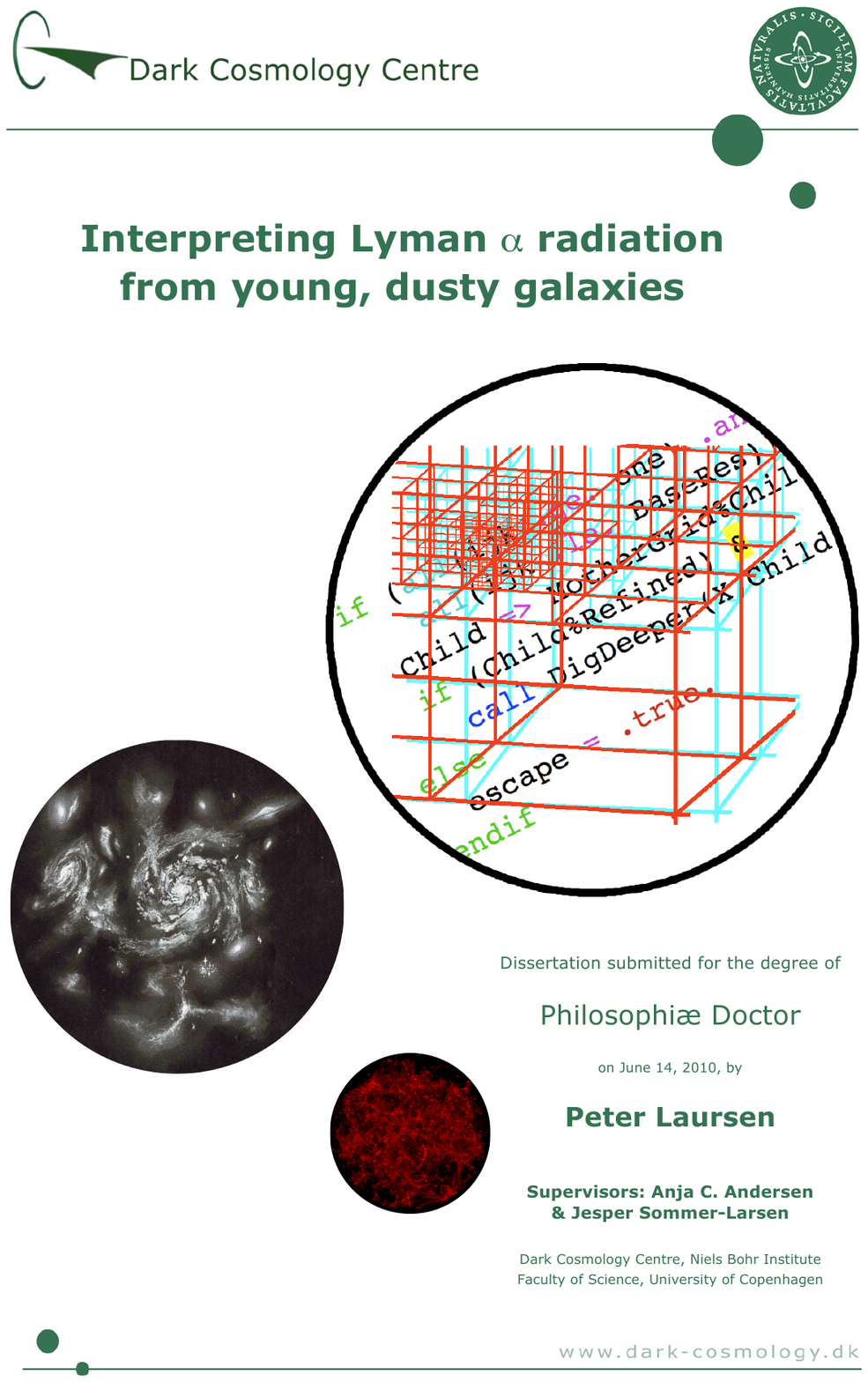}
  \end{figure}

\thispagestyle{empty}
\ \\
\vspace*{14cm}
\ \\
\hspace*{-3.5cm}
\thispagestyle{empty}
\begin{minipage}[!h]{17cm}
\KUcol{%
\emph{Cover art:}\vspace{3mm}
}

\KUcol{%
The front page illustrates the essence of this work, under the directives given
by the University Design Guidelines, which seem to be very preoccupied with
bubbles.\vspace{1mm}
}

\KUcol{%
The \emph{lower bubble} displays a snapshot from a cosmological simulation at
the time when the Universe was two and a half billion years old.
The region is some 13 million lightyears across.
Roughly one thousand galaxies float around in this volume, organized in
beautiful filamentary structures, separated by huge voids of
nothingness.\vspace{2mm}
}

\KUcol{%
Zooming in on one of the little dots, the \emph{middle bubble} shows a pencil
drawing of a few galaxies, slightly smeared out with my finger to create the
effect of gaseous nebulae, and subsequently color-inverted.\vspace{2mm}
}

\KUcol{%
The \emph{upper bubble} shows the closest zoom-in, symbolizing the fact that
this work is predominantly numerical. In the simulations, the physical
properties of the galaxies are represented by millions of numbers arranged in a
three-dimensional irregular matrix of cells. Two different
projections of this matrix are shown in \red{red} and \textcolor{cyan}{cyan},
respectively, generating
an anaglyph image which can be viewed with a pair of 3D glasses (although in
the printed version the cyan may come out a bit too dark).
In between the cells is seen a few lines from the computer code.
}
\end{minipage}
%
%
\newpage
\frontmatter
\ \\
\ \\
\ \\

\subsection*{\textsf{Abstract}}
\textsf{%
The significance of the Ly$\alpha$ emission line as a probe of the
high-redshift Universe has long been established. Originating mainly in the
vicinity of young, massive stars and in association with accretion of large
bulks of matter, it is ideal for detecting young galaxies, the fundamental
building blocks of our Universe.
Since many different processes shape the spectrum and the spatial distribution
of the Ly$\alpha$ photons in various ways, a multitude of physical properties
of galaxies can be unveiled.
}\vspace{1mm}

\textsf{%
However, this also makes the interpretation of Ly$\alpha$ observations
notoriously difficult. Because Ly$\alpha$ is a resonant line, it scatters on
neutral hydrogen, having its path length from the source to
our telescopes vastly increased, and taking it through regions of unknown
physical conditions.
}\vspace{1mm}

\textsf{%
In this work, a numerical code capable of calculating realistically the
radiative transfer of Ly$\alpha$ is presented. The code is capable of performing
the radiative transfer in an arbitrary
and adaptively refined
distribution of Ly$\alpha$ source emission, temperature and velocity field of
the interstellar and intergalactic medium, as well as density of neutral and
ionized hydrogen, and, particularly important, dust.
}\vspace{1mm}

\textsf{%
Accordingly, it is applied to galaxies simulated at high resolution, yielding 
a number of novel and interesting results, most notably the escape fractions
of Ly$\alpha$ photons, the effect of dust, and the impact of the transfer
through the intergalactic medium.
}\vspace{1mm}

\textsf{%
Furthermore, the remarkable detection of Ly$\alpha$ emission from a so-called
``damped Ly$\alpha$ absorber'' --- a special type of objects thought to be the
progenitor of present-day's galaxies --- is presented, and the potential of
the code for interpreting observations is demonstrated.
}\vspace{1mm}
\ \\
\ \\
\ \\
\ \\
\ \\
\ \\
\ \\
``\emph{It would seem that large digital computers could be applied very
profitably to this problem.}''\\
\ \\
Donald E.~Osterbrock (1962), on the complications of Ly$\alpha$ radiative
transfer.
\newpage

\ \\
\newpage

\ \\
\ \\
\ \\

\begin{center}
\subsection*{Acknowledgments}
\end{center}

I am very appreciative of having had the opportunity to work at the Dark
Cosmology Centre, with its inspiring atmosphere.
Although all my colleagues have contributed to making my days joyous, some have
played a more direct part in accomplishing my research:

Obviously, this thesis would not have been possible without the aid and support
from my supervisors Anja and Jesper. Your faith in my abilities have been
overwhelming, verging on the irrational.
The same must be said about Johan, who has also been very eager to include me
in various projects.
Steen has always been prompt to help me with numerical and statistical
issues.
Giorgos was very patient with me during reduction of data, and was a great
traveling partner and art co-investor.
I have benefited considerably from Bo's knowledge on various observational
aspects of Ly$\alpha$, and from Brian's perpetual sysadmin support.
It is a pleasure to thank all of you.
\textcolor{white}{Og tak til Morten for p{\o}lse.}

Last, but not least, in fact most, I am grateful to
Frk.~Lehmann for your truly invaluable support and for taking care of
everything, in particular through the past couple of months, and to
\href{http://www.dark-cosmology.dk/~pela/Pics/Sinus/Sigurd.png}
{\black{Sigurd Sinus}},
for making me remember the essential stuff in life, such as watching excavators
and throwing rocks in the water.

\newpage 

\tableofcontents

\vspace{3.9mm}
{\bf Bibliography}
\vspace{-8mm}\begin{flushright}{\bf \pageref{bib}}\end{flushright}

  {\bf Index}
  \vspace{-8mm}\begin{flushright}{\bf \pageref{ind}}\end{flushright}

\newpage

\mainmatter

\partL{Theoretical background}\label{theoback}

\chapter{Introduction}
\label{cha:intro}

\init{S}{ince the emergence of} a deeper consciousness, man\-kind has strived
to understand and explain the place we inhabit in the cosmos, along with the
numerous enchanting phenomena that surround us. The magnificent and
eternal sky, occasionally strewn with transient events, cannot help but
mesmerize and impose a spirit of inquiry on the beholder.

Possessing only pure reason and the eye, it is not surprising that our
ancestors ascribed most natural phenomena to the will of devine beings whenever
something seemed \emph{too} fantastic.
Without in any way denying the existence of such creatures,
perhaps the most stunning fact about Nature is that, at least to a very
large degree, it is comprehensible within the realm of known physical laws.
With the advent of modern telescopes and sensitive detectors we began to grasp
the vastness of the Universe, with our Milky Way only being one among at least
billions, and most likely an infinite amount of galaxies, organized in
beautiful filamentary structures throughout the cosmos.
Galaxies are thus the building blocks of our Universe, and to understand the
physics governing the formation, structure, and evolution of these
enigmatic conglomerations of stars, gas, and dust, as well as more exotic
components such as black holes and dark matter,
is a magnificent challenge of astronomy, revealing ultimately the necessary
conditions for the genesis of life itself.

Astronomers, unlike other physicists, do not have the privilege of being able
to perform experiments in a lab. Instead we either have to stick with whatever
experiment Nature performs for us, and then simply \emph{observe}, applying
our knowledge of physics in order to interpret our observations. Thanks to the
finite speed of light, however, a privilege that we \emph{do} have is our
ability to look back in time, and thus by observing different cosmic epochs
gain a picture of the evolution of the Universe and its contents.
Nevertheless, since most cosmological phenomena happen on timescales much
longer than the human lifetime, we are always confined to ``snapshots'' of the
observed objects.

To actually learn in much greater detail how things happen, we have another
possibility, namely to \emph{simulate} on a computer how we think Nature might
operate.
Although in principle a computer cannot do anything humans are not also able
to, they do it a billion times faster.
Over the years, many elaborate numerical codes have been constructed that
reproduce and predict the Universe on all scales, from dust agglomeration, over
planetary and stellar formation, to simulations of cosmological
volumes\footnote{That is, volumes that are more or less representative of the
Universe as a whole.}.
Although these codes include an steadily increasing number of physical
processes, numerical resolution, etc., and although they are able to predict
various
observables, many of them fail to account for the fact that the observed light
may be rather different from the light that was emitted.

Except for meteorites and dust grains in the neighborhood of Earth, the only
means of gaining information from the Universe is by observations of
light\footnote{At least until we manage to detect gravitational waves.},
be it cooling radiation from huge collapsing gas reservoirs forming galaxies,
spectral lines from
planetary nebulae, or thermal radiation from heated dust. However, once
radiation leaves its origins, it may still be subject to various physical
processes altering its intensity, direction, and spectral distribution. If we
do not understand these processes, we may severely misinterpret the predictions
of the models, when comparing to observations.

This is particularly true in the case of \emph{resonantly scattered} photons,
i.e.~photons that are absorbed and emitted in another direction with more
or less almost the same frequency. In this case, the photons will follow a
complicated path from it is emitted to it is detected, taking it through
regions of space that may be physically quite different from the those lying
between its point of emission and the observer.

\section{A brief history of Lyman $\alpha$}
\label{sec:lyahist}\index{Ly$\alpha$|textbf}

The most ``famous'' resonant line is Ly$\alpha$, the energy of which
corresponds to the energy difference between the ground state and the first
excited state of neutral hydrogen. Because this is the most frequent transition,
and because hydrogen constitutes more than 90\% of all elements in the
Universe by number, the Ly$\alpha$ line is often the most prominent line
emerging from astrophysical objects.
In \sec{sources}, the various physical processes
that may result in emission of Ly$\alpha$ are discussed in detail,
but in anticipation of
events we note that they are mainly produced by gas surrounding young, massive
stars, and, to a lesser extent, by the cooling of collapsing gas.
In their classic paper, \citet{par67} suggested how detection of young galaxies
would be feasible, using the Ly$\alpha$ line. Nevertheless, for almost three
decades only a few such Ly$\alpha$ emitters (LAEs, i.e.~galaxies
detected from their emission in Ly$\alpha$, either by
narrowband photometry or spectroscopy, see also \sec{LAEs}) were discovered
\citep[see, e.g.,][]{djo92}. Several theories were proposed to explain the high
amount of null-results, e.g.~suppression of the Ly$\alpha$ line due to metals
\citep{mei81}, absorption by dust\index{Absorption!By dust}
\citep{har88} and lower-than-expected
formation of massive stars \citep{val93}.

However, as surveys eventually were able to go deeper, and as searching wide
regions on the sky became feasible, large numbers of high-redshift, star-forming
galaxies were discovered. Notable surveys include
the Large Area Lyman Alpha survey\index{LALA}
\citep[LALA; e.g.,][]{rho00} and
the Subaru Deep Field survey\index{Subaru} \citep[e.g.,][]{tan05}, and together
with more recent surveys, gradually a census of the nature of LAEs at a wide
range of redshifts has been provided (e.g. \citet{gua10} at $z=2.1$,
\citet{nil09} at $z=2.3$, \citet{wan09} at $z=4.5$, \citet{ouc10} at $z=6.6$,
and for small, spectroscopically unconfirmed samples even farther, with
\citet{hib10} and \citet{til10}).
The current LAE redshift record holder \citep[$z = 6.96$;][]{iye06} is now five
years ``old'', demonstrating the arduousness of detecting galaxies thus
distant.
Galaxies up to $z \sim 8$ \citep{bou04} has been detected via the Lyman-break
technique (see \sec{LBGs}), but have not been spectroscopically
confirmed\footnote{Since the submission of this thesis the record has broken,
with a spectroscopically confirmed LAE at $z = 8.55$ \citet{leh10}.}
(both of these techniques are
discussed in the next chapter), but in time deeper observations will be
realized with, e.g., the Ultra-VISTA\index{Ultra-VISTA} project
(launched this spring, reaching $z=8.8$; PIs Dunlop, Le Fevre, Franx, \& Fynbo)
and the James Webb Space Telescope\index{JWST}
\citep[to be launched in 2013; e.g.,][]{gar06}.

For this reason, Ly$\alpha$ has become one of the most 
important diagnostic tools for exploring the high-redshift Universe.
Since it is an ultraviolet (UV) wavelength, the atmosphere is
opaque to Ly$\alpha$ emitted at redshifts smaller than $\sim$1.5, but hereafter
it is often easily detectable, and at $z \gtrsim 2.1$, it becomes the
strongest emission line in the optical-NIR window. 

Since it is often the only line visible, it is crucial for determining
redshifts \citep[e.g.,][]{hu02a,hu02b,kod03}.
From the shape of the line profile, equivalent width and offset from other
emission and absorption lines, information about morphology, kinematics
and underlying stellar population of the host galaxy can be gained;
\citet{kun98} used P Cygni features in the Ly$\alpha$ line to infer the
presence of strong, galactic outflows, while the equivalent width has been used
as evidence for an unusually strong ionizing continuum, possibly caused by
first generation (Pop\,III) stars \citep{mal02}.
Also, strong Ly$\alpha$ emission from distant
quasars are used to map the cosmic web of sheets and filaments in the
intervening Universe through Ly$\alpha$ \emph{absorption},
resulting in the so-called Ly$\alpha$ forest
\citep[e.g.,][see also \sec{LAF}]{wei03}.

\section{Motivation}
\label{sec:moti}

\begin{figure}[!t]
\centering
\includegraphics [width=0.48\textwidth] {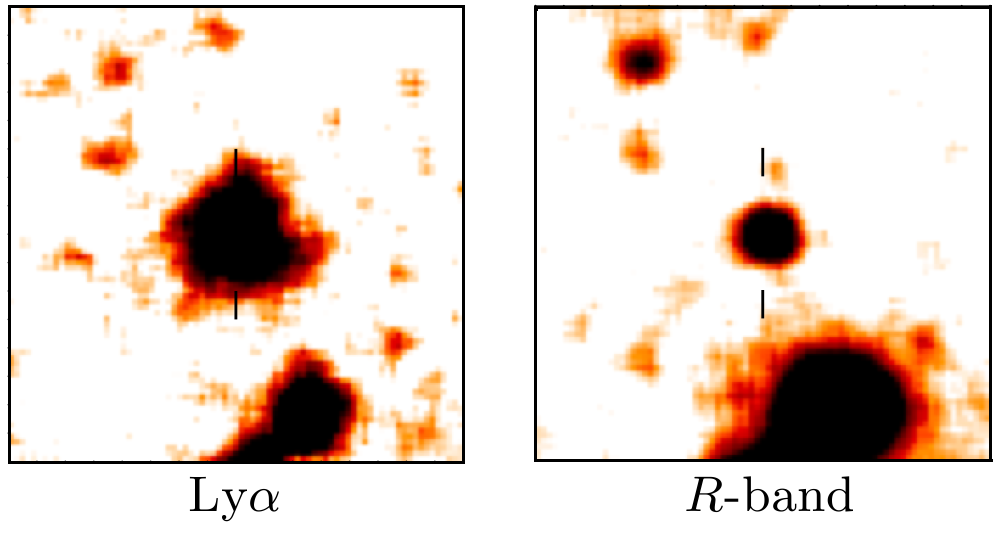}
\includegraphics [width=0.35\textwidth] {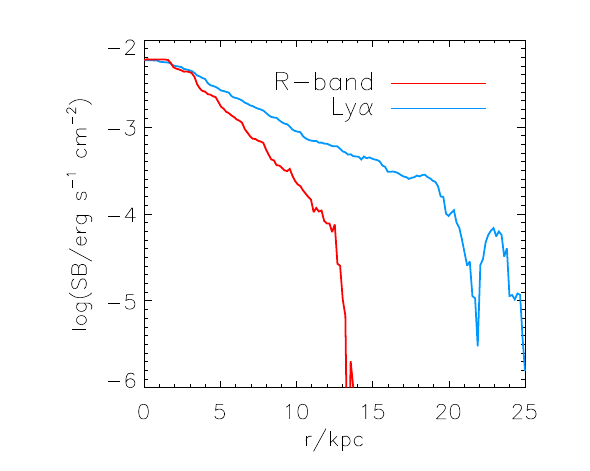}
\caption{{\small The Ly$\alpha$ emitter LEGO2138\_29 (redshift $z = 2.86$)
                 observed in Ly$\alpha$ \emph{left} and in the $R$-band
                 \emph{middle}.
                 \emph{Right} panel shows the azimuthally averaged
                 SB profiles, where the SB of the $R$-band
                 \emph{red} is
                 normalized such that its maximum coincides with the maximum of
                 the Ly$\alpha$ SB \emph{blue}. The source is clearly much more
                 spatially extended in Ly$\alpha$ than in the $R$-band.
                 Courtesy of \citet{fyn03}.}}
\label{fig:LEGO}
\end{figure}

Through the past years it has become possible to actually resolve
observationally these young Ly$\alpha$ emitting galaxies. In several cases,
the galaxies have been\index{Ly$\alpha$ emitters!Extendedness}
found to be significantly more extended on the sky when observed in Ly$\alpha$
as opposed to optical bands \citep[e.g.,][]{mol98,fyn01,fyn03,sai06}.
Specifically, \fig{LEGO} displays surface brightness (SB) maps of the LAE
LEGO2138\_29 \citep{fyn03}, observed in Ly$\alpha$ and in
the $R$-band, as well
as their SB profiles, i.e.~the SB of the objects averaged over the azimuthal
angle.
Because
the galaxies are still in a proto-phase, they should be surrounded by an
envelope of accreting and partially neutral gas, and thus this phenomenon may
be the result of the photons having to scatter their way out of the galaxies.

\begin{figure}[!t]
\centering
\includegraphics [width=0.90\textwidth] {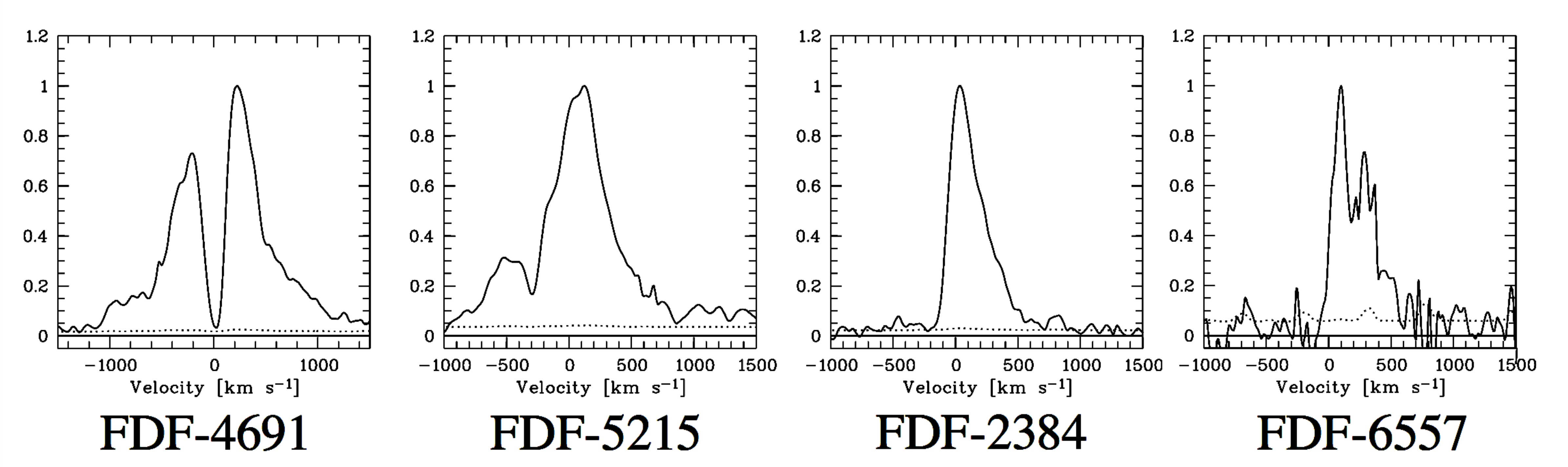}
\caption{{\cap Four examples of Ly$\alpha$ line profiles, obtained with a
               resolution of $R \simeq 2000$.
               \emph{Panel 1} shows a beautiful double-peaked profile.
               In \emph{panel 2}, the blue peak is diminished, but still
               clearly visible.
               \emph{Panel 3} is an archetypical Ly$\alpha$ profile, with the
               blue peak completely gone. Its asymmetry is a clear signature
               of resonant scattering processes.
               The profile in \emph{panel 4} is more noisy, and it is difficult
               to say whether we see one peak, as in panel 3, or two peaks
               blended by insufficient resolution.
               From \citet{tap07}.}}
\label{fig:LAEspec}
\end{figure}

Spectroscopically, LAEs also exhibit an interesting feature: As will be
discussed in (much!) further detail in \cha{ResScat}, due to the high opacity
of neutral hydrogen for a photon in the line center, in general the radiation
will have to diffuse in frequency to either the red or the blue side of the
line center, and should thus escape in a double-peaked profile. Although this
has indeed been observed in several cases \citep[e.g.][]{yee91,ven05,tap07},
more often the Ly$\alpha$ profile are characterized by an asymmetric profile
resembling the red peak of the anticipated double-peak\index{Double-peaks}
profile
\citep[e.g][]{nil07,gro09}. \Fig{LAEspec} shows some typical LAE line
profiles.
In some cases this may simply be an issue of insufficient spectral resolution,
but galactic outflows or absorption in the intergalactic medium (IGM) could
also be the cause.

To cast light upon the physical mechanisms that might govern these phenomena
and thus learn more about the LAEs themselves, the complex
problem of Ly$\alpha$ RT needs to be solved. 
The puzzle is further complicated by the possible presence of dust in the
galaxies, the effect of which is yet quite poorly constrained.
Since photons are only observed from the location where their last scattering
took place, it is difficult to say exactly where it was produced, how long a
distance it has been traveling, as well as through which sort of regions.

The problem of Ly$\alpha$ RT under various astrophysical
circumstances has already been dealt
with, both analytically \citep[e.g.,][]{ost62,ada72,har73,neu90,loe99} and
numerically \citep[e.g.][]{ahn00,ahn01,zhe02,can05,tas06a,dij06a,han06,ver06}. 
Unfortunately, due to analytical and computational limitations, most of the works
have had to consider only very simplified constellations of gas, not capturing
the full intricacy of realistic astrophysical situations. A number of studies
have used cosmological simulations to predict the appearance and emergent
spectrum from LAEs
\citep[e.g.,][]{far01,fur03,fur05,bar04,gne04,can05,le05,le06}. However, with
a few important exceptions \citep{can05,tas06a,lau07,kol10,fau10},
traditionally the radiative transfer (RT) of Ly$\alpha$ radiation has been
treated as absorption instead of scattering, simply
modyfying the observed intensity by a factor of $e^{-\tau}$, where $\tau$ is
the optical depth of neutral hydrogen lying between the emitter and the
observer. For moderately thin media, this may be an acceptable approximation,
since the radiation is scattered out of the direction to the observer, but in
the dense regions in the vicinity of the emitter, radiation that is initally
emitted in a different direction may be scattered \emph{into} the direction of
the observer.


\section{Structure}
\label{sec:struct}

Toward these ends, a numerical code, capable of predicting the diffusion of
Ly$\alpha$ radiation in real and frequency space, as well as the effects of
dust,
is constructed and applied to simulated galaxies resulting from a fully
cosmological simulation.
An early, more primitive version of the code has already adressed the question
of the extendedness of LAEs \citep{lau07}, and the present code expands on this
work. The dissertation is structured as follows:\vspace{1mm}

In the remainder of Part \ref{theoback} the project is put into context by
describing in sufficient detail
the theoretical background and the equations necessary for understanding the
requirement for,
and the development of, the numerical RT code that is the
mainstay of this work.
The different types of galaxies that will be investigated are described,
the basic theory of resonant scattering and dust, and a
brief account of the IGM.\vspace{1mm}

Part \ref{NumRT} presents first the underlying cosmological simulations to
which the RT code is applied, and subsequently the details of code itself are
explained.
In \Cha{tests} the code is extensively tested against a numbers of
situations for which analytical solutions exist.
Also in this chapter, various convergence tests and parameter studies are
discussed, which are probably better understood after having seen the results.
Finally, the scheme for performing RT in the IGM is described.\vspace{1mm}

Each of the first four chapters of Part \ref{science} elaborate on the tangible
fruit of
three years' labor --- three papers written \citepalias{lau09a,lau09b,lau10a},
and one co-written \citepalias{fyn10}, by the author
of this thesis. Longer sections of these chapters are taken directly from the
papers, but much is rearticulated, expanded, and/or reshuffled to create a
coherent presentation. Three of the chapters represent theoretical results
obtained from numerical simulations, while the fourth report on related
observational achievements.
The main conclusions are summarized and discussed in \Cha{post}, where also
future prospects are considered.\vspace{1mm}

The appendices in Part \ref{appendix} features a quantum mechanichal derivation
of the hydrogen cross section, illuminating under which circumstances it is
valid. Subsequently, the abstracts of each of the four papers are
provided.\vspace{1mm}

If you are currently reading the .pdf version of the thesis, the references in
the bibliography are hyperlinks taking you to the online NASA ADS database.
A version including the full papers can be downloaded from the URL\\
\href{http://www.dark-cosmology.dk/~pela/PhD/PhD.html}
{www.dark-cosmology.dk/\~{}pela/PhD/PhD.html}.



\chapter{Galaxies in the early Universe}
\label{cha:gals}\index{Galaxy formation}

\init{T}{he graceful, revolving arms} of spiral galaxies,
the symmetry,
the beautiful colors and patterns,
the colossal masses and sizes,
impels the astronomer to unveil their origin.
How did such marvelous objects come into existence?
How did they evolve?
What are they made of?
In order to answer these questions, physical theories and observational data
must be united to construct a comprehensive picture.

\begin{figure}[!t]
\centering
\includegraphics [width=0.90\textwidth] {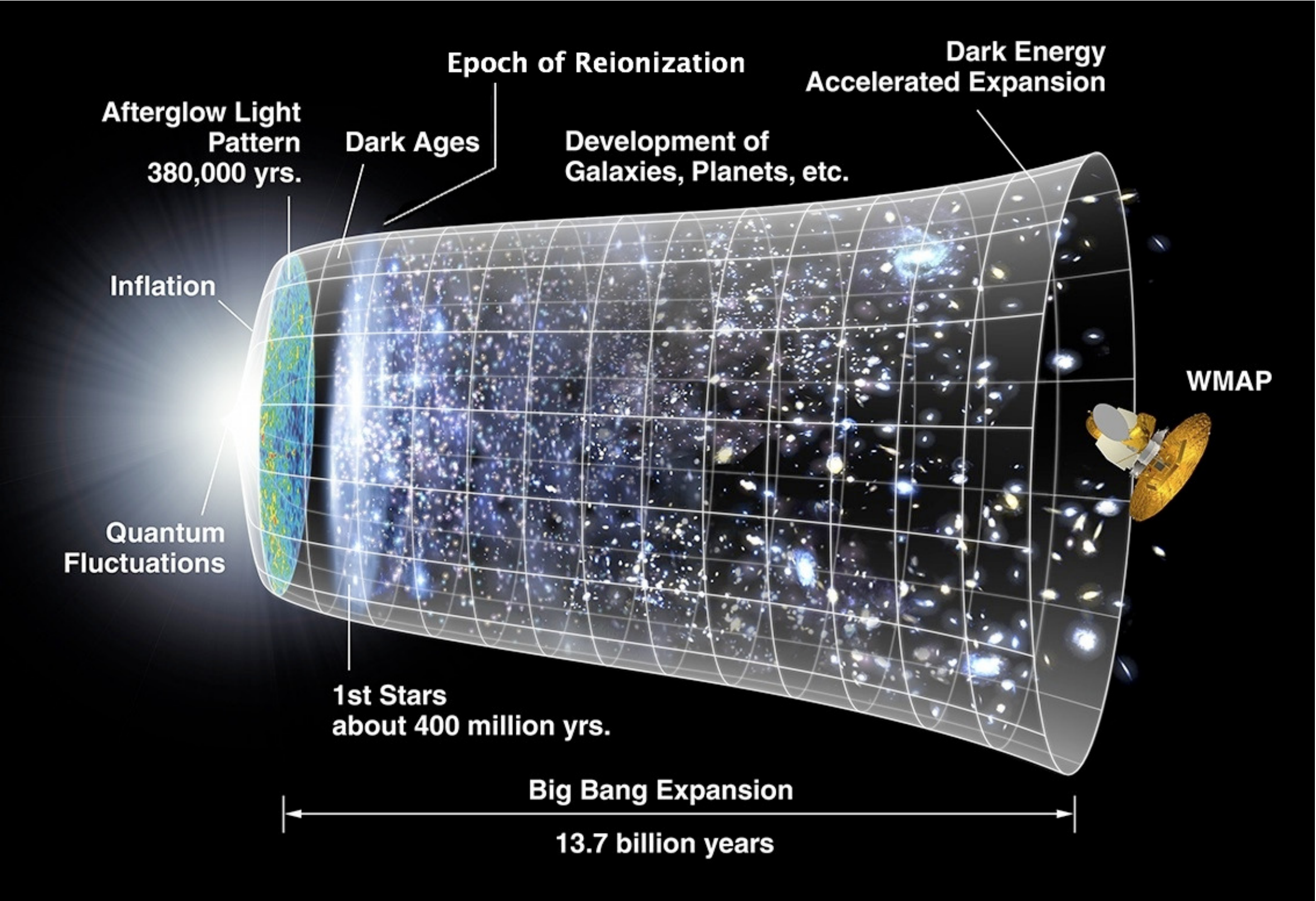}
\caption{{\cap Timeline of the Universe: Two spatial dimensions, plus time
         increasing from left to right.
         The Universe was born 13.7 Gyr ago, and has expanded ever since.
         From the recombination of hydrogen 380\,000 years
         after the Big Bang, to its reionization aproximately half a Gyr later,
         the Universe was opaque to UV radiation, and
         since the first stars emit primarily in this wavelength region,
         observing the very first galaxies is a challenging task.
         Credit: NASA/WMAP Science Team.}}
\label{fig:WMAPtimeline}
\end{figure}

The formation of a galaxy can be outlined as follows: primordial quantum
fluctuations
existing immediately after the
creation of the Universe --- the Big Bang --- and being blown up to
astronomical sizes by the inflation, grow in amplitude with time, due to the
gravitational attraction of matter. Since dark matter outbalances baryonic
matter by a factor of five, the dynamics and structure formation of both are
initially governed by dark matter.\index{Dark matter}
Eventually, some 400 Myr after the Big Bang, the gas falling into the huge
potential wells created by the dark matter becomes dense enough that it may
collapse further, giving birth to stars. The very first stars --- ``Pop\,III''
\index{Pop\,III stars}\index{Stars!Pop\,III} --- are thought to be extremely
massive and luminous, since the pristine,
metal-free gas is not able to ignite fusion before several hundreds of solar
masses of gas have accreted.
The massive amount of radiation, and the feedback imposed by the exploding
stars generate galactic superwinds that blow out from the proto-galaxy with
speeds of thousands of km s$^{-1}$.
At this time, the gas in the Universe is largely neutral, and the hard UV
radiation from the stars cannot travel far before ionizing the hydrogen.
At first the high density of the gas means that it quickly recombines, making
it virtually impossible for any radiation with an energy higher than that of
the hydrogen ionization potential to escape. This era in the history of the
Universe is referred to as \emph{the Dark Ages}.\index{Dark Ages}
Eventually, the continuous emission of ionizing radiation, together with the
expansion of the Universe diluting the gas, makes it possible to create
``bubbles'' of ionized hydrogen around the galaxies that grow in size and
ultimately overlap, making the Universe largely ionized.
This\index{Epoch of Reionization} ``Epoch of Reionization'', described
further in \sec{EoR} marks the end of the Dark Ages \citep[e.g.][]{fer03}, and
is thought to take place
during a relatively short period of time, cosmologically speaking.
The exact course, however,  is still quite far from being well understood.
\Fig{WMAPtimeline} summarizes the above discussion in graphics.

Observing galaxies in their infancy is a challenging task: apart from the
opaqueness of the IGM, their mere distance makes them so faint
that they easily drown in the diffuse background radiation of star light.
Moreover, at redhifts $z \gtrsim 1.5$, the angular diameter of an
object begins to \emph{increase}, thus lowering its SB,
making it even fainter.

Looking for such faint blobs in an image, making sure that it is not only noise,
or irrelevant foreground objects, clearly requires special techniques.
Whereas distant galaxies was once only spotted in visible wavelengths on
photographic
plates imaging everything along the line of sight, modern filters, CCDs, and
spectrographs now make it possible to confine the search to specific regions
in space, time, and wavelength.

Obviously the physical properties of galaxies do not depend on the means
by which they are observed, but different techniques impose different selection
criteria on the observed galaxy population. If one searches for
galaxies in a particular wavelength region, one will tend to find a special
kind of galaxies, namely those that are bright in this region.
Merging
various galaxy populations is a major goal of galaxology. In the present
work we will be studying three distinct populations, viz.~the already
introduced LAEs, along with the so-called \emph{Lyman-break galaxies}, and
\emph{damped Ly$\alpha$ absorbers}.
To get a notion of their individual selection criteria
and their overall physical properties, the following three sections describe
these three populations.

\section{Ly$\alpha$ emitters}
\label{sec:LAEs}\index{Ly$\alpha$ emitters|textbf}

Ly$\alpha$ emitters (LAEs) are simply galaxies observed in Ly$\alpha$,
achieved either through spectroscopy or through narrowband imaging.
Since a redshift of $\sim$1.5 is needed to shift the ultraviolet wavelength of
Ly$\alpha$ into the optical atmospheric window, ground-based observations
generally  yield high-redshift LAEs
\citep[e.g.][]{cow98,fyn01,ven05,iye06,nil09}.
Nearby LAEs are only observable from space, with the
International Ultraviolet Explorer\index{IUE}
satellite \citep[IUE; e.g][]{mei81,ter93},
the Galaxy Evolution Explorer\index{GALEX}
\citep[GALEX; e.g.][]{deh08},
and the Hubble Space Telescope\index{HST}
\citep[HST; e.g][]{hay07,ate08,ost09}.
Narrowband imaging has the advantage that many LAE candidates can be found
simultaneously, but to investigate a non-vanishing volume of space the width
of the filter cannot be too small, and thus the redshift determination is not
very accurate. Spectroscopically confirmed LAEs cannot only pinpoint their
redshift precisely, but also make sure that an object is in fact an LAE and not
an interloper emitting strongly at another wavelength. Common contaminators are
H$\alpha/\beta$ and \fion{O}{ii/iii} emitters located at lower redshifts.
On the other hand spectroscopy is very time-consuming, usually requiring each
object to be targeted individually for several hours.

\subsection{Sources of emission}
\label{sec:sources}\index{Ly$\alpha$!Emission}

A Ly$\alpha$ photon is emitted from a hydrogen atom in the first excited state.
Several physical processes may result in the atoms being in this state, but can
in most cases be divided into recombinations following the ionization of the
hydrogen, or collisions with other atoms. In the context of galaxies, three
distinct processes are the main contributors to Ly$\alpha$.
These are described below, along
with a fourth agent that may in some cases outshine the others --- a quasar.

\subsubsection{Stellar sources}
\label{sec:stellar}

The most significant source to Ly$\alpha$ radiation in young galaxies is
recombing hydrogen following ionization by the hard UV radiation from
massive stars. At the onset of star formation, UV radiation from young, hot,
massive stars will ionize the surrounding neutral hydrogen (\ion{H}{i}). When
the protons and electrons
recombine, a fraction of the recombinations will result in the emission of a
Ly$\alpha$ photon. \citet{spi78} found that, at $T = 10^4$ K, a typical
temperature of the \ion{H}{ii}/\ion{H}{ii} regions, approximately 38\% of
the recombinations go directly to the ground state. The result is then just
another ionizing UV photon. Eventually, a recombination will go to a higher
state, subsequently cascading down the various states, generating one or more
photons with
insufficient energy to ionize hydrogen. If the result is
a photon with more energy than the Ly$\alpha$ photon, it may excite a hydrogen
atom. In the end, the effect will be an atom excited to the first state.
However, if this state is the $2S$ state, the most probable decay is via an
intermediate state, resulting in the emission of two photons, each with less
energy than a Ly$\alpha$ photon. Only in the $2P$ state can the atom decay
directly to
the ground state, emitting a Ly$\alpha$ photon (see \sec{tran}).
\citet{spi78} found that, at
$T = 10^4$ K, ultimately 68\% of the recombinations are accompanied by the
emission of a Ly$\alpha$ photon. This fraction is only mildly sensitive to
temperature; for $5\times10^3$ K ($2\times10^4$ K) the
fraction is 70\% (64\%).

This is the scenario described by \citet{par67}, who argued that as much as
10\% of the total luminosity of the galaxy may be emitted as Ly$\alpha$.


\subsubsection{Gravitational cooling}
\label{sec:gravcool}\index{Gravitational cooling}\index{Cooling radiation}

Galaxies are formed from gas falling from the IGM into the deep potential wells
created by mostly dark matter, collapsing under its own gravity.
As the gas becomes more dense, atomic collisions becomes more
frequent, heating the gas to several million degrees, which cools and emits
Ly$\alpha$, i.e.~the potential energy of the gas is released as cooling
radiation. \citet{far01} find
that, at high redshifts, most of this radiation is emitted by gas with
$T < 20\,000$ K, and consequently $\sim$50\% in Ly$\alpha$ alone. 


\subsubsection{UV background radiation}
\label{sec:UVB}\index{UV background}

UV radiation that is not absorbed in the interstellar medium (ISM) escapes its
host galaxy. Whether emanating from massive
stars or from quasars, in this way the Universe was reionized at a
redshift around $z\sim 6$--11, and the
IGM was filled up with a ubiquitous, \emph{metagalactic} UV background
(UVB). This
field pierces through some or all of the neutral hydrogen in the young galaxies.
For column densities $\Nhi$ less
than about $10^{17.2}$ \cmsq, the optical depth for a UV photon is of the
order unity, meaning that the whole system can be penetrated. In most galactic
systems, however, $\Nhi$ is so large that
the inner regions becomes self-shielded from the UV field. Systems with
$10^{17.2}$ \cmsq $< \Nhi < 10^{20.3}$ \cmsq are referred
to as Lyman limit systems (LLSs)\index{Lyman limit systems}, while systems with
$\Nhi > 10^{20.3}$ \cmsq are called damped Ly$\alpha$ absorbers
(DLAs)\index{Damped Ly$\alpha$ absorbers}, since for
column densities higher than this the damping wings of the absorption line
profile becomes apparent.

Although the Ly$\alpha$ photons resulting from the UVB are
only about 1\% of those resulting from massive stars, the mechanism is the
same --- the neutral gas is photo-ionized, recombines, and produces Ly$\alpha$
radiation.

Moreover, the UV field can also photo-heat non-self-shielded gas,
which subsequently cools, radiating Ly$\alpha$ \citep{fur05}.


\subsubsection{Quasars and superwinds}
\label{sec:QSOs}\index{Quasars}\index{Superwinds}

A particular class of Ly$\alpha$ emitting objects, known as ``Ly$\alpha$
blobs'' (LABs)\index{Lya blobs@Ly$\alpha$ blobs} has been known now for a
decade \citep{fyn99,ste00,mat04,nil06}.
Although LABs are extremely large (a few tens to $>$150 kpc) and very
Ly$\alpha$-luminous (up to $5\times10^{43}$ \ergs), they exhibit little or no
continuum radiation.
Hence, a probable mechanism behind these objects could be cooling radiation
\citep{hai00,far01,dij06a,dij06b}, although the phenomenon has
also been attributed to hidden quasars and superwinds:

Many, if not all, galaxies go through a phase in their early life in which
gas accreting onto a central, supermassive black hole results in a
active galactic nucleus (AGN), ejecting tremendous amounts of energy in two
jets along the axis of rotation. A sub-class of AGNs are the quasars, or QSOs.
The jets may then ionize surrounding gas, possibly producing Ly$\alpha$
radiation \citep{hai01,wei04,wei05}

Even if no quasar is present, superwinds from starbursts and supernovae (SNe)
sweeping up surrounding material may have the same effect
\citep{tan00,tan01,mor04,wil05}.




\subsection{Physical characteristics}
\label{sec:laechar}\index{Ly$\alpha$ emitters!Physical characteristics}

As with many other galaxy ``types'', the term ``LAEs'' represents a selection
method, not a physically distinct category of galaxies.
Nevertheless, one may speak of a \emph{typical} LAE as a fairly small, quite
young, relatively highly star-forming galaxy or proto-galaxy, of rather low
metallicity and dust contents.
However, large deviations from this exist.

\subsubsection{Ages}
\label{sec:age}

The ages of LAEs, i.e.~the time that has passed since the onset
of star formation, has typically been found to be of the order 100 Myr
\citep{lai07,lai08}, but ages of $\sim$10 Myr \citep{gaw07,pir07,fin08} and
$\sim$1 Gyr (\citealt{nil07}, and \citealt[][when including objects also
detected with IRAC]{lai08}) have also been found.


\subsubsection{Stellar masses}
\label{sec:mass}

LAEs are generally small systems. Their stellar masses can be obtained, or at
least constrained, through spectral energy distribution (SED) fitting.
Such fitting have found typical masses
of $M_\star/\Msun \sim 10^8$-$10^9$ \citep{gaw06a,lai07,fin07,nil07}.


\subsubsection{Star formation rates}
\label{sec:SFR}\index{Star formation rate!Ly$\alpha$ emitters}

One of the essential characteristics of a galaxy is its star formation rate
(SFR), i.e.~the amount of gas mass converted into stars per unit time.
Typical
LAE SFRs lie in the range 1--10 $M_\odot$ yr$^{-1}$
\citep{cow98,hu98,gro07,nil07}. This may be compared to the Milky Way SFR of
around 1 $M_\odot$ yr$^{-1}$ \citep[e.g.][]{rob10}),
but taking into account their low masses reveals that LAEs have the highest
\emph{specific} SFR (sSFR; the SFR divided by stellar mass) of any type of
galaxy, a signature of their young ages \citep{cas06}.


\subsubsection{Metallicities and dust contents}
\label{sec:Zdust}

Perhaps the most debated attribute of LAEs is their dust contents and its
influence on their observability. Due to the
resonant scattering of the Ly$\alpha$ photons increasing the path out of the
galaxies by a large and unknown factor, Ly$\alpha$ may be much more vulnerable
to dust than continuum radiation. Accordingly, LAEs have been thought to be
relatively
free of dust. The color excess $E(B-V)$, defined as the difference
between the extinction in the $B$ band and the $V$ band, is a measure of the
amount of dust in a galaxy, since it quantifies the reddening of the continuum.
At $z \sim 3$, \citet{gro07} find $E(B-V)$'s lying approximately in the range
0.01 to 0.1. \citet{ver08} find similar values by fitting line profiles using
a radiative transfer model similar to the one deveoped in this work.

At somewhat higher redshifts, $z\sim$5, marginal evidence for slightly higher
extinction have been reported \citep{lai07,fin07}, although also similar values
are found \citep{pir07}. 

Since dust is made of metals, the amounts of dust must in some way be
correlated with metallicity. In general, approximately half of the produced
metal deplete onto dust, with only marginal evidence for a redshift evolution
\citep[see, e.g., Fig.~1 of][]{pei99}.


\subsubsection{Clustering properties}
\label{sec:clust}

Galaxies are thought in some way to trace the underlying dark matter
distribution \citep[e.g.][]{kai84,bar86}. This distribution is of course
inherently impossible to observe,
but one way to learn about the enigmatic dark matter\index{Dark matter}
is by studying the
clustering properties of galaxies, usually quantified through the two-point
correlation function. However, due to the complexities involved in both the
formation and the radiative transfer of Ly$\alpha$, no consensus has yet been
reached on whether or not LAEs are a good tracer. To substantiate this issue,
probably much larger volumes needs to be surveyed \citep{ors08}.

As LAEs are, in general, smaller systems than other high-redshift
populations such as Lyman-break galaxies (see next section), sub-mm galaxies
(SMGs), and distant red galaxies (DRGs), they provide an opportunity of probing
the faint end of the bolometric luminosity function (LF). Other populations are
heavily
biased toward having strong continua, thus implying very massive galaxies, in
turn implying that they probe preferentially the overpopulated regions of the
Universe like galaxy glusters.



\subsubsection{Evolution}
\label{sec:evol}

Looking at the LF of LAEs, no significant evolution seems to
occur from $z \sim 6$ to $z \sim 3$
(\citealt{ouc03,mai03,mal04,van05,shi06}, but see also \citealt{kas06})

Comparing $z = 2.25$ LAEs in the COSMOS\index{COSMOS} field with LAEs
at $z \sim 3$,
corresponding to a time interval of roughly one Gyr, \citet{nil09} studied the
evolution of LAEs.
They found that the sample at the lower redshift appears to be characterized
by redder objects as well as an increase in the ratio
between SFRs inferred from UV and those inferred from Ly$\alpha$.
Both of these factors are a
signature of the LAEs becoming more dusty, which is in accord with what may be
expected (more on this in \cha{dust}).
Moreover, the distribution of equivalents widths (EWs) becomes narrower with
time. This is consistent with the higher UV-to-Ly$\alpha$ SFRs, since the EW is
proportional to the ratio of the Ly$\alpha$ flux to UV flux, similar to the
SFRs. 

Finally, the fraction af LAEs containing AGN was found to increase. This is also
expected, since $z \sim 2$ corresponds to the peak of the AGN number density
distribution \citep[e.g.][]{wol03b}.




\section{Lyman-break galaxies}
\label{sec:LBGs}\index{Lyman-break galaxies|textbf}

\begin{figure}[!t]
\centering
\includegraphics [width=0.70\textwidth] {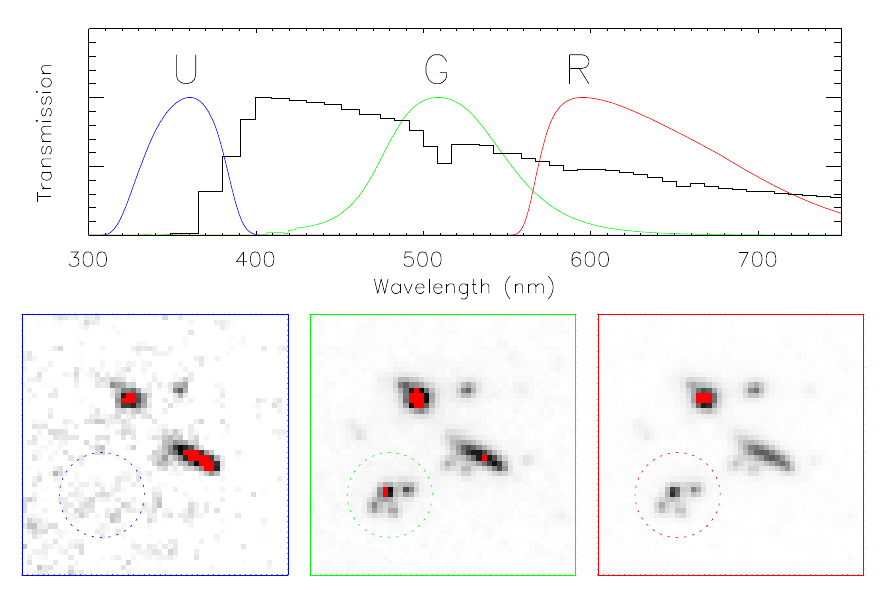}
\caption{{\cap Demonstration of the LBG technique.
               \emph{Top:} Due to the presence of neutral hydrogen, the
               spectrum of a galaxy exhibits a ``break'' at a rest wavelengths
               of 912 {\AA}, in this case redshifted to around 400 {\AA},
               i.e.~corresponding to a redshift of $\sim 3.4$.
               Almost no flux hence enters the $U$-filter (central wavelength
               3600 {\AA}), whereas the galaxy is easily detected in the
               $G$-filter (central wavelength 5100 {\AA}). The actual images are
               seen in the \emph{bottom} plot.
               The redshift of the galaxy can in this case be constrained to lie
               between $z_{\mathrm{lo}} \sim 3600/900 - 1 = 3$ and
               $z_{\mathrm{hi}} \sim 5100/900 - 1 \sim 4.5$.
               To further constrain
               the redshift, spectroscopic follow-up is needed.
               \emph{Credit: J. P. U. Fynbo.}}}
\label{fig:LBG}
\end{figure}

The \emph{Lyman-break} is the sharp drop in intensity in the spectrum of a
galaxy, located at a (rest) wavelength of 912 {\AA}. This wavelength matches the
ionization potential of neutral hydrogen, so photons of shorter wavelengths
will have a high probability of being absorbed before escaping the galaxy.
Observing a region of the sky in several filters may thus reveal an image of
a galaxy in the ``redder'' filters, whereas the sky seems empty in the ``bluer''
filters. \Fig{LBG} shows an example of this.

Since large areas can be surveyed for many galaxies simultaneously, this
technique significantly moved forward the frontier of galaxy surveys.
While first mentioned by \citet{mei76}, it was not employed to detect
high-redshift galaxies until almost two decades later, the first being
\citet{ste93} (although it was used to put constraints on the abundances of
galaxies at $z > 3$ a few years earlier \citep{guh90,son90}).
By now, the total sample a galaxies detected in this way
--- appropriately dubbed Lyman-break galaxies (LBGs), or drop-out galaxies ---
probably approaches 10\,000
\citep{ste03,mad96,pet01,bun04,sta04,ouc04a,ouc04b,wad06}.

LBGs tend to be somewhat older and more massive than LAEs, with stellar masses
lying in the range $10^9$ to $10^{11} \Msun$.
Their SFRs are typically higher, reaching several hundred $\Msun$ yr$^{-1}$
\citep[e.g.][]{rig06,sha01}.
Considerable amounts of dust are observed to reside in the LBGs
\citep[e.g.][]{saw98,cal01,tak04,rig06}, with extinction from
$A_V \simeq 0.3$--0.5 \citep{sha01,verm07} to $A_V \simeq 1$--2 \citep{pap01}.
Their clustering properties appear to be similar to those of LAEs
\citep{ste98,gia98,ouc04b}.


\section{Damped Ly$\alpha$ absorbers}
\label{sec:DLAs}\index{Damped Ly$\alpha$ absorbers|textbf}

The LAEs and LBGs described above, together with other populations such as
SMGs and DRGs, all have in common that
they are selected by virtue of their emission. Damped Ly$\alpha$ absorbers
(DLAs), in contrast, are identified by their ability to cause broad absorption
lines in the spectra of bright background sources.
The ``bright background source'' will usually be a quasar, but may also be more
transient sources such as a SN or gamma-ray burst (GRB).

\begin{figure}[!t]
\centering
\includegraphics [width=0.90\textwidth] {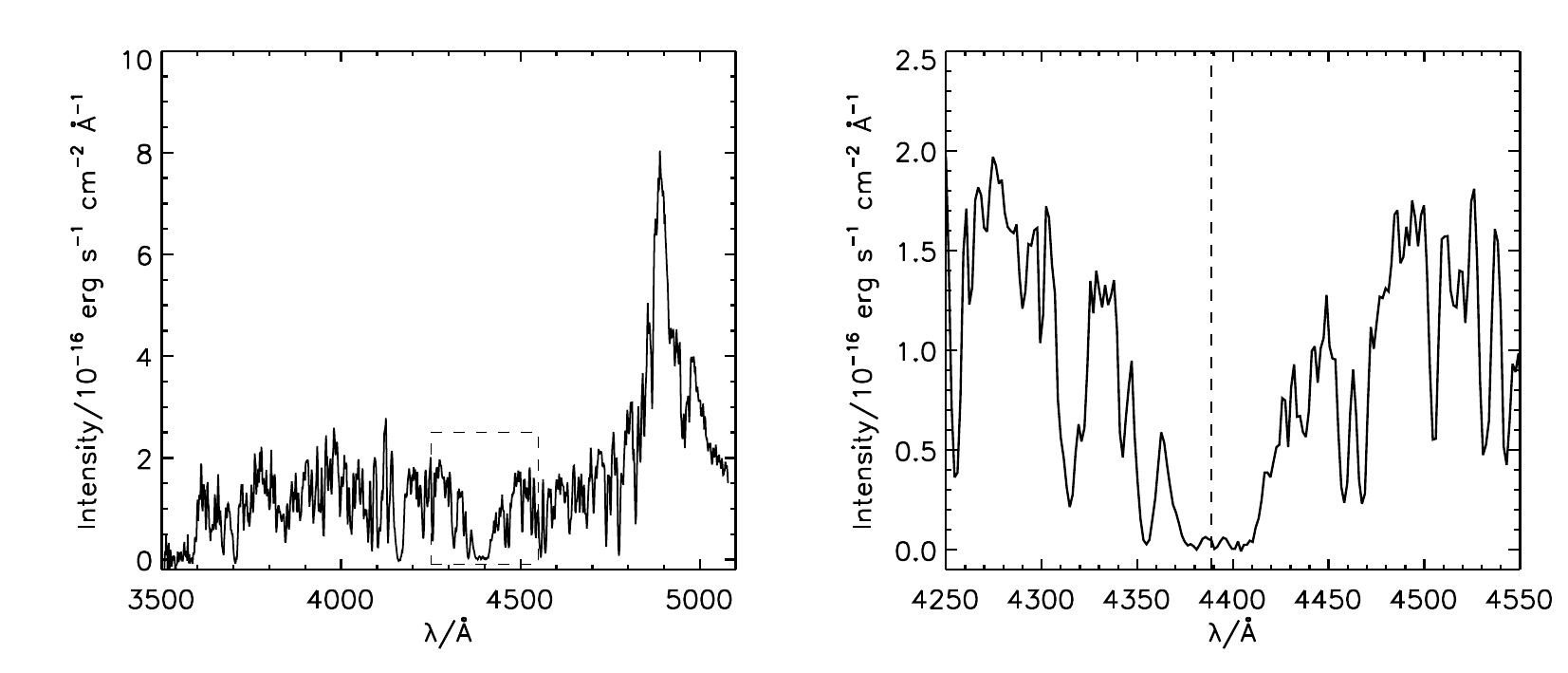}
\caption{{\cap Full spectrum of the quasar Q2348-011 at $z = 3.0$ (\emph{left})
               and a zoom-in on the region around a DLA situated at $z = 2.6$.
               This spectrum was acquired at the Nordic Optical Telescope by
               the author and collaborators, and will be discussed further in
               \cha{DLAs}.}}
\label{fig:DLA}
\end{figure}

As will be discussed in further detail in \sec{LAF}, the IGM consists
of diffuse clouds of neutral hydrogen of various column densities. As the
intense light from a source travels through the IGM, it suffers a cosmological
redshift by the expansion of the Universe. If a sufficient amount of neutral
hydrogen happens to be located at the place where a given wavelength region
intially with a wavelength bluer than that of Ly$\alpha$ has been redshifted
to 1216 {\AA}, the spectrum will be subjected to an absorption line.
For clouds of column densities $\Nhi \gtrsim 10^{20.3}$ \cmsq, the
absorption
feature becomes so broad that the damping wings of line profile, i.e.~the part
of the profile caused by natural broadening as opposed to thermal broadening,
becomes visible. An example of such a system, the DLA, is seen in \fig{DLA}.
Sometimes the absorption feature itself is labelled ``DLA'', but usually the
term refers to the system responsible of the absorption.

At $z \sim 3$, most of the neutral gas in the Universe is contained within
DLAs \citep{ome07}.
The fact that DLAs are in their very nature
self-shielded against the ionizing UVB implies that the gas is able to cool
sufficiently to initiate star formation. This makes them obvious candidates for
present-day galaxies \citep{wol86}.

Due to the high luminosity of the background quasars, detecting galaxies
associated with DLAs is, in general, a strenuous task. Both the scarce number
of observations \citepalias[e.g.][]{fyn10} and numerical simulations
\citep[e.g.][]{pon08} indicate that a characteristic impact parameter for the
line of sight through the hydrogen cloud responsible for the absorption
is, at most, $\sim$10 kpc, corresponding at a redshift of $z\sim3$
to the order of an arcsecond.
Consequently, even under very good seeing conditions,
detection of a galaxy against a bright quasar will be extremely difficult.
For this reason, not much is known about the underlying galaxy population
resulting in DLAs, if at all there is such a ``population''.
The number of known DLAs now exceeds one thousand \citep{pro05,not09}.
By selection they
are very gas rich. Since DLAs are always accompanied by narrow metal lines at
the same redshift, at least some star formation is expected to
have been occurring for a while \citep{vla99,wol03a}.
However, inferred metallicities\index{Metallicity} are usually
quite low ([X/H]\footnote{Here, X denotes a generic term for various metals,
and the metallicity [X/Y] is a convenient way of writing log(X/Y) $-$
log(X$_\odot$/Y$_\odot$). Usually metals that are believed not to deplete
significantly to dust are used, e.g.~Zn \citep{rot95}.} $\sim -1$
\citep{vla02,sav00}, and even down to $-2$ \citep{pet97,pet99}, indicating that
the systems must be very young.



\chapter{Resonant scattering}
\label{cha:ResScat}\index{Resonant scattering|textbf}
 
\init{I}{t has been stated that} ``\emph{galaxies are to astronomy what atoms are
to physics}'' \citep{san61}. However, whereas the physics of atoms can easily be
studied with no knowledge of galaxies, the converse is certainly not true.
To gain more intuition
about the concept of resonant scattering and the conditions under which the
derived atomic
cross section --- crucial to the developed code --- is valid,
the proper quantum mechanical derivation of the cross section will now be
reviewed.
For a more thorough derivation, see, e.g., \citet{bra03}.

We will then proceed to go deeper into the actual scattering process and
finally discuss the spatial motion of the photons, establishing how
radiative transfer is a journey, or diffusion, in both real and frequency
space.

\section{Atomic cross section}
\label{sec:Xsec}

The cross sectional area of the scattering hydrogen atom is dependent on the
frequency $\nu$
of the photon. Only if its energy matches closely
the energy difference $E_f - E_i$ between the initial
state $\psi_i$ and final
state $\psi_f$ of the hydrogen atom --- if it is in \emph{resonance} ---
does it have a large probability of exciting the electron. Traditionally, this
cross section is expressed as
\begin{equation}
\label{eq:XsecClas}
\sigma_\nu = \frac{\pi e^2}{m_e c} f_{if} \phi_{if}(\nu),
\end{equation}
where $-e$ and $m_e$ is the charge and the mass of the electron, respectively,
$f_{if}$ is the \emph{oscillator strength}\index{Oscillator strength}, and
$\phi_{if}(\nu)$ is the normalized line profile of the corresponding
transition, giving the probability of interaction as a function of frequency.
For the Ly$\alpha$ transition, $i=1$ and $f=2$.

Historically, the electron-proton system was viewed as an oscillating spring,
and the oscillator strength was introduced as an empirically determined
quantity for optical transitions with bound electrons. However, it can also be
derived analytically by considering the Hamiltonian of the interaction of the
atom with a photon. This is done in \app{quant}, yielding
the absorption cross section for the Ly$\alpha$ transition to be
\index{Cross section!Hydrogen|textbf}
\begin{equation}
\label{eq:signuintext}
\sigma_\nu = f_{12} \frac{\pi e^2}{m_e c}
                   \frac{\Delta\nu_L/2\pi}{(\nu-\nu_0)^2 + (\Delta\nu_L/2)^2},
\end{equation}
where
\begin{equation}
\label{eq:f12intext}
f_{12} = 0.4162
\end{equation}
is the oscillator strength,
\begin{equation}
\label{eq:nu0intext}
\nu_0 = 2.466\times10^{15} \textrm{ s}^{-1}
\end{equation}
is the line center frequency, and
\begin{equation}
\label{eq:DnuL}
\Delta\nu_L = 9.936\times10^7 \textrm{ s}^{-1}
\end{equation}
the natural line width.

It is of conceptual importance to understand that the cross section ---
contingent on the assumption that an ensemble of non-interacting particles is
incident upon the scatterer ---  must be interpreted in a statistical sense,
whether described classically or quantum mechanically. In the classical limit,
each incident particle can be assigned an impact parameter, i.e.~a distance to
the axis parallel to which it approaches the scatterer. It is the essence of
scattering cross section measurements that no effort be made to determine the
actual path of an individual particle.

\subsection{Thermal broadening of the line profile}
\label{sec:therm}\index{Thermal broadening}\index{Line profile!Thermal}

The result obtained in \eq{signuintext} gives the probability distribution
function (PDF) for a hydrogen atom absorbing a photon of frequency $\nu$.
However,
so far it was assumed that the scattering atom is at rest. If the atom is
moving with (non-relativistic) velocity $\mathbf{v}_{\textrm{atom}}$, the
frequency of the photon will be Doppler shifted in the reference frame of the
atom. To first order, a Lorentz transformation yields\index{Lorentz
transformation}
\begin{equation}
\label{eq:lornu}
\nu' = \nu \left( 1 - \frac{\mathbf{\hat{n}}_i \cdot \mathbf{v}_{\textrm{atom}}}
                          {c} \right),
\end{equation}
where $\mathbf{\hat{n}}_i$ is a unit vector representing the direction of the
incident photon.

The total velocity of the atoms is a sum of several contributions, which can be
divided into two categories; macroscopic and
microscopic. Macroscopic velocity is the total bulk motion of the gas,
e.g.~the turbulent motion of the fluid elements and/or an overall expansion of a
gas cloud. This
is most easily taken into account by a Lorentz transformation between reference
frames (numerically between adjacent cells of the simulation, see
Sec.~\ref{sec:Lorcell}) using \eq{lornu}. Microscopic velocities can
be due to pressure,
i.e.~atom collisions. This was verified in \sec{tran} to be insignificant in the
environments studied in the present work. However, it could easily be
implemented in the model, since collisions give rise to a Lorentzian profile
\citep[e.g.,][]{eme96}. \Eq{DnuL} would then be replaced by
\index{Half-life!Collisional}
\begin{equation}
\label{eq:pres}
\Delta\nu_{L\mathrm{,nat.+coll.}}
  = \frac{E_{\mathrm{2,nat.+coll.}}}{h_{\mathrm{Pl}}}
  = 2\pi \left( \frac{1}{t_{1/2}} + \frac{1}{t_{1/2,\textrm{coll.}}} \right),
\end{equation}
where $h_{\mathrm{Pl}}$ is Planck's constant,
\begin{equation}
\label{eq:t12}
t_{1/2} = 1.60\times10^{-9} \textrm{ s}
\end{equation}
is the natural half-life of the excited state (compare with \sec{tran})
and $t_{1/2,\textrm{coll.}}$ is the half-life of the excited state due to the
possibility of the electron being de-excited by collisions.

The second microscopic effect is the random thermal velocity of the gas. For a
gas at some temperature $T$,
these velocities follow a Maxwellian distribution,
i.e.~a Gaussian distribution in three mutually perpendicular directions.
Equating the kinetic energy of the individual particles to their thermal
energy, the thermal velocity dispersion (times $\sqrt{2}$) is
\begin{eqnarray}
\label{eq:vth}
v_{\textrm{th}} & = & \left( \frac{2 k_B T}{m_H} \right)^{1/2}\\
                & = & 12.85 \, T_4^{1/2} \textrm{ km s}^{-2},
\end{eqnarray}
where $k_B$ is Boltzmann's constant, $m_H$ is the mass of the hydrogen atom,
and $T_4 = T/(10^4 \textrm{ K})$.
From Eqs.~\ref{eq:lornu} and \ref{eq:vth} the associated Doppler frequency
shift $\Delta\nu_D$ of the frequency distribution is then
\begin{eqnarray}
\label{eq:DnuD}
\Delta\nu_D & = &\frac{v_{\mathrm{th}}}{c} \nu_0\\
            & = & 1.057\times10^{11} \, T_4^{1/2} \textrm{ s}^{-1}.
\end{eqnarray}

To simplify notation, frequency will be parametrized through\index{x@$x$}
\begin{equation}
\label{eq:x}
\boxed{
x \equiv \frac{\nu - \nu_0}{\Delta\nu_D}.
}
\end{equation}
In terms of these quantities, with $\phi(\nu)d\nu = \phi(x)dx$, the thermal
line profile is then
\begin{equation}
\label{eq:Gx}
\mathcal{G}(x) = \frac{1}{\sqrt{\pi}} e^{-x^2},
\end{equation}
while the natural line profile is
\index{Lorentzian}
\begin{equation}
\label{eq:Lx}
\mathcal{L}(x) = \frac{a}{\pi} \frac{1}{x^2 + a^2},
\end{equation}
where\index{Damping parameter}\index{a@$a$}
\begin{equation}
\label{eq:a}
\boxed{
a \equiv \frac{\Delta\nu_L}{2\Delta\nu_D}
}
\end{equation}
is (half) the relative line width, or the ``damping parameter''.

The resultant line profile in the reference
frame in which the gas is on average in rest (the Lagrangian frame of the fluid
element) is a Voigt profile, i.e.~a convolution of the
Lorentzian and the Gaussian:\index{Voigt profile}
\index{Line profile!Total (Voigt)}
\begin{eqnarray}
\label{eq:Vx}
\nonumber
\mathcal{V}(x) & = & \int_{-\infty}^{\infty} \mathcal{L}(x-y) 
                                            \mathcal{G}(y)\,dy\\
               & = & \frac{1}{\sqrt{\pi}\Delta\nu_D}
                     H(a,x),
\end{eqnarray}
where\index{Voigt function}
\begin{equation}
\label{eq:Haxtheo}
H(a,x) = \frac{a}{\pi} \int_{-\infty}^{\infty}
         \frac{e^{-y^2}}{(x-y)^2 + a^2}\,dy
\end{equation}
is the Voigt function, and $\int_{-\infty}^{\infty} \mathcal{V}(x)\,dx = 1$,
since the same is true for $\mathcal{L}$ and $\mathcal{G}$, and convolution
conserves normalization.

Thus, we may write the final cross section of the hydrogen atom as
\index{Cross section|textbf}\index{sigmax@$\sigma_x$}
\begin{equation}
\label{eq:sigxtheo}
\boxed{
\sigma_x = f_{12} \frac{\sqrt{\pi}e^2}{m_e c \Delta\nu_D} H(a,x).
}
\end{equation}
To get a notion of the magnitude of the cross section,
for a temperature of $10^4$ K it evaluates in the line center
to\index{Cross section!In the line center}
\begin{equation}
\label{eq:sig0}
\sigma_0 = 5.898 \times 10^{-14} \textrm{ cm}^2,
\end{equation}
four orders of magnitude larger than the \ion{H}{i} cross section at the
Lyman limit.




\section{Scattering}\index{Scattering!Theory}
\label{sec:theoscat}

\subsection{Frequency shift}
\label{sec:freq}
\index{Frequency shift}

It must be emphasized that the discussed broadening of the line and the
corresponding uncertainty in energy does \emph{not} imply that a photon of a
given energy can be absorbed, and subsequently re-emitted with a different
energy. Indeed, this would be possible had the energy of the ground state been
associated with an uncertainty in energy as well. However, since the lifetime
of this state is effectively infinite, its energy is well-defined.
Except for a very small recoil effect, which will be
discussed in Sec.~\ref{sec:recoil}, the scattering is
coherent in the reference frame of the atom. However, to an external observer,
any motion of the atom will, in general, add a Doppler shift to the photon.

\begin{figure}[!t]
\centering
\includegraphics [width=0.50\textwidth] {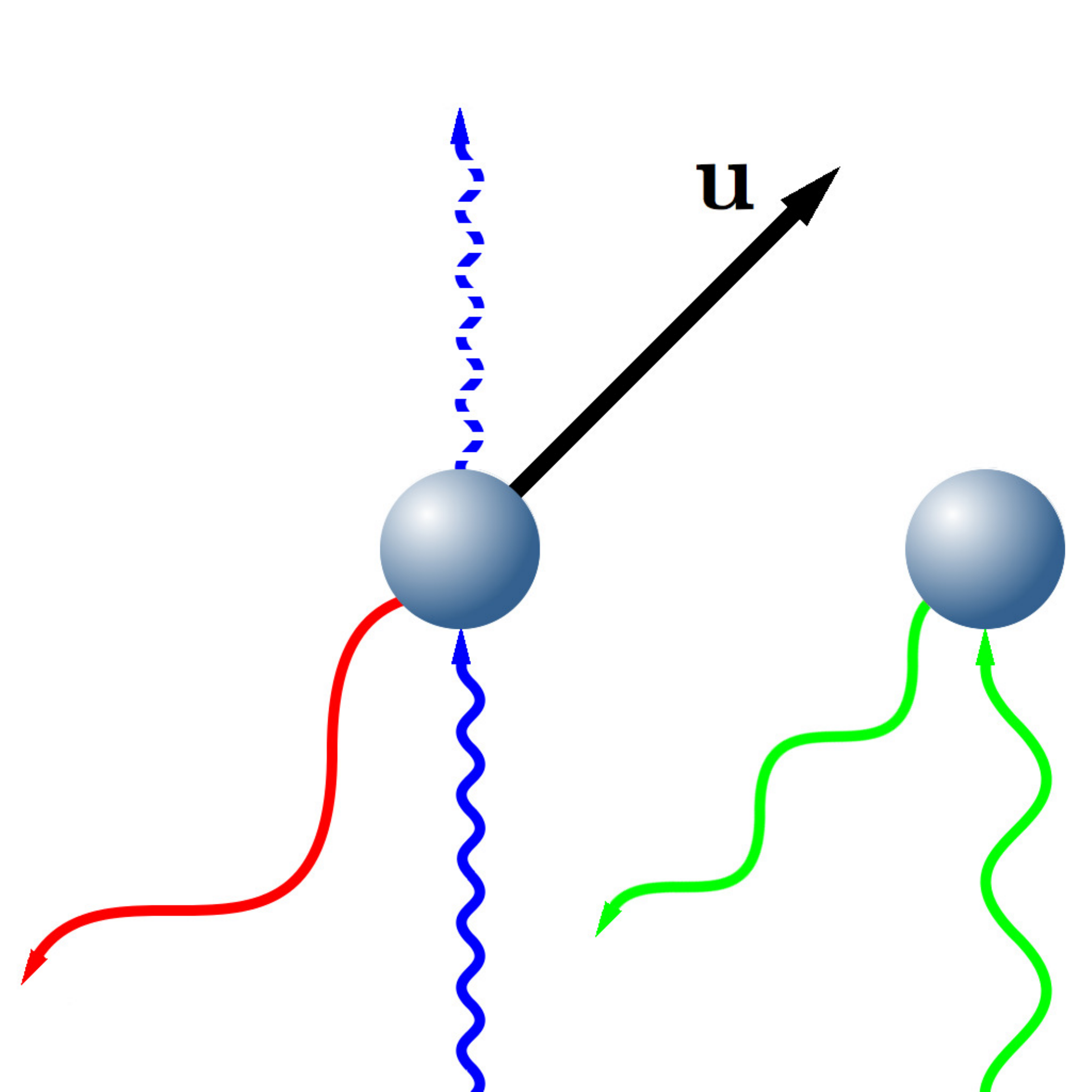}
\caption{{\cap Illustration of the mechanism responsible for the frequency
               shift of a scattered photon. In the reference frame of an
               external observer (\emph{left}), a photon blueward of the line
               center
               (\emph{blue}) is scattered by an atom receding in such manner
               that the
               component of its velocity $\mathbf{u}$ along the direction of
               the photon matches closely the frequency $x$. In the reference
               frame of the atom (\emph{right}), the photon then seems close to
               the line center (\emph{green}). Except for a minute recoil
               effect, the
               photon leaves the atom with the same frequency. However, to the
               external observer, if the photon is scattered in a direction
               opposite the atom's motion (\emph{red}), it will be redshifted.
               Only if by chance it is scattered in the exact same direction
               (\emph{dashed blue}), its frequency remains unaltered.
               For $|x| \gg 1$, the number of atoms with sufficiently high
               velocities is so
               small that the photon is most likely to be scattered by a
               low-velocity atom. In this case, no matter in which direction
               the photon is scattered the motion of the atom will not shift
               the frequency significantly.}}
\label{fig:RefFrame}
\end{figure}

Measuring the velocity of the atom in terms of the thermal velocity,
\begin{equation}
\label{eq:utheo}
\mathbf{u} = \frac{\mathbf{v_{\mathrm{atom}}}}{v_{\mathrm{th}}},
\end{equation}
allows us to compare directly frequency and velocity, so that \eq{lornu} reads
\begin{equation}
\label{eq:lorx}
x' = x - \mathbf{u} \cdot \mathbf{\hat{n}}_i,
\end{equation}
where $x'$ is the frequency as measured in the reference frame of the atom.
Figure \ref{fig:RefFrame} shows a qualitative interpretation of how the
Doppler shift arises.

Consequently, to track the photons in both real and frequency space, it is
important to know both the direction in which the photon is scattered
\emph{and} the exact velocity of the scattering atom.

One may ask if it makes sense to refer to the photon as being the same before
and after the scattering event, since in fact it ceases to exist (for about
$10^{-9}$ seconds), and since, to the external observer, its energy has changed.
However, even though it does not exist for a short while, in the reference
frame of the atom --- which is just as good as any other reference frame ---
the photon retains a ``memory'' of its energy before the event. Its state
of existence is thus not independent of its former being, and the fact that its
intrinsic properties seem to change is just a consequence of us not moving
along with the atom. Hence, from any philosophical and, in particular,
pragmatic point of view, we propose that a photon be one and the same photon,
until it is destroyed by some other physical process.

\subsection{Phase function}
\label{sec:phasea}\index{Phase function|textbf}\index{Polarization}

When light is scattered, it may be polarized. A full discussion of the concept
of polarization is beyond the scope of this thesis (albeit it would certainly
be interesting --- and feasible --- to numerically probe the polarization of
the scattered light from young galaxies). However, we note that polarization is
strongly connected with the direction of scattering, in that the radiation
will be more polarized, the closer to $90^{\circ}$ the scattering is
\citep[e.g.,][]{cha50}. Thus,
isotropic scattering will on average result in unpolarized light.

\subsubsection{Core scattering}
\label{sec:corescat}

The probability distribution of the directions of scattering is given by the
\emph{phase function} $W(\theta)$,
where $\theta$ is the angle between the direction
vectors $\mathbf{\hat{n}}_i$ and $\mathbf{\hat{n}}_f$
of the incident and outgoing
photon, respectively (for reasons of symmetry, the scattering must always be
isotropic in the azimuthal direction and hence independent of $\phi$).
\citet{ham40} found that $W(\theta)$ is determined by
the multipole order of the emitted radiation and the difference in total
angular momenta $J$ of the three involved states, i.e.~the initial,
intermediate and final state. Specifically, in the dipole approximation
(see \sec{abs}) the phase function is,
\begin{equation}
\label{eq:Wmu}
W(\theta) \propto 1 + \frac{R}{Q} \cos^2\theta,
\end{equation}
where $R/Q$
is the degree of polarization for $90^\circ$ scattering.

For resonant scattering, the initial and final state are the same, while
the intermediate state corresponds to the excited state.
For one-electron atoms, the spin quantum number is $s = \pm1/2$. Since $\ell$ 
can assume values up to $n-1$, the total angular momentum $J = \ell + s$
for the ground state
is always 1/2, while the first excited state can have
$J=1/2$ or $J=3/2$.
Thus, the involved states are the ground state $1S_{1/2}$ and the three
excited states $2S_{1/2}$, $2P_{1/2}$ and $2P_{3/2}$.
As discussed in \app{quant},
we ignore the possibility of being excited to the $2S_{1/2}$ state.
It is then found \citep{ham40} that for the $2P_{1/2}$ transition, with
$J=1/2$ and $\Delta J_{if} = \Delta J_{fi} = 1$,
\begin{eqnarray}
\label{eq:RQ12}
\nonumber
\frac{R}{Q} & = & \frac{(2J - 1) (2J + 3)}{12J^2 + 12J + 1}\\
            & = & 0,
\end{eqnarray}
while for the $2P_{3/2}$ transition, with $J = 3/2$, $\Delta J_{if} = 1$ and
$\Delta J_{fi} = -1$

\begin{eqnarray}
\label{eq:RQ32}
\nonumber
\frac{R}{Q} & = & \frac{(J + 1) (2J + 3)}{26J^2 - 15J - 1}\\
            & = & \frac{3}{7}.
\end{eqnarray}

The spin multiplicity of each state is $2J+1$, so the probability of being
excited to the $2P_{3/2}$ state is twice that of the $2P_{1/2}$ state. Thus,
changing variable to $\mu = \cos\theta$ so that $\mu \in [-1,1]$,
with a probability of 1/3 the photon is scattered according to\index{Phase
function!Core scatterings}
\begin{equation}
\label{eq:W12}
\boxed{
W_{\mathrm{core,}1/2}(\mu) = \frac{1}{2},
}
\end{equation}
i.e.~isotropically, while with a probability of 2/3 according to
\begin{equation}
\label{eq:W32}
\boxed{
W_{\mathrm{core,}3/2}(\mu) = \frac{7}{16}
                              \left( 1 + \frac{3}{7}\mu^2 \right).
}
\end{equation}
Note that Eqs.~\ref{eq:W12} and \ref{eq:W32} are normalized to unity, not
to $1/4\pi$.

\subsubsection{Wing scattering}
\label{sec:wingscat}\index{Wing scattering}

As indicated by the subscript in Eqs.~\ref{eq:W12} and \ref{eq:W32}, the
derived probability distributions are only valid in the ``core'' of the
line profile, i.e.~close to the line center. Investigating polarization of
scattered light from the Sun, \citet{ste80} found that in the
``wings'', the two lines of
$2P_{1/2}$ and $2P_{3/2}$ interfere quantum mechanically, making the scattering
behave like that of a classical oscillator, i.e.~pure Rayleigh scattering.
\index{Rayleigh scattering}\index{Scattering!Rayleigh}
In this case the direction follows a dipole distribution\index{Dipole
distribution}, with 100\% polarization
at $90^\circ$ scattering. Thus, the normalized phase function is\index{Phase
function!Wing scatterings}
\begin{equation}
\label{eq:Wwing}
\boxed{
W_{\mathrm{wing}}(\mu) = \frac{3}{8} \left( 1 + \mu^2 \right).
}
\end{equation}
\ \\
The transition $x_{\mathrm{cw}}$ from core to wing scattering is not
well-defined, but can be taken to be the value of $x$ where the Gaussian and
the Lorentzian contributes about equally much to the profile, i.e.~where
\index{Core/wing transition}
\begin{equation}
\label{eq:LGx}
\frac{1}{\sqrt{\pi}} e^{-x_{\mathrm{cw}}^2}
                                      \sim \frac{a}{\pi x_{\mathrm{cw}}^2}
\end{equation}

The solution to this equation can be approximated as
\begin{equation}
\label{eq:xcw}
x_{\mathrm{cw}}(a) = 1.59 - 0.60 \log a - 0.03 \log^2 a.
\end{equation}
The difference in the phase functions is not very important (this is verified
by tests), but since the value of $x_{\mathrm{cw}}$ is calculated anyway
in the process of determining the velocity of the scattering atom (\sec{u_II})
and in the acceleration scheme described in \sec{coreskip}, the exact value
is used.


\subsection{Recoil effect}
\label{sec:recoil}\index{Recoil effect|textbf}

By conservation of momentum, except for the case when the photon is scattered
in the same direction as the incident photon, at each scattering the photon
must transfer some energy to the atom. \citet{fie59} found that, on average,
the photon loses a fractional energy of
\begin{equation}
\label{eq:recoil}
\boxed{
g = \frac{h \Delta\nu_D}{2 k_B T}.
}
\end{equation}

Except for very low temperatures, compared to the Doppler shift this effect is
negligible, at least for a single scattering. However, as a photon easily
undergoes millions or even hundreds of millions of scatterings, one might
suspect that the cumulative energy loss would eventually shift all photons 
to the red side of the line. This was suggested by \citet{kah62}. But as will
be discussed in Sec.~\ref{sec:RT}, when a photon is in the wing, there is a
bias toward being scattered back toward the center, so that any
systematic redshift will be counteracted. In this way, \eq{recoil}
can be understood as reflecting the thermalization of photons around the
frequency $\nu_0$.
Still, even though \citet{ada71} argues that the effect of recoil \emph{is} negligible (at least in the case of
a homogeneous, static medium of uniform temperature), since the calculation is
quite rapid it is not omitted in the developed code.


\subsection{Atom velocity}
\label{sec:utheo}\index{Atom velocity|textbf}

\begin{figure}[!t]
\centering
\includegraphics [width=0.60\textwidth] {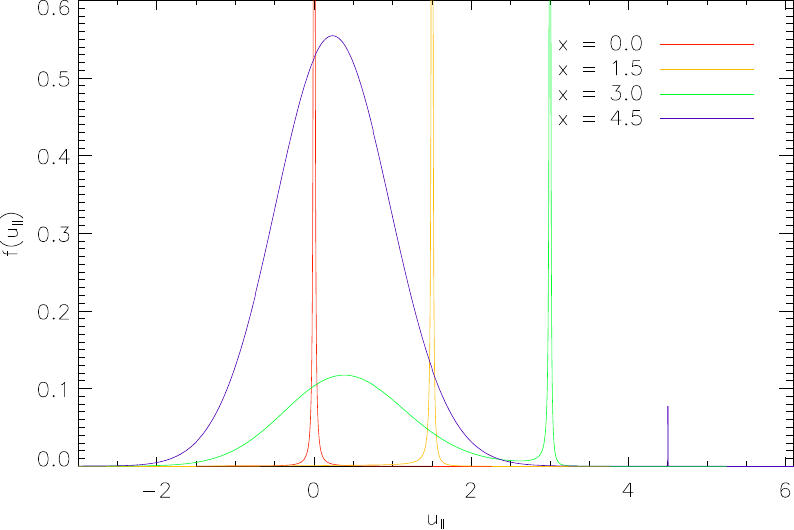}
\caption{{\small Probability function as given by \eq{fupartheo} for a
                 number of different frequencies. For $x=0$, $f(u_{||})$ simply
                 resembles a normal Lorentz profile, but for larger values of
                 $x$, the Gaussian part becomes increasingly significant.}}
\label{fig:fupartheo}
\end{figure}

As mentioned earlier, the velocities of the atoms follow a Gaussian probability
distribution in three directions. However, due to the frequency dependency of
the scattering cross section, the velocity distribution $f$ of the atoms
responsible for the scattering of a particular photon is quite different. Since
motion perpendicular to $\mathbf{\hat{n}}_i$ does not contribute to any
Doppler shift, a natural basis for $\mathbf{u}$ is
\begin{equation}
\label{eq:ubasis}
\mathbf{u} =
\left( 
\begin{array}{l}
u_{||}\\
u_{\perp,1}\\
u_{\perp,2}
\end{array}
\right),
\end{equation}
where $u_{||}$ is the velocity parallel to $\mathbf{\hat{n}}_i$ and
$u_{\perp,i}$, with $i=1,2$,  are the --- mutually orthogonal --- velocities
perpendicular to $\mathbf{\hat{n}}_i$. In this basis, $u_{\perp,i}$
still follow a Gaussian distribution, i.e.
\begin{equation}
\label{eq:Gu}
\boxed{
\mathcal{G}(u_{\perp,i}) = \frac{1}{\sqrt{\pi}} e^{-u_{\perp,i}^2}.
}
\end{equation}
In the direction parallel to $\mathbf{\hat{n}}_i$, the PDF
$\mathcal{G}(u_{||})$ must be convolved with the probability 
$\mathcal{L}(x-u_{||})$ of the atom being able to scatter the photon. The
resulting, normalized probability distribution
\begin{equation}
\label{eq:fupartheo}
\boxed{
f(u_{||}) = \frac{a}{\pi H(a,x)} \frac{e^{-u_{||}^2}} {(x-u_{||})^2 + a^2}
}
\end{equation}
can be seen in Fig.~\ref{fig:fupartheo} for a number of incident frequencies.
This distribution
reflects the fact that for a photon of frequency $x$, being scattered by an
atom with $u_{||} = x$ is highly favored, so that in the reference frame of the
atom the photon appears to be exactly at resonance. Since for large values of
$|x|$ the 
number of such atoms reduces as $e^{-x^2}$, in this case the photon is more
likely to be scattered by an atom to which it appears to be far in the
wing.\\[5mm]

From Eqs.~\ref{eq:lorx} and \ref{eq:g} and with $\mathbf{\hat{n}}_f$
given by the proper phase function, for a frequency $x_i$ of the incident
photon, the frequency $x_f$ of the scattered photon is then
\index{Frequency shift}
\begin{equation}
\label{eq:xf}
\boxed{
x_f = x_i - u_{||} + \mathbf{\hat{n}}_f \cdot \mathbf{u}
    + g (\mathbf{\hat{n}}_i \cdot \mathbf{\hat{n}}_f - 1).
}
\end{equation}
%



\section{Radiative transfer}
\label{sec:RT}

\subsection{Optical depth}
\label{sec:tautheo}
\index{Optical depth|textbf}

We will now look deeper into the spatial transfer of radiation.
Consider first a beam of radiation of initial intensity $I_0$, traveling
through an isothermal hydrogen cloud of homogeneous density $\nhi$.
For a small distance $dr$, effectively the atoms are covering a fractional area
$\nhi \sigma_x$, so the change $dI$ in intensity is
\begin{equation}
\label{eq:dI}
dI = -I_0 d\tau,
\end{equation}
where
\begin{equation}
\label{eq:taudef}
d\tau \equiv \nhi \sigma_x dr
\end{equation}
is the optical depth covered by the beam. If the gas is mixed with dust of
density $n_{\mathrm{d}}$ and cross section $\sigma_{\mathrm{d}}$, \eq{taudef}
is replaced by
\begin{equation}
\label{eq:taudefDUST}
\boxed{
d\tau \equiv (\nhi \sigma_x + n_{\mathrm{d}} \sigma_{\mathrm{d}}) dr.
}
\end{equation}

Integrating \eq{dI} along the path of the beam, covering a total
optical depth $\tau$ reduces the intensity to
\begin{equation}
\label{eq:I_tau}
I(\tau) = I_0 e^{-\tau}.
\end{equation}
from which we see that the optical depth traversed by a single photon is
governed be the probability distribution
\begin{equation}
\label{eq:P_tau}
\boxed{
P(\tau) = e^{-\tau},
}
\end{equation}
which is properly normalized in the interval $[0,\infty[$. This distribution
has an average of $\langle\tau\rangle = 1$.


\subsection{Photon peregrination}
\label{sec:pere}
\index{Radiative transfer}

The first attempts to predict the diffusion of Ly$\alpha$ was made under the
assumption of coherent scattering in the observers frame \citep{amb32,cha35}.
Several physical quantities which may or may not be directly observable have
been subject to interest, e.g.~the average number $N_{\mathrm{scat}}$ of
scatterings required to escape the medium (to determine the possibility of the
photon being destroyed by collisions or dust) and the shape of the emergent
spectrum.
Due to the complexity of the problem, the physical configurations investigated
have traditionally been photons emitted from the center of a homogeneous,
isothermal cloud of either spherical symmetry or, in particular, of infinite
extension in two directions and finite extension in one (a plane-parallel
``slab'').
Denoting
by $\tau_0$\index{Optical depth!In the line center}\index{tau0@$\tau_0$}
the optical depth for a photon in the line center from the
initial point of emission to the edge of the gaseous cloud, from pure random
walk considerations $N_{\mathrm{scat}}$ would be expected to be of the order
$\tau_0^2$. Accordingly, the medium would not have to
be very optically thick for destruction processes of Ly$\alpha$ to
become significant. 

\citet{hen40} and \citet{spi44} acknowledged the fact that scattered photons
undergo a change in frequency due to thermal Doppler broadening of the
scattering atoms. Relying on these considerations,
\citet{zan49,zan51} argued that, in each scattering, the frequency of the
Ly$\alpha$
photon would undergo \emph{complete redistribution}\index{Complete
redistribution} over the Doppler line
profile, i.e.~there is no correlation between $x_i$ and $x_f$, and the
probability that $x < x_f < x+dx$ is $\phi(x_f)dx$. In this picture,
the photon still executes a random walk, but at each scattering there is a small
possibility that it will be redistributed so far into the wing as to render the
medium optically thin and thus allow escape.
This reduces $N_{\mathrm{scat}}$ significantly, and the result was later
verified numerically for intermediate optical depths ($\tau_0 \sim 10^4$) by
\citet{koe56}.

Still based on the assumption of isotropic scattering,
\citet{unn52a,unn52b} calculated an ``exact redistribution''\index{Exact
redistribution} formula
$q(x_i,x_f)$, giving the probability distribution of $x_f$ as a function of
$x_i$. With this result,
\citet{ost62} found that in the wings, the rms frequency shift
$(\Delta x)_{\mathrm{rms}}$ per scattering is
\begin{equation}
\label{eq:rms}
\boxed{
(\Delta x)_{\mathrm{rms}} = 1,
}
\end{equation}
and the mean shift $\langle \Delta x \rangle$ per scattering is
\begin{equation}
\label{eq:meanx}
\boxed{
\langle \Delta x \rangle = -1/|x|,
}
\end{equation}
i.e.~the photon has a tendency to drift toward the line center.
Thus, a photon at frequency
$x \gg 1$ will execute a nearly random walk in frequency, returning to the
core in $N_{\mathrm{scat,ret.}} \sim x^2$ scatterings. Simultaneously, the
photon is undergoing a
random walk in real space, but barely diffuses spatially. Only when eventually
it is shifted to $x \ge x_1$, where $x_1$ is the frequency that renders the
optical depth unity, the photon will escape in a \emph{single longest flight}.
The probability $w(x_1)$ of escape per scattering is
\begin{equation}
\label{eq:Pesctheo}
w(x_1) \sim \int_{x_1}^\infty q(x_i,x_f) dx_f,
\end{equation}
from which he found that $N_{\mathrm{scat}} \propto \tau_0$ for moderate
optical depths. However, he
argued that for some limiting large optical depth --- which he was not able to
calculate due to the lack of ``sufficiently large digital computers'' --- $x_f$
is so large that the photon will execute a random walk also in real space,
whence in this case $N_{\mathrm{scat}} \propto \tau_0^2$.

Nonetheless, applying the method of \citet{fea64}, \citet{ada72} found
numerically that also for extremely large
optical depths ($\tau_0$ up to $10^8$), $N_{\mathrm{scat}} \propto \tau_0$.
Although he could not
prove it rigorously, he was able to give a heuristic argument on physical
grounds for this behavior. The essential point in \citeauthor{ost62}'s
argument was that during each excursion to the wing, the
photon would travel a distance in space much smaller than the size of the
medium. But since $\phi(x)$ varies quite slowly in the wings, the mean free
path $\tau_{\mathrm{mfp}}$, measured in line center optical depths, is
\begin{eqnarray}
\label{eq:mfp}
\nonumber
\tau_{\mathrm{mfp}} & \sim & \frac{1}{\phi(x)}\\
                    & \sim & \frac{x^2}{a}.
\end{eqnarray}
Since
\begin{equation}
\label{eq:invprob}
N_{\mathrm{scat}} = \frac{1}{w(x_1)},
\end{equation}
the rms distance $\tau_{\mathrm{rms}}$ traversed
before returning to the core is
\begin{eqnarray}
\label{eq:drms}
\nonumber
\tau_{\mathrm{rms}} & \sim & \tau_{\mathrm{mfp}} \sqrt{N_{\mathrm{scat,ret.}}}\\
                    & \sim & \frac{x^3}{a}.
\end{eqnarray}
Now, if the photon escapes in a \emph{single longest excursion}, then the
frequency $x_m$
where the average photon escapes must satisfy
$\tau_{\mathrm{rms}} \sim \tau_0$, so that
\begin{equation}
\label{eq:xmada}
x_m \sim (a\tau_0)^{1/3},
\end{equation}
which is in fact the frequency where the emergent spectrum takes its maximum
value. Assuming for simplicity complete redistribution, so that $x_f$ is given
by $\phi(x) \sim e^{-x^2}$, \citet{ada72} realized the important fact that
$\phi(x)dx$ does
not, as previously assumed, give the probability for a \emph{given photon} to
be scattered into the interval $[x,x+dx]$, since the photon will scatter there
$\sim x^2$ times before returning to the core. This implies that the
probability that a given photon scatter into this
frequency interval for the first time is $[\phi(x)/x^2]dx$. Hence, from
\eq{Pesctheo}, the probability of escape is
\begin{eqnarray}
\label{eq:PescFR}
\nonumber
w(x_m) & \sim & \int_{x_m}^\infty \frac{\phi(x)}{x^2}dx\\
       & \sim & \frac{a}{x_m^3},
\end{eqnarray}
so that, from Eqs.~\ref{eq:invprob} and \ref{eq:PescFR},
\begin{equation}
\label{eq:Nsada}
N_{\mathrm{scat}} \sim \tau_0.
\end{equation}

Inspired by \citet{unn55}, utilizing the Eddington approximation --- which
implies that the radiation field is everywhere nearly isotropic, but with a
small net outward flow --- and
expanding the redistribution function as formulated by \citet{hum62} to
second order, \citet{har73} obtained a diffusion equation for the angular
averaged intensity $J(\tau,x)$ within a (non-absorbing) slab of extremely
large optical depths (defined\footnote{Note that in \citeauthor{har73}'s papers, as well
as most coeval authors', the optical depth at frequency $x$ is defined as
$\tau_x=\tau_0 \phi(x)$, whereas in our definition $\tau_x=\tau_0 H(a,x)$.
Since $H(a,x)=\sqrt{\pi}\phi(x)$, this implies that
$\tau_{\mathrm{Harrington}} = \sqrt{\pi}\tau_{\mathrm{us}}$. This definition
has been chosen to follow more recent studies.} as
$a\tau_0 \ge 10^3 / \sqrt{\pi}$, or $\tau_0 \ge 1.2\times10^6$ for
$T = 10^4$ K). For the sake of completeness we here give the equation, which
may be written as
\begin{equation}
\label{eq:diffeq}
\frac{\partial^2 J}{\partial\tau^2} + \frac{\partial^2 J}{\partial\varsigma^2}
 = -3\phi(x)\frac{E(\tau,x)}{4\pi},
\end{equation}
where $E(\tau,x)$ is the rate of photon generation per unit mean optical depth,
per unit area, per unit Doppler width, and
$\varsigma \equiv \sqrt{2/3}\int_0^x [1/\phi(x')]dx'$.
With the photons emitted isotropically from a central source
emitting 1 photon per unit time, i.e.~$1/4\pi$ photons per unit time per
steradian, such that $E(\tau,x)=(3/2)^{1/2}\delta(\tau)\delta(\varsigma)$, 
an initial frequency $x_{\mathrm{inj}} = 0$,
and scatterings assumed to be dominated by
isotropic wing scatterings, he was able to solve \eq{diffeq} and
obtain an expression for the emergent spectrum. \citet{neu90} gave a more
general solution to the problem, allowing for the destruction of photons and
the injection at any initial optical depth in the slab, with arbitrary initial
frequency. For centrally\footnote{\citeauthor{neu90} assumed that the photons
are emitted from a thin layer inside the slab, parallel to the surface.
However, for reasons of symmetry, we may as well assume that they are emitted
from a single point.} emitted radiation in a non-absorbing medium, the solution
to \eq{diffeq} at the surface, i.e.~at $\tau = \pm\tau_0$, is
\index{Neufeld solution|textbf}
\begin{equation}
\label{eq:neufeld}
\boxed{
J(\pm\tau_0,x) = \frac{\sqrt{6}}{24} \frac{x^2}{\sqrt{\pi}a\tau_0}
 \frac{1}{\cosh\big[\sqrt{\pi^3/54}\, (x^3 - x_{\mathrm{inj}}^3)/a\tau_0 \big]}.
}
\end{equation}
With perhaps some injustice, we will refer to \eq{neufeld} as the
``Neufeld solution'', even when $x_{\mathrm{inj}} = 0$, in which case it reduces to the
result of \citet{har73}. The profile is normalized to $1/4\pi$ and exhibits
two bumps, symmetrically centered on $x=0$ and drifting further
apart for increasing $a\tau_0$. Note that it solely
depends on the product $a\tau_0$, and that the physical size of the gaseous
system does not enter the equation. A higher optical depth is compensated for 
by a lower value of $a$, i.e.~a higher temperature since
$a \propto \Delta\nu_D^{-1} \propto T^{-1/2}$. The physical explanation for
this is that the more dense the medium is, the further into the wing the
photons have to drift, while a higher temperature will make the medium less
opaque to radiation, since in the wings $\phi(x) \propto a$.

Setting $\partial J/\partial x = 0$, we obtain a transcendental equation for
the maximum of the profile with the solution \citep{har73}
\begin{equation}
\label{eq:xm}
\boxed{
x_m = \pm 1.066 (a\tau_0)^{1/3}.
}
\end{equation}
Moreover, the average number of scatterings can be shown from
\eq{neufeld} to be \citep{har73}
\begin{equation}
\label{eq:Nscat}
\boxed{
N_{\mathrm{scat}} = 1.612 \tau_0.
}
\end{equation}
Notice the close agreement between these results for $x_m$ and
$N_{\mathrm{scat}}$, and the
results obtained by \citet{ada72} given by Eqs.~\ref{eq:xmada} and
\ref{eq:Nsada}.

\citet{dij06a} derived an expression similiar to \eq{neufeld}, but
for spherical symmetry. In this case the emergent spectrum is\footnote{Note
that \eq{dijkstra} differs from the result given by
\citet{dij06a} by a factor of 1/2, since their profile is normalized to $1/2\pi$
instead of $1/4\pi$. This convention was chosen to be able compare more easily
with the Neufeld solution.}
\begin{equation}
\label{eq:dijkstra}
\boxed{
J_{\mathrm{sph}}(\pm\tau_0,x) = \frac{\sqrt{\pi}}{4\sqrt{6}}
                                 \frac{x^2}{a\tau_0}
 \frac{1}{1 + \cosh\big[\sqrt{2\pi^3/27}\, x^3/a\tau_0 \big]},
}
\end{equation}
with the maximum occurring at
\begin{equation}
\label{eq:xmsph}
\boxed{
x_{m\mathrm{,sph}} = \pm 0.92 (a\tau_0)^{1/3}.
}
\end{equation}
and an average number of scatterings approximately one-half of that of the
slab.

%

Furthermore, the spectrum for an
homologously expanding (as in Hubble flow) or contracting (as in a
gravitational collapse) medium, but with no thermal motion, was examined
analytically by \citet{loe99}.

Evidently, all of the configurations considered so far are highly idealized
compared to realistic, astrophysical situations, but for more general
geometries and velocities analytic solutions are not obtainable.
Nevertheless, they provide valuable and
at least qualitative insight into the characteristics of young galaxies,
H\textsc{i} envelopes surrounding hot stars, etc.
Moreover, they offer direct means of testing numerical methods, and with the
analytical approximations obtained in this section, we have a firm
basis for testing the developed code.




\chapter{Cosmic dust}
\label{cha:dust}\index{Dust|textbf}

\init{O}{nce solely an obscuring nuisance} to astronomers,
infrared (IR) astronomy revealed
cosmic dust to be not only interesting in itself, but also to play an
important role in many astrophysical processes.
For example, to form molecular hydrogen, a third agent is needed to carry away
the released energy. Since the probability for three hydrogen atoms to
meet simultaneously is extremely small, the most efficient way
is by first sticking to a dust grain, crawl stochastically across its surface,
and finally meet another atom and form H$_2$, transferring the excess
energy to the dust grain \citep{gou63}. Many other molecules, especially
organic, are believed to be catalyzed by grain surface chemistry
\citep[e.g.][]{tur90,cec01,bis07}.

In the final stages of the life of a star, radiation pressure drives out large
amounts of gas from the surface of the star. In this process the dust aids the
mass loss by absorbing the radiation, being accelerated, and sweeping up gas
on its way away from the star \citep[see, e.g.,][]{hof09}.

Dust absorbs light and converts it to IR radiation. Since such long wavelengths
are only little affected by other absorption processes, observing astrophysical
objects in IR is sometimes easier, or even the only possibility.
This is the underlying mechanism of the SMG selection method.

Moreover, dust grains stick together to form rocks which, in turn, form planets
which are probably needed to form life.

From dust samples from air- and spaceborn plate collectors, the physical
properties of \emph{local} dust is reasonably well-constrained. However,
extrapolating this to the rest of the Universe is most likely a bit too bold.
One of the incentives for implementing dust in the numerical simulations of
Ly$\alpha$ RT was to learn about the dust itself. Unfortunately it turned out
that in the context of Ly$\alpha$, different types of dust give rise to
vitually identical observables. However, since this means that the outcome is
not very sensitive to the exact nature of the modeled dust, the \emph{effect}
of dust in the early Universe can thus be ascertained with relatively high
confidence.

The remainder of this chapter discusses the basic theory of dust, necessary
for understanding the constructed scheme for the effects of dust in the
numerical code.

\section{Effect on radiative transfer}
\label{sec:effRT}\index{Absorption!By dust}

Whereas absorption processes in gas in many cases are well-known,
the effect of dust on the RT is still an intensely debated subject.
In contrast to gas, laboratory experiments with dust are extremely
complex, partly due to the complications involved in replicating the physical
environments of the ISM, partly due to our limited
knowledge vis-\`a-vis what actually constitutes cosmic dust.

In the present-day Universe, most dust is formed\index{Dust!Formation} in the
atmospheres of stars on
the asymptotic giant branch\index{Asymptotic giant branch}
(AGB)\index{Stars!AGB} of the Hertzsprung-Russel diagram; the dying
phase of stars less massive than $\sim$8 $M_\odot$
\citep[e.g.][]{hof07,mat08,gai09}.
In these environments the gas is sufficiently cool, yet sufficiently dense
that molecules may form and stick together to form dust grains.

However, there is observational evidence that dust is also abundantly present
in the early Universe \citep[e.g.][]{sma97,ber03,str07,cop09}.
Since the time to reach
the AGB phase is of the order of 1 Gyr,
something else may have provided the
ISM with dust at these epochs. A promising candidate is SNe\index{Supernovae},
the ejecta of which are believed to exhibit favorable conditions
for the formation of dust for a short period of time, approximately 600 days
after the explosion \citep[e.g.][]{kot09}.

For the Milky Way (MW), as well as  the Small and Large Magellanic Clouds
\index{Small Magellanic Cloud}%
\index{Large Magellanic Cloud}%
(SMC; LMC), the dust extinction curves\index{Extinction curve},
i.e.~the extinction of light as a function of wavelength,
are fairly well established \citep[e.g.][]{bia96,nan81,pre84,pei92}, and from
the observed color excess $E(B-V)$ one may then derive the total extinction.
The term ``extinction'' refers to removal of light from the line of sight,
be it due to absorption or scattering, and may be characterized by the number
$A_\lambda$ of magnitudes by which the observed light from an object is
diminished.
For more distant galaxies one is usually obliged to assume similar extinction
curves. Since the stellar population of
the SMC is younger than that of the LMC, an SMC extinction curve might be
expected to describe better the dust in high-redshift galaxies, and
has indeed proved to be a good fit in GRB host
galaxies \citep[e.g.][]{jak04} and quasar host galaxies
\citep[e.g.][]{ric03,hop04}.
Note, however, that the prominent feature at
2175 {\AA},\index{2175 {\AA} feature} characteristic of the
LMC and MW extinction but the SMC, has been detected in a few cases also at
high redshift \citep{jun04,ell06,sri08,eli09}.

Extinction curves are obtained by comparing the flux received from a pair of
identical (i.e.~same spectral class) stars --- one obscured by dust and the
other unobscured. Measuring at a range of wavelengths, one gets the shape of
the curve.
The overall normalization of extinction curves comes from the observed
property that the extinction is found to be very close to proportional with the
column density $N_{\mathrm{H}}$ of hydrogen \citep[e.g.][]{boh78}.
Typically, one combines measurements of 
$\Nhi$ (and $N_{\textrm{{\scriptsize H}}_2}$)
with the extinction in the $V$ band, $A_V$. In this way, one then knows how
much light is extinguished when traveling a given physical distance in space.

However, for light that does not travel directly from the source to the
observer, as is the case for resonantly scattered lines like Ly$\alpha$, the
situation becomes more complicated. Not only does the total distance covered by the
photons increase by a large and a priori unknown factor, but the photons
received from a given point on the sky may also have traveled through
physically different environments, in turn implying an unknown
and possibly highly increased probability of being absorbed by
dust.\footnote{In principle the Ly$\alpha$ photon may also be destroyed by
other mechanisms, e.g.~by collisionally induced transition of the excited atom
from the 2$P$ state to the 2$S$ state and subsequent two-photon emission, or
pumping
of the nearby $B$-$X$ 1-2 $P$(5) and 1-2 $R$(6) electronic transition of H$_2$
\citep{neu90}. However, under almost all conditions encountered in the ISM,
these processes can be safely ignored (see also \app{quant}).}

\subsection{Ly$\alpha$ escape fraction}
\label{sec:fesctheo}\index{Ly$\alpha$!Escape}
\index{Escape fractions!Ly$\alpha$}

For this reason, the observed fact that Ly$\alpha$ radiation nonetheless
\emph{does} escape has long puzzled astronomers.
Many astrophysical and cosmological key questions depend upon precise
measurements of the luminosities of distant galaxies;
in particular, SFRs and SFR histories, as well as LFs,
are crucially contingent on the amount of assumed luminosities.
A fundamental problem in this context is naturally the question of how large a
fraction of the emitted light actually escapes the galaxy.
If an unknown fraction of the emitted light is absorbed, either by gas or by
dust, the inferred quantity of interest clearly will be subject to large
uncertainties or, at best, a lower limit.

The fact that Ly$\alpha$ line profiles
are often seen to exhibit a P Cygni-like profile has led to the suggestion
that high-velocity outflows\index{Outflows} of gas are needed to enable escape
\citep{kun98,ost09,ate08}, and in fact \citet{dij09} showed that the pressure
exerted on the ISM by Ly$\alpha$ itself can drive large gas masses out of the
galaxies. However, at high redshifts many galaxies are still
accreting matter, which should result in an increased \emph{blue} peak.
Since this is rarely observed, the shape could be caused by other mechanisms,
e.g.~IGM absorption.

The angle under which a galaxy is viewed may also affect the amount of
observed radiation. Ionizing UV radiation could create ``cones'' of low neutral
hydrogen density emanating from the star-forming regions through which the
Ly$\alpha$
can escape \citep{ten99,mas03}. Even without these ionized cones, scattering
effects alone may cause an anisotropic escape of the Ly$\alpha$; tentative
evidence for this was found with the early version of the code \citep{lau07}.

Another commonly repeated scenario is a multi-phase\index{Multi-phase medium}
 medium, where the dust is locked up in cold clouds so that the
photons primarily travel in an ionized, dustless medium \citep{neu91,han06}.
Since continuum radiation travels through the cloud, it would be attenuated
more by the dust. This could explain the high Ly$\alpha$ equivalent widths
occasionally observed in LAEs \citep[e.g.][]{mal02,rho03,shi06}.

Previous attempts to determine Ly$\alpha$
escape fractions from high-redshift galaxies have mainly been
trying to match observed Ly$\alpha$ luminosities with expected, and different
methods obtain quite different results.
In fact, as is evident from the following discussion, there seems to be no
general consensus on the value, the scatter, or even the order of magnitude of
$f_{\mathrm{esc}}$.

\citet{le05,le06} found very good agreement between galaxies simulated with
the galaxy formation model {\sc galform} and observational data at
$z = 3$--6, using a constant escape fraction of $f_{\mathrm{esc}} = 0.02$ and
assuming no IGM absorption.
\citet{dav06} obtained similar results by matching the Ly$\alpha$ LF
of galaxies from their cosmological smoothed particle hydrodynamics
simulation to the data of
\citet{san04}, although \citet{nag08} argued that the data are based on a small
sample and that the simulation box size is too small. Matching the simulated
Ly$\alpha$ LF to the observed one by \citet{ouc08}, \citet{nag08} themselves
obtain $f_{\mathrm{esc}} \simeq 0.1$, although the preferred scenario is not
that a certain fraction of the Ly$\alpha$ radiation escapes, but rather that
a certain fraction of LAEs are ``turned on'' at a given time (the so-called
``duty cycle scenario'').
In a similar way, \citet{day09} find somewhat higher escape fractions at
$z\sim5.7$ and $\sim$6.5 ($f_{\mathrm{esc}} \sim 0.3$), which they use for
predicting the LF of LAEs at $z\sim7.6$.

\citet{ver08}, using the Monte Carlo Ly$\alpha$
radiative transfer code {\sc MCLya}
\citep{ver06} and assuming a shell-like structure of gas, found a large range
of escape fractions by fitting calculated spectra to observed ones of $z\sim3$
LBGs, with $f_{\mathrm{esc}}$ ranging from $\sim$0 to $\sim$1.

\citet{gro07} compared inferred Ly$\alpha$ and rest-frame UV continuum SFRs of
a large sample of LAEs from the MUSYC\index{MUSYC} \citep{gaw06a}
survey and argue that an
escape fraction of $\sim$$1/3$ is needed to explain the discrepancy,
although \citet{nil09} pointed out that a missing $(1+z)$-factor probably
explains the difference.
Matching  Ly$\alpha$-inferred SFRs to SED modeling of
observed LAEs, \citet{gaw06b} found an $f_{\mathrm{esc}}$ of $\sim$0.8, with a
lower limit of 0.2. While SED fitting may not be the most accurate way of
estimating SFRs, aiming to match these observations, \citet{kob07} obtain
similar results theoretically by incorporating the effects of galactic
outflows.

To calculate the SFR from Ly$\alpha$, case B recombination is assumed,
i.e.~that the recombinations take place in optically thick regions. In this
case \citet{ost89} showed that the ratio between
emitted Ly$\alpha$ and H$\alpha$ radiation is 8.7.
Additionally assuming solar element abundances, a \citet{sal55} initial mass
function (IMF) with mass limits 0.1 and 100 $\Msun$,
and that star formation has been going on for $\sim$$10^8$ yr,
the calculated H$\alpha$ luminosity can be converted to a SFR using the
relation \citep{ken98}\index{Star formation rates!From H$\alpha$}
\begin{equation}
\label{eq:ken98}
\frac{\mathrm{SFR}}{\Msun\mathrm{\,yr}^{-1}} = 
\frac{L_{\mathrm{H}\alpha}}{1.3\times10^{41}\mathrm{\,erg\,s}^{-1}}.
\end{equation}
This relation is widely used, even though obviously the assumption may be far
from valid.

If available, Ly$\alpha$ may also be compared directly to H$\alpha$.
Usually, both lines are not obtained simultaneously, however. Instead one may
compare the average luminosities of two samples of galaxies observed in each
line, i.e.~compare the LFs.
The problem is obviously that a sample of LAEs will usually be biased against
objects emitting strongly in Ly$\alpha$, and similarly for H$\alpha$ emitters.
\citet{hay10b} overcome this issue by performing a double-blind survey
targeting the GOODS-S\index{GOODS} field \citep{gia04}. They find that
a Ly$\alpha$ escape fraction of $(5.3\pm3.8)$\% is needed to make the two
obtained LFs match.



\section{What characterizes cosmic dust?}
\label{sec:dusttheo}\index{Dust!Properties}

Four quantities characterize what impact the dust grains will have on the
propagating Ly$\alpha$ photons:
the \emph{density};
the (wavelength-dependent) \emph{cross section} of interaction;
the \emph{albedo} giving the probability that a photon incident on a dust
grain will be scattered rather than absorbed;
and finally the \emph{phase function} defining the direction
into which a non-absorbed photon is scattered.
To understand how this is implemented in the the code, these quantities will be
discussed below.

Dust grains are built up from metals, and thus the dust
density is expected to scale with gas metallicity in some fashion.
Metals are created in
dying stars, i.e.~in AGB stars and SNe. For sufficiently dense and cold
environments, the neutral metals form molecules which eventually stick together
to form dust. No formal definition of the distinction between large molecules
and dust grains exists, but may be taken to be of the order of $\sim$500 atoms
or so.

Depending on the abundances of the individual metals, as well as the physical
conditions, a variety of different types of dust may be produced, with
regards to both composition and structure, and hence with different scattering
properties. Much effort has been put into unraveling the nature of cosmic dust,
in particular in explaining the 2175 {\AA}\index{2175 {\AA} feature} bump.
This feature is generally
attributed to carbonaceous materials, e.g.~graphite, diamonds, and/or
polycyclic aromatic hydrocarbons,
but still the precise nature remains unknown.

For grain sizes much smaller than the wavelength $\lambda$ of the light, the
exact shape
of the particle is not significant and the scattering can be calculated as
Rayleigh scattering. For grains larger than $\lambda$, Mie theory provides a
solution assuming
spherical geometry. However, a significant fraction of interstellar dust is
expected
to be comprised by particles of sizes comparable to the wavelength of
Ly$\alpha$.
In principle, the result of a photon interacting with a dust grain may be
calculated analytically by solving Maxwell's equations, on the basis of the
geometry of the particle and its optical properties, i.e.~the dielectric
functions. This is possible in the case of simple geometries
such as spheres and spheroidals \citep{mie08,van57,boh83}.
More general shapes and composites can be modeled by discretizing
the grain into a large number of dipoles; the so-called \emph{discrete dipole
approximation} \citep{pur73,dra88},
but for the complex and, more importantly,
uncertain or unknown shape of realistic grains, this is not feasible.

Had we full knowledge of the relevant properties of dust, a distribution of the
various species could be calculated in
simulated galaxies, and the radiative transfer could then be realized by
computing the total optical depth of the ISM as a sum
of all contributors, and determining for each scattering the kind of particle
responsible for the scattering.
However, lacking a sound theory of the formation of dust grains, in particular
in the high-redshift Universe, we take a
different approach: although the exact nature of cosmic dust is not known,
the average extinction --- and hence the cross-sectional area --- of dust as a
function of wavelength is known for many different sightlines through the SMC
and the LMC \citep[e.g.][]{gor03}.
Since the metallicity of the Magellanic Clouds is fairly well
known, the extinction curve of the SMC (or LMC) can be scaled to the
metallicity of the gas at each point in the simulated galaxies, thus yielding
the extinction in the simulations.

\subsection{Cross section}
\label{sec:xsecd}\index{Dust!Cross section}\index{Cross section!Dust}

Observationally, the extinction $A_V$ in the $V$ band is found to have a
surprisingly constant proportionality with the column density of hydrogen from
sightline to sightline within the MW \citep[e.g.][]{boh78}.
Similar results, but with
different normalizations, are found for the SMC and the LMC \citep{gor03}.
Accordingly, the cross section $\sigma_{\mathrm{d}}(\lambda)$ of dust may be
conveniently expressed as an effective cross section \emph{per hydrogen atom},
thus eliminating any assumptions about the size distribution, shape,
composition, etc., and merely relying on observed extinction curves.
The optical depth $\tau_{\mathrm{d}}$ of dust when traveling a distance $r$
through a region of hydrogen density $n_{\mathrm{H}}$ is then
\begin{equation}
\label{eq:taud}
\tau_{\mathrm{d}} = n_{\textrm{{\scriptsize H}}} r \sigma_{\mathrm{d}}
                  = N_{\textrm{{\scriptsize H}}}   \sigma_{\mathrm{d}}.
\end{equation}

The quantity usually measured is $A_\lambda/N_{\mathrm{H}}$,
and the cross section is then 
\begin{equation}
\label{eq:sigdth}
\sigma_{\mathrm{d}} = \frac{\ln 10}{2.5} \frac{A_\lambda}{N_{\mathrm{H}}}
\end{equation}
We use the fit to the SMC or LMC extinction curves proposed by \citet{pei92},
which is an extension of the \citet{mat77}-model. The fit is a sum of six terms
(Drude profiles) representing a background, a
far-ultraviolet (FUV), and a far-infrared (FIR) extinction, as well as the
2175 {\AA},\index{2175 {\AA} feature} the 9.7 $\mu$m, and the 18 $\mu$m
extinction features.  Based on newer data from \citet{wei01}, \citet{gne08}
adjusted the fit and added a seventh term to account for the narrow, asymmetric
FUV peak in the dust extinction.

\begin{figure}[!t]
\centering
\includegraphics [width=0.70\textwidth] {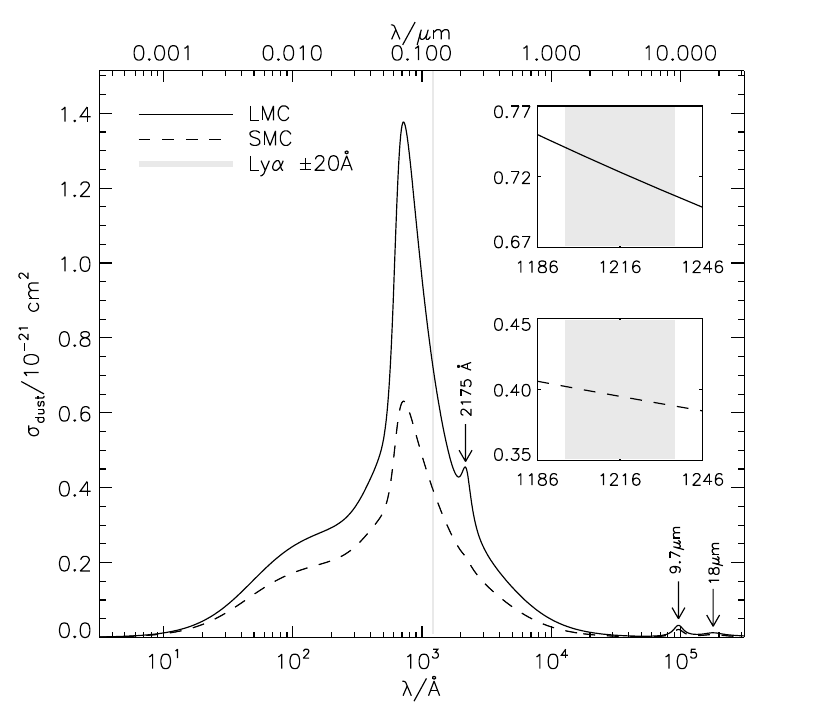}
\caption{{\cap Dust cross section fits to the observed extinction curves of the
         LMC (\emph{solid}) and the SMC (\emph{dashed}).
         The difference in amplitude is mainly due to the SMC being less
         metal-rich than the LMC. The vertical, gray-shaded area is the region
         inside which the (rest-frame) Ly$\alpha$ line is expected to fall.
         The two inlets show a zoom-in of this region on the extinction curves
         (\emph{top}: LMC, \emph{bottom}: SMC), demonstrating the linearity
         across the Ly$\alpha$ line.}}
\label{fig:Xsec}
\end{figure}
Figure \ref{fig:Xsec} shows these fits. The two inlets show that the extinction
curves are very close to being linear in the vicinity of the Ly$\alpha$ line.
In fact, in this region it is an excellent approximation to write the
cross section as

 

%
\begin{equation}
\label{eq:sigd}
\boxed{
\sigma_{\mathrm{d}}/10^{-21}\textrm{ cm}^2 = 
\left\{ \begin{array}{lll}
0.395 + 1.82\times10^{-5}\,T_4^{1/2}\, x & \textrm{ for the SMC}\\
\\
0.723 + 4.46\times10^{-5}\,T_4^{1/2}\, x & \textrm{ for the LMC}.
\end{array} 
\right.
}
\end{equation}
where as usual $x \equiv (\nu-\nu_0)/\Delta\nu_{\mathrm{D}}$, and
$T_4 \equiv T / 10^4$ K.
Note that $T$ only enters
\eq{sigd} to account for the temperature dependency of $x$;
$\sigma_{\mathrm{d}}$ itself is independent of $T$.


\subsection{Number density}
\label{sec:dens}

\begin{figure}[!t]
\centering
\includegraphics [width=0.70\textwidth] {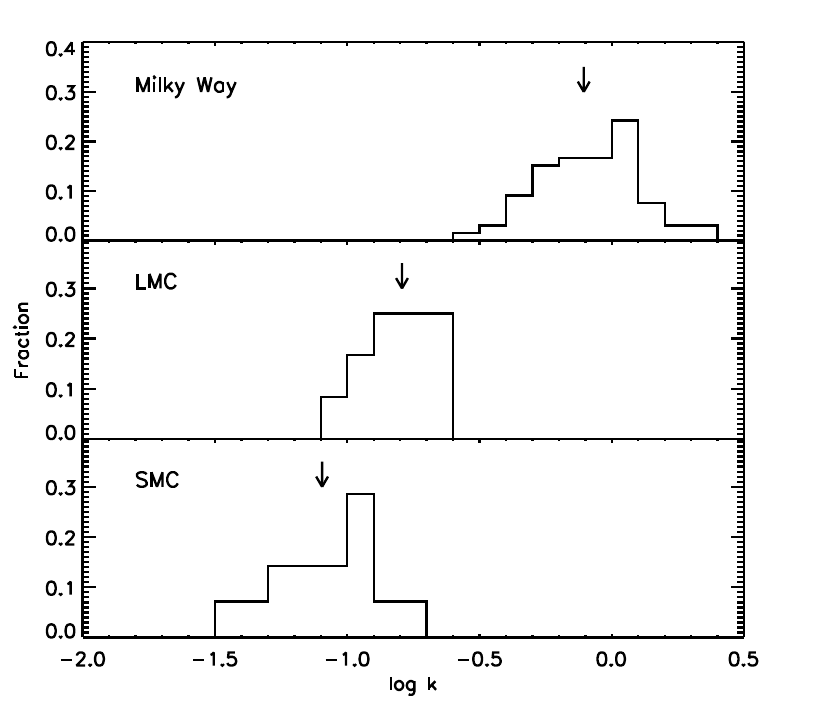}
\caption{{\cap Distributions of dust-to-gas ratio along individual lines of
               sight within the MW (\emph{top}), LMC (\emph{middle}), and SMC
               (\emph{bottom}). Here, $k$ is defined as
               $k = 10^{21}(\tau_B/\Nhi)$ cm$^{-2}$, where $\tau_B$ is the
               extinction optical depth in the $B$ band. The arrows indicate
               the mean dust-to-gas ratio in each galaxy. Although the three
               histograms differ substantially in the means, they have similar
               logarithmic dispersion.
               The plot is taken from \citet{pei92}.}}
\label{fig:pei92}
\end{figure}

The reason for the variability of extinction with galaxy, and the
non-variability with sightline, is to a large degree the different overall
metallicities of the galaxies. Although differences do exist within the
galaxies, as seen in \fig{pei92} the differences in dust-to-gas ratio are
larger from galaxy to galaxy \citep{pei92}.

In most of the calculations we will use an SMC
curve, but as shown in \sec{parstud} the result are not very different if
an LMC curve is used.

Because the cross section is expressed as a cross section per hydrogen atom,
the relevant quantity is not dust density, but hydrogen density.
However, since in general the metallicity at a given location in a
simulated galaxy differs
from that of the Magellanic Clouds, the amplitude of the extinction will also
differ. Assuming that extinction scales with metallicity, a corresponding
pseudo number density $n_{\mathrm{d}}$ of dust at a given location of hydrogen
density $n_{\textrm{H}}$ and metallicity $Z_i$ of element $i$ can then be
calculated as
\begin{equation}
\label{eq:ndpre}
n_{\mathrm{d}} \sim n_{\textrm{H}} \frac{\sum_i Z_i}{\sum_i Z_{i,0}},
\end{equation}
where $Z_{i,0}$ is the average metallicity of element $i$ in the galaxy
the extinction curve of which is applied.
Obviously, $n_{\mathrm{d}}$ is not a true dust number density, but merely a
rescaled hydrogen number density.

On average, the SMC metallicities of the different elements are deficient
relative to Solar values by 0.6
dex \citep[e.g.][]{wel97}, while the LMC is deficit by 0.3 dex
\citep[e.g.][]{wel99}.
Small metal-to-metal deviations from this exist, but, as will be shown in
\sec{parstud}, scaling $Z_i$ to the metallicity of the
individual metals, using values from \citet{rus92}, does not change the
outcome significantly.

The reason that \eq{ndpre} is not expressed as a strict equality is
that we have so far neglected to differentiate between neutral and ionized
hydrogen.
Dust grains may be destroyed in a number of ways, e.g.~through collisions with
other grains,
sputtering due to collisions with ions, sublimation or evaporation, or even
explosions due to UV radiation
\citep[e.g.][]{gre76}.
These scenarios are all expected to become increasingly important for
hotter environments.
Accordingly, studies of the interstellar abundances of dust have usually
assumed that ionized regions contribute negligibly to the dust density,
and merely concerned themselves with measuring
column densities of neutral hydrogen, i.e.~H{\sc i} + H$_2$.
Moreover, many metallicity measurements are derived from low-resolution spectra
not capable of resolving and characterizing various components of the ISM.
As discussed in \sec{iongas}, dust is also observed in regions
that are primarily ionized, and since the bulk of the Ly$\alpha$ photons is
produced in the proximity of hot stars with a large intensity of ionizing UV
radiation, even a little dust associated with the ionized gas might affect the
results.

Hence, we assume that the amount of dust scales with the total amount of
neutral hydrogen \emph{plus} some fraction
$f_{\mathrm{ion}}$ of the ionized hydrogen,
and \eq{ndpre} should then be
\begin{equation}
\label{eq:nd}
\boxed{
n_{\mathrm{d}} = (\nhi
               +  f_{\mathrm{ion}}
                  n_{\textrm{{\scriptsize H}{\tiny \hspace{.1mm}II}}})
                  \frac{\sum_i Z_i}{\sum_i Z_{i,0}} 
}
\end{equation}

Again it is emphasized that this is not a physical number density of dust grains
but with this expression, the total optical depth of gas and dust as seen by a 
photon traveling a distance $r$ is
\begin{equation}
\label{eq:Ntot}
\tau_{\mathrm{tot}} = r (\nhi
                         \sigma_x
                    + n_{\mathrm{d}} \sigma_{\mathrm{d}}).
\end{equation}

In principle, the summation term in \eq{nd} should also include
a term accounting for the fact that the
dust-to-metal
ratio\index{Dust!Dust-to-metal ratio}
$f_{\mathrm{dm}}$
in a given cell may be different from that for which the empirical
data exist. In the Milky Way and the Magellanic Clouds,
$f_{\mathrm{dm}} \simeq 1$
for most metals, i.e.~roughly 1/2 of the metals is condensed to dust grains.
The depletion patterns in high-redshift galaxies are not well
constrained, but no measurements suggest that it should be substantially
different from the local Universe. In fact \citet{pei99} interpret the
depletion patterns of Cr and Zn measured in DLAs by
\citet{pet97} as giving $f_{\mathrm{dm}} \simeq 1$ out to $z \lesssim 3$.
Similarly, fitting depletion patterns of eight elements in GRB host galaxies,
\citet{sav03} find $f_{\mathrm{dm}} \simeq 1$.
To recap, we make no assumptions about how metals deplete to dust other than it
is not appreciably different from the present epoch.

\subsubsection{Dust in ionized gas}
\label{sec:iongas}\index{Dust!Destruction}

Ionized gas is found in a number of physically distinct locations throughout
the Universe.
Compact\index{HII regions@\ion{H}{ii} regions}\index{Strz@Str\"omgren spheres}
H{\sc ii} regions, or Str\"omgren spheres, surround young, hot stars, while
more diffuse H{\sc ii} is a part of the ISM. Larger H{\sc ii}
``bubbles'' are formed around regions of massive star formation due not only
to ionizing radiation from the stars but also to the energy deposited in the
ISM from supernova feedback. Outside the galaxies, the
IGM is predominantly ionized out to redshifts of at least $z \sim 5$--6.
Observations show or indicate the presence of dust in all of these media.
While generally lower than in the neutral gas, inferred
dust-to-gas\index{Dust!Dust-to-gas ratio} mass
ratios ($f_{\mathrm{dg}}$) in
ionized gas span a range from roughly equal to the typically assumed MW ISM
value of $\sim$0.01, to upper limits of $\sim$$10^{-4}$ times lower than this.

Based on 45--180 $\mu$m (FIR) spectroscopy, \citet{aan01} found the Galactic
H{\sc ii} region S125 to be strongly depleted of dust, with a dust-to-gas ratio
of $f_{\mathrm{dg}} \le 10^{-6}$, while \citet{smi99}, using mid-infrared (MIR)
imaging and
spectroscopy, inferred a dust-to-gas ratio of the Galactic H{\sc ii} region
RCW 38 of $10^{-5}$ to $10^{-4}$.
On the other hand, using FIR spectroscopy \citet{chi86} found 12 H{\sc ii}
regions to be dust-depleted by ``only'' a factor of 10 relative to the MW ISM
(i.e.~$f_{\mathrm{dg}} \sim 10^{-3}$), while from FIR photometry,
\citet{har71} found the median dust-to-ionized-gas ratio of seven H{\sc ii}
regions to be close to 0.01.

For the more diffuse H{\sc ii} gas that constitutes part of the ISM, most
obtained
extinction curves in a sense already include the contribution of H{\sc ii} to
$n_{\mathrm{d}}$, although its quantity is
not revealed when measuring H{\sc i} column densities. Hence, any value of
$f_{\mathrm{ion}}$ for the diffuse ISM
larger than 0 would account twice for the ionized gas. 

The dominant destruction mechanism of dust is probably shock waves, associated
with, e.g., high-velocity clouds and SN winds \citep{dra79a,dra79b}.
However, since SNe are thought to be the prime creator of dust at high
redshifts, the H{\sc ii} bubbles in the vicinity of massive star-forming
regions cannot be entirely devoid of dust, and observational evidence of dust
related to SN remnants (SNR) and starburst regions does indeed exist.
Using MIR imaging, \citet{bou06}
determined the dust-to-gas ratio of SN 1987A to be $\sim$$5\times10^{-3}$.
Somewhat lower results are found in Kes 75
\citep[$\sim$$10^{-3}$ from FIR and X-ray,][]{mor07} and in Kepler's SN
\citep[$\sim$$10^{-3}$ from IR and bremsstrahlung,][]{con04}.
On larger scales, the hostile environments imposed by the SNe and ionizing
radiation will reduce the dust density in starburst regions. Fitting continuum
SEDs, \citet{con03} found that
$10^{-4} \lesssim f_{\mathrm{dg}} \lesssim 10^{-2}$
in various starburst regions in a sample of seven luminous infrared galaxies.
However, such regions are not ionized to the same level as compact H{\sc ii}
regions and SNRs, and as argued in the case of the diffuse H{\sc ii}, the
scaling of dust with H{\sc i} to some extend already accounts for the
H{\sc ii}.

Various feedback processes are also responsible for expelling a non-vanishing
amount of metals and
dust into the IGM, although inferred dust-to-gas ratios tend to be small:
from IR-to-X-ray luminosities, \citet{gia08} inferred a dust-to-gas ratio of a
few to 5 times $10^{-4}$, as did \citet{che07} by comparing photometric and
spectroscopic properties of quasars behind SDSS clusters.
Higher values \citep[dust-to-H{\sc i} $\sim0.05$ in the M81 Group,][]{xil06} ---
possibly expelled from the starburst galaxy M82 --- and lower values
\citep[$f_{\mathrm{dg}}\sim10^{-6}$ in the Coma cluster and even less in five
other Abell clusters,][]{sti02}
are also found.
Additionally, sputtering by the hot
halo gas may tend to destroy primarily small grains, leading to a flattening of
the extinction curve in the UV; at the Ly$\alpha$ wavelength, this may reduce
the average cross section by a factor of 4--5 \citep{agu01}.

In summary, the factor $f_{\mathrm{ion}}$ is a practical way of modeling the
destruction of dust in physically ``hostile'' environments.
For simplicity, in the RT code we will not distinguish between H{\sc ii} in
various regions but merely settle on an average dust-to-gas ratio of ionized
gas of $\sim10^{-4}$; that is we set
$f_{\mathrm{ion}} = 0.01$. In
\sec{parstud}, other values of $f_{\mathrm{ion}}$ are investigated and it is
found that using 0.01,
the resulting escape fractions lie approximately midway between those
found when using $f_{\mathrm{ion}} = 0$ (corresponding to the \emph{complete}
destruction of dust in regions where hydrogen is ionized) and
$f_{\mathrm{ion}} = 1$ (corresponding to no destruction of dust at all).
Moreover, these extreme values do not seem to change $f_{\mathrm{esc}}$
by more than $\sim25$\%.



\subsection{Albedo}
\label{sec:alb}\index{Albedo}

When a photon interacts with a dust grain, it may be either absorbed or
scattered. The efficiency with which the dust grain
scatters radiation is dependent on the composition (material, shape, etc.)
of the dust and on the wavelength of the incident photon. If the photon is not
scattered (i.e.~emitted with the same wavelength as the incident photon), it is
absorbed. In this case it is
converted into heat and re-emitted at a later time as IR radiation.
Expressing the total cross section as a sum of a scattering cross section
$\sigma_{\mathrm{s}}$ and an absorbing cross section $\sigma_{\mathrm{a}}$,
such that $\sigma_{\mathrm{d}} = \sigma_{\mathrm{s}} + \sigma_{\mathrm{a}}$,
the albedo $A$ of the dust is defined as
\begin{equation}
\label{eq:alb}
A = \frac{\sigma_{\mathrm{s}}}{\sigma_{\mathrm{d}}}.
\end{equation}

The albedo of dust has been investigated observationally from reflection
nebulae \citep[e.g.][]{cal95} and diffuse galactic light \citep[e.g.][]{lil76}.
At the Ly$\alpha$ wavelength, $A$ lies approximately between $0.3$ and $0.4$
for various size distributions fitted to the LMC and SMC, assuming that the
dust is made mainly of graphite and silicates \citep{pei92,wei01}.
We adopt the model-derived value of $A = 0.32$ \citep[from][]{li01}. The albedo
of this model matches observed values over a wide range of wavelengths.
In \sec{parstud}, the impact of using other values is investigated.


\subsection{Phase function}
\label{sec:phase}\index{Phase function!Dust}

If a photon is not absorbed, it is scattered. As with scattering on hydrogen,
the probability distribution of deflection angles $\theta$
from its original path is given by the phase function. For reasons of
symmetry, the scattering must be symmetric in the azimuthal angle $\phi$
(unless the grains are collectively oriented in some preferred direction due
to, e.g., magnetic field lines), but
in general this is not the case in $\theta$. In fact, dust is often
observed to be considerably forward scattering
[e.g.~in reflection nebulae \citep{bur02}, diffuse galactic light
\citep{schi01}, and interstellar clouds \citep{wit90}].
This asymmetric scattering may be described by the \citet{hen41} phase function 
\begin{equation}
\label{eq:Phg}
P_{\mathrm{HG}}(\mu) = \frac{1}{2} \frac{1 - g^2}{(1 + g^2 - 2g\mu)^{3/2}},
\end{equation}
where $\mu = \cos\theta$, and $g = \langle \mu \rangle$ is the asymmetry
parameter. For $g = 0$, \eq{Phg} reduces to \eq{W12} (isotropic scattering),
while $g = 1$ $(-1)$ implies complete forward (backward) scattering.
$g$ is a function of
wavelength, but for $\lambda$ close to that of Ly$\alpha$, \citet{li01} found
that $g = 0.73$. Again, other values are investigated in \sec{parstud}.




\chapter{The intergalactic medium}
\label{cha:IGMtheo}
\index{Intergalactic medium|textbf}
\index{Interstellar medium|textbf}

\init{A}{lbeit extremely dilute}, space between galaxies is not entirely empty.
In the \emph{intracluster} medium (ICM), i.e.~the space between galaxies bound
together in a cluster, number densities are typically of the order $10^{-3}$ to
$10^{-4}$ \cmcb \citep[e.g.][]{fab94}.
For comparison, the density in the ISM is roughly equal to one hydrogen atom
per cm$^3$, while star-forming regions exceed $10^2$, and even up to $10^6$
atoms per cm$^3$ in the densest molecular clouds \citep[e.g.][]{fer01}.
The clusters are not randomly dispersed in the IGM, but are connected via
huge sheets and filaments of gas of $\sim$ten times lower densities.
However, this is still much denser than the average of the Universe, the bulk
of the gas lying in the immense voids of densities of $10^{-7}$ to $10^{-6}$
\cmcb, stretching several tens to hundreds of Mpc across.
To put things into perspective, a sphere the size of the Earth-Moon system in
the IGM would contain an amount of matter suitable to fill up a cup.

Whereas the temperature of the ISM is generally of the order $10^4$ K, and
one or two orders of magnitude lower in molecular clouds, the rarefied IGM
is easily heated to much higher temperatures.
When gas falls from the voids onto the filaments it heats up, reaching
temperatures of $10^{5}$ to $10^{7}$ K. Even higher temperatures of
$\sim$$10^{8}$ K are reached as it falls into the clusters.
The filaments are themselves connected in ``knots'', where the largest of all
clusters are found.

In this
thesis, the IGM is taken to mean everything outside the galaxies, although
sometimes the term is used only for the ICM.

\section{The Ly$\alpha$ forest}
\label{sec:LAF}
\index{Lya forest@Ly$\alpha$ forest}

The physical state of the IGM can be probed by looking at absorption lines in
the spectrum of a bright source whose intrinsic spectrum is well-known.
Since hydrogen constitutes
the vast majority of the elements, a particularly
popular line is obviously the Ly$\alpha$ line.
As light travels through the
expanding Universe, it gets redshifted, implying that
wavelengths blueward of the Ly$\alpha$ line center are eventually
shifted into resonance. If for a given
wavelength this happens in the vicinity of a sufficient amount of neutral
hydrogen, the spectrum experiences an absorption line (although strictly
speaking the
photons are not absorbed, but rather scattered out of the line of sight).
This results in the so-called Ly$\alpha$ forest \citep[LAF;][]{lyn71,sar80}.

The different absorption features in the LAF should not be thought of as arising
from a number of individual clouds of neutral hydrogen along the line of sight,
as was originally thought. Rather, as has been substantiated
observationally \citep[e.g.][]{bec94,din94,fan96},
theoretically   \citep[e.g.][]{rau95,bi97,hui97}, and
numerically     \citep[e.g.][]{zha95,her96,mir96},
the features are
mostly due to continuously distributed, relatively smooth, low-density gas
regions, with the line widths dictated by the Hubble flow across them, smoothed
further by thermal broadening.

As the transmission of radiation is sensitive to the ionization state of the
IGM, the LAF has been used observationally to put constraints on the so-called
Epoch of Reionization, described in the following section. However,
many other interesting problems can be constrained by looking at the LAF.
For instance, the primordial fluctuations will leave an imprint in the LAF
of later epochs \citep{cro98}.
Prior to decoupling at $z \sim 1100$, acoustic waves in the photon/baryon plasma
--- the so-called \emph{baryon acoustic oscillations} (BAOs) --- shape the power
spectrum on scales of $\sim$150 Mpc. In principle these should also be
observable in the LAF, thereby probing the expansion history of the Universe and
thus providing information about dark energy \citep{eis05}.
Constraints can also be put on the maximum amount of hot dark matter (HDM)
allowed, since too much HDM erases structure on small scales.
Additionally, the abundance of deuterium can be measured, and as the absorbing
systems are generally of low metallicity, this will probably be ``unprocessed''
deuterium, thus gauging nucleosynthesis.


\section{The Epoch of Reionization}
\label{sec:EoR}
\index{Epoch of Reionization|textbf}
\index{Reionization|textbf}

\begin{figure}[!t]
\centering
\includegraphics [width=0.90\textwidth] {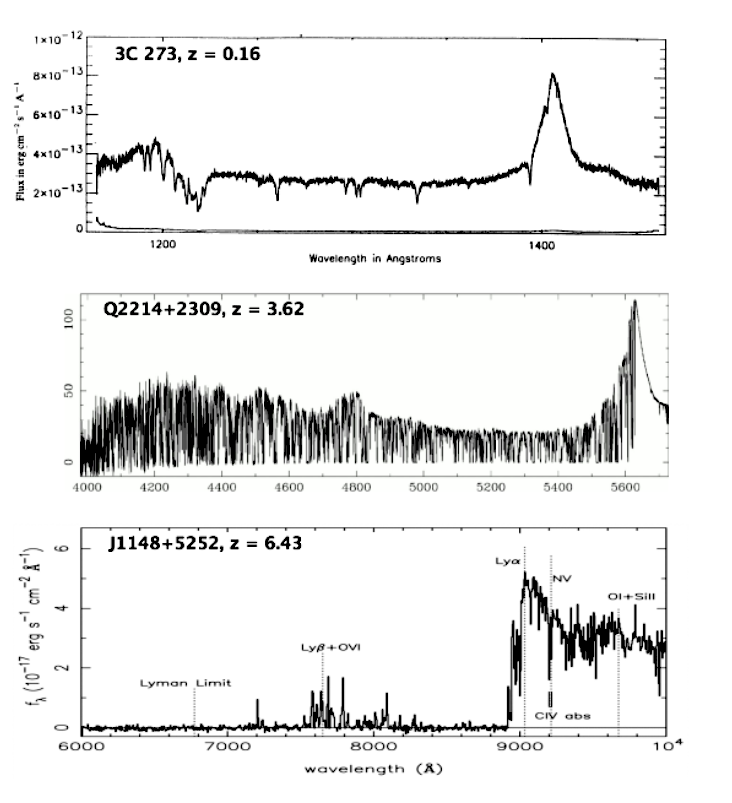}
\caption{{\cap Spectra of quasars at three different epochs. The present-day
               Universe is almost completely ionized, transmitting essentially
               the full spectrum (\emph{upper panel}).
               While the Universe at $z \sim 3.6$ is largly ionized, many
               diffuse \ion{H}{i} region still persists, resulting in the
               Ly$\alpha$ forest (\emph{middle panel}).
               At even higher redshift, close to the Epoch of Reionization,
               all radiation
               blueward of the Ly$\alpha$ line is absorbed, leading to the
               Gunn-Peterson trough (\emph{lower panel}).
               The spectra are taken from \citet{bra93}, \citet{rau98}, and
               \citet{got06}, respectively.
               }}
\label{fig:QSOtriad}
\end{figure}

Inspecting the spectra of quasars at successively higher redshift, one notices
that the Ly$\alpha$ absorption lines become increasingly copiuos, eventually
overlapping, until at a redshift of $z \sim 6$ the spectrum is rendered
completely black blueward of the Ly$\alpha$ line.
This missing flux is the \emph{Gunn-Peterson trough}, predicted
theoretically by \citet{gun65}, but only observed more than three decades
later \citep{bec01}.
\Fig{QSOtriad} shows this evolution in quasar spectra.

Observations of a large number of quasars
show that the Universe was largely opaque to radiation blueward of Ly$\alpha$
at $z \gtrsim 6$ \citep{son04,fan06}.
Around this redshift, the Universe underwent its second major change of state,
the first being the ``recombination`` at $z \sim 1100$ where the cosmic
microwave background (CMB) was released. Since the recombination,
the Universe had been largely neutral and thus opaque to all wavelengths
blueward of Ly$\alpha$. Accordingly, this era is referred to as the \emph{dark
ages}.\footnote{Sometimes the era \emph{before} recombination is denoted the
dark ages. At this time the Universe was even more opaque since light
scattered on free electrons, the cross section of which is independent of
wavelength. However, at this time the Universe was 3000 K and above, so in fact
it was not dark, but filled with a yellowish glow.}
Somewhen relatively shortly before $z \sim 6$, during the
\emph{Epoch of Reionization} (EoR), the Universe became transparent, and a
paramount puzzle in modern cosmology is the question of when and how the
hydrogen, and later helium, of the Universe was reionized.
The EoR marks a comprehensive change
of the physical state of the gaseous Universe, and to understand the cause,
as well as the course, of this phenomenon is a challenging task.
Besides being a compelling event in itself, it also has profound
implications for the interpretation of observations and theoretical
cosmological models, not only due to the increased transparency of the
IGM, but also because of the accompanying rise in IGM temperature.
While reionization is in one way or another directly caused by the appearance
of the earliest luminous sources, it also in turn affects subsequent
structure formation \citep{cho06}.

The very first stars that formed,
Pop\,III,\index{Pop\,III stars}\index{Stars!Pop\,III}
are now thought to be
too few and too ephemeral to produce enough ionizing photons to sustain
reionization \citep{mei05}. Quasars\index{Quasars}
provide a massive amount of photons, but tend to appear too late for ionization
of hydrogen (although they may be important for the reionization of helium).
It is generally believed that the main source for the reionization of hydrogen
is massive star formation in galaxies,
and much effort is being put into ascertaining how
easily ionizing photons can escape their host galaxies in order to be able to
ionize the IGM \citep{raz07,gne08,raz09}.
A plausible scenario is that galaxies gradually ionize their immediate
surroundings such that they are encompassed in large ``bubbles'' of ionized
gas, thus resulting in a highly inhomogeneous ionization structure of the IGM.
Eventually these bubbles percolate and overlap, until ultimately almost all of
the IGM is ionized \citep[e.g][]{gne00}.

Different probes exist for scrutinizing the EoR,
yielding different and not readily mergable results.
Regardless of the physical mechanism responsible for the reionization, it is
likely that it did not happen at a specific moment in time, but rather over an
extended period, the EoR, although this period may have been quite brief.
In addition to measuring the evolution of quasar spectra,
two promising methods exist:

\subsection{The cosmic microwave background}
\label{sec:CMB}
\index{Cosmic microwave background}
\index{Thomson optical depth}

After the CMB was released at the ``surface of last scattering'', it traveled
freely through the neutral Universe.
However, when electrons were released by the reionization a small fraction of
the CMB photons Thomson scattered on these free electrons, introducing small
anisotropies in their polarization. These anisotropies has been studied with
the WMAP\index{WMAP} satellite, allowing the total optical depth
$\tau_e$ of electrons to be deduced.
In a short interval $dt$ of time, a photon travels a
distance $c\,dt$, through which the optical depth is
\begin{eqnarray}
\label{eq:dtaue}
\nonumber
d\tau_e & = & n_e(z)\, \sigma_{\mathrm{T}}\, c\, dt\\
        & = & n_e(z)\, \sigma_{\mathrm{T}}\, c\, \frac{1}{(1+z) H(z)}\, dz,
\end{eqnarray}
where $\sigma_{\mathrm{T}} = 6.65\times10^{-25}$ cm$^2$ is the Thomson
scattering
cross section, and $H(z)$ is the Hubble parameter at redshift $z$, given by
\eq{Hz}. That is, given an electron density
history, the resulting total $\tau_e$ can be calculated by integrating
\eq{dtaue}. In realitiy, the situation is the converse, however.
A total optical depth is
measured, or at least inferred from the CMB polarization maps, and to convert
this to an EoR, an \emph{instant} reionization is typically assumed, i.e.~that
the neutral fraction $x_{\mathrm{neu}}$ of the Universe went from 1 to 0 at a
given redshift, $z_{\mathrm{reion}}$\footnote{The term ``$z_{\mathrm{reion}}$''
thus refers to a characteristic redshift for the EoR, whereas the term
``$\zre$'' in the present work is used for the redshift at which the UVB
initiates is the cosmological simulations, the EoR occurring slightly later.}.
The latest measured value of $\tau_e$ is
0.088, in this way corresponding to $z_{\mathrm{reion}} = 10.5$ \citep{jar10}.


\subsection{The 21 cm line}
\label{sec:21cm}
\index{21 cm line}

The other method relies on the temperature dependence of the forbidden hydrogen
21 cm line. When the first sources of light appeared and a Ly$\alpha$
background was established, the spin temperature $T_{\mathrm{S}}$ of the gas
coupled to the kinetic temperature $T_{\mathrm{k}}$ through the
Wouthuysen-Field effect \citep{wou52,fie58}.
A small fraction of the Ly$\alpha$ photons exchanged energy
with the hydrogen atoms through the scattering, causing a spin flip between the
two hyperfine levels $F = 0$ and $F = 1$.
If the IGM has expanded adiabatically since it was thermally decoupled from the
CMB at $z \sim 200$, then $T_{\mathrm{k}} < T_{\mathrm{CMB}}$ implying that
this mechanism is supposed to manifest itself as absorption at $21(1+z)$ cm.
If X-rays have heated the neutral IGM efficiently, however, it will be
observable as an emission signal \citep{mad97}.
Either way, the signal will thus be a signature of the first sources of light,
and hence of the EoR.

While this probe of the high-redshift IGM has been considered for many years
\citep[e.g.][]{hog79,sco90} no observations have yet confirmed its efficiency,
although this will hopefully change in the near future with the new generation
of radio telescopes such as the Low Frequency Array
(LOFAR)\footnote{http://www.lofar.org}\index{LOFAR},
the Primeval Structure Telescope
(PaST)\footnote{http://web.phys.cmu.edu/\~{}past}\index{PaST},
and the Square Kilometre Array
(SKA)\footnote{http://www.skatelescope.org}\index{SKA}.




\partL{Numerical radiative transfer}\label{NumRT}

\chapter{Numerical background}
\label{cha:alg}

\init{C}{hapter \ref{cha:ResScat} dealt with} the theoretical aspects of
resonant line RT, and various analytical solutions.
An entirely different approach to the problems of RT is the
so-called Monte Carlo method.
The principal
achievement of the present work is the development and application of a
numerical code, relying on this technique, capable of performing
Ly$\alpha$ RT in the most realistic way possible.
This chapter describes the algorithms and numerical simulations anterior to
the RT, while the developed Ly$\alpha$ RT code is
described in the following chapters.

\section{Monte Carlo RT}
\label{sec:MC}
\index{Monte Carlo|textbf}

The Monte Carlo (MC) technique is a class of
computational algorithms, suitable for simulating the behavior of various
physical and
mathematical systems. It distinguishes itself from other simulation methods
(such as cosmological simulations) by being stochastic, i.e.~in some manner
nondeterministic, as opposed to deterministic algorithms.
Being especially useful in studying systems with a large number of coupled
degrees of freedom, the basic idea of this technique is that every time the
fate of some process --- be it physical, mathematical, financial, or
whatever --- is governed by a probability distribution, the outcome is
determined randomly by using a random number (or, more often, a pseudorandom
number supplied by a random number generator).

MC methods were originally practiced under more generic names such as
``statistical sampling''. The first to apply such methods, at least in the
literature, was \citet{lec33}, calculating $\pi$ from the number of randomly
thrown needles falling inside some area. \citet{kel01} used numbers written on
pieces of papers, drawn from a bowl, to numerically evaluate certain integrals,
and in the 1930s Enrico Fermi, though never publishing anything on the subject,
used numerical experiments that would now be called MC calculations to
study the behavior of the newly discovered neutron.

\citet{met49} published the first paper on the MC technique, dubbing
it so after the famous casino in Monaco to honor Stanis{\l}aw Ulam's uncle,
a passionate gambler \citep{ula91}. It has been used for solving RT
problems since the early 1960s \citep{fle63} and for resonant scattering RT
a few years later \citep{aue65}. In essence, a large number of photons are
followed as they diffuse randomly in real and frequency space from their
locations of emission until they escape the medium. The optical depth that a
photon reaches before being scattered, the velocity of the scatterer, the
direction into which the photon is scattered,
and several other physical quantities
are determined from the various probability distributions by which they are
governed.

Nevertheless, though conceptually simple, the demand for strong
computer power until quite recently restricted this technique to deal with
more or less the same idealized configurations that had already been dealt with
analytically. Thus, the majority of previous attempts to
model RT in astrophysical situations have been based on
strongly simplified configurations.

By far, the most work on the subject has been concerned with the emerging
spectrum from an isothermal,
homogeneous medium of plane-parallel or spherical symmetry
\citep[e.g.][]{aue65,ave68,pan73,ahn01,ahn02,zhe02}. Some allow for isotropic
velocities \citep[e.g.][]{car72,nat86,loe99,dij06a} and for a density gradient
\citep{bar10}, and some include
simple models for dust \citep[e.g.][]{bon79,ahn00,han06,ver06}.
However, even though the results of this work
have improved our knowledge tremendously on many physical processes, they do
not capture the complexity and diversity of realistic, astrophysical situations
where velocities can be quite chaotic, and densities and temperatures can vary
by many orders of magnitude over relatively small distances.

A few codes have been constructed and applied to
arbitrary distributions of physical parameters resulting from cosmological
simulations \citep{can05,tas06a,lau07,kol10,fau10}.
The next chapter describe the construction a similar code ---
{\sc MoCaLaTA}\footnote{In a somewhat pathetic attempt to come up with a cool
acronym, \textsc{MoCaLaTA} stands for
{\bf Mo}nte
{\bf Ca}rlo
{\bf L}yman
{\bf a}lpha
{\bf T}ransfer with
{\bf A}MR.}.
Although the work carried out in this thesis is largely inspired by the above
discussed, it distinguishes itself in several ways from earlier works:
the two most important features of \textsc{MoCaLaTA} are its adaptive grid and
its realistic treatment of dust.
The adaptive mesh refinement (AMR) allows for investigation of arbitrarily
detailed systems, the resolution only being limited by the underlying
cosmological simulation (or, in extreme cases, by computer memory), while the
dust is absolutely necessary to rely on the resulting spectra and images,
for instance allowing us to compute escape fractions.
Furthermore,
in addition to studying the emergent spectrum and SB
distribution, the effect of viewing the system from different angles
is investigated. Also, the wavelength and redshift dependent RT through the IGM
is treated.

\subsection{Random numbers}
\label{sec:ran}

Evidently, a key concept of an MC code is the generation of \emph{random
numbers}. Since the output of any deterministic computation is inherently
predictable, truly random numbers cannot be generated by a computer, but
several algorithms exist that for virtually any purpose come sufficiently
close to. Most such pseudorandom number generators provide a series of numbers
randomly distributed in the interval $[0,1]$. Such a number is called a
univariate.
The developed code makes use of the pseudorandom number generator \verb+ran1+
\citep{pre92}, providing a series of numbers that does not repeat itself before
at least $2\times10^{18}$ callings.

The general technique for generating random numbers with an arbitrary PDF that
is analytically integrable,
\emph{and} the integral of which is invertible, is as follows:

As a case in point, we will explicate how to generate a random value of an
optical depth $\tau$, governed by the PDF found in \eq{P_tau}:
\begin{equation}
\label{eq:Pt}
P(\tau) = e^{-\tau}.
\end{equation}
Since \eq{Pt} is properly normalized in the interval $[0,\infty[$, finding a
$\tau$ randomly distributed under this function corresponds to integrating
\eq{Pt} from 0 to $\tau$ until the area under the function is equal
to some univariate $\mathcal{R}$:
\begin{eqnarray}
\label{eq:Rt}
\mathcal{R} & = & \int_0^{\tau} e^{-\tau'}\,d\tau'\\
            & = & -e^{-\tau} + 1.
\end{eqnarray}
If the PDF is not normalized, it should be so before integrating.
Next, this expression for 
$\mathcal{R}$ is inverted to give us $\tau = \tau(\mathcal{R})$:
\begin{equation}
\label{eq:tRttt}
\tau(\mathcal{R}) = -\ln(1-\mathcal{R}),
\end{equation}
or, equivalently,
\begin{equation}
\label{eq:tR}
\boxed{
\tau(\mathcal{R}) = -\ln\mathcal{R}.
}
\end{equation}

If the PDF is not analytically integrable, it can be integrated numerically,
split up into bins. The appropriate bin is then found by cumulating the values
in the bins until the univariate is reached.



\section{Smoothed particle hydrodynamics}
\label{sec:SPH}
\index{Smoothed particle hydrodynamics}

Smoothed particle hydrodynamics (SPH) is a class of computational algorithms
used for simulating the flow of fluids. Introduced over three decades ago
\citep{luc77,gin77}, it is now applied in a wide range of fields of physics,
including astrophysics, aerodynamics, oceanography, and volcanology. The basic
principle of SPH is to represent a fluid by $N_{\mathrm{par}}$ discrete particles
containing the physical parameters of interest.
In contrast to mesh-based, or Eulerian, methods where derivatives are
evaluated at fixed points in space, SPH is a Lagrangian technique, i.e.~the
coordinates move with the fluid. Hence, compared to regularly spaced gridding,
SPH allows for a huge dynamical range in density, since little computational
power is spent in regions of low density.

The parameters of a particle are ``smeared out'' over a finite spatial distance
by a kernel function $W$ that decreases monotonically away from the
particle, out to some distance where it vanishes. This distance is given by
(twice) the \emph{smoothing length} $\h$ of the particle, and
is usually dependent on the local density such that $\h$ is small in dense
regions and large in rarefied regions. The kernel could for instance be a
Gaussian. More commonly, however, a cubic spline is used.
The value $A(\mathbf{r})$ of any given quantity $A$ in any given
point $\mathbf{r}$ is then
\begin{equation}
\label{eq:SPH}
A(\mathbf{r}) = \sum_j \frac{A_j}{\rho_j} m_j
                W(|\mathbf{r} - \mathbf{r}_j|, \h),
\end{equation}
where the summation is over the nearest $N_{\mathrm{nb}}$ neighbors, and
$A_j$, $\rho_j$, $m_j$, and $\mathbf{r}_j$ are the quantity of interest, the
associated density, the mass, and the position of the $j$'th particle,
respectively.
The optimal choice for $N_{\mathrm{nb}}$ is somewhat controversal (and
dependent on
the specific simulation); in the present simulation, $N_{\mathrm{nb}} = 50$
is used. When $\h$ varies in time and space, the choice of which particle's
$\h$ to use in \eq{SPH} becomes ambiguous; in the \emph{scatter} interpretation
$\h$ refers to the $j$'th particle, whereas in the \emph{gather} interpretation
$\h$ refers to particle closest to $\mathbf{r}$. There is no a priori reason
to favor one over another; a compromise may be to use the mean of the two
smoothing lengths, or, even better, to use the mean of the to kernels
\citep{her89}.

The motion of the particles is found by solving the Euler equation including
the appropriate forces between them (or the Navier-Stokes equation, if
viscosity is included);
in pure N-body simulations, e.g.~DM-only simulations, the only relevant
force is gravity. The large scale structure of the Universe is dominated by
dark matter, but on galactic scales, hydrodynamics needs to be taken into
account, resulting in pressure forces between particles. Thus, in SPH
simulations two different types of particles are used: collisionless dark
matter particles and collisional gas particles.
Since each of the $N_{\mathrm{par}}$ particles influence the
other $N_{\mathrm{par}} - 1$, brute force calculations of their mutual
attraction results in $\mathcal{O}(N_{\mathrm{par}}^2)$ coupled differential
equations.
 Interpolating particles onto a mesh and
 Fourier transforming the gravitational potential converts the partial
 differential equations to multiplications with Green's functions, making the
 number of operations scale like
 $\mathcal{O}(N_{\mathrm{par}}\log N_{\mathrm{par}})$ instead; this scheme is
 called the Particle-Mesh (PM) method \citep[e.g.][]{efs85}.
An alternative way, requiring
 also
only
$\mathcal{O}(N_{\mathrm{par}}\log N_{\mathrm{par}})$, is a hierarchical tree
where forces between nearby particles are calculated as a direct sum, but
particles increasingly farther away are treated in increasingly larger
collective groups \citep{barn86}.

To avoid numerical singularities, the gravitational force $F$ between two
particles of masses $m_i$ and $m_j$, separated by a distance $r_{ij}$, is
calculated as $F = G m_i m_j / (r_{ij}^2 + \epsilon^2)$, where $G$ is the
gravitational
constant and $\epsilon$ is the (small) \emph{gravity softening length}.
\index{Gravity softening length}

A plethora of additional physical processes may then be implemented, such as
viscosity, thermal conduction, star formation (converting gas particles into
collisionless star particles), chemical evolution, stellar feedback (converting
the stars back to gas and injecting energy into the surrounding medium), RT
of ionizing radiation, etc.


\section{Underlying cosmological simulations}
\label{sec:cosmo}
\index{Cosmological simulations}

The cosmologic simulations used in this study are conducted using an
N-body/hydro\-dynamical TreeSPH code.
The simulations are first carried out at low resolution,
but in a large spherical volume of space with open boundary conditions.
Subsequently, interesting
galaxy-forming regions are resimulated at high resolution.
Typically, resimulations are performed at 8$\times$ higher mass resolution,
but also ultrahigh resolution (64$\times$) simulations are executed.

The spherical hydrosimulations are themselves resimulations at 8$\times$ the
resolution of DM-only simulations, run with $128^3$ DM particles with periodic
boundary conditions. In the hydrosimulations, all of the original DM particles
are then split into a DM particle
and a gas (SPH) particle according to an adopted universal baryon fraction of
$f_b = 0.15$, in line with recent estimates.

The simulations are started at an
initial redshift $z_i = 39$, at which time there is only DM and
gas particles. The latter eventually evolves partly into star particles,
while, in turn, star particles can become gas particles again.
Both a \citet{sal55} and a \citet{kro98} IMF has been
considered. A standard, flat $\Lambda$CDM cosmology is assumed,
with $\Omega_m = 0.3$, $\Omega_\Lambda = 0.7$, and $h = H_0/100$ km $^{-1}$
Mpc$^{-1} = 0.7$. Two models with different values of the rms linear
density fluctuation $\sigma_8$\index{s8@$\sigma_8$} on scales of
$8 h^{-1}$ Mpc are examined:
one with $\sigma_8 = 0.74$ and one with $\sigma_8 = 0.9$. These values bracket
the latest WMAP-inferred value of $\sim$0.8 \citep{jar10}.
The comoving diameter of the simulated volumed is
$D_{\mathrm{box}} = 10 h^{-1}$ Mpc.

In addition to H and He, the code also
follows the chemical evolution of C, N, O, Mg, Si, S, Ca, and Fe, using the
method of \citet{lia02a,lia02b}. This algorithm invokes in a non-instantaneous
fashion the effects of
supernovae of type II and type Ia, and mass loss from stars of all masses.
Star formation spawns feedback processes, manifesting itself in galactic
superwinds. In the simulations, these winds are realized using the
``conservative'' entropy equation solving scheme \citep{spr02} (rather than
thermal energy), improving the shock resolution over classical SPH schemes.

The Ly$\alpha$ emission is produced by the three different processes described
in \sec{sources};
from recombinations in photoionized regions around massive stars
(responsible for $\sim$90\% of the total Ly$\alpha$ luminosity),
gravitational cooling of infalling gas ($\sim$10\%), and a metagalactic UV
background (UVB) photoionizing the external parts of the galaxy ($\sim$1\%).

The UVB\index{UV background} field is assumed to be that given by \citet{haa96},
where the
gas is treated as optically thin to the UV radiation until the mean free path
of a UV photon at the Lyman limit becomes less than 0.1 kpc, at which point the
gas is treated as optically thick and the UV field is ``switched off''.
Motivated by the steep decline in the transmission of the IGM blueward of the
Ly$\alpha$ line around $z \sim 6$ (discussed in \sec{LAF}), the original
\citeauthor{haa96} UVB is assumed to switch on at a redshift of
$z_{\mathrm{re}} = 6$. To comply with the
results of WMAP, which predicts a somewhat earlier UVB, a different set of
models was also run, in which the intensity
curve was ``stretched'' to initiate at $z \sim 10$.
When these two different versions are discussed together, the model with
$\zre = 10$\index{zre@$\zre$} will be referred to as ``early'' reionization,
while $\zre = 6$ will be referred to as ``late'' reionization.
\index{Reionization}

The masses of SPH, star and DM particles were
$m_{\rm{gas}} = m_\star = 7.3\times10^5$ and
$m_{\rm{DM}}  = 4.2\times10^6$ $h^{-1}$ M$_{\odot}$, and the
gravity softening lengths were
$\epsilon_{\rm{gas}} = \epsilon_\star = 380$ and
$\epsilon_{\rm{DM}} = 680$ $h^{-1}$pc. The
gravity softening lengths were fixed in physical coordinates from $z=6$
to $z=0$, and in comoving coordinates at earlier times.

\begin{figure}[!t]
\centering
\includegraphics [width=1.00\textwidth] {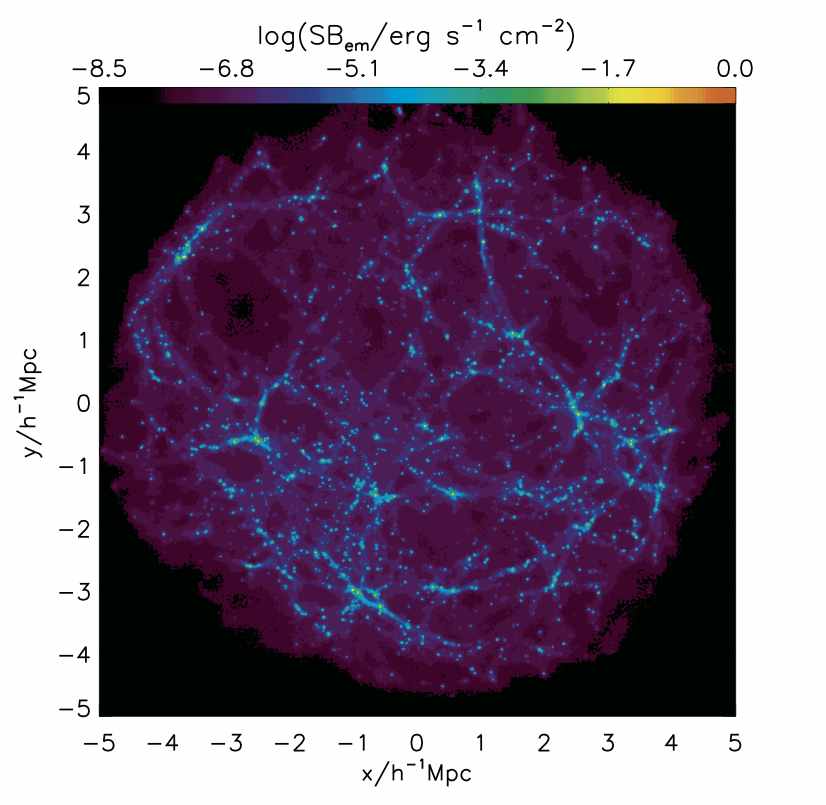}
\caption{{\cap Snapshot of a cosmological simulation at a redshift of $z = 2.5$.
               Color coding indicates Ly$\alpha$ emissivity, and distances are
               given in comoving coordinates. In this particular simulation,
               $\zre = 10$ and $\sigma_8 = 0.74$ was used.}}
\label{fig:S29COSMO}
\end{figure}

For a more thorough description of the code, the reader is referred to
\citet{som03} and \citet{som06}. A snapshot of one of the simulations at
$z = 2.5$ is seen in \fig{S29COSMO}.

Nine individual galaxies are extracted from
the cosmological simulation at redshift $z = 3.6$ --- at which time the
Universe was 1.8 Gyr old --- to be used for the
Ly$\alpha$ RT. These galaxies
are representative of typical galaxies in the sense that they
span three orders of magnitudes in mass, the most massive
eventually evolving into a disk galaxy with circular speed
$V_{\mathrm{c}} \simeq 300$ \kms at $z = 0$.
The numerical and physical properties of these galaxies are
listed in \tab{num} and \tab{phy}, respectively.
\def\ryknum{-2cm} 
\def\rykphy{-3cm} 
\begin{table}[!t]
\begin{center}
{\sc Characteristic quantities of the simulations}
\end{center}
\hspace{\ryknum}
\begin{tabular}{lccccccccc}
\hline
\hline
Galaxy                                               &          S33sc    &            K15       &           S29     &            K33        &         S115        &          S87   &          S108  &         S115sc  &          S108sc \\
\hline
$N_{\mathrm{p,tot}}$                                 & 1.2$\times$$10^6$ &   2.2$\times$$10^6$  & 1.1$\times$$10^6$ &  $1.2$$\times$$10^6$  & $1.3$$\times$$10^6$ & 1.4$\times$$10^6$ & 1.3$\times$$10^6$ & 1.3$\times$$10^6$  & 1.3$\times$$10^6$  \\
$N_{\mathrm{SPH}}$                                   & 5.5$\times$$10^5$ &   1.0$\times$$10^6$  & 5.1$\times$$10^5$ &  $5.5$$\times$$10^5$  & $6.4$$\times$$10^5$ & 7.0$\times$$10^5$ & 6.3$\times$$10^5$ & 6.4$\times$$10^5$  & 6.3$\times$$10^5$  \\
$m_{\mathrm{SPH}}$,$m_{\mathrm{star}}$               & 5.4$\times$$10^5$ &   9.3$\times$$10^4$  & 9.3$\times$$10^4$ &  $9.3$$\times$$10^4$  & $1.1$$\times$$10^4$ & 1.2$\times$$10^4$ & 1.2$\times$$10^4$ & 2.6$\times$$10^3$  & 1.5$\times$$10^3$  \\
$m_{\mathrm{DM}}$                                    & 3.0$\times$$10^6$ &   5.2$\times$$10^5$  & 5.2$\times$$10^5$ &  $5.2$$\times$$10^5$  & $6.6$$\times$$10^4$ & 6.5$\times$$10^4$ & 6.5$\times$$10^4$ & 1.4$\times$$10^4$  & 8.1$\times$$10^3$  \\
$\epsilon_{\mathrm{SPH}}$,$\epsilon_{\mathrm{star}}$ &     344           &         191          &     191           &       191             &      96             &    96             &        96         &     58             &     48             \\
$\epsilon_{\mathrm{DM}}$                             &     612           &         340          &     340           &       340             &      170            &   170             &       170         &    102             &     85             \\
$l_{\mathrm{min}}$                                   &      18           &         10           &      10           &        10             &        5            &     5             &         5         &      3             &      2.5           \\
\hline
\end{tabular}
\hspace{\ryknum}
\caption{{\cap Total number of particles ($N_{\mathrm{p,tot}}$), number
               of SPH particles only ($N_{\mathrm{SPH}}$), masses ($m$),
               gravity
               softening lengths ($\epsilon$), and minimum smoothing lengths
               ($l_{\mathrm{min}}$) of dark matter (DM), gas (SPH), and star
               particles used in the simulations. Masses are measured in
               $h^{-1}M_\odot$, distances in $h^{-1}$pc}}
\label{tab:num}
\end{table}
\begin{table}[!t]
\begin{center}
{\sc Physical properties of the simulated galaxies}
\end{center}
\hspace{\rykphy}
\begin{tabular}{lccccccccc}
\hline
\hline
Galaxy                                &           S33sc        &            K15         &           S29        &            K33         &          S115          &          S87           &          S108         &          S115sc         &          S108sc        \\
\hline
SFR/$M_\odot$ yr$^{-1}$               &         70             & 16                     &           13         & 13                     & 0.5                    &        0.46            &        1.62           &  $3.7$$\times$$10^{-3}$ & $1.7$$\times$$10^{-3}$ \\
$M_*/M_\odot$                         & $3.4$$\times$$10^{10}$ & $1.3$$\times$$10^{10}$ & $6.0\times10^{9}$    & $6.5$$\times$$10^{9}$  & $2.5$$\times$$10^{8}$  & $1.8$$\times$$10^{8}$  & $4.9$$\times$$10^{8}$ &  $2.0$$\times$$10^{7}$  & $5.9$$\times$$10^{6}$  \\
$M_{\mathrm{vir}}/M_\odot$            & $7.6$$\times$$10^{11}$ & $2.8$$\times$$10^{11}$ & $1.7\times10^{11}$   & $1.3$$\times$$10^{11}$ & $2.5$$\times$$10^{10}$ & $2.1$$\times$$10^{10}$ & $2.6$$\times$$10^{9}$ &  $4.9$$\times$$10^{9}$  & $3.3$$\times$$10^{8}$  \\
$r_{\mathrm{vir}}$/kpc                &      63                & 45                     &       39             & 35                     & 20                     &   19                   &     10                &        12               &          5             \\
$[$O/H$]$                             &    $-0.08$             &     $-0.30$            &   $-0.28$            &     $-0.40$            &    $-1.22$             &    $-1.28$             &   $-0.51$             &     $-1.54$             &       $-1.64$          \\
$V_{\mathrm{c}}(z=0)$/\kms            &   300                  & 245                    &    205               & 180                    & 125                    &           132          &       131             &        50               &            35          \\
$L_{\mathrm{Ly}\alpha}$/\ergs         & 1.6$\times$$10^{44}$   & $4.5$$\times$$10^{43}$ & 2.9$\times$$10^{43}$ & $2.5$$\times$$10^{43}$ & $1.3$$\times$$10^{42}$ & 1.1$\times$$10^{42}$   & 2.6$\times$$10^{42}$  & 4.9$\times$$10^{40}$    & 3.2$\times$$10^{39}$   \\
$L_{\nu,\mathrm{UV}}$/\ergs Hz$^{-1}$ & 5.0$\times$$10^{29}$   & $6.7$$\times$$10^{28}$ & 9.3$\times$$10^{28}$ & $5.5$$\times$$10^{28}$ & $3.6$$\times$$10^{27}$ & 3.3$\times$$10^{27}$   & 1.2$\times$$10^{28}$  & 2.6$\times$$10^{25}$    & 1.2$\times$$10^{25}$   \\
\hline
\end{tabular}
\hspace{\rykphy}
\caption{{\cap Star formation rates (SFRs),
               stellar masses ($M_*$),
               virial masses ($M_{\mathrm{vir}}$),
               virial radii ($r_{\mathrm{vir}}$),
               metallicities ($[$O/H$]$),
               circular velocities ($V_{\mathrm{c}}$),
               Ly$\alpha$ luminosities ($L_{\mathrm{Ly}\alpha}$),
               and UV luminosities ($L_{\nu,\mathrm{UV}}$)
               for the simulated galaxies.
               All quoted values correspond to a redshift of
               $z = 3.6$, except $V_{\mathrm{c}}$ which is given for $z = 0$.}}
\label{tab:phy}
\end{table}

\subsection{Ionizing UV radiative transfer}
\label{sec:uvrt}
\index{UV radiative transfer}

To model the propagation of ionization fronts realistically, \citet{raz06,raz07}
employed the following RT scheme to post-process the cosmological simulation:
first, the physical properties of the SPH particles are interpolated from the
50 nearest neighboring particles onto an adaptively refined grid of base
resolution $128^3$ cells, with
dense cells recursively subdivided into eight cells
until no cell contains more than ten particles.
\emph{The resulting adaptively refined grid is the same that will be used for
the Ly$\alpha$ RT}.

Around each stellar source, a system of $12\times4^{n-1}$ radial rays
($n = 1,2,\ldots$ being the angular resolution level) is constructed that split
either as one
moves farther away from the source or as a refined cell is entered.
Once a radial ray is refined angularly, it stays refined at larger distances
from the source, even when leaving the high-resolution region. In each
cell, the photoreaction number and energy rates due to
photons traveling along ray segments passing through that cell are accumulated.
These rates are then used to update temperature and the ionization state of
hydrogen and helium, which in turn are used to calculate the Lyman continuum
(LyC) opacities used in the RT. In addition
to stellar photons, ionization and heating by LyC
photons originating outside the computational volume is accounted for with the
FTTE scheme \citep{raz05} assuming the \citeauthor{haa96} UVB, modified to match
the particular reionization model.

Since the ratios of \ion{H}{i}, \ion{He}{i}, and \ion{He}{ii} densities varies
from cell to cell, the UV RT cannot be conducted monochromatically but has to be
done as multi-frequency transfer. The UV photons are separated in three bands,
$[13.6,24.6[$ eV, $[24.6,54.4[$ eV, and $[54.4,\infty[$, assuming a mixture of
stellar and quasar spectra which would provide the UVB.
In each cell the angle-averaged intensity is added to the chemistry solver to
compute the ionization equilibrium.





\chapter{{\sc MoCaLaTA}}
\label{cha:mocalata}

\init{A}{ firm basis has now} been established for understanding the
development and the structure of the developed MC Ly$\alpha$ RT code
{\sc MoCaLaTA}.
In short the structure of {\sc MoCaLaTA} is as follows:
at the heart of a simulation lies an \emph{adaptively refined grid of
cells}, with each cell containing a set of physical parameters characterizing
that particular point in space. Being adaptively refined means that one may
have extremely high resolution only in the places where it is required,
retaining low resolution elsewhere. In this way the need for memory is
tremendously reduced, compared to achieving the same resolution with a regular
grid\footnote{Throughout this thesis we will use the terms cell, grid point and
mesh point more or less interchangably; a mesh- or grid point can be thought of
as being the center of a cell.}.

The physical parameters may be set ``by hand'', as in the case of a homogeneous,
isothermal sphere, or may be a ``snapshot'' from a cosmological simulation.
If the underlying cosmological simulation is an AMR code, like
AP$^3$M \citep{cou91}, ENZO
\citep{osh04} and ART \citep{kra97}, the original grid is simply
used for the Ly$\alpha$ RT.
If the underlying cosmological simulation is particle based,
like Gadget \citep{spr05} and the code that we will make use
of \citep[][see also \sec{cosmo}]{som03},
first the physical parameters of interest must be interpolated onto the mesh.
Cells may be subdivided into eight subcells which, in turn, may be
further refined. The refinement criterion is usually taken to be density, but
can in principle be any condition, e.g.~density gradient, velocity, etc.

Thus, in each cell we will have values for
the temperature $T$,
the number density $\nhi$ of neutral hydrogen,
the number density $\nd$ of dust,
the three dimensional bulk velocity
$\mathbf{v}_{\mathrm{bulk}}$,
as well as the luminosity $L_{\mathrm{Ly}\alpha}$.
$n_{\mathrm{d}}$ is itself calculated on the basis of the metallicity of the
eight different metals and the ionization state of hydrogen in the cell.

A photon is then launched with a probability of being launched from a given
cell proportional to the luminosity in that cell. Depending on the optical
depth of the gas and dust lying along the path of the photon,
it then travels some
distance before it either scatters on a hydrogen atom or a dust grain, is
absorbed by dust, or escapes the computational box.
If scattered, it changes direction and continues its journey.

At each point of scattering the probability that the photon is emitted in the
direction of a virtual observer \emph{and} escapes through the intervening
column of
gas is calculated and added as a weight to the pixel element of a three
dimensional array corresponding to the frequency and the projected position of
the photon.

The whole process is then repeated until the photon escapes the computational
box, and subsequently repeated for the other photons until the output
converges, and adding more photons does not alter the output
significantly.\vspace{1mm}

{\sc MoCaLaTA} is written in the general-purpose, procedural, imperative
programming
language Fortran (95). Generally, to avoid numerical errors, calculations are
carried out in double precision, providing approximately 16 digits of precision.
To
save memory, however, the large arrays storing the physical parameters are kept
in single precision. Since each photon's path is independent of the others',
the simulation can be distributed on several CPUs, and the code has thus been
parallelized using OpenMP.

\section{Emission of photons}
\label{sec:em}

The ratio of $L_{\mathrm{Ly}\alpha}$ of a given cell to the total luminosity
$L_{\mathrm{tot}}$ of all cells determines the probability of a photon being
emitted from that particular cell. The initial position
$\mathbf{x}_{\mathrm{em}}$ of the
photon is a random location in the cell.

Irrespective of the emission being due to recombination or collisional
excitation and subsequent decay, the photons are emitted isotropically.
Accordingly, an initial direction vector $\mathbf{n}_i$ can be found as
\begin{equation}
\label{eq:n_i}
\mathbf{n}_i = \left( 
\begin{array}{c}
2\mathcal{R}_1 - 1\\
2\mathcal{R}_2 - 1\\
2\mathcal{R}_3 - 1
\end{array}
\right)
\end{equation}
where $\mathcal{R}_1, \mathcal{R}_2, \mathcal{R}_3$ are three different
univariates. In order not to favor the corners of the cube surrounding
$\mathbf{x}_i$, if $n_i^2 > 1$ (i.e.~if $\mathbf{n}_i$ lies outside a unit
sphere) it is rejected and three new univariates are drawn.
Otherwise it is accepted and normalized to get the inital \emph{unit} direction
vector $\mathbf{\hat{n}}_i$.

In the reference frame of the emitting
atom, the photon is injected with a frequency $x_{\mathrm{nat}}$, given by the
distribution $\mathcal{L}(x)$ (Eq.~\ref{eq:Lx}).
The atom, in turn, has a velocity $\mathbf{v}_{\mathrm{atom}}$ in the reference
frame of the gas element
drawn from a thermal profile of Doppler width $\Delta\nu_{\mathrm{D}}$.
Measuring atom velocities in terms of Doppler widths,
$\mathbf{u} = \mathbf{v}_{\mathrm{atom}}/v_{\mathrm{th}}$, each component $u_i$
is then distributed according to $\mathcal{G}(u_i)$, given by
\eq{Gx}.

To first order in $v/c$, \eq{n_i} is valid in all relevant reference frames,
and a Lorentz transformation to the reference frame of
the gas element then yields the initial
frequency
\begin{equation}
\label{eq:x_i}
x_i = x_{\mathrm{nat}} + \mathbf{u} \cdot \mathbf{\hat{n}}_i.
\end{equation}

For photons emitted in the dense, star-forming regions, it makes no difference
whether $x_i$ is calculated in the above manner or simply set
equal to zero. However, when studying large volumes of space, a nonvanishing
fraction of the Ly$\alpha$ photons may be produced through cooling radiation,
which also takes place well away from the star-forming regions of the galaxy.
In these environments, whereas the probability of a photon with $x = 0$
escaping is still
extremely small, being injected one or two Doppler widths away from line center
may allow the photon to escape directly.


\section{Propagation of the radiation}
\label{sec:prop}

\subsection{Optical depth}
\label{sec:tau}
\index{Optical depth}

The optical depth $\tau$ covered by the photon before scattering is governed by
the PDF given by \eq{P_tau}. Thus, instead of
calculating the probability of being scattered for each single atom encountered
along the way, we can determine the optical depth from \eq{P_tau} and
subsequently convert it into a physical distance $r$ using \eq{taudefDUST}.
The method of generating a random value of $\tau$ was discussed in \sec{ran}.


\subsection{Gas and dust cross section}
\label{sec:sigma}
\index{Cross section}

\subsubsection{Neutral hydrogen}
 
Recall that in the rest frame of the gas element, the scattering cross section
$\sigma_x$ of H\,\textsc{i} given by \eq{sigxtheo}:
\begin{equation}
\label{eq:sigx}
\sigma_x = f_{12} \frac{\sqrt{\pi}e^2}{m_e c \Delta\nu_D} H(a,x),
\end{equation}
where
\begin{equation}
\label{eq:Hax}
H(a,x) = \frac{a}{\pi} \int_{-\infty}^{\infty}
         \frac{e^{-y^2}}{(x-y)^2 + a^2}\,dy.
\end{equation}
is the Voigt function.

\begin{figure}[!t]
\centering
\includegraphics [width=0.90\textwidth] {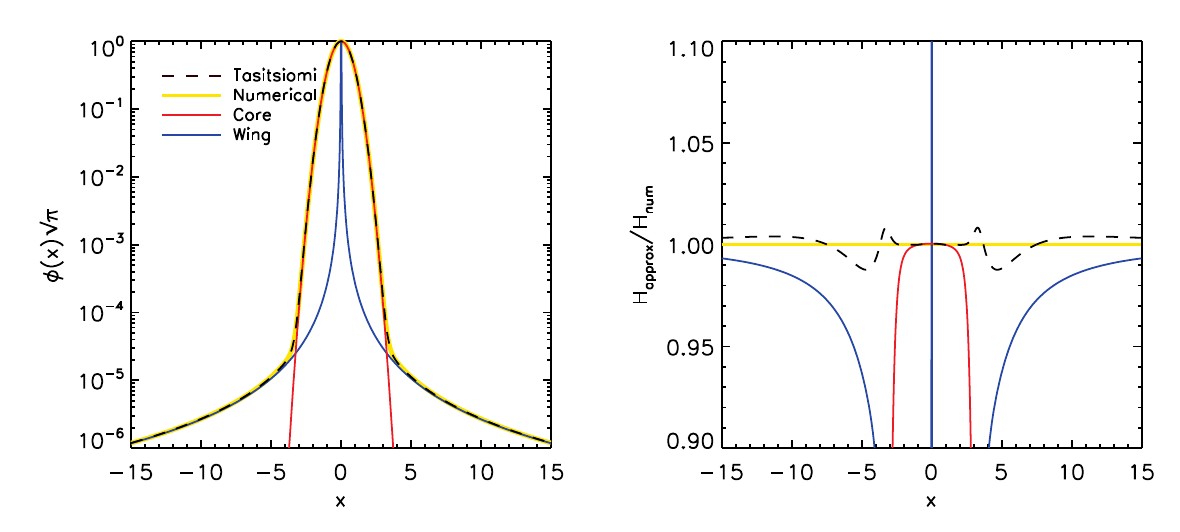}
\caption{{\small Comparison of the approximation given by \eq{Haxtas}
                 (\emph{dashed}) with the numerical solution (\emph{yellow}) of
                 the Voigt
                 function for a temperature of $T = 10\,000$ K. Also shown are
                 the core (Gaussian, \emph{red}) and the wing (Lorentzian,
                 \emph{blue})
                 approximation given by \eq{Haxappr}. In the right
                 panel the discrepancy between the solutions stand out more
                 clearly. It is clear that in the core/wing approximation,
                 the exact transition between the two becomes crucial.}}
\label{fig:Voigt}
\end{figure}

Unfortunately, \eq{Hax} is not analytically integrable. Indeed, it 
can be integrated numerically, but due to the broad wings of the
Lorentzian this is not practical, especially since one would have to this for
a fine grid covering all the anticipated values of $a$ and $x$.
A common way to come around this is to simply use
\begin{equation}
\label{eq:Haxappr}
H(a,x) \simeq \left\{ \begin{array}{ll}
e^{-x^2}                & \textrm{in the core}\\
\frac{a}{\sqrt{\pi}x^2} & \textrm{in the wings},
\end{array}
\right.
\end{equation}
or to expand $H(a,x)$ in $a$. However, \citet{tas06a} offers an
analytical fit which is an excellent approximation for temperatures $T > 2$ K,
and may be written as\index{Voigt function!Approximation}
\begin{equation}
\label{eq:Haxtas}
\boxed{
H(a,x) = q\sqrt{\pi} + e^{-x^2},
}
\end{equation}
where
\begin{equation}
\label{eq:q}
q = \left\{ \begin{array}{ll}
0                                     & \textrm{for } z \le 0\\
\left(1 + \frac{21}{x^2}\right) 
          \frac{a}{\pi(x^2 + 1)} P(z) & \textrm{for } z > 0,
\end{array}
\right.
\end{equation}
with
\begin{equation}
\label{eq:z}
z = \frac{x^2 - 0.855}{x^2 + 3.42},
\end{equation}
and
\begin{equation}
\label{eq:Pz}
P(z) = 5.674z^4 - 9.207z^3 + 4.421z^2 + 0.1117z.
\end{equation}
In Fig.~\ref{fig:Voigt} the Voigt function as given by \eq{Haxtas} is 
compared with the ``exact'' solution resulting from a 4th order numerical
integration (Simpson's method), and with the approximation given by
\eq{Haxappr}.

\subsubsection{Dust}
\label{sec:nd}

The total cross section (per hydrogen atom) of the dust grains is given by
\eq{sigd}:
\begin{equation}
\label{eq:sigd2}
\sigma_{\mathrm{d}}/10^{-21}\textrm{ cm}^2 = 
\left\{ \begin{array}{lll}
0.395 + 1.82\times10^{-5}\,T_4^{1/2}\, x & \textrm{ for the SMC}\\
\\
0.723 + 4.46\times10^{-5}\,T_4^{1/2}\, x & \textrm{ for the LMC}.
\end{array} 
\right.
\end{equation}

\subsection{Spatial displacement}
\label{sec:r}
\index{Radiative transfer}

With $\tau$, $\sigma_x$, and $\sigma_{\mathrm{d}}$ as given by
Eqs.~\ref{eq:tR} and \ref{eq:sigx}, and \ref{eq:sigd2}, respectively,
we can now determine the physical distance $r$
that the photon will travel before being scattered:
\begin{equation}
\label{eq:r}
r = \frac{\tau}{\nhi \sigma_x + n_{\mathrm{d}} \sigma_{\mathrm{d}}}.
\end{equation}
The new --- not necessarily final, as we will see --- position of the photon is
then
\begin{equation}
\label{eq:x_f}
\mathbf{x}_f = \mathbf{x}_i + r\hat{\mathbf{n}}_i,
\end{equation}
where $\mathbf{x}_i = \mathbf{x}_{\mathrm{em}}$.

If $\mathbf{x}_f$ lies inside the cell, the photon is scattered. However,
because in general the physical conditions vary from cell to cell, if the final
location is outside the initial cell we must redo the calculations above
inside the neighboring cell. Towards this end, we consecutively check whether
$\mathbf{x}_f$ is outside the six planes constituting the boundaries of the 
cell and, if so, determine the exact position of the intersection
$\mathbf{x}_{\mathrm{cut}}$ of the photon's trajectory with the plane.

Specifically, let $x_+$ be the $x$-value of the plane including the face of the
cell lying in the positive $x$-direction and define $x_-$, $y_{\pm}$ and
$z_{\pm}$ accordingly. Then the distance from $\mathbf{x}_i$ to the face is
\begin{equation}
\label{eq:delx}
\delta x = \frac{x_+ - x_{i,1}}{\hat{n}_{i,1}},
\end{equation}
where $x_{i,1}$ is the $x$-component of $\mathbf{x}_i$, etc., and should not be
confused with the initial frequency $x_i$.
If $x_{f,1} > x_+$, the photon has crossed the plane and is ``pulled back'' 
to the point of intersection
\begin{equation}
\label{eq:x_cut}
\mathbf{x}_{\mathrm{cut}} = \mathbf{x}_i + \delta x \,\mathbf{\hat{n}}_i.
\end{equation}
However, only if $y_- \le x_{\mathrm{cut,2}} < y_+$ and 
$z_- \le x_{\mathrm{cut,3}} < z_+$, the photon actually left the cell through
the $x_+$-face; otherwise, it must have left through one of the other faces
which we
then check. Whatever the case, when $\mathbf{x}_{\mathrm{cut}}$ has been 
determined, we set $\mathbf{x}_f = \mathbf{x}_{\mathrm{cut}}$ and let the
photon continue its journey, redoing the steps above, but with the physical
parameters of the new cell. The optical depth of this particular part of the
journey has already been determined, however, so the new optical depth
$\tau'$ that will enter the equations is reduced by the amount spent in the
previous cell:
\begin{equation}
\label{eq:tau_new}
\tau' = \tau_{\mathrm{orig.}} - |\mathbf{x}_{\mathrm{cut}} - \mathbf{x}_i|\,
        ( \nhi\, \sigma_x )_{\mathrm{prev.cell}}.
\end{equation}

In contrast to a  regular grid, in an AMR grid a given cell will not in general
have a unique neighbor.
The cells are structured in a nested grid, where a refined cell is the
``parent'' of eight ``child'' cells which, in turn, may or may not be refined.
The new host cell of the photon is then determined by walking
up and down the hierarchical tree structure.

\subsection{Lorentz transformation between adjacent cells}
\label{sec:Lorcell}
\index{Lorentz transformation}

Obviously, the frequency $\nu$ of the photon is not altered upon entering the
neighboring cell. However, because of the temperature dependence of
$\Delta\nu_D$,
$\nu$ is no longer represented by the same value of $x$. Moreover, since $x$ 
was defined relative to
the gas bulk motion, we must perform a Lorentz transformation in order to 
express $x$ relative to the bulk motion in the new cell.

To an external observer, the frequency of the photon is measured to be
\begin{equation}
\label{eq:x_ext}
x_{\mathrm{ext}} = x + \mathbf{u}_{\mathrm{bulk}} \cdot \mathbf{\hat{n}}_i,
\end{equation}
where
$\mathbf{u}_{\mathrm{bulk}} = \mathbf{v}_{\mathrm{bulk}}/v_{\mathrm{th}}$.
Denoting with primed and unprimed variables values in the ``new'' and the 
``old'' cell, respectively, from \eq{x_ext} and the definition of $x$,
\begin{equation}
\label{eq:eq}
(x' + \mathbf{u}_{\mathrm{bulk}}' \cdot \mathbf{\hat{n}}_i) \Delta\nu_D' =
(x + \mathbf{u}_{\mathrm{bulk}} \cdot \mathbf{\hat{n}}_i) \Delta\nu_D,
\end{equation}
or
\begin{equation}
\label{eq:x_newcell}
\boxed{
x' = (x + \mathbf{u}_{\mathrm{bulk}} \cdot \mathbf{\hat{n}}_i) 
     \frac{\Delta\nu_D}{\Delta\nu_D'}
   - \mathbf{u}_{\mathrm{bulk}}' \cdot \mathbf{\hat{n}}_i.
}
\end{equation}

In this way the photon is transferred through the medium until either the 
initially assigned $\tau$ is ``spent'' and the photon is scattered/absorbed,
or it escapes the computational box.


\section{Scattering}
\label{sec:codescat}
\index{Scattering!In code}

Once the location of the scattering event has been determined, another
univariate $\mathcal{R}$ determines whether the photon hits a hydrogen atom or
a dust grain by comparing it to the ratio
\begin{eqnarray}
\label{eq:varrho}
\nonumber
\varrho & = & \frac{\nd \sigma_{\mathrm{d}}}
              {\nhi \sigma_x + n_{\mathrm{d}} \sigma_{\mathrm{d}}}\\
        & = & \frac{\tau_{\mathrm{d}}}{\tau_x + \tau_{\mathrm{d}}}.
\end{eqnarray}

\subsection{Atom velocity}
\label{sec:u}
\index{Atom velocity}

If $\mathcal{R} > \varrho$, the interaction is caused by hydrogen, in which
case the
velocity $\mathbf{u}$ of the atom responsible is generated.

\subsubsection{Parallel velocity}
\label{sec:u_II}

The probability distribution of velocities parallel to the incident direction
of the photon is given by \eq{fupartheo}:
\begin{equation}
\label{eq:fupar}
f(u_{||}) = \frac{a}{\pi H(a,x)} \frac{e^{-u_{||}^2}} {(x-u_{||})^2 + a^2}
\end{equation}
To pick a random parallel velocity $u_{||}$, in principle all we have to do is
to integrate \eq{fupar}
from $-\infty$ to $u_{||}$ until we reach some univariate $\mathcal{R}$. The
problem is that \eq{fupar} is not
analytically integrable and that agaian numerical integration is not practical.

The solution to this problem is based on the \emph{rejection method}
\index{Rejection method}\citep{pre92}: choosing a random number
with a given distribution can be interpreted geometrically as choosing a random
point $P(x,y)$ in two dimensions uniformly distributed in the area under the
function graphing the distribution.
Since the function is not integrable, we do not have a direct way of
choosing $P$. However, we can draw a function $g(u_{||})$ that
lies everywhere \emph{above} the original function $f(u_{||})$, and
that \emph{is} integrable. We will call this function the \emph{comparison
function}. If a random point chosen uniformly under this function
lies  under $f(u_{||})$ as well, we will accept the
corresponding value of $u_{||}$; if not, we will reject it and draw a
new $P$ (in fact, this is similar to the method we used in three dimensions to
generate the inital direction vector, \eq{n_i}).

To determine whether $P$ lies within the original probability
distribution, a uniform deviate between 0 and $g(u_{||})$ is drawn as
the $y$-value of $P$. If this number lies also between 0 and
$f(u_{||})$, we keep the $x$-value of $P$ as $u_{||}$.
Alternatively, we can pick the second random number between
zero and one, and then accept or reject $u_{||}$ according to whether
it is respectively less or greater than the ratio
$f(u_{||})/g(u_{||})$.

For a given frequency and temperature, the factor $a/\pi H(a,x)$ in
\eq{fupar} is constant. Thus, we want to find a random value of
$u_{||}$ with the distribution
\begin{equation}
\label{eq:fu}
f(u_{||}) \propto \frac{e^{-u_{||}^2}}{(x-u_{||})^2 + a^2},
\end{equation}
where $a$ and $x$ are given. The distribution being 
invariant with respect to the transformation
$(x,u_{||}) \to (-x,-u_{||})$, we can operate with a positive $x$, and
then multiply the result with the sign of the original $x$.

The comparison function can be chosen as
\begin{equation}
\label{eq:gu0}
g(u_{||}) \propto \frac{1}{(x-u_{||})^2 + a^2},
\end{equation}
since this is integrable, invertible, and
everywhere larger than $f(u_{||})$ (except at $u_{||} = 0$, where
$g=f$). A first univariate $\mathcal{R}_1$ then gives us a value of 
$u_{||}$, which is accepted if a second univariate $\mathcal{R}_2$ is
smaller than $f/g = e^{-u_{||}^2}$.

Unfortunately, when $x \gtrsim 2$ the fraction of rejected velocities
becomes inexpediently large. To increase the fraction of acceptance,
\citet{zhe02} make use of \emph{two} comparison
functions, applying to two domains separated by some parameter $u_0$:
\begin{equation}
  \label{eq:g}
  g(u_{||}) \propto \left\{ \begin{array}{llll}
                    g_1(u_{||}) & = &
        \frac{1}{(x-u_{||})^2 + a^2} & \textrm{for } u_{||} \le u_0\vspace{2mm}\\
                    g_2(u_{||}) & = &
     \frac{e^{-u_0^2}}{(x-u_{||})^2 + a^2} & \textrm{for } u_{||}  >  u_0
                     \end{array} \right.,
\end{equation}
in which case the acceptance fractions are $e^{-u_{||}^2}$ and
$e^{-u_{||}^2}/e^{-u_0^2}$ for $g_1$ and $g_2$, respectively.

\begin{figure}[!t]
\centering
\includegraphics [width=0.70\textwidth] {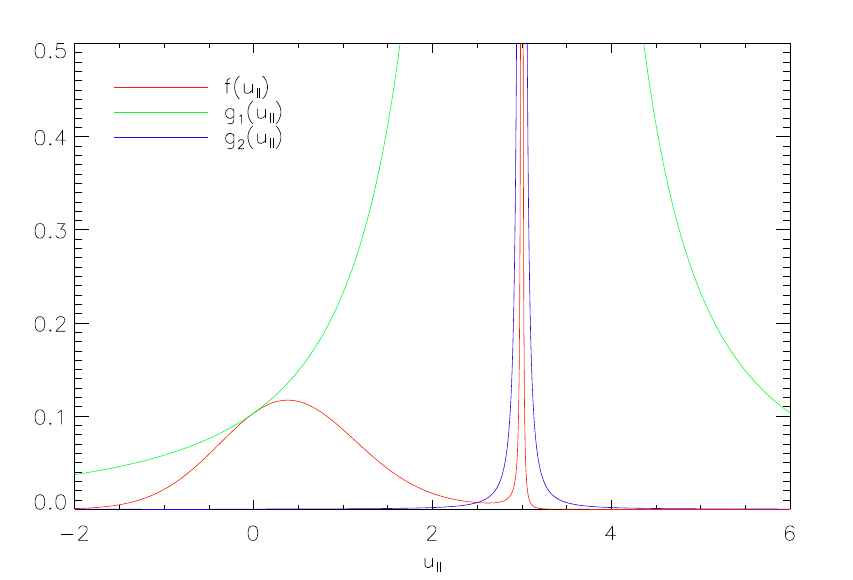}
\caption{{\small Probability distribution $f(u_{||})$ of parallel velocities
                 $u_{||}$ (\emph{red}) for $x=3$. Also shown are the two
                 comparison functions $g_1(u_{||})$ (\emph{green}) and
                 $g_2(u_{||})$ (\emph{blue}), with a separating
                 parameter of $u_0 = 2.75$. A random point $P(x,y)$ under $g$ 
                 is determined, and the corresponding value of $u_{||}$
                 accepted if $P$ lies under $f$ as well.}}
\label{fig:gu}
\end{figure}
Defining a number $p$ as the relative area under the first part of the
comparison function
\begin{eqnarray}
  \label{eq:p}
\nonumber
  p & = & \frac{\int_{-\infty}^{u_0} g(u_{||}) \, du_{||}}
               {\int_{-\infty}^{+\infty} g(u_{||}) \, du_{||}}\\
\nonumber
    & = & \frac{\int_{-\infty}^{u_0} g(u_{||}) \, du_{||}}
               {\int_{-\infty}^{u_0} g_1(u_{||}) \, du_{||} +
                \int_{u_0}^{+\infty} g_2(u_{||}) \, du_{||}}\\
    & = & \frac{\theta_0 + \pi/2}
               {(1 - e^{-u_0^2})\theta_0 + (1 + e^{-u_0^2})\pi/2},
\end{eqnarray}
where
\begin{equation}
  \label{eq:theta0}
  \theta_0 = \tan^{-1} \left( \frac{u_0 - x}{a}  \right),
\end{equation}
we determine which domain to use by comparing $p$ with a univariate
$\mathcal{R}_1$.

Then $u_{||}$ is generated through
\begin{equation}
\label{eq:uII}
  u_{||} = a \tan \theta + x,
\end{equation}
where $\theta$ is a random number uniformly distributed in
$[-\pi/2,\theta_0]$ and $[\theta_0,\pi/2]$ for $\mathcal{R}_1 \le p$ and 
$\mathcal{R}_1 > p$, respectively.

Finally, a second univariate $\mathcal{R}_2$ is compared with
the corresponding fraction of acceptance, thus determining whether the
generated value of $u_{||}$ is accepted or rejected.

The last obstacle in determining $u_{||}$ is how to find an appropriate value of
$u_0$, since the number of accepted values of $u_{||}$ is highly dependent on
$u_0$.
For the wide range of temperatures and frequencies involved we find that a
satisfactory average acceptance-to-rejection ratio of order unity is
achieved for
\begin{equation}
\label{eq:u0}
u_0 = \left\{ \begin{array}{ll}
0                           & \textrm{for }
                              0 \phantom{.2}\le x < 0.2\\
x - 0.01 a^{1/6} e^{1.2x}   & \textrm{for }
                              0.2 \le x < x_{\mathrm{cw}}(a)\\
4.5                         & \textrm{for }
                              \phantom{0.2| \le } x \ge x_{\mathrm{cw}}(a)
\end{array}
\right.
\end{equation}
as the value $u_0$ separating the two comparison functions,
an $x_{\mathrm{cw}}$  defines the boundary between the core and the wings of
the Voigt profile, given by \eq{xcw}.

\subsubsection{Perpendicular velocity}
\label{sec:u12}

Velocities perpendicular to $\mathbf{\hat{n}}_i$ follow a regular Gaussian
which unfortunately is also not integrable. However, \citet{box58} showed that
\emph{two independent} deviates can be generated simultaneously from this
distribution from
two univariates $\mathcal{R}_1$ and $\mathcal{R}_2$ in the following way:
\begin{equation}
\label{eq:u12BM}
\left. \begin{array}{lll}
u_{\perp,1} & = & 
\left(- \ln \mathcal{R}_1 \right)^{1/2} \cos2\pi\mathcal{R}_2\\
u_{\perp,2} & = &
\left(- \ln \mathcal{R}_1 \right)^{1/2} \sin2\pi\mathcal{R}_2.\\
\end{array}
\right.
\end{equation}

In this particular part it is possible to speed up the code tremendously.
The acceleration scheme is described in \sec{acc}.


\subsubsection{Change of basis}
\label{sec:basis}

The axes defining $u_{\perp,1}$ and $u_{\perp,2}$ can be in any directions, as
long as
$\mathbf{\hat{u}}_{\perp,1} \perp \mathbf{\hat{u}}_{\perp,2}
\perp \mathbf{\hat{u}}_{||} \equiv \mathbf{\hat{n}}_i$. To be specific, we
generate $\mathbf{\hat{u}}_{\perp,1}$ by projecting $\mathbf{\hat{n}}_i$ onto
the $xy$-plane, re-normalizing and rotating by $\pi/2$ about the $z$-axis,
resulting in
\begin{equation}
\label{eq:uhat1}
\mathbf{\hat{u}}_{\perp,1} = \frac{1}{(\hat{n}_{i,1}^2+\hat{n}_{i,2}^2)^{1/2}}
\left( 
\begin{array}{c}
- \hat{n}_{i,2} \mathbf{\hat{x}}\\
\phantom{-}  \hat{n}_{i,1} \mathbf{\hat{y}}\\
  0
\end{array}
\right),
\end{equation}
where $\mathbf{\hat{x}}$, $\mathbf{\hat{y}}$ are the unit direction vectors of
the ``lab'' system (the external observer), while $\hat{\mathbf{u}}_{\perp,2}$
is generated through the cross product
\begin{equation}
\label{eq:uhat2}
\mathbf{\hat{u}}_{\perp,2} = 
\mathbf{\hat{n}}_i \times \hat{\mathbf{u}}_{\perp,1}
\end{equation}

With the atom velocities in the three directions given by Eqs.~\ref{eq:uII}
and \ref{eq:u12dij}, the total velocity $\mathbf{u}$ of the atom in the lab
frame is then
\begin{equation}
\label{eq:u}
\boxed{
\mathbf{u} = u_{\perp,1} \mathbf{\hat{u}}_{\perp,1}
           + u_{\perp,2} \mathbf{\hat{u}}_{\perp,2}
           + u_{||}      \mathbf{\hat{n}}_i.
}
\end{equation}
%


\subsection{Re-emission}
\label{sec:reem}

The direction vector $\mathbf{\hat{n}}_f$ of the scattered photon is determined
by the
proper phase function; for core scatterings, according to the discussion in
Sec.~\ref{sec:phase} a first univariate establishes if the intermediate state
is $2P_{1/2}$ (probability: 1/3) or $2P_{3/2}$ (probability: 2/3). In the first
case the photon is scattered isotropically so that, as in the case of emission,
$\mathbf{\hat{n}}_f$ is given by
\eq{n_i}. Otherwise it is scattered according to the distribution
given by \eq{W32}. Wing scatterings follow the distribution given by
\eq{Wwing}. Integrating these functions from $\mu = -1$ until a
univariate $\mathcal{R}$ is reached and inverting yields the only real solution
\begin{equation}
\label{eq:mu32}
\mu = \varphi^{-1/3} - \varphi^{1/3},
\end{equation}
where
\begin{equation}
\label{eq:fmu}
\varphi = \left\{ \begin{array}{ll}
\frac{1}{7} \left( 14 - 24\mathcal{R}
              + (245 - 672\mathcal{R}
                     + 576\mathcal{R}^2)^{1/2} \right) & \textrm{in the core}\\
2 - 4\mathcal{R} + (5 - 16\mathcal{R} + 16\mathcal{R}^2)^{1/2}
                                                       & \textrm{in the wings.}
\end{array}
\right.
\end{equation}
Since the scattering is azimuthally isotropic, we may draw two more variates
$\mathfrak{r}_1,\mathfrak{r}_2 \in [-1,1]$, rejecting them if
$\mathfrak{s} \equiv \mathfrak{r}_1^2 + \mathfrak{r}_2^2 > 1$, and
normalize them to
\begin{eqnarray}
\label{eq:rnorm}
\mathfrak{r}_1'& = & \left( \frac{1-\mu^2}{\mathfrak{s}} \right)^{1/2}
                     \mathfrak{r}_1\\
\mathfrak{r}_2'& = & \left( \frac{1-\mu^2}{\mathfrak{s}} \right)^{1/2}
                     \mathfrak{r}_2.
\end{eqnarray}
The direction of the outgoing photon can then be represented by
\begin{equation}
\label{eq:nhatf}
\boxed{
\mathbf{\hat{n}}_f =
\left( 
\begin{array}{c}
\mathfrak{r}_1' \mathbf{\hat{u}}_{\perp,1}\\
\mathfrak{r}_2' \mathbf{\hat{u}}_{\perp,2}\\
\mu \mathbf{\hat{n}}_i
\end{array}
\right).
}
\end{equation}

\subsection{Interaction with dust}
\label{sec:interdust}

If $\mathcal{R} \le \varrho$, the interaction
is caused by dust. In this case a second univariate is compared to the
albedo of the dust grain, dictating whether the photon is absorbed, thus
terminating the journey of this particular photon, or scattered, in which case
it is re-emitted in a random direction given by the \citeauthor{hen41} phase
function (\eq{Phg}):
\begin{equation}
\label{eq:Phg2}
P_{\mathrm{HG}}(\mu) = \frac{1}{2} \frac{1 - g^2}{(1 + g^2 - 2g\mu)^{3/2}}.
\end{equation}

To generate a new direction from this distribution, again the rejection method
described above is applied.



\section{Simulating observations}
\label{sec:obs}

Following the scheme described in Secs.~\ref{sec:em} through
\ref{sec:codescat},
the photon is trailed as it scatters in real and frequency space, until 
eventually it escapes the computational box. Subsequently, this procedure is
repeated for the remaining $n_{\mathrm{ph}} - 1$ photons. Each time a photon
leaves the box, we can sample its frequency and its point of last scattering,
in this way yielding the spectrum and a three dimensional image of the
extension of the Ly$\alpha$ radiation.

\subsection{Surface brightness maps and spectra}
\label{sec:SBsp}

However, since in general the morphology of a galaxy may
very well cause an anisotropic luminosity, it is more interesting to see how
the system would appear when observed from a given angle.
Because the number of photons escaping in a
particular direction is effectively zero, following \citet{yus84} we calculate
instead \emph{for each scattering} and for each photon the probability of
escaping the medium in the direction of the
observer, or, in fact, six different observers situated in the positive and
negative directions of the three principal axes, as
\begin{equation}
\label{eq:w}
w = W(\mu) e^{-\tau_{\mathrm{esc}}},
\end{equation}
where $W(\mu)$ is given by the proper phase function (Eqs.~\ref{eq:W12},
\ref{eq:W32} or \ref{eq:Wwing}), $\mu$ is given by the angle between
$\mathbf{\hat{n}}_i$ and the direction of the observer, and
$\tau_{\mathrm{esc}}$ is the optical
depth of the gas and dust lying between the scattering event and the edge of
the computational box (integrated through the intervening cells).

This probability is added as a
weight to an array of three dimensions; two spatial and one spectral. The two
spatial dimensions can be thought of as a CCD, consisting of
$N_{\mathrm{pix}}^2$ pixels each suspending a solid angle
$\Omega_{\mathrm{pix}}$
of the computational box. Behind each pixel element is a one-dimensional
spectral array with $N_{\mathrm{res}}$ bins. 

Since flux leaving the source diminishes as $1/d_L^2$, where
$d_L$ is the luminosity distance, the total surface
brightness
SB$_{\mathrm{pix}}$ of the area covered by the pixel, measured in energy per
unit time, per unit area \emph{at the location of the observer},
per unit solid angle suspended by the pixel is then
\begin{equation}
\label{eq:SBpix}
\boxed{
\textrm{SB}_{\mathrm{pix}} = \frac{L_{\mathrm{tot}}}
                                  {d_L^2 \Omega_{\mathrm{pix}}}
                             \frac{1}{n_{\mathrm{ph}}}
   \sum_{\mathrm{ph.,scat.}} W(\mu) e^{-\tau_{\mathrm{esc}}},
}
\end{equation}
where the sum is over all photons and all scatterings. Note that
\eq{SBpix} does not contain a factor $1/4\pi$, due to the fact that
the phase functions are normalized to unity.

In a flat Universe, the luminosity distance of an object at redshift $z$ is
given by \citep[e.g.,][]{wei72}
\begin{equation}
\label{eq:dL}
d_L = \frac{c}{H_0} (1+z) \int_0^z \frac{dz'}{H(z')/H_0}
\end{equation}
(which must be integrated numerically), where $H_0$ is the present Hubble
parameter, and
\begin{equation}
\label{eq:Hz}
\frac{H(z)}{H_0} = \Big[ \Omega_m (1+z)^3
                        + \Omega_k (1+z)^2
                        + \Omega_\Lambda \Big]^{1/2},
\end{equation}
with $\Omega_m$ and $\Omega_\Lambda$ the (dark and baryonic) matter and dark
energy density parameter, respectively. For a non-zero curvature parameter
$\Omega_k \equiv 1 - \Omega_m - \Omega_\Lambda$, the integral in
\eq{dL} must be replaced by
\begin{eqnarray}
\label{eq:Ok}
\frac{1}{\sqrt{\phantom{|}\Omega_k\phantom{|}}} \sinh\left(
         \sqrt{\Omega_k} \int_0^z \frac{dz'}{H(z')/H_0} \right)
         & \textrm{for } & \Omega_k > 0\\
\frac{1}{\sqrt{|\Omega_k|}} \sin\left(
         \sqrt{|\Omega_k|} \int_0^z \frac{dz'}{H(z')/H_0} \right)
         & \textrm{for } & \Omega_k < 0.
\end{eqnarray}

\Eq{SBpix} is the SB that an observer would measure
at a distance $d_L$ from the galaxy. Hence, this is the interesting quantity
for comparing with actual observations. Theorists tend to be more concerned
with the intrinsic SB, i.e.~the flux measured by a hypothetical observer at the
location of the source. The conversion is given by
\begin{equation}
\label{eq:SBtrad}
\textrm{SB}_{\mathrm{there}} = \frac{\Omega d_L^2}{A}
                                   \textrm{SB}_{\mathrm{here}},
\end{equation}
where $A$ is the area suspended by the solid angle $\Omega$. The ratio between
the two defines the angular diameter distance
\begin{equation}
\label{eq:dA}
d_A^2 = \frac{A}{\Omega}.
\end{equation}
Since the SB of a receding object decreases as $(1+z)^4$ and the angular area
as $(1+z)^2$, $d_A$ is related to $d_L$ by
\begin{equation}
\label{eq:dLdA}
d_L = (1+z)^2 d_A.
\end{equation}
From Eqs.~\ref{eq:dA} and \ref{eq:dLdA}, \eq{SBtrad} can be
written
\begin{equation}
\label{eq:SBtrad2}
\textrm{SB}_{\mathrm{there}} = (1+z)^4 \textrm{SB}_{\mathrm{here}},
\end{equation}
or, if $\Omega$ is measured in arcsec$^2$ rather than steradians
\begin{equation}
\label{eq:SBtas}
\boxed{
\textrm{SB}_{\mathrm{there}} = 206265^2 (1+z)^4 \textrm{SB}_{\mathrm{here}}.
}
\end{equation}

Finally, the 3D array can be collapsed along the frequential direction to give
a ``bolometric'' SB map (i.e.~all wavelengths in the vicinity of the Ly$\alpha$
line), along the two spatial directions to give the integrated
spectrum, or along all directions to give the total flux received from the
source. Since in fact a full spectrum is obtained for each pixel, it is also
possible to simulate long-slit spectroscopy, giving frequency as a function of
position of a selected part of the image.



\section{Acceleration schemes}
\label{sec:acc}

\subsection{Core-skipping scheme}
\label{sec:coreskip}

In very dense regions of the gas, as
long as the photon is in the core, the optical depth is so enormous that each
scattering is accompanied by a negligible spatial shift. Only when by chance
the photon encounters an atom with a large perpendicular velocity will it be
scattered out of the core, i.e.~beyond $|x| = x_{\mathrm{cw}} \sim 3$, and be
able to make a long journey. Since the probability $P$ of this happening is
\begin{eqnarray}
\label{eq:Pwing}
P & \simeq & \frac{2}{\sqrt{\pi}} \int_{x_{\mathrm{cw}}}^\infty e^{-x^2}dx\\
  & \sim   & 1 - \textrm{erf}(3)\\
  & \sim   & 10^{-5},
\end{eqnarray}
and since complete redistribution is a fair approximation in the core, it
will take of the order of $10^5$ scatterings before entering the wing. These
scatterings are insignificant in the sense that they do not contribute to any
important displacement in neither space nor frequency. Hence, we may as well
skip them altogether and go directly to the first scattering that pushes the
photon into the wing. This can be achieved by drawing $u_{\perp,1}$,
$u_{\perp,2}$ from a \emph{truncated} Gaussian that favors atoms with high
velocities \citep{ave68}. The total perpendicular velocity
$u_\perp = (u_{\perp,1}^2 + u_{\perp,2}^2)^{1/2}$ follows a two-dimensional
Maxwellian $\mathcal{M}(u_\perp)$ where the direction of $\mathbf{u}_\perp$
is isotropically distributed in the plane. Replacing for \eq{u12BM}
\citep{dij06a},
\begin{equation}
\label{eq:u12dij}
\boxed{
\left. \begin{array}{lll}
u_{\perp,1} & = & \left( x_{\mathrm{crit}}^2 - \ln \mathcal{R}_1 \right)^{1/2}
                  \cos2\pi\mathcal{R}_2\\
u_{\perp,2} & = & \left( x_{\mathrm{crit}}^2 - \ln \mathcal{R}_1 \right)^{1/2}
                  \sin2\pi\mathcal{R}_2,\\
\end{array}
\right.
}
\end{equation}
corresponds to drawing $u_\perp$ from the distribution
\begin{equation}
\label{eq:M}
\mathcal{M}(u_{\perp}) = \left\{ \begin{array}{ll}
0 & \textrm{for } |u_{\perp}| \le x_{\mathrm{crit}}\\
2 u_\perp e^{-(u_\perp^2 - x_{\mathrm{crit}}^2)}
  & \textrm{for } |u_{\perp}|  >  x_{\mathrm{crit}}
\end{array}
\right.
\end{equation}
where $x_{\mathrm{crit}}$
is the critical value of $x$ within which scatterings can be neglected.

The value $x_{\mathrm{crit}}$ is not simply equal to $x_{\mathrm{cw}}$, since
for a non-dense medium, a core scattering can in fact be associated with a
considerable spatial journey, while in extremely thick clouds even scatterings
in the inner part of the wing may be neglected. Moreover, the exact value of
$x_{\mathrm{crit}}$ is actually quite important; a high value can decrease the
computational execution time by \emph{several orders of magnitude}, while a
too high value will push the photons unnaturally far out in the wings,
skewing the result. From the Neufeld solution we know that the important
parameter is the product $a\tau_0$. Correspondingly, we expect
$x_{\mathrm{crit}}$ to be a function of the value of $a\tau_0$ in the current
cell. Indeed, from numerous tests it is found that we can use
\begin{equation}
\label{eq:xcrit}
\boxed{
x_{\mathrm{crit}} = \left\{ \begin{array}{ll}
0                               & \textrm{for } a\tau_0 \le 1\\
0.02 e^{\xi \ln^\chi\! a\tau_0} & \textrm{for } a\tau_0 > 1,
\end{array}
\right.
}
\end{equation}
where $(\xi,\chi) = (0.6,1.2)$ or $(1.4,0.6)$ for $a\tau_0 \le 60$ or
$a\tau_0 > 60$,
respectively, without affecting the final result.
Of course, if the photon is already in the wing, the proper Gaussian is used.

\subsubsection{Dust absorption in the core}
\label{sec:dustcore}

With a dusty medium, we must investigate the possibility that the photon
would have been destroyed, had we \emph{not} used this acceleration scheme,
i.e.~the probability $P_{\mathrm{abs}}(x_{\mathrm{crit}})$ of absorption for a
photon initially in the core, before escaping the frequency interval
$[-x_{\mathrm{crit}},x_{\mathrm{crit}}]$.
Ultimately, we will determine this probability numerically, but to interpret
the result, we will first investigate the scenario analytically.
In the following calculation, factors of order unity will be omitted.

The probability per interaction that a photon with frequency $x$ be absorbed is
\begin{equation}
\label{eq:Pabsi}
p_{\mathrm{abs}}(x) = \frac{\tau_{\mathrm{a}}}
                           {\tau_{\mathrm{d}} + \tau_x}
                 \sim \frac{1}{1 + \phi(x)\tau_0/\tau_{\mathrm{a}}},
\end{equation}
since $\tau_{\mathrm{a}} \sim \tau_{\mathrm{d}}$ and
$\tau_x \sim \phi(x)\tau_0$.
Here, $\tau_{\mathrm{a,d,}x\mathrm{,0}}$ corresponds to the optical depth of
this particular part of the journey.
The number $dN(x)$ of scatterings taking place when the frequency of the
photon is close to $x$ is
\begin{equation}
\label{eq:dN}
dN(x) = N_{\mathrm{tot}} \phi(x) dx,
\end{equation}
where $N_{\mathrm{tot}}$ is the total number of scatterings before the
photon exits $[-x_{\mathrm{crit}},x_{\mathrm{crit}}]$,
i.e.~the total number of scatterings
skipped. Here we have assumed \emph{complete redistribution} of the frequency,
i.e.~there is no correlation between the frequency of the photon before and
after the scattering event. This is a fair approximation in the core
\citep{unn52b,jef60}. However, as discussed in \sec{pere}, once the photon
is in the wing it has a tendency to stay there, only slowly drifting toward
the line center with a mean shift per scattering
$\langle \Delta x \rangle = -1/|x|$.

For the purpose of the current calculation, the Voigt profile is
approximated by a Gaussian in the core and a power law in the wing, such that
\begin{equation}
\label{eq:phiapp}
\phi(x) \sim \left\{ \begin{array}{ll}
e^{-x^2} & \textrm{ for } x < x_{\mathrm{cw}}\\
\frac{a}{x^2}           & \textrm{ for } x \ge x_{\mathrm{cw}},
\end{array}
\right.
\end{equation}
where $x_{\mathrm{cw}}$ marks the value of $x$ at the transition from core to
wing (\eq{xcw}).

At each scattering, the probability of escaping the region confined by
$x_{\mathrm{crit}}$ is
\begin{eqnarray}
\label{eq:Pesc}
\nonumber
p_{\mathrm{esc}}(x_{\mathrm{crit}}) & = & 2 \int_{x_{\mathrm{crit}}}^\infty
                                          \phi(x) dx\\
& \sim & \left\{ \begin{array}{ll}
    \mathrm{erfc}\, x_{\mathrm{crit}} &
    \textrm{\, \, \, for } x_{\mathrm{crit}} < x_{\mathrm{cw}}\\
\frac{a}{x_{\mathrm{crit}}}   &
    \textrm{\, \, \, for } x_{\mathrm{crit}} \ge x_{\mathrm{cw}},
\end{array}
\right.
\end{eqnarray}
where erfc is the complimentary error function.

Using \eq{dN}, the total probability of being absorbed can be
calculated as
\begin{equation}
\label{eq:Pabst}
P_{\mathrm{abs}}(x_{\mathrm{crit}}) = \int_0^{x_{\mathrm{crit}}}
                                       p_{\mathrm{abs}}(x) dN(x).
                   = N_{\mathrm{tot}} \int_0^{x_{\mathrm{crit}}}
                                       p_{\mathrm{abs}}(x) \phi(x) dx.
\end{equation}
The total number of scatterings before escape if the photon is not absorbed
is $N_{\mathrm{tot}} \sim 1/p_{\mathrm{esc}}$.
For $x_{\mathrm{crit}} < x_{\mathrm{cw}}$, from Eqs.~\ref{eq:Pabsi} and
\ref{eq:Pesc}, \eq{Pabst} then evaluates to
\begin{equation}
\label{eq:Pabs1}
P_{\mathrm{abs}}(x_{\mathrm{crit}}) \sim \frac{1}
                                          {{\mathrm{erfc}} x_{\mathrm{crit}}}
   \int_0^{x_{\mathrm{crit}}}
   \frac{dx}{e^{x^2} + \tau_0/\tau_{\mathrm{a}}}.
\end{equation}
The exponential integral and the factor $1/\textrm{erfc}\, x_{\mathrm{crit}}$
are of the same order, but since the factor
$\tau_0/\tau_{\mathrm{a}} \sim \sigma_0/\sigma_{\mathrm{a}}$ is of the order
$10^8$, \eq{Pabs1} will usually be negligible.

In the case of $x_{\mathrm{crit}} \ge x_{\mathrm{cw}}$,
\begin{equation}
\label{eq:Pabs2}
P_{\mathrm{abs}}(x_{\mathrm{crit}}) \sim
     \int_0^{x_{\mathrm{crit}}} \frac{dx}
     {a/\phi(x) + a\tau_0/\tau_{\mathrm{a}}}.
\end{equation}
This integral can be evaluated separately for the intervals
$[0,x_{\mathrm{cw}}[$ and $[x_{\mathrm{cw}},x_{\mathrm{crit}}]$. For the first,
the result is usually negligible, as in the case with \eq{Pabs1}.
The second integral yields
\begin{equation}
\label{eq:Pabs22}
P_{\mathrm{abs}}(x_{\mathrm{crit}}) \sim \frac{1}{\mathfrak{t}}
(\tan^{-1}\frac{x_{\mathrm{crit}}}{\mathfrak{t}} -
 \tan^{-1}\frac{x_{\mathrm{cw}}}  {\mathfrak{t}}),
\end{equation}
where
\begin{equation}
\label{eq:ainacc}
\mathfrak{t} \equiv (a\tau_0/\tau_{\mathrm{a}})^{1/2}.
\end{equation}

\begin{figure}[!t]
\centering
\includegraphics [width=0.70\textwidth] {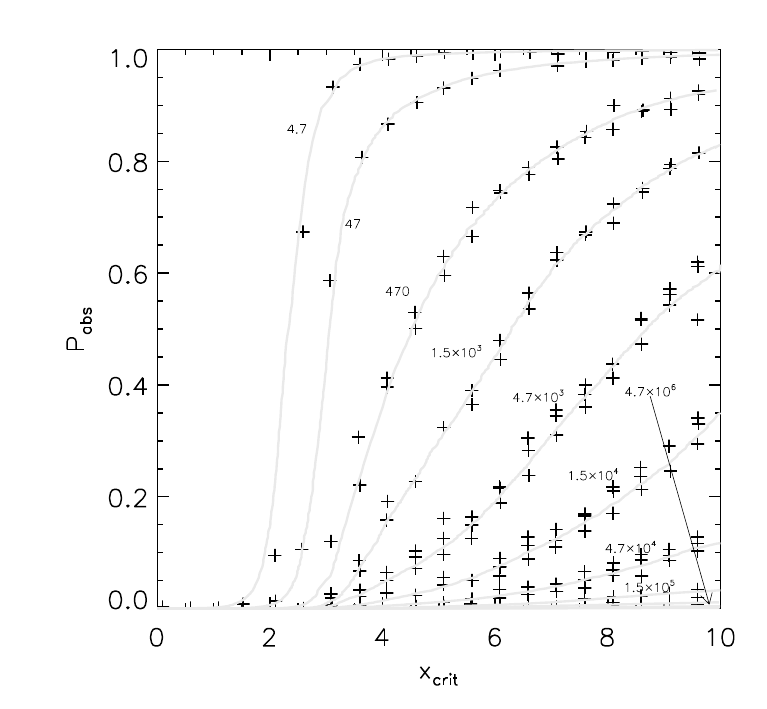}
\caption{{\cap Probability of absorption $P_{\mathrm{abs}}$ before escaping the
             region of the line confined by the value $x_{\mathrm{crit}}$, for a
             series of different values of $a\tau_0/\tau_{\mathrm{a}}$ (labeled
             at the corresponding lines, and obtained with various combinations
             of $a$, $\tau_0$, and $\tau_{\mathrm{a}}$).}}
\label{fig:Pabs}
\end{figure}

For $x_{\mathrm{crit}} \ge x_{\mathrm{cw}}$, the assumption of complete
redistribution becomes very inaccurate, as the photon spends considerably
more time in the wings, with a larger probability of being destroyed. However,
Eqs.~\ref{eq:Pabs22} and \ref{eq:ainacc} reveal a signature of the behavior of
$P_{\mathrm{abs}}(x_{\mathrm{crit}})$, namely that it has a ``$\tan^{-1}$-ish''
shape, and that it scales not with the
individual parameters $a$, $\tau_0$, and $\tau_{\mathrm{a}}$, but with their
interrelationship as given by the parameter $\mathfrak{t}$. To know exactly the
probability of absorption,
a series of Monte Carlo simulations were carried out
for a grid of different temperatures, gas densities, and dust densities. The
results, which are stored as a look-up table, are shown in Fig.~\ref{fig:Pabs}.
Indeed, the same fit applies approximately to different $T$,
$\nhi$, and $n_{\mathrm{d}}$
giving equal values of $\mathfrak{t}$.

Whenever the acceleration scheme is applied, a bilinear interpolation over
$\log\mathfrak{t}$ and $x_{\mathrm{crit}}$ determines the appropriate value
of $P_{\mathrm{abs}}$. A univariate is then
drawn and compared to $P_{\mathrm{abs}}$, thus
determining if the photon is absorbed or allowed to continue its journey.
Note that under most physically realistic conditions, only low values of
$P_{\mathrm{abs}}$ are actually met. However, when invoking the acceleration
scheme many times, the probability of absorption may become significant.



\subsection{Semi-analytical scheme}
\label{sec:semianal}

Most of the computing time is spent in the very dense cells. Since each
cell is in fact a cube of homogeneous and isothermal gas with zero velocity
gradient (henceforward a ``uniform cube''),
if an analytical Neufeld-equivalent solution for the distribution of
frequencies
exists, it would be possible to skip a great number of scatterings and thus
speed up the code further.

The slab solution is an alternate series which can be written in closed form.
Unfortunately, this is not feasible for the cube solution, but under certain
approximations, \citet{tas06b} found that it is still possible to write it as
an alternate series. The problem is that, whereas for the slab the terms
quickly die off, the same is not true for the cube. In fact she found that to
achieve an accuracy better than 3\%, one must exceed 30 terms.

Hence, it seems more convenient to seek a ``Neufeld-based'' approximation.
Since for the cube, the radiation can escape from six faces rather than just
two, we may expect the emergent radiation to be described by a function similar
to the slab solution, but using a lower value of $a\tau_0$.

\subsubsection{Emergent Spectrum}
\label{sec:x_cube}

\begin{figure}[!t]
\centering
\includegraphics [width=0.80\textwidth] {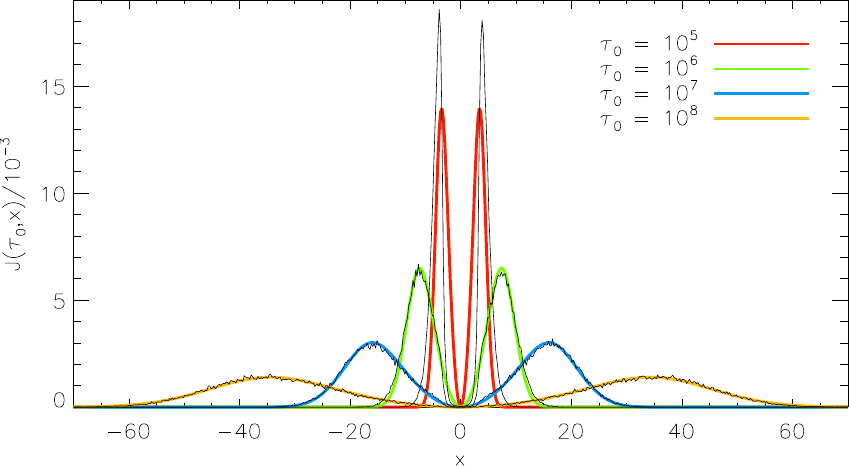}
\caption{{\cap Emergent spectra of a uniform cube of different optical
         depths. A damping parameter of $a = 0.00047$ has been used.
         Neufeld profiles are fitted to the spectra using
         $\eta a\tau_0$, with $\eta = 0.71$ for all $\tau_0$.}}
\label{fig:x_cube}
\end{figure}

Toward these ends, a series of simulations is run in which photons are emitted
isotropically from the centers of cubes of constant --- but different ---
temperature and density,
and zero bulk velocity. The distance from the center to each face is $z_0$.
We will investigate optical depths $\tau_0 = 10^5$, $10^6$, $10^7$, and $10^8$
(measured along the shortest path from center to face). In
all simulations, $n_{\mathrm{ph}} = 10^5$, and different temperatures are
tested. A Neufeld profile is then fitted to the emergent spectrum, using
$\eta a \tau_0$
as the independent variable, where $\eta$ is the parameter to be determined.
A priori, we have no reason to believe that the same value of $\eta$, if any,
should be able to describe all optical depths. However, it is found that, save
for the lowest optical depth ($\tau_0 \sim 10^5$),
excellent fits are obtained using\index{eta@$\eta$}
\begin{equation}
\label{eq:eta}
\boxed{
\eta = 0.71.
}
\end{equation}
This is seen in Fig.~\ref{fig:x_cube}.

\subsubsection{Directionality of the Emergent Photons}
\label{sec:dir_cube}

In realistic,
cosmological simulations, the direction with which the photons exit
the cell is also important. Since in the limit $\tau_0 \to \infty$, any finite
size step \emph{not} perpendicular to the surface will just shift to position
of the photons in the parallel direction, for extremely optically thick slabs,
the photons should have a
tendency to exit perpendicular to the surface. In this case, \citet{phi86}
found that the directionality of the emergent radiation approaches that of
Thomson scattered radiation from electrons, with
intensity
\begin{equation}
\label{eq:ImuI0}
\frac{I(\mu)}{I(0)} = \frac{1}{3} \left( 1 + 2\mu  \right),
\end{equation}
where $\mu = \cos\theta$, with $\theta$ the angle between the outgoing
direction $\hat{\mathbf{n}}_f$ of the photon and the normal to the surface.

Since the number of photons emerging at $\mu$ is $\propto I(\mu)\mu\,d\mu$, the
probability $P(\le\mu)$ of exiting the slab with $\mu \le \mu'$ is
\citep{tas06b}
\begin{eqnarray}
\label{eq:Pltmu}
\nonumber
P(\le\mu') & = &\frac{\int_0^{\mu'} (1+2\mu)\mu\,d\mu}
                    {\int_0^1      (1+2\mu)\mu\,d\mu}\\
          & = & \frac{\mu'^2}{7} \left( 3 + 4\mu' \right).
\end{eqnarray}

\begin{figure}[!t]
\centering
\includegraphics [width=0.90\textwidth] {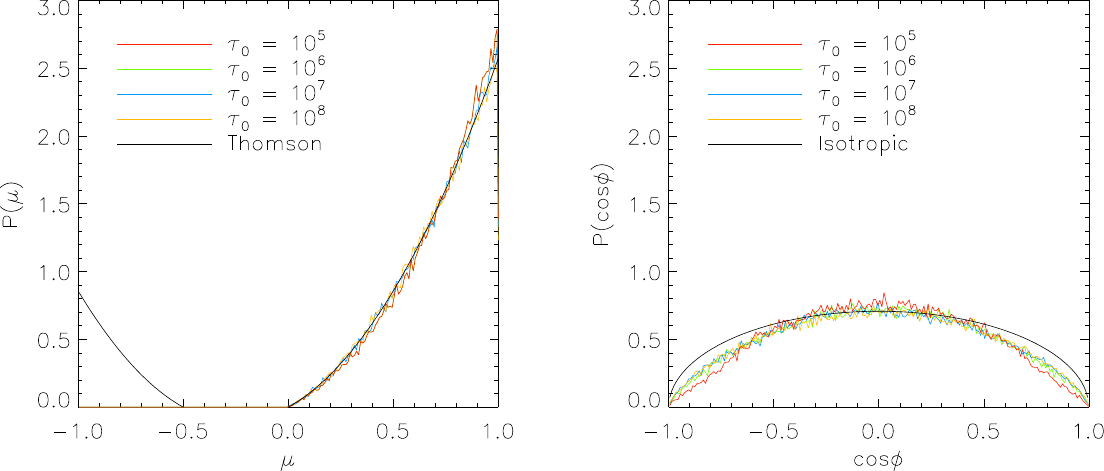}
\caption{{\cap Directionality of the photons emerging from a uniform cube
         for different values of $\tau_0$. In
         the direction perpendicular to the face of the cube (\emph{left}),
         $\hat{\mathbf{n}}_f$ follows the distribution given by
         \eq{Pmucube}, while in the azimuthal direction
         (\emph{right}) there is a slight deviation from isotropy.}}
\label{fig:mu_cube}
\end{figure}

We confirm that this is also an excellent description for a cube
(\fig{mu_cube}).

The probability distribution is found by differentiating \eq{Pltmu}
and recognizing that $\mu$ must be positive for the photon to escape:
\begin{equation}
\label{eq:Pmucube}
P(\mu) =  \left\{ \begin{array}{ll}
           \frac{6}{7} (\mu + 2\mu^2) & \textrm{for 0 $< \mu \le$ 1}\\
           0                          & \textrm{otherwise},
\end{array}
\right.
\end{equation}

Since \eq{Pmucube} is valid for all six faces of the cube, the
azimuthal angle $\phi$ parallel to the face cannot, as in the case of a slab,
be evenly distributed in $[0,2\pi]$ \citep{tas06b}. However, as can be seen
from Fig.~\ref{fig:mu_cube}, the deviation from uniformity is quite small, and
can probably be neglected. Furthermore, it seems less pronounced, the higher
the optical depth.

\subsubsection{Point of Escape}
\label{sec:exitcube}

\begin{figure}[!t]
\centering
\includegraphics [width=0.50\textwidth] {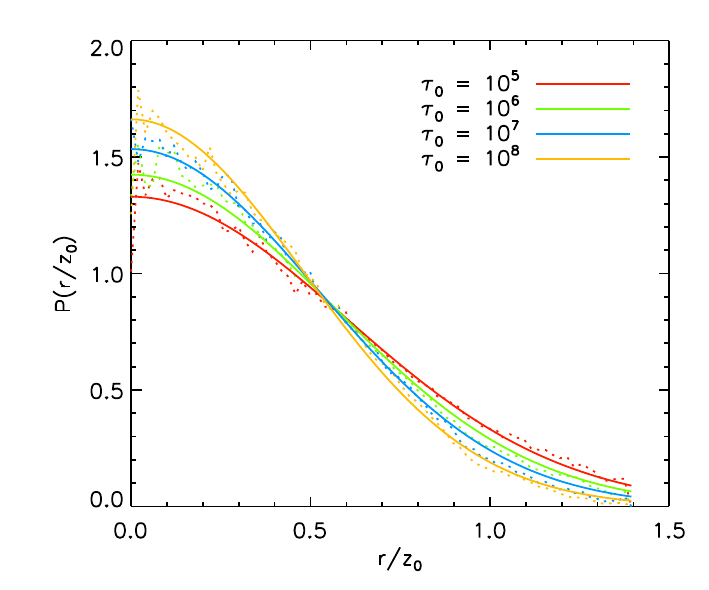}
\caption{{\cap Probability distribution (solid lines) of the exiting point for
         photons emerging from a uniform cube of side length $2 z_0$, as a
         function of distance $r$ from the center of the face,
         normalized to $z_0$ , for different values of $\tau_0$.
         The distributions have been calculated as best fits to the
         corresponding simulated SB
         profiles (dotted lines), as given by \eq{SBcube}.}}
\label{fig:SBcube}
\end{figure}

The final parameter characterizing the photons escaping the cube is the point
$\mathbf{x}_{\mathrm{esc}}$ where it crosses the face. Figure \ref{fig:SBcube}
shows the azimuthally averaged SB profiles of the emergent radiation as a
function of distance from the center of the face, for different optical depths.
It is found that the SB profile is fairly well described by
a truncated Gaussian
\begin{equation}
\label{eq:SBcube}
\textrm{SB}(r/z_0) =  \left\{ \begin{array}{ll}
 \frac{2}{\sqrt{2\pi}\sigma_{\mathrm{SB}}}
 e^{-(r/z_0)^2/2\sigma_{\mathrm{SB}}^2}& \textrm{for $0\le r\le z_0\sqrt{2}$}\\
           0                           & \textrm{for $r > z_0\sqrt{2}$}.
\end{array}
\right.
\end{equation}
%
The dispersion $\sigma_{\mathrm{SB}}$ of the SB decreases very slowly with
optical depth, and can be written as
$\sigma_{\mathrm{SB}} = 0.48 - 0.04 \log \tau_0/10^8$.
However, in the context of a cell-based structure, one might state that it is
meaningless to discuss differences in position on scales smaller than the size
of a cell, and it is found that final results are not altered by simply setting
$\sigma_{\mathrm{SB}} = 0.5$.

\subsubsection{Implementation of the Cube Solution}
\label{sec:impl}

\begin{figure}[!t]
\centering
\includegraphics [width=0.70\textwidth] {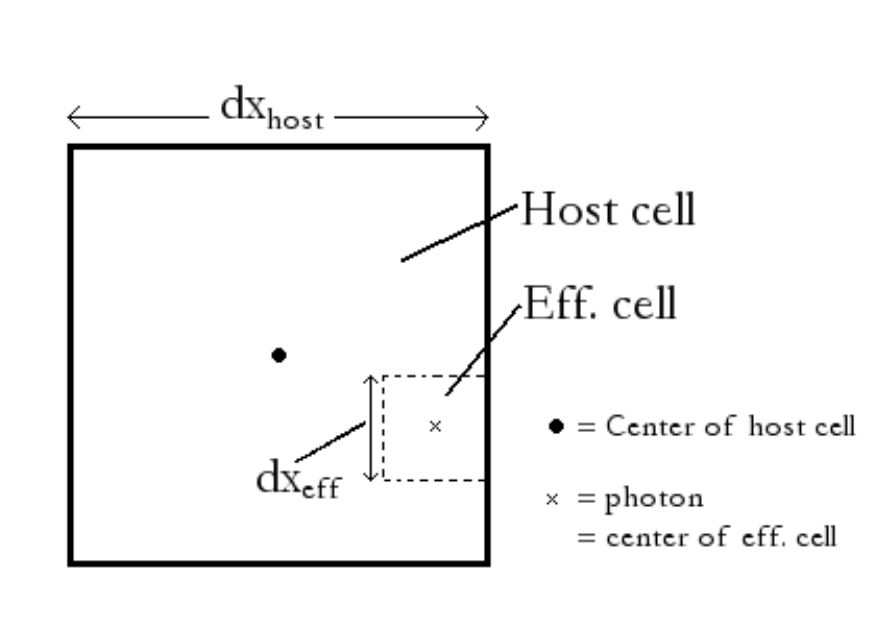}
\caption{{\cap Geometry of the ``effective cell inside cell''-configuration.
               The ``radius'' of the effective cell is
               $z_0 = dx_{\mathrm{eff}}/2$.}}
\label{fig:EffCell}
\end{figure}

With the probability distributions of frequency, direction and position for the
photons escaping the cell, we are now able to accelerate the code further:
every time a photon finds itself in a host cell of $a\tau_0$ higher than some
given threshold, which to be conservative we define as
$a\tau_0 \gtrsim 2\times10^3$, an \emph{effective cell} with the photon in the
center is built, with ``radius'' $z_0$ equal to the distance from the photon to
the nearest face of the host cell (see \fig{EffCell}).
Since the effective cell is always completely circumscribed by the host cell,
its physical parameters are equal to those of its host cell.

If the value of $a\tau_0$ in the effective cell, $(a\tau_0)_{\mathrm{eff}}$, is
below the threshold (i.e.~if the photon is too close to the face
of the host cell), the normal scheme is used. Otherwise, the photon is assigned
a new frequency according to the effective Neufeld distribution:
drawing a univariate
$\mathcal{R}$ and setting this equal to the Neufeld-equivalent cube
solution\footnote{Of course normalized to unity instead of the usual
$1/4\pi.$} integrated from $-\infty$ to $x$ yields (after some algebra)
\begin{eqnarray}
\label{eq:intJcube}
\nonumber
\mathcal{R} & = & \int_{-\infty}^{x_f} J_{\mathrm{cube}}(\tau_0,x)dx\\
            & = & \frac{2}{\pi} \tan^{-1}
                  e^{\sqrt{\pi^3/54}
                     (x_f^3-x_i^3) / 
                     \eta (a\tau_0)_{\mathrm{eff}}}.
\end{eqnarray}
Inverting the above expression, the frequency $x$ of the photon then becomes
\begin{equation}
\label{eq:x_cube}
x_f = \left( \sqrt{\frac{54}{\pi^3}} \eta (a\tau_0)_{\mathrm{eff}}
    \ln \tan \frac{\pi\mathcal{R}}{2}  + x_i^3 \right)^{1/3}.
\end{equation}

Finally, the direction and the position of the photon is determined from
Eqs.~\ref{eq:Pmucube} and \ref{eq:SBcube} (with a probability of escaping
from a given face equal to $1/6$), whereafter it continues its journey.

\subsubsection{Dust absorption in a uniform cube}
\label{sec:dustcube}

With the inclusion of dust, we must calculate the possibility of the photon
being absorbed in such a cube. From the above, we might expect that replacing
$a\tau_0$ by $\eta a\tau_0$ and $\tau_{\mathrm{a}}$ by $\eta\tau_{\mathrm{a}}$
in the slab-relevant equation for the escape fraction (\eq{neufesc})
yields the relevant solution.
In fact, even better fits can be achieved by simultaneously replacing the
square root
by an exponantiation to the power of 0.55. That is, every time the
semi-analytical acceleration scheme is invoked, a univariate is drawn and
compared to the quantity
\begin{equation}
\label{eq:cube}
f_{\mathrm{esc}} = \frac{1}
   {\cosh\left( \zeta'
         \left[ \eta^{4/3}
                (a\tau_0)^{1/3}
                (1-A)\tau_{\mathrm{d}}
         \right]^{0.55}
         \right)},
\end{equation}
determining whether or not the photon should continue its journey.

\begin{figure}[!t]
\centering
\includegraphics [width=0.60\textwidth] {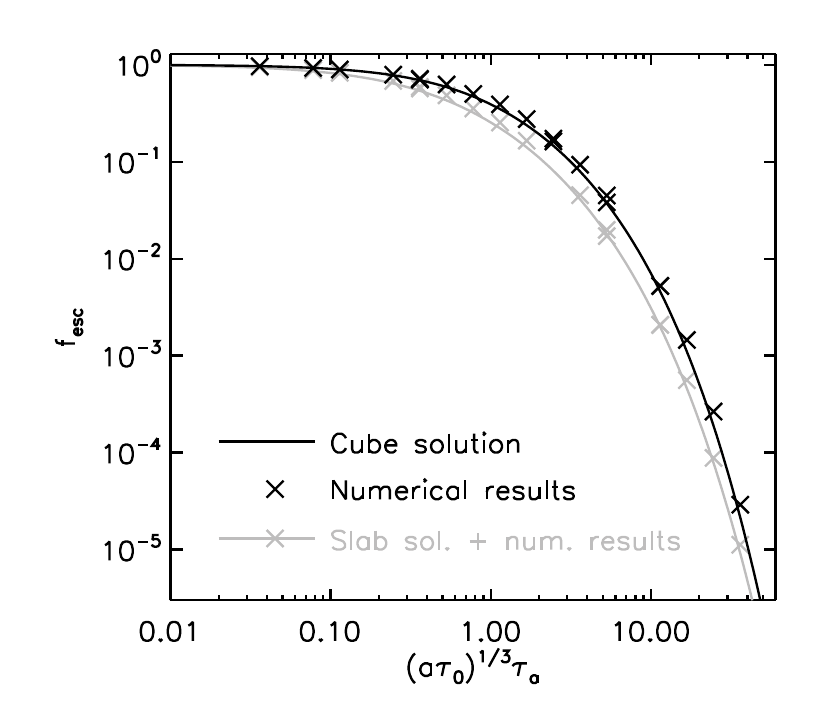}
\caption{{\cap Escape fractions (\emph{black crosses}) of photons emitted from
         the center of a cube of
         damping parameter $a$, line center optical depth $\tau_0$, and dust
         absorption optical depth $\tau_{\mathrm{a}}$, measured from the center
         to the face, compared to the analytical solution (\emph{solid black})
         given by \eq{cube}. For comparison, the equivalent results for the
         slab (\emph{gray}; the same as in \fig{neufesc}) are also displayed.}}
\label{fig:cube}
\end{figure}

Figure \ref{fig:cube} shows the calculated escape fractions from a number of
cubes of different physical properties.

The tests performed in
\sec{analtest} and many of the results of Chapters \ref{cha:conseq} and
\ref{cha:dusteffect} were performed both
with and without this acceleration scheme, all agreeing to a few percent within
statistical errors.



\subsection{Luminosity-boosting scheme}
\label{sec:boost}

Since the vast majority of the photons are emitted within a relatively small
volume of the total computational domain, many photons are needed to reach good
statistics in the outer regions. In order to reach convergence faster, the
probability of emitting photons from low-luminosity cells can be artificially
boosted by some factor $1/w > 1$, later corrected for by letting the emitted
photon only
contribute with a weight $w$ to statistics (spectra, SB profiles, escape
fractions).

This factor is calculated for the $i$'th cell as
\begin{equation}
\label{eq:lumw}
w_i = \left(\frac{L_i}{L_{\mathrm{max}}}\right)^{1/b},
\end{equation}
where $L_i$ is the original luminosity of the cell, $L_{\mathrm{max}}$ is the
luminosity of the most luminous cell (not to be confused with
$\mathcal{L}_{\mathrm{max}}$), and $b$ is a ``boost buffer'' factor that
determines the magnitude of the boost; for $b = 1$, all cells will have an
equal probability of emitting a photon while for $b \to \infty$ the
probability approaches the original probability.

The optimal value of $b$ depends on the quantity and physical region of
interest. Since photons are absorbed primarily in the central regions,
$f_{\mathrm{esc}}$ calculations will usually converge fastest using
$b \to \infty$, and since after all most photons are received from this region
as well, the same counts for the spatially integrated spectrum. If one wishes
to investigate the SB or the spectrum of the outer regions, $b$ should not
simply be set equal to unity, however, since a significant fraction of the
photons received from here are photons originating in the central parts and
later being scattered in the direction of the observer. In this case, faster
convergence can be reached with $b \sim 1.5$ and up to a few tens.




\chapter{Tests}
\label{cha:tests}

\init{B}{efore the developed code} is applied to yield any scientific results,
it is important to test it in several ways.
The analytical solutions derived in \cha{ResScat} serves as a way of testing
the code.
Without loss of continuity, this chapter can be read at a later time. However,
several of the tests are a premise for the semi-analytical acceleration scheme
described in \sec{semianal}.
After the code has been applied, the sensitivity of the obtained results on the
underlying cosmological simulations must also be tested. This is done in
\sec{convtest}. Lastly, in \sec{parstud} the model for the effects of dust is
tested by varying the different parameters in the model.
These two sections are probably better read \emph{after} having read the
chapters on the results.

\section{Testing against analytical solutions}
\label{sec:analtest}

\begin{figure}[!t]
\centering
\includegraphics [width=0.70\textwidth] {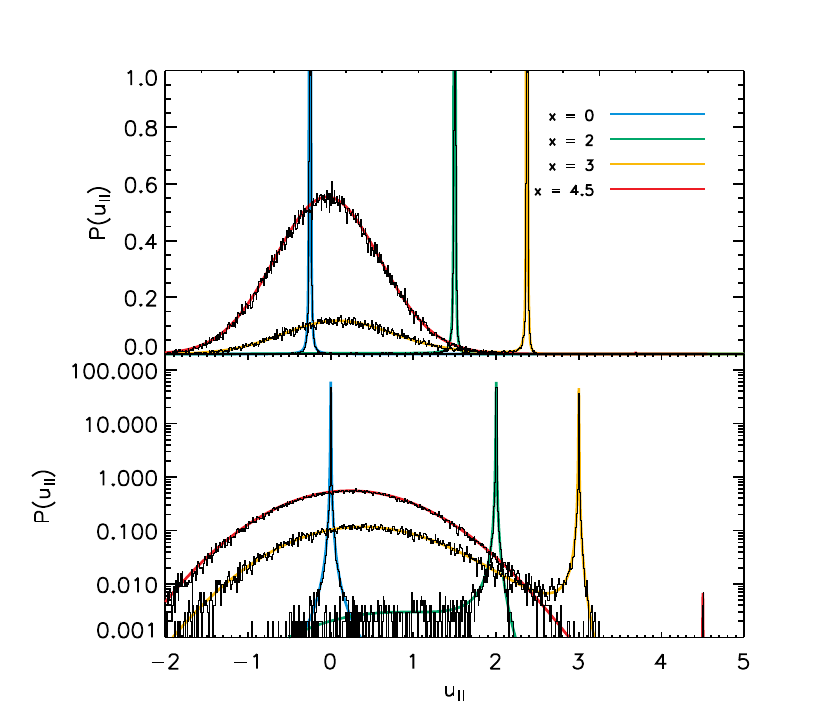}
\caption{{\cap Probability distribution $P(u_{||})$ of parallel velocities
         $u_{||}$ of the scattering atom
         for four values $x$ of the frequency of the incoming photon, as
         generated from \eq{fu}. For photons in the line center
         ($x = 0$, \emph{blue}), $P(u_{||})$ resembles
         the natural line broadening function. For successively larger, but
         relatively low frequencies ($x = 2$, \emph{green}, and $x = 3$,
         \emph{yellow}), the photon still has a
         fair chance of being scattered by an atom to which it appears
         close to resonance. For larger frequencies ($x = 4.5$, \emph{red}),
         however, atoms
         with sufficiently large velocities are so rare that the
         distribution instead resembles a regular Gaussian, slightly shifted
         toward $u_{||} = x$. The method for
         generating $u_{||}$ is quite good at resolving the resonance
         peak. This is particularly visible in the logarithmic plot
         (bottom panel).}}
\label{fig:uII}
\end{figure}

The various probability distribution generators of {\sc MoCaLaTA} were tested
against
their analytical solutions (in the case such solutions exist; otherwise against
numerical integration). Here, only the result for the parallel velocities
$u_{||}$ (Fig.~\ref{fig:uII}) is showed.

\subsection{Individual scatterings}
\label{sec:indiv}

\begin{figure}[!t]
\centering
\includegraphics [width=0.90\textwidth] {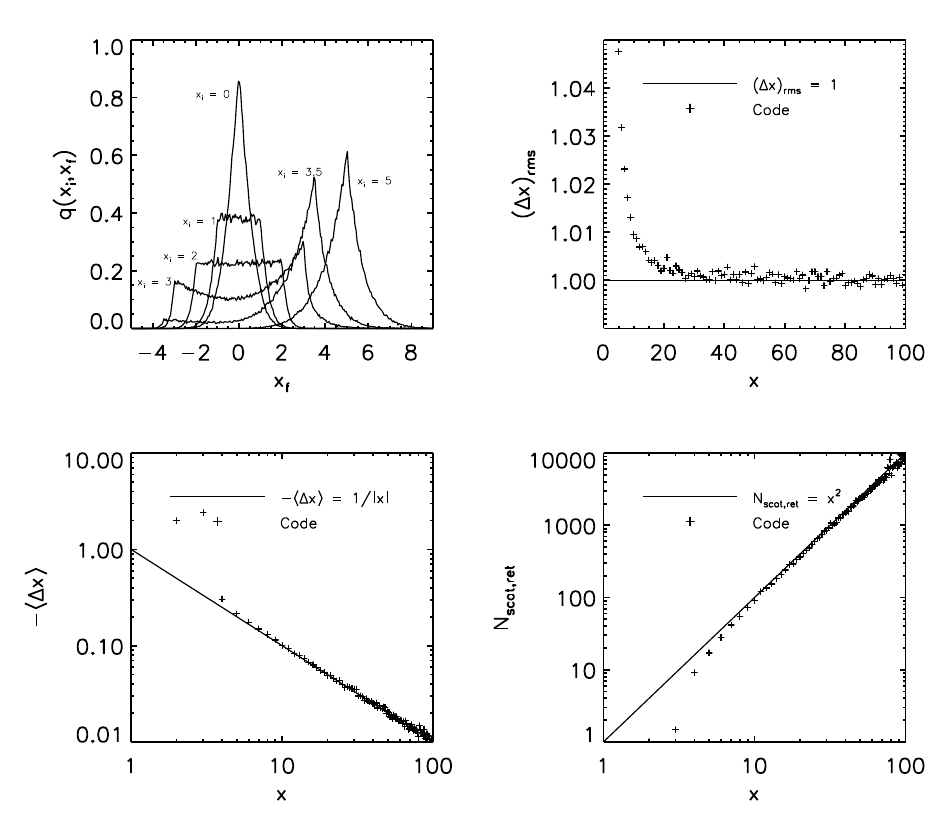}
\caption{{\cap Tests of the relation between the frequency $x_i$ of the incoming
         photon, and the frequency $x_f$ of the outgoing photon
         (\emph{top left}).
         For photons close to the line center, frequencies are distributed more
         or less uniformly over the line profile. For larger $x$, frequencies
         close to the incoming frequency are preferred, but also frequencies
         of opposite sign.
         The distribution follows that predicted by \citet{hum62}.
         For even
         larger frequencies, photons are less likely to be scattered by atoms
         to which they are at resonance, and the outgoing frequency is then
         only a few Doppler widths away from the ingoing. For sufficiently
         large $x$, the rms shift $(\Delta x)_{\mathrm{rms}} \rightarrow 1$
         (\emph{top right}), the mean shift
         $\langle\Delta x \rangle \rightarrow -1/|x|$ (\emph{bottom left}), and
         the average number of scattering needed to return to the core
         $N_{\mathrm{scat,ret.}} \rightarrow x^2$ (\emph{bottom right}), as predicted
         by \citet{ost62}.}}
\label{fig:ost}
\end{figure}

To test the individual scatterings, Fig.~\ref{fig:ost} shows the relation
between the frequency of the incident and of the scattered photon, compared
with the exact redistribution function as formulated by \citet{hum62}.
Furthermore, the rms and mean shift, and the average number of scattering
before returning to the core for wing photons are shown.
For $x \rightarrow \infty$, the values
are seen to converge to the results derived by \citet{ost62}, and given by
Eqs.~\ref{eq:rms} and \ref{eq:meanx}.
%


\subsection{Neufeld solution}
\label{sec:neufeld}

\begin{figure}[!t]
\centering
\includegraphics [width=0.65\textwidth] {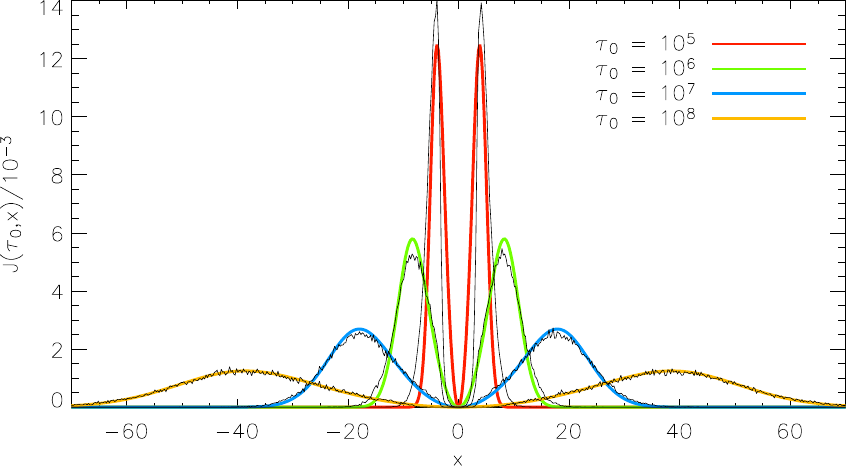}
\caption{{\cap Emergent spectrum of photons injected in the line center in an
         isothermal and homogeneous slab of gas, for different values
         of line center optical depth $\tau_0$ from the center of the slab to
         the surface, compared with the
         corresponding Neufeld solutions. For increasingly optically
         thick media, the photons must diffuse in frequency further and
         further from the line center in order to escape the medium.
         For all simulations, $T = 10^4$ K (corresponding to $a = 0.00047$) and
         $n_{\mathrm{ph}} = 10^5$
         was used. The analytical solution becomes increasingly more
         accurate as $\tau_0 \to \infty$.}}
\label{fig:neufeld}
\end{figure}

\begin{figure}[!t]
\centering
\includegraphics [width=0.60\textwidth] {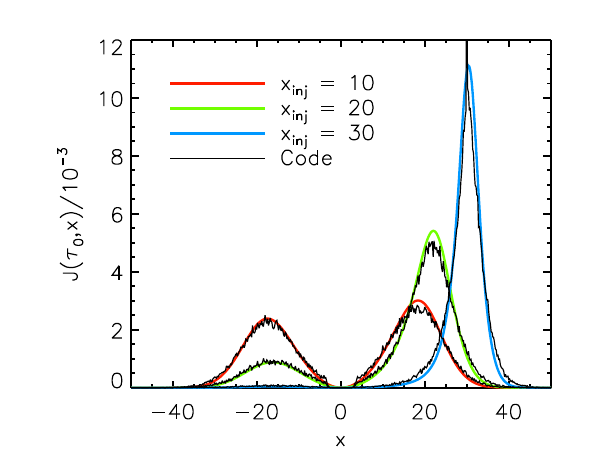}
\caption{{\cap Emergent spectrum of $10^5$ photons injected with different
         initial frequencies $x_{\mathrm{inj}}$ in a slab of line
         center optical depth $\tau_0 = 10^7$ and temperature
         $T = 10^4$ K (corresponding to $a\tau_0 = 4700$), compared with the
         corresponding Neufeld solutions.}}
\label{fig:x_inj}
\end{figure}

The most basic confirmation of the reliability of the code is a test of the
Neufeld solution. Hence, a simulation of a slab (i.e.~with the $x$- and
$y$-dimension set to infinity) is run in which the bulk velocity of the
elements is set to zero, while the temperature and hydrogen density are
constant in such a way as to give the desired line center optical depth
$\tau_0$ from the center of the slab to the surface. Base cells are refined
in arbitrary locations, to an arbitrary level of refinement. Also,
$W(\theta) =$ constant is used and the recoil term in \eq{xf} is
omitted to match the assumptions made by Neufeld. The result for different
values of $\tau_0$ is shown in Fig.~\ref{fig:neufeld}, while the result of
varying the initial frequency is shown in Fig.~\ref{fig:x_inj}.
For the lowest
optical depth ($\tau_0 = 10^5$, corresponding to $a\tau_0 = 47$ at $T = 10^4$
K), the fit is not very accurate. However, this is not an artifact caused by,
say, an insufficient number of photons in the simulation,
 but merely reflects the
fact that the Neufeld solution is no longer valid when the optical depth becomes
too low (at low optical depths, the transfer of photons is no longer dominated
by wing scatterings, where the line profile can be approximated by a power law).


\subsection{Maximum of the emergent spectrum}
\label{sec:max}

\begin{figure}[!t]
\centering
\includegraphics [width=0.60\textwidth] {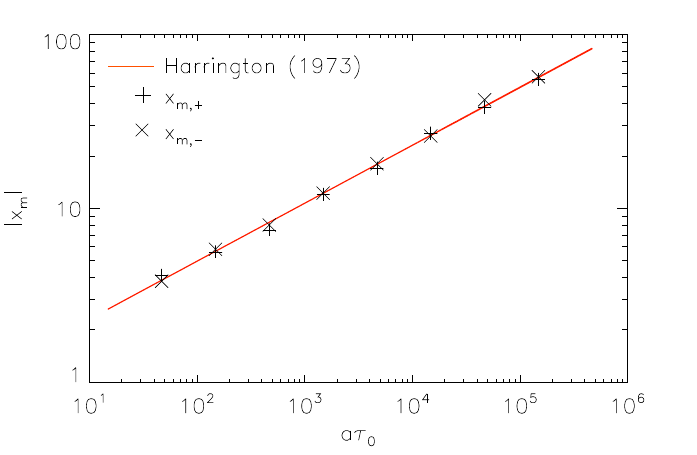}
\caption{{\small Absolute values $x_m$ of $x$ for which the emergent spectrum
                 of a slab
                 takes its maximum, as a function of $a\tau_0$ of the slab,
                 compared with the analytical solution (red) given by
                 \eq{xm}. Values are shown separately for the
                 positive (+) and the negative ($\times$) peak, denoted by
                 $x_{m,+}$ and $x_{m,-}$, respectively.}}
\label{fig:xm}
\end{figure}

Figure \ref{fig:xm} shows the values of $x$ for which the emergent spectrum
takes its maximum, compared with the analytical solution
\citep[\eq{xm},][]{har73}.
%


\subsection{Average number of scatterings}
\label{sec:Nscat}

\begin{figure}[!t]
\centering
\includegraphics [width=0.60\textwidth] {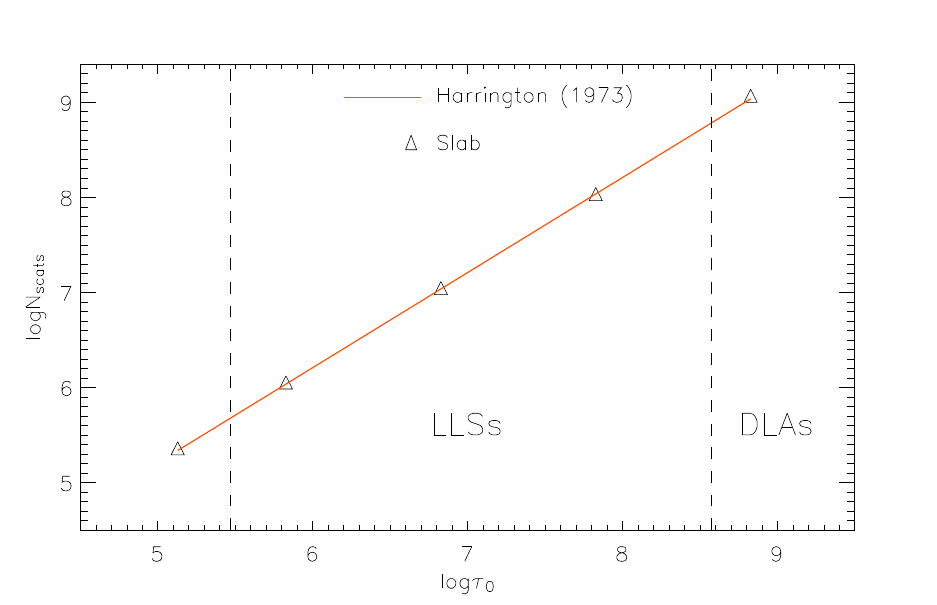}
\caption{{\small Average number of scatterings $N_{\mathrm{scat}}$
                 (\emph{triangles}) for different line center optical depths
                 $\tau_0$, compared with the analytical solution (\emph{red
                 line}) given by \eq{Nscat}. The \emph{dashed
                 lines} indicate the regions of optical depths for
                 LLSs and DLAs. While a
                 temperature of $T = 10$ K was used for simulations, the number
                 of photons varied from $10^5$ to $\sim10^3$ for the lowest and
                 highest optical depths, respectively.}}
\label{fig:Nscat}
\end{figure}

Figure \ref{fig:Nscat} shows the average number of scatterings
$N_{\mathrm{scat}}$.
Of course, in this case
a non-accelerated version of the code (i.e.~$x_{\mathrm{crit}} = 0$) was used,
since we are interested in the true number of scatterings. To get a feeling for
the physical significance of the optical depths, the region of
$\tau_0$ is divided into the domains of
LLSs and DLAs, characterized by limiting neutral hydrogen column densities of
$\Nhi = 10^{17.2}$ \cmsq and $\Nhi = 10^{20.3}$ \cmsq, respectively.
%


\subsection{Gas bulk motion}
\label{sec:Vbulk}
\index{Outflows}

To test if the implementation of the bulk velocity scheme produces reliable
results, we inspect the emergent spectrum of a sphere subjected to isotropic,
homologous expansion or collapse. Thus, the velocity
$\mathbf{v}_{\mathrm{bulk}}(\mathbf{r})$ of a
fluid element at a distance $\mathbf{r}$ from the center is set to
\begin{equation}
\label{eq:vr}
\mathbf{v}_{\mathrm{bulk}}(\mathbf{r}) = \mathcal{H} \mathbf{r},
\end{equation}
where the Hubble-like parameter $\mathcal{H}$ is fixed such that the velocity
increases linearly from 0 in the center to a maximal absolute velocity
$v_{\mathrm{max}}$ at the edge of the sphere ($r = R$):
\begin{equation}
\label{eq:vmax}
\mathcal{H} = \frac{v_{\mathrm{max}}}{R},
\end{equation}
with $v_{\mathrm{max}}$ positive (negative) for an expanding (collapsing)
sphere.

\begin{figure}[!t]
\centering
\includegraphics [width=0.85\textwidth] {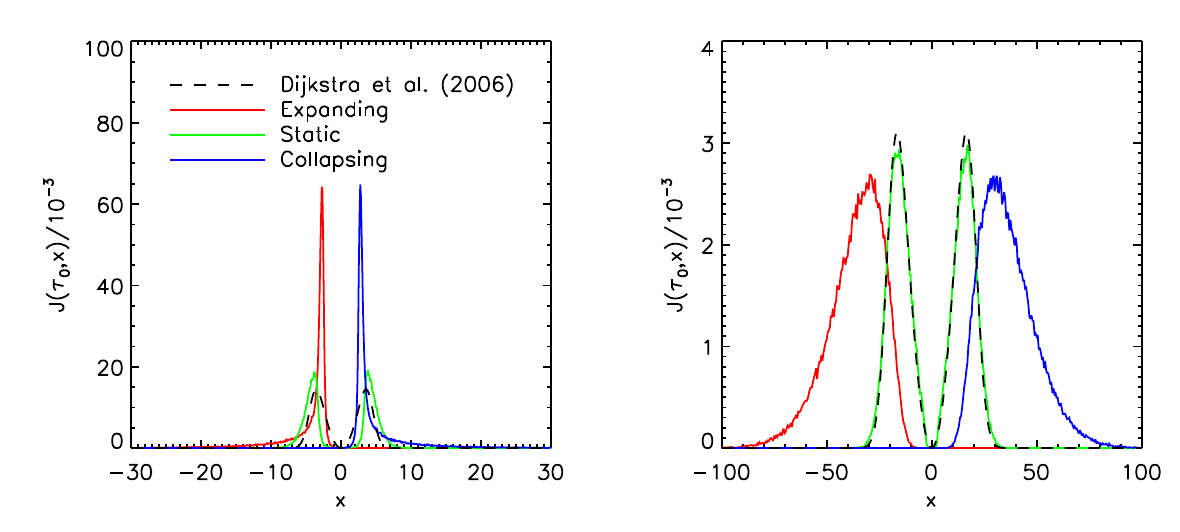}
\caption{{\cap Emergent spectrum from an isothermal ($T = 10^4$ K) and
         homogeneous sphere of gas undergoing isotropic expansion (\emph{red})
         or contraction (\emph{blue}) in such a way that the velocity at the
         edge of the sphere is $v_{\mathrm{max}} \pm200$ km s$^{-1}$.
         Left panel shows the result for a column density $\Nhi$ from the
         center to the edge of $2\times10^{18}$ \cmsq,
         corresponding to $\tau_0 = 1.2\times10^5$ and characteristic
         of a typical LLS. Right panel shows the result for
         $\Nhi = 2\times10^{20}$ \cmsq ($\tau_0 = 1.2\times10^7$),
         characteristic of a typical DLA. Also shown is the result from
         a simulation with $v_{\mathrm{bulk}} = 0$ (\emph{black dashed}),
         and the
         analytical solution for the static sphere (\emph{green}) as given
         by \citet{dij06a}. For the LLS, $\tau_0$ is clearly
         too small to give an accurate fit.}}
\label{fig:LLSDLA}
\end{figure}

For $T \neq 0$ K, no analytical solution for the spectrum exists.
Qualitatively, we expect an expansion to cause a suppression of
the blue wing and an enhancement of the red wing of the spectrum. The reason
for this is that photons blueward of the line center that
would otherwise escape the medium, are shifted into resonance in the reference
frame of atom lying closer to the edge, while red photons escape even more
easily. Conversely, a collapsing sphere will exhibit an enhanced blue wing and
a suppressed red wing. This is indeed seen in Fig.~\ref{fig:LLSDLA}.
Another
way to interpret this effect is that photons escaping an expanding cloud are,
on the average, doing work on the gas, thus losing energy, and vice versa for
a collapsing cloud.

\begin{figure}[!t]
\centering
\includegraphics [width=0.60\textwidth] {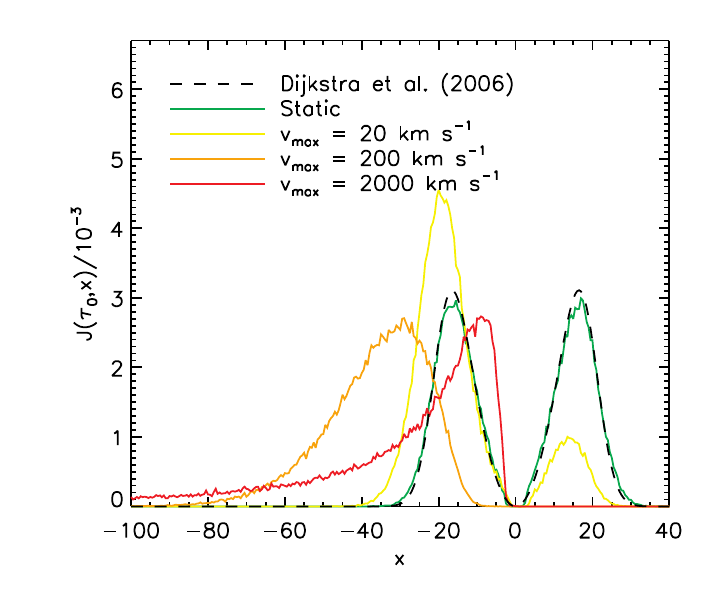}
\caption{{\small Emergent spectrum from an isothermal ($T = 10^4$ K) and
                 homogeneous sphere of hydrogen column density
                 $\Nhi = 2\times10^{20}$ \cmsq (a DLA) undergoing isotropic
                 expansion with different maximal velocities $v_{\mathrm{max}}$
                 at the edge of the sphere. For increasing $v_{\mathrm{max}}$,
                 the peak of the profile is pushed further away from the line
                 center. However, if $v_{\mathrm{max}}$ becomes too large, the
                 medium becomes optically thin and the peak moves back towards
                 the center again.}}
\label{fig:expand}
\end{figure}

In Fig.~\ref{fig:expand}, results for a sphere of gas expanding at different
velocities are shown.
For increasing $v_{\mathrm{max}}$, the
red peak is progressively enhanced and displaced redward of the line center.
However, above a certain threshold value the velocity gradient becomes so large
as to render the medium optically thin and allow less redshifted photons to
escape, making the peak move back toward the line center again.

The results matches closely those found by previous authors
\citep{zhe02,tas06a,ver06}.


\subsection{Escape fraction}
\label{sec:fesc}\index{Absorption!By dust}

\begin{figure}[!t]
\centering
\includegraphics [width=0.70\textwidth] {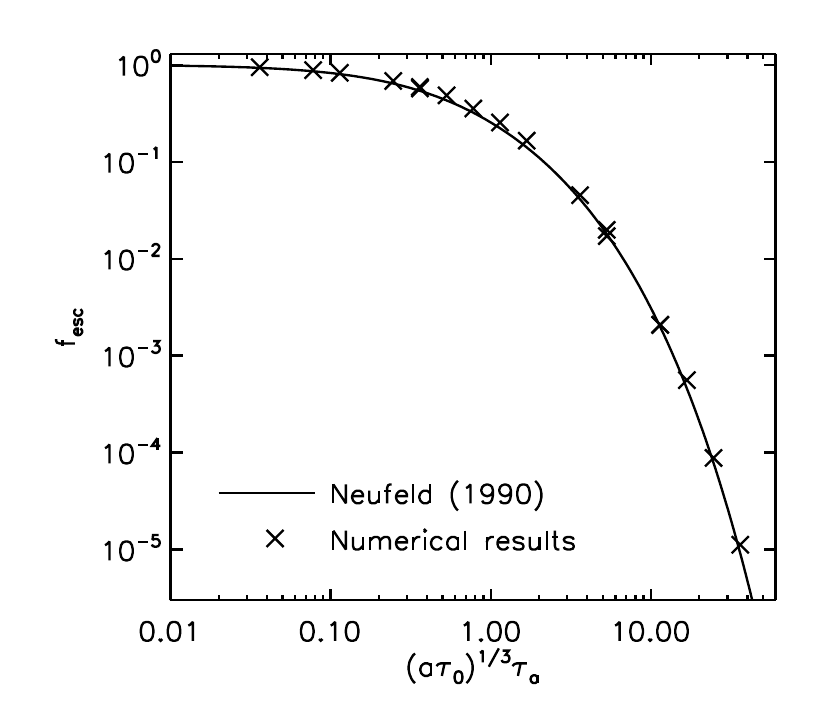}
\caption{{\cap Escape fractions $f_{\mathrm{esc}}$ of photons emitted from the
         center
         of a semi-infinite slab of gas damping parameter $a$, hydrogen optical
         depth $\tau_0$, and dust absorbing optical depth $\tau_{\mathrm{a}}$,
         compared to the analytical solution in \eq{neufesc}.}}
\label{fig:neufesc}
\end{figure}

\citet{neu90} provided an analytical expression for the escape fraction of
photons emitted from inside a slab
of an absorbing medium. The solution, which is
valid for very high optical depths ($a\tau_0 \gtrsim 10^3$, where $\tau_0$ is
the optical depth of neutral hydrogen from the center to the surface of the
slab) and in the limit $(a\tau_0)^{1/3} \gg \tau_{\mathrm{a}}$, where
$\tau_{\mathrm{a}}$ is the absorption optical depth of dust, is
\begin{equation}
\label{eq:neufesc}
f_{\mathrm{esc}} = \frac{1}
   {\cosh\left[ \zeta'
                \sqrt{(a\tau_0)^{1/3}\tau_{\mathrm{a}}}\right]},
\end{equation}
where $\zeta' \equiv \sqrt{3} / \zeta\pi^{5/12}$,
with $\zeta \simeq 0.525$ a fitting parameter. Figure \ref{fig:neufesc}
shows the result of a series of such simulations, compared to the analytical
solution.

Various AMR grid configurations were tested. Of course, the physical parameters
such as $\nhi$, $\nd$, and $T$ of a cell do not depend on the level of
refinement, but the acceleration schemes discussed in \sec{acc} do.



\section{Convergence tests}
\label{sec:convtest}

This section, as well as \sec{parstud} where the dust model is tested, makes
numerous references to the results obtained in Chapters \ref{cha:conseq} and
\ref{cha:dusteffect}, and may advantageously be read subsequent to those chapters.

\subsection{Resolution and interpolation scheme}
\label{sec:resinterp}

\begin{figure}[!t]
\centering
\includegraphics [width=1.00\textwidth] {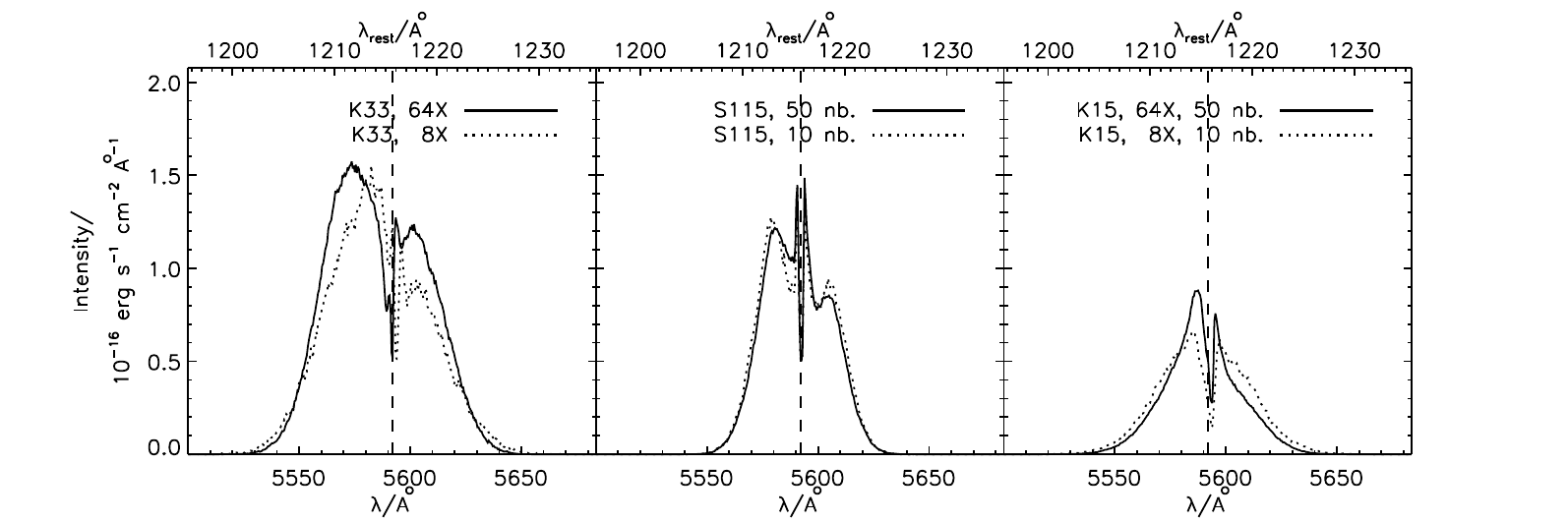}
\caption{{\cap Comparison spectra for the resolution test and the interpolation
         test.
         \emph{Left} panel shows the spectrum escaping in the negative
         $y$-direction of K33, simulated at high (\emph{solid curve}) and
         intermediate (\emph{dotted}) resolution.
         \emph{Middle} panel shows the spectrum escaping in the negative
         $x$-direction of S115, simulated at ultrahigh (64$\times$)
         resolution, but interpolating the physical parameters onto the AMR
         grid using the 50 nearest neighboring particles (\emph{solid}) and the
         10 nearest neighbors (\emph{dotted}).
         \emph{Right} panel shows the spectrum escaping in the
         negative $z$-direction of K15, simulated at high resolution and
         interpolating from 50 neighbors (\emph{solid}), compared with
         intermediate resolution/10 neighbors (\emph{dotted}). The differences
         do not change the results qualitatively. Note that in the
         middle (right) panel, the intensity has been multiplied (divided) by
         a factor of 10 in order to use the same scale as for all three
         galaxies.}}
\label{fig:SpectraComp}
\end{figure}

\begin{figure}[!t]
\centering
\includegraphics [width=1.00\textwidth] {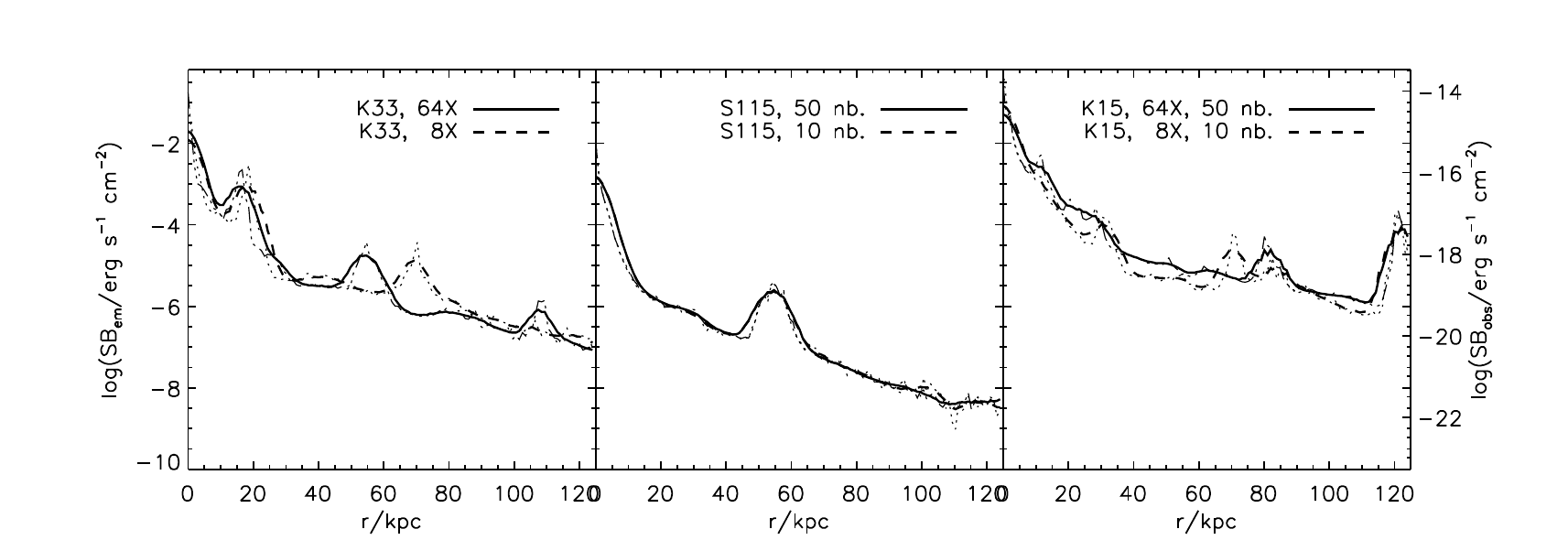}
\caption{{\cap Comparison SB profiles of the six models
         from \fig{SpectraComp} at a seeing of $0\farcs8$ (\emph{solid} and
         \emph{dashed} curves),
         overplotted on the true SB profiles (\emph{dotted}).
         While performing the simulations at
         different resolutions may shift some of the luminous regions somewhat
         spatially, the maximum SB and the overall slope remain virtually
         unaltered. Moreover, modifying the number of neighboring particles
         used for the interpolation scheme seems unimportant.}}
\label{fig:SBprofComp}
\end{figure}

In order to check the impact of the resolution of the cosmological simulation
on the results of the Ly$\alpha$ RT, the above RT calculations were also
carried out on the galaxies extracted directly from the cosmological
simulations, i.e.~at 8$\times$ lower resolution than the resimulated galaxies.
Furthermore, the procedure by which the
SPH particles are interpolated onto the grid was tested by using the 10
nearest neighboring particles instead of the usual 50 particles, and
running similar RT calculations on the output grids.
Although in both cases the results changed somewhat, there
seems to be no general trend.
Due to the slightly different evolution of the lo-res galaxies (being a
different simulation), the precise
configuration of stars and gas clouds will not necessarily be the same, and
thus luminous peaks in the SB maps cannot be expected to coincide exactly.
However, the maximum SBs appear to agree to within a few tens of percents, as
do the slopes and the overall amplitudes of the SB profiles.
The outcome of three such simulations, performed \emph{without} dust, can be
seen in Figures \ref{fig:SpectraComp} and \ref{fig:SBprofComp}.

\begin{figure}[!t]
\centering
\includegraphics [width=1.00\textwidth] {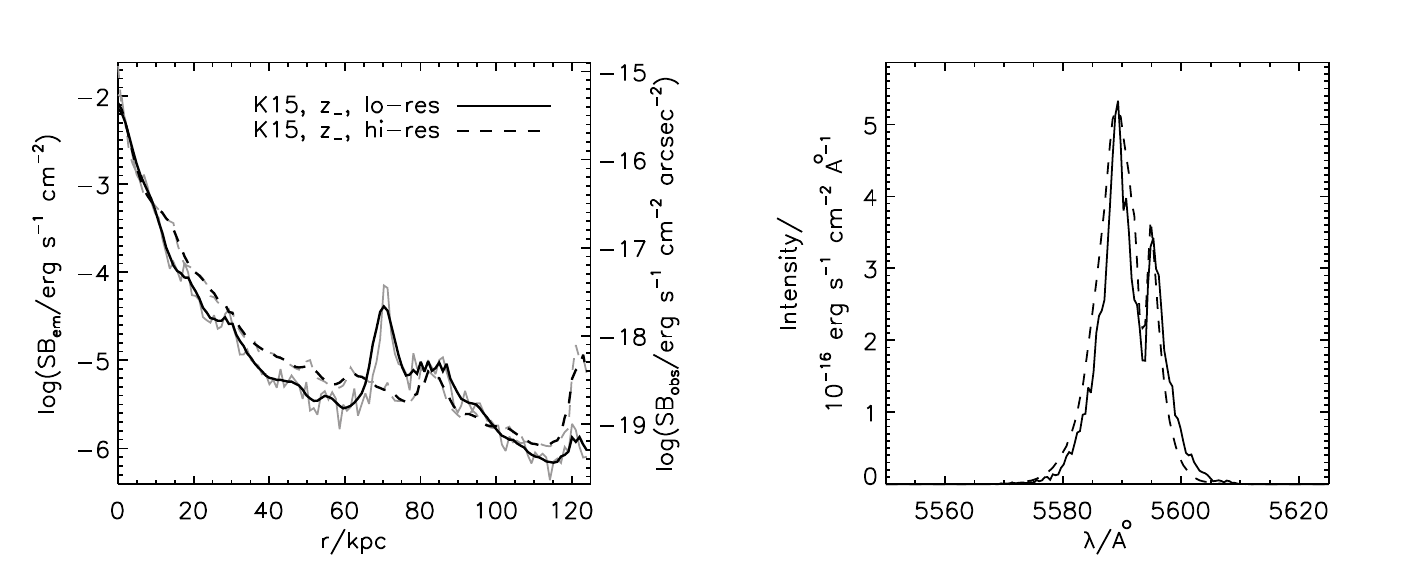}
\caption{{\cap Comparison of the SB profile (\emph{left})
         and emergent spectrum (\emph{right}) of the radiation escaping the
         galaxy K15 when performed on galaxies simulated at low
         (\emph{solid}) and high (\emph{dashed}) resolution. Gray lines show
         the true profile and black lines show the profile convolved with a
         seeing of $0\farcs5$. While the SB in
         the central regions appear to agree nicely, the fact that we are
         comparing two different simulations make luminous regions in the
         outskirts appear somewhat shifted.}}
\label{fig:lowres}
\end{figure}

The convergence proves stable also when dust is added to the calculations.
The resulting SB profile and spectrum for K15 are seen in \fig{lowres}.
The resulting escape fraction is a bit lower than for the hi-res
simulation, but less than 5\%.
%


\subsection{AMR structure}
\label{sec:AMRtest}

To investigate the significance of the AMR structure, simulations were also
carried out in which the structure was progressively desolved, level by level.
In K15, K33, and S115, the
maximum level $\mathcal{L}$ of refinement is 7, where $\mathcal{L} = 0$
corresponds to the unrefined base grid of $128^3$ cells. Eight cells of
$\mathcal{L} = \ell$ are desolved to $\mathcal{L} = \ell - 1$ by taking the
average of the physical parameters. Since temperature reflects the internal
energy of a body of gas, and since the combined velocity is given by momentum,
$T$ and $\mathbf{v}_{\mathrm{bulk}}$ are weighted by the respective cell masses.
\begin{figure}[!t]
\centering
\includegraphics [width=0.65\textwidth] {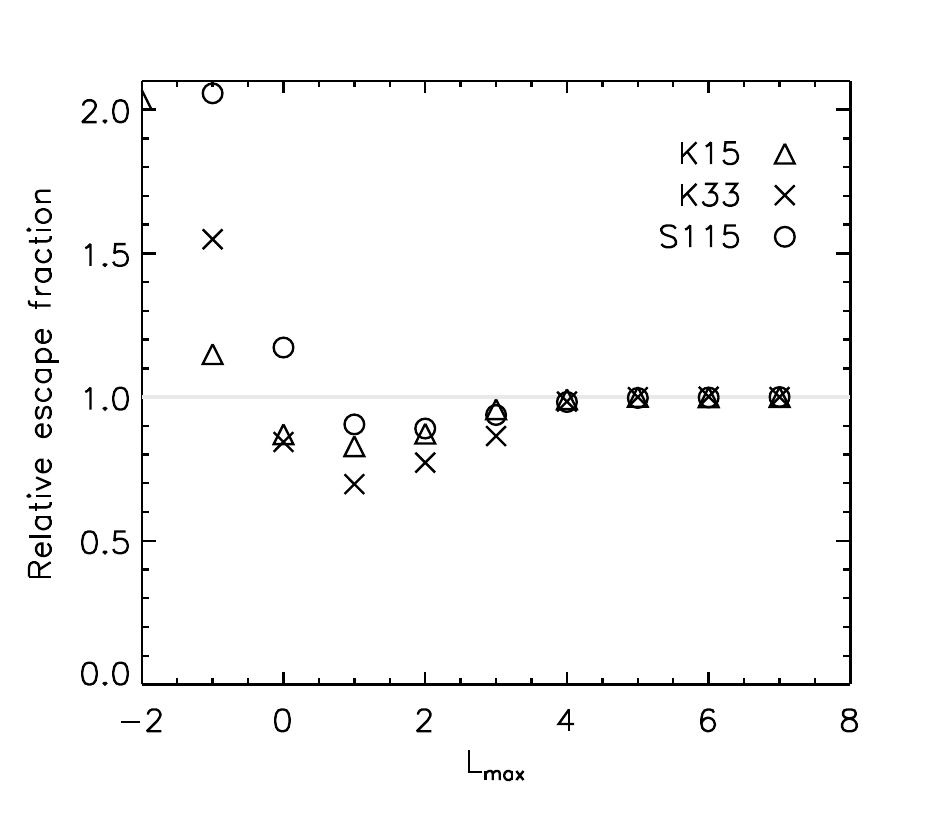}
\caption{{\cap Relative escape fractions $f_{\mathrm{esc}}$ from the three
         galaxies K15, K33, and S115 as a function of the
         maximum level $\mathcal{L}_{\mathrm{max}}$ of AMR refinement.
         $\mathcal{L}_{\mathrm{max}} = 0$ corresponds to having only the
         $128^3$ base grid, while $\mathcal{L}_{\mathrm{max}} < 0$ corresponds
         to desolving the base grid. For increasingly lower resolution,
         $f_{\mathrm{esc}}$ drops due to the low-density paths being smeared
         out with high-density regions. Eventually, however, when the
         resolution is so course that the central star- and gas-rich regions
         are mixed with the surrounding low-density region, $f_{\mathrm{esc}}$
         increases rapidly.}}
\label{fig:Lmax}
\end{figure}

Figure \ref{fig:Lmax} shows the resulting escape fractions of these three
galaxies. Desolving the first
few levels does not alter $f_{\mathrm{esc}}$ notably, indicating that the
galaxies are sufficiently resolved. However, eventually we see the importance of
the AMR structure: with insufficient resolution, the clumpiness of the central,
luminous ISM is lost, ``smoothing out''
the low-density paths that facilitate escape,
and consequently $f_{\mathrm{esc}}$ decreases. When the resolution becomes even
worse, the central regions are averaged with the surrounding low-density gas,
so that most of the photons are being emitted from medium dense cells,
resulting in a small probability of scattering on neutral hydrogen, and hence
a small probability of being absorbed by dust.


\subsection{UV RT scheme}
\label{sec:UVRTtest}

\begin{figure}[!t]
\centering
\includegraphics [width=0.90\textwidth] {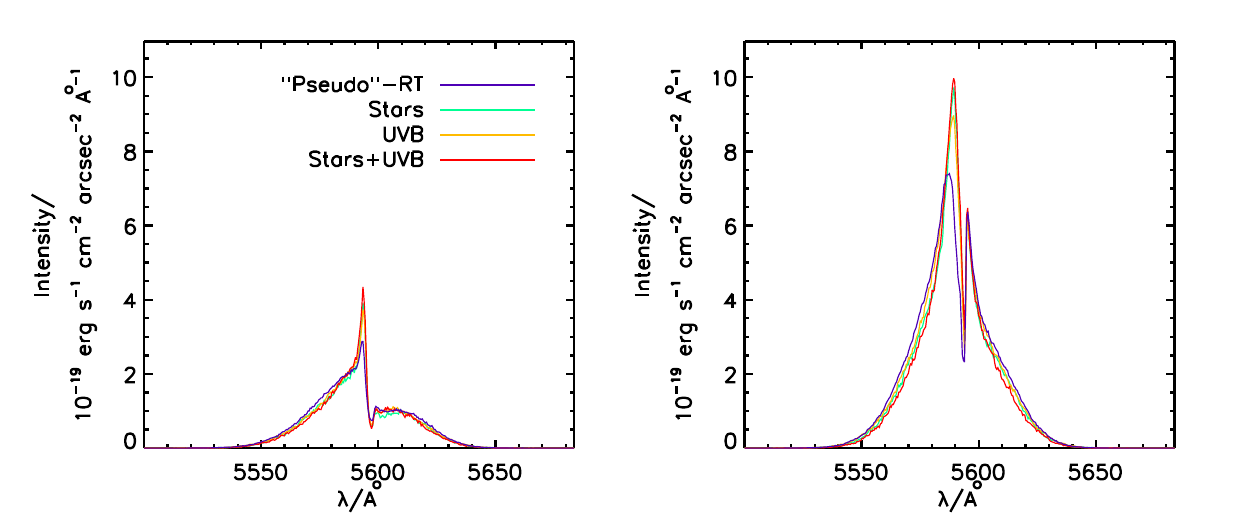}
\caption{{\cap Emergent spectrum of the galaxy K15, as seen when observing the
         sheet-like structure in which the galaxy is embedded edge-on
         (\emph{left}) and face-on (\emph{right}).
         \emph{Blue lines} show the
         spectrum for the model without the improved UV RT, while \emph{green},
         \emph{yellow}, and \emph{red} lines show the spectrum when treating
         the UV RT
         properly for the stellar sources only, the UV background only, and
         both, respectively.
         The only real difference is seen in the blue peak of the spectra,
         which is a bit higher for the improved models.}}
\label{fig:AlexSpec}
\end{figure}

Finally, to see the effect of the improved UV RT scheme (\sec{uvrt}),
\fig{AlexSpec} shows the emergent spectrum from K15 for the different schemes
(excluding the effects of dust).
Both edge-on and face-on spectra are sensitive
to the changes in ionization of predominantly low-density regions
which is computed with LyC radiative transfer.
As discussed in \sec{neufeld}, the quantity that determines the
shape of the spectrum is the product $a\tau_0 \propto \nhi / T$
(at a fixed physical size).
Generally, a higher temperature will also imply a lower density of neutral
hydrogen, and vice versa, and thus we might expect
$\nhi / T$ to change rapidly to higher or lower values for the improved scheme.
However, for high density cells the change in the ratio
$\nhi / T$ is minor when invoking the improved scheme, in most
cases of order unity. Only in low-density cells is this ratio considerably
altered, but since $>90$\% of all scatterings take place in high-density
cells, the overall effect is small. The only notable difference is seen in
the inner part of the spectrum, which is exactly the part that is created by
the low-density regions, since photons near the line center cannot escape from
high-density regions.



\section{Dust parameter study}
\label{sec:parstud}

\begin{figure}[!t]
\centering
\includegraphics [width=0.65\textwidth] {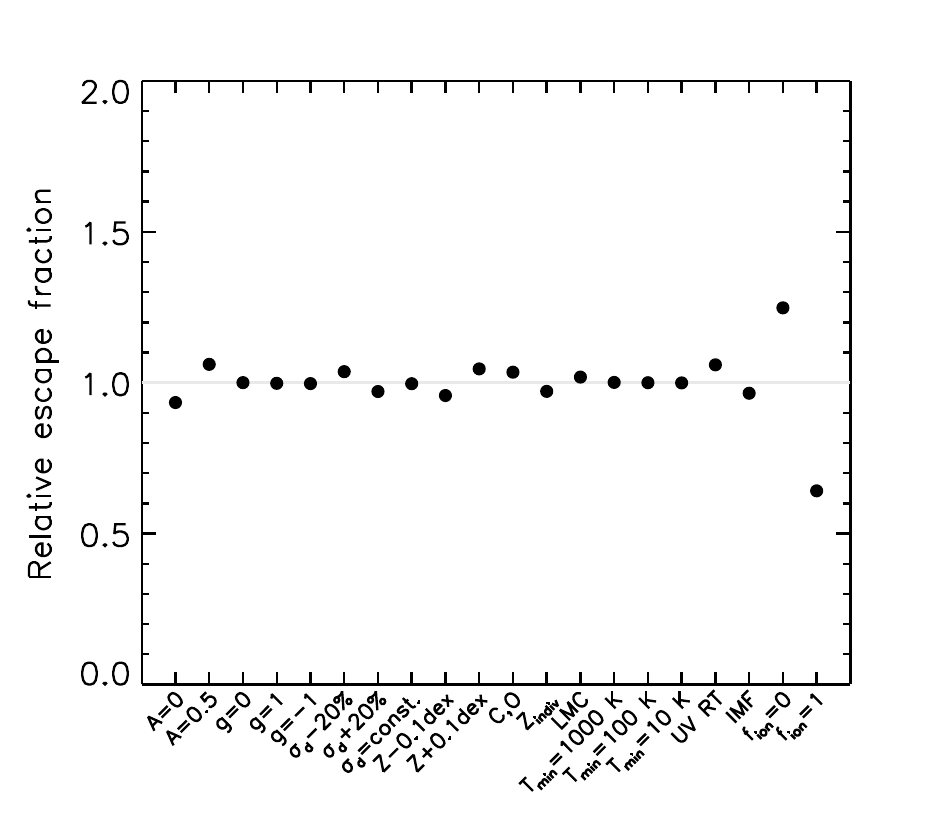}
\caption{{\cap Relative escape fraction from the galaxy K15 as a ``function'' of
         model, i.e.~simulation where all parameters but one are equal to that
         of the benchmark model used for the simulations in \cha{dusteffect}.
         See text for an explanation of abscissa labels.
         Except for the factor $f_{\mathrm{ion}}$, the
         chosen benchmark model appears quite robust to varying other
         parameters.}}
\label{fig:allpar}
\end{figure}

The adopted model of dust clearly involves a multitude of assumptions, some
more reasonable than others. To inspect the dependency of the outcome on the
values of the parameters, a series of simulations of K15 was run,
varying the below discussed values. The resulting escape fractions compared to
that of the ``benchmark'' model, used for the simulations in
\cha{dusteffect}, are shown in \fig{allpar}.

\emph{The albedo} $A$ of the dust grains. The chosen value of 0.32 is bracketed
by the values 0.5 and 0, i.e.~somewhat more reflective and completely
black, respectively. As expected, the higher the albedo, the higher the escape
fraction, but note that even completely black dust reduces $f_{\mathrm{esc}}$
by less than $10$\%. This is because the bulk of the
photons is absorbed in the very dense environments, where scattering off of
one grain in many cases just postpones the absorption to another grain.

\emph{The scattering asymmetry parameter} $g$. The three cases $g = 0,1,-1$
are tested, corresponding to isotropic scattering, total forward scattering,
and total backscattering. The difference from the benchmark model value of
$0.73$ is virtually
nonexistent; the fact that most of the scatterings take place in the
dense environments makes the transfer be dominated by scattering on hydrogen.

\emph{The dust cross section} $\sigma_{\mathrm{d}}$. \citet{fit07} showed that
the variance of the extinction curves (in the MW, normalized to $A_V$) is
approximately 20\% at the Ly$\alpha$ wavelength.
Decreasing (increasing) the dust cross section by this quantity increases
(decreases) the escape fraction as
expected, but not by more than $\sim$$\pm5$\% percent.
Also, a constant cross section
with $\sigma_{\mathrm{d}} = \sigma_{\mathrm{d}}|_{x=0}$ was tested, but with no
notable effect.

Likewise, a variance of \emph{the reference metallicity} $Z$ is present from
sightline to sightline in the Magellanic Clouds, probably at least 0.1 dex.
Using a smaller (larger) reference $Z$ makes the metallicity in the simulations
comparatively larger (smaller), with a larger (smaller) dust density as a
result and hence a smaller (larger) escape fraction. Since the escaping photons
represent different sightlines in the galaxies, it is fair to use the average
$Z$ (for the individual metals) in \eq{nd}, but to investigate the
sensitivity on the reference metallicity, simulations with $Z_{\mathrm{SMC}}$
increased (reduced) by 0.1 dex was run, resulting in a $-5$\% ($+5$\%) change
in $f_{\mathrm{esc}}$.

Letting $n_{\mathrm{d}}$ scale with the metallicity of only C and O instead
of the total metallicity does not alter $f_{\mathrm{esc}}$ much either. This
model is relevant since C and O probably are the main constituents of dust.

Although the metallicity of the Magellanic Clouds is smaller than that of the
MW, the \emph{relative} abundances between the various elements are more or
less equal. Small deviations do exist, however, but letting $n_{\mathrm{d}}$
scale with the metallicity of the individual metals does not change the results
much (point labeled ``$Z_{\mathrm{indiv}}$'' in Fig.~\ref{fig:allpar}).


\emph{Dust type}. Using an LMC extinction curve instead of SMC results in a
slightly (few \%) larger escape fraction, since the quantity
$\tau_{\mathrm{d}} \propto
n_{\mathrm{d}} \sigma_{\mathrm{d}} \propto
\sigma_{\mathrm{d}}/Z_0$ is roughly 10\% lower for the LMC than for the SMC.

\emph{The minimum temperature} $T_{\mathrm{min}}$ of the simulations. The
cosmological simulation includes cooling of the gas to $\sim$$10^4$ K. Since
the temperature affects the RT of Ly$\alpha$, the
temperature of the cells with $T \simeq 10^4$ K was artificially lowered to
$10^3$, $10^2$ (approximately the temperature of the cold neutral medium),
and $10$ K (approximately the
temperature of a molecular cloud), to see if not including sufficient cooling
could affect the results. However, as is seen in Fig.~\ref{fig:allpar}, the
difference is insignificant.

\emph{The ionizing UV RT scheme}. As found in \sec{UVRTtest}, implementing a
more
realistic UV RT scheme does not alter the outcome of the non-dusty Ly$\alpha$
RT significantly. When including dust, as seen from Fig.~\ref{fig:allpar}
the improved RT results in a slightly increased $f_{\mathrm{esc}}$, although
less than 10\%. The reason
is that this RT is more efficient than the ``old'' RT scheme at ionizing the
neutral gas in the immediate vicinity of the stars and, accordingly, at
lowering the dust density. However, in these regions the gas density is so high
that ionization in most cases is followed by instantaneous recombination, and
hence the physical state of the gas in the case of the improved RT is not
altered significantly.

\emph{The initial mass function}.
Since a
Salpeter IMF is more top-heavy than a Kroupa IMF, i.e.~produces relatively
more massive stars per stellar mass, it also yields a higher metallicity
and hence a higher absorption by dust. However, the increased feedback from the
massive stars serves as to counteract star formation, and these
two effects more or less balance each other. The ratio of the
Kroupa-to-Salpeter feedback energy is 0.617, while for the yield the ratio
is 0.575 (for oxygen). The result, as is seen from \fig{allpar}, is only a
slightly smaller escape fraction.

\begin{figure}[!t]
\centering
\includegraphics [width=0.65\textwidth] {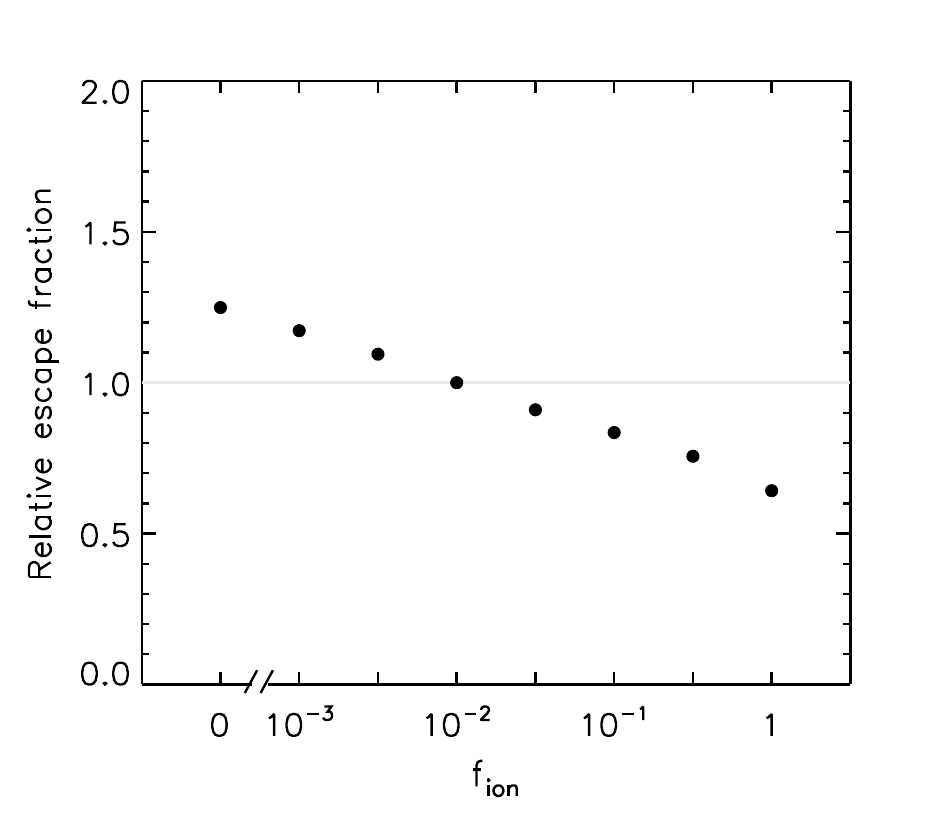}
\caption{{\cap Relative variation in escape fraction $f_{\mathrm{esc}}$ from
         the galaxy K15, as a function of
         $f_{\mathrm{ion}}$, the fraction of
         ionized hydrogen that contributes to the density of dust
         (\eq{nd}). Note the discontinuity on the abscissa axis.
         Even a little dust in the ionized region can
         affect $f_{\mathrm{esc}}$ quite a lot.}}
\label{fig:subnosub}
\end{figure}

\emph{The fraction of ionized hydrogen}
$f_{\mathrm{ion}}$ contributing to the dust
density. Insufficient knowledge about the dust contents of ionized gas is by
far the greatest source of uncertainty in $f_{\mathrm{esc}}$, as is seen in
the figure. This effect is further investigated in Fig.~\ref{fig:subnosub},
where the relative escape fraction from K15 for different values of
$f_{\mathrm{ion}}$ is shown.

From the figure, it is seen that even a small amount of dust associated with
the ionized hydrogen can affect the escape fraction quite a lot. The reason is
that most of the scatterings and the absorption take place in the dense region
where also the star formation is high. In these regions supernova feedback
shockwaves will recurrently sweep through the ISM, heating and ionizing the
medium without significantly lowering the density.
For $f_{\mathrm{ion}} = 1$, the
calculated dust density of these regions is not affected, while
for $f_{\mathrm{ion}} = 0$, this effect
renders the gas virtually dustless, resulting in highly porous medium with
multiple possibilities for the photons of scattering their way out of the
dense regions.

The resulting escape fraction of the benchmark model lies approximately midway
between the two
extrema and if nothing else, $f_{\mathrm{esc}}$ can be regarded as having an
uncertainty given by the result of these extrema, i.e.~$\sim$20\%.
Nevertheless, $f_{\mathrm{ion}} = 0.01$ seems a realistic value,
cf.~the discussion in \sec{iongas}.



\chapter{Intergalactic radiative transfer}
\label{cha:IGMRT}

\init{T}{he Ly$\alpha$ RT described in the} previous chapters in principle only
 predicts the spectrum of
radiation that one would observe if located in the vicinity of the galaxies.
Obviously, in reality the radiation has to travel trough the IGM afterwards.
At a redshift
of $\sim 3.6$, the IGM is largely ionized, and the spectra escaping the
galaxies are, in general, not very different from what would be observed at
Earth. However, even a very small amount of neutral hydrogen may influence the
observations.
To investigate just \emph{how} large an impact the IGM exerts on the radiation,
full IGM RT has to be computed. Furthermore, at higher redshifts where
the IGM is generally more neutral and more dense, neglecting the effect would
lead to severely erroneous results.

In principle this could be achieved by performing first the ``galactic'' RT in
the high-resolution resimulations and subsequently continuing the RT in the
low-resolution cosmological volume from the location of the individual
galaxies. However, although the physics of scattering in galaxies and that of
scattering in the IGM is not inherently different, the different physical
conditions imposes a natural division of the two schemes: in the dense gas of
galaxies, photons are continuously scattered in and out of the line of sight,
whereas in the IGM, once a photon is scattered out of the line of sight, it
is ``lost'', becoming part of the background radiation. The probability of a
background photon being scattered \emph{into} the line of sight, on the other
hand, is vanishingly small.

In order to disentangle galactic from intergalactic effects,
and, more importantly, to investigate the general effect of the IGM instead of
merely the IGM lying between us and the nine resimulated galaxies, we take a
different approach: the transmission properties of the IGM are studied by
calculating the normalized spectrum $\Flam$ --- the \emph{transmission function}
--- in the vicinity of the Ly$\alpha$ line, as an average
of a large number of sightlines cast through the cosmological volume, and
originating just outside a large number of galaxies.

As the red part of the spectrum is only shifted farther away from resonance,
IGM absorption tends to skew
the line and not simply diminish it by some factor. In most earlier studies of
the line profiles of high-redshift galaxies, the IGM has either been
ignored \citep[e.g.][]{ver06,ver08},
taken to transmit the red half and remove the blue part \citep[e.g.][]{fin08},
or to influence the line uniformly by a factor
$e^{-\langle\tau\rangle}$, where $\langle\tau\rangle$ is the average optical
depth of the Universe \citep[e.g.][]{bru03,mei05,fau08,ryk09}.

Consider a source emitting the normalized spectrum
$\F_{\mathrm{em}}(\lambda) \equiv 1$ for all $\lambda$.
In an idealized, completely homogeneous universe undergoing completely
homologous expansion (``ideal Hubble flow'') and with the absorption profiles
having a negligible width, the observed spectrum would simply be a step
function, with $\Flam = \F_{\mathrm{red}} = 1$ for $\lambda > \lambda_0$, and
$\F(\lambda) = \F_{\mathrm{blue}} < 1$ for $\lambda \le \lambda_0$.
Three factors contribute to make $F$ differ from a step function:

Firstly, the Ly$\alpha$ line is not a delta function, but has a finite width.
For high densities of neutral hydrogen, the damping wing of the profile may
even extend significantly into the red part of the spectrum, such that in the
vicinity of the line center also part of $\F_{\mathrm{red}}$ is less than one.

Secondly, the IGM is highly inhomogeneous, leading to large variations
in $\F$ blueward of $\lambda_0$.
This is the reason for the LAF seen at intermediate to high redshifts.
However, when we consider the average effect of the IGM and let $\F$ be the
average of many sightlines $\F_{\mathrm{blue}}$ should still be a constant
function of wavelength.
Nevertheless,
in the proximity of galaxies the gas density is higher than far from the
galaxies; on the other hand, in these regions the stellar ionizing UV radiation
may reduce the neutral hydrogen density.
Consequently, wavelengths just blueward of the line center may not on average
be subject to the same absorption as farther away from the line.
Regardless of which of
these two effects is more important, the correlation of the IGM with the source
cannot be neglected.

Finally the expansion is not exactly homologous, since peculiar velocities of
the gas elements will cause fluctuations around the pure Hubble flow.
Considering again not individual sightlines but the average effect of the IGM,
these fluctuations are random and cancel out; on average, for every gas element
that recedes from a galaxy faster than the Hubble flow and thus causes an
absorption line at a slightly bluer wavelength, another gas element does the
opposite.
Hence, the average transmission in a ``realistic'' universe is the same as in a
universe where there are no peculiar velocities.
However, in the proximity of overdensities, the extra mass results in a
``retarded'' expansion of the local IGM.
When expansion around a source is somewhat slower than that of the rest
of the Universe, on average matter in a larger region will be capable of
causing absorption,
since the slower the expansion, the farther the photons will have to travel
before shifting out of resonance.

This chapter focuses on two different aspects of the intergalactic
transmission. First, to see how the Ly$\alpha$ line is affected by the IGM, we
calculate a ``transmission function'' given by the average, normalized flux
$\Flam$.
Specifically, $\F$ is calculated by taking the median value in
each wavelength bin of many sightlines, originating just outside a large number
of galaxies (where ``just outside'' will be defined later).
The standard deviation is defined by the 16 and 84 percentiles.
From the above discussion we may expect that $\F$ be characterized by a red
part $\F_{\mathrm{red}} \simeq 1$, and a blue part $\F_{\mathrm{blue}} < 1$,
but with a non-trivial shape just blueward of the line center, since the IGM in
the vicinity of the sources is different than far from the sources.
We will also examine the average transmission $\T = \T(z)$ of the
IGM, defined by first calculating for each individual sightline the fraction
of photons that are transmitted through the IGM in a relatively large
wavelength interval well away from the line center, and then taking the median
of all sightlines (again with the 16 and 84 percentiles defining the
standard deviation).
This quantity is sensitive to the overall ionization state of the IGM, and has
therefore been
used observationally to put constraints on the EoR \citep{bec01,djo01}.

\section{Preparations}
\label{sec:prep}

\subsection{Cosmological models}
\label{sec:models}

The basics of the cosmological simulation were described in \sec{cosmo}.
For the IGM RT, three different models will be investigated:
Model 1 has the \citeauthor{haa96} UVB field initiating at $\zre = 10$,
and has density fluctuations given by $\sigma_8 = 0.74$;
Model 2 has $\zre = 10$ and $\sigma_8 = 0.9$;
Model 3 has $\zre = 6$ and $\sigma_8 = 0.74$.
In all cases, the improved UV RT is applied afterwards (\sec{uvrt}).


\subsection{Galaxy selection criteria}
\label{sec:select}

Galaxies are located in the simulations as described in \citet{som05}.
To make sure that a given identified structure is a real galaxy, the
following selection criteria are imposed on the sample:
\begin{enumerate}
  \item To ensure that a given structure ``$i$'' is not just a substructure of
        a larger structure ``$j$'', if the center of structure $i$ is situated
        within the virial radius of $j$, it must have more stars than $j$.
  \item The minimum number of star particles must be at least
        $N_{\star\mathrm{,min}} = 15$. This corresponds to a minimum stellar
        mass of $\log(M_{\star\mathrm{,min}}/\Msun) = 7.2$.
  \item The circular velocity, given by 
        $V_c = \sqrt{G M_{\mathrm{vir}} / r_{\mathrm{vir}}}$ must be
        $V_c \ge 35$ km s$^{-1}$.
\end{enumerate}

The IGM in the vicinity of large galaxies is to some extend different from the
IGM around small galaxies. The more mass gives rise to a deeper gravitational
potential, enhancing the retarded Hubble flow. On the other hand, the larger
star formation may cause a larger bubble of ionized gas around it.
To investigate the difference in transmission, the sample of accepted galaxies
is divided into three subsamples, denoted ``small'', ``intermediate'', and
``large''. To use the same separating criteria at all redshifts, instead of
separating by mass --- which increases with time due to merging and accretion
--- the galaxies are separated according to their circular velocity, which
does not change significantly over time.
The thresholds are defined somewhat arbitrarily as $V_1 = 55$ km s$^{-1}$
(between small and intermediate galaxies) and $V_2 = 80$ km s$^{-1}$ (between
intermediate and large galaxies).

\begin{figure}[!t]
\centering
\includegraphics [width=1.10\textwidth] {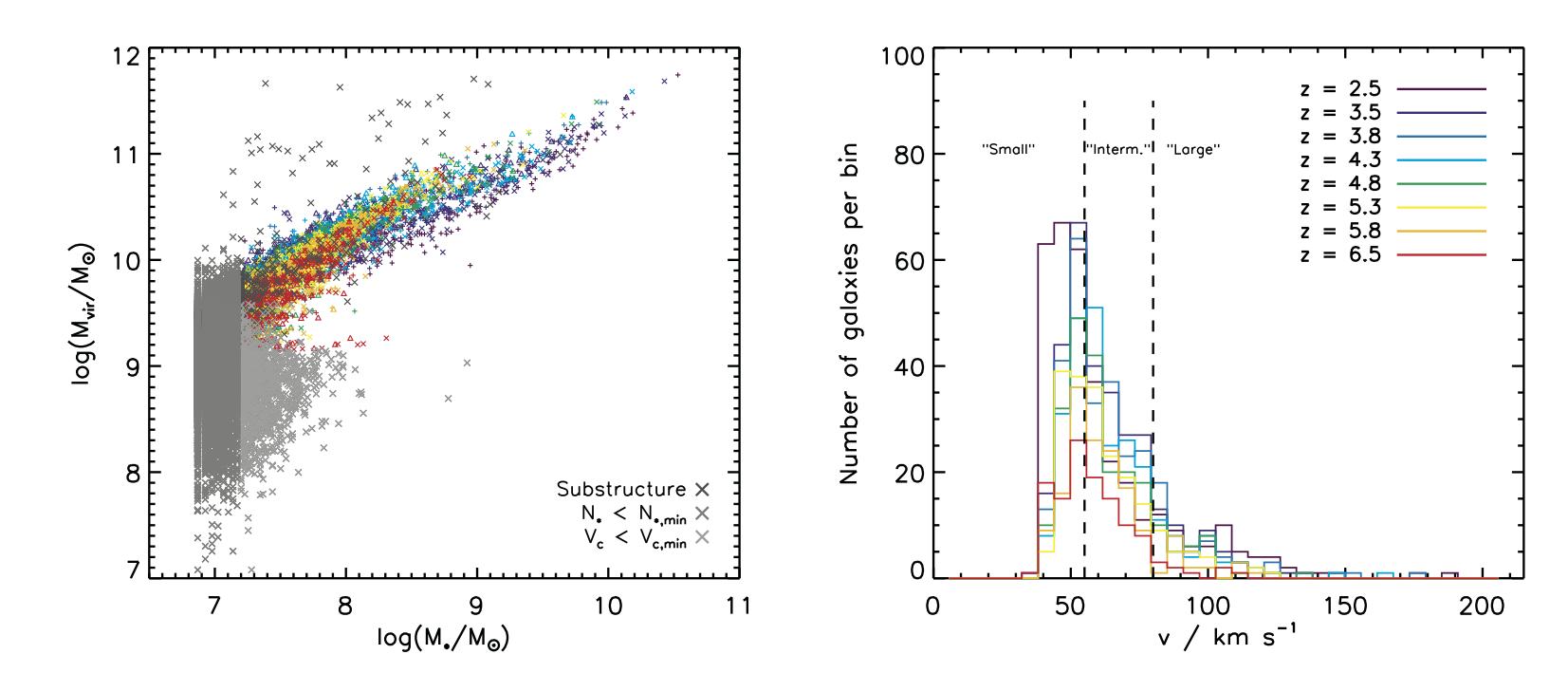}
\caption{{\cap \emph{Left:} Scatter plot of virial masses $M_{\mathrm{vir}}$
         vs.~stellar masses $M_\star$ for the full, unfiltered sample of
         galaxies. The colors signify redshift, with more red meaning higher
         redshift (exact values are seen in the right plot's legend).
         The data points of Model 1, 2, and 3 are shown with \emph{triangles},
         \emph{plus signs}, and \emph{crosses}, respectively.
         The galaxies that are rejected are overplottet with \emph{gray}
         crosses, with \emph{dark}, \emph{medium}, and \emph{light} gray
         corresponding to rejection criterion 1, 2, and 3, respectively.
         \emph{Right:} Distribution of circular velocities $V_c$ for the
         accepted galaxies in Model 1.
         The distributions for Model 2 and 3 look similar, although Model 2
         has more galaxies (see \tab{Ngals}).}}
\label{fig:Filter}
\end{figure}

The final sample of galaxies is seen in \Fig{Filter}, which shows the relation
between stellar and virial mass, and the distribution of circular velocities.

The exact number of galaxies in the three models studied in this work is seen
in \tab{Ngals}.
\begin{table}[bth]
\begin{center}
{\sc Number of galaxies in the simulations}
\end{center}
\centering
\begin{tabular}{cccccccc}
\hline
\hline
Model  & $\zre$ & $\sigma_8$ &  $z$  & Total & Small & Intermediate & Large\\
\hline
 1.    &   10   &   0.74     &  2.5  &  343  &  207  &       80     &  56 \\
       &        &            &  3.5  &  309  &  138  &      127     &  44 \\
       &        &            &  3.8  &  283  &  125  &      119     &  39 \\
       &        &            &  4.3  &  252  &  102  &      115     &  35 \\
       &        &            &  4.8  &  225  &  111  &       86     &  28 \\
       &        &            &  5.3  &  201  &   98  &       85     &  18 \\
       &        &            &  5.8  &  154  &   75  &       62     &  17 \\
       &        &            &  6.5  &  121  &   70  &       44     &   7 \\
       &        &            &       &       &       &              &      \\
 2.    &   10   &   0.9      &  3.5  &  405  &  207  &      128     &  70 \\
       &        &            &  4.3  &  384  &  165  &      150     &  69 \\
       &        &            &  4.8  &  341  &  150  &      129     &  62 \\
       &        &            &  5.3  &  324  &  145  &      125     &  54 \\
       &        &            &  5.8  &  277  &  121  &      106     &  50 \\
       &        &            &  6.5  &  252  &  126  &       92     &  34 \\
       &        &            &       &       &       &              &      \\
 3.    &    6   &   0.74     &  3.5  &  325  &  162  &      122     &  41 \\
       &        &            &  3.8  &  318  &  165  &      117     &  36 \\
       &        &            &  4.3  &  293  &  150  &      110     &  33 \\
       &        &            &  4.8  &  250  &  138  &       84     &  28 \\
       &        &            &  5.3  &  204  &  101  &       85     &  18 \\
       &        &            &  5.8  &  160  &   75  &       67     &  18 \\
       &        &            &  6.5  &  126  &   69  &       50     &   7 \\
\hline
\end{tabular}
\caption{{\cap ``Small'', ``intermediate'', and ``large'' galaxies are defined
               as having circular velocities
               $V_c < 55$ km s$^{-1}$,
               55 km s$^{-1} \le V_c < 80$ km s$^{-1}$, and
               $V_c \ge 80$ km s$^{-1}$, respectively.}}
\label{tab:Ngals}
\end{table}
%



\section{{\sc IGMtransfer}}
\label{sec:IGMRTsims}
\index{Intergalactic medium!Radiative transfer}

For the IGM RT the same nested grid is used as for the ionizing
UV RT. The transmission properties of the IGM are studied by calculating the
normalized spectrum $\Flam$ in the vicinity of the Ly$\alpha$ line, as an
average of a high number of sightlines cast through the simulated cosmological
volume. The code used for these calculations --- dubbed {\sc IGMtransfer} ---
has been made publically available \citep{lau10b} and can be downloaded from
the URL
\href{http://www.dark-cosmology.dk/~pela/IGMtransfer.html}
{www.dark-cosmology.dk/\~{}pela/IGMtransfer.html}.

The resulting value of $\Flam$ at wavelength $\lambda$ for a given sightline is
\begin{equation}
\label{eq:F} 
\Flam = e^{-\tau(\lambda)}.
\end{equation}
The optical depth $\tau$ is the sum of contributions from all the cells
encountered along the line of sight:
\begin{equation}
\label{eq:tau}
\tau(\lambda) = \sum_i^{\mathrm{cells}} 
                n_{\textrm{{\scriptsize H}{\tiny \hspace{.1mm}I}},i}
                \,r_i
                \,\sigma(\lambda + \lambda v_{||,i}/c).
\end{equation}
Here, 
$n_{\textrm{{\scriptsize H}{\tiny \hspace{.1mm}I}},i}$ is the density of
neutral hydrogen in the $i$'th cell,
$r_i$ is the distance covered in that particular cell,
$v_{||,i}$ is the velocity component of the cell along the line of sight,
and
$\sigma(\lambda)$ is the cross section of neutral hydrogen.
Due to the resonant nature of the transition, the largest contribution at a
given wavelength will arise from the cells the velocity of which corresponds
to shifting the wavelength close to resonance.

\begin{figure}[!t]
\centering
\includegraphics [width=0.90\textwidth] {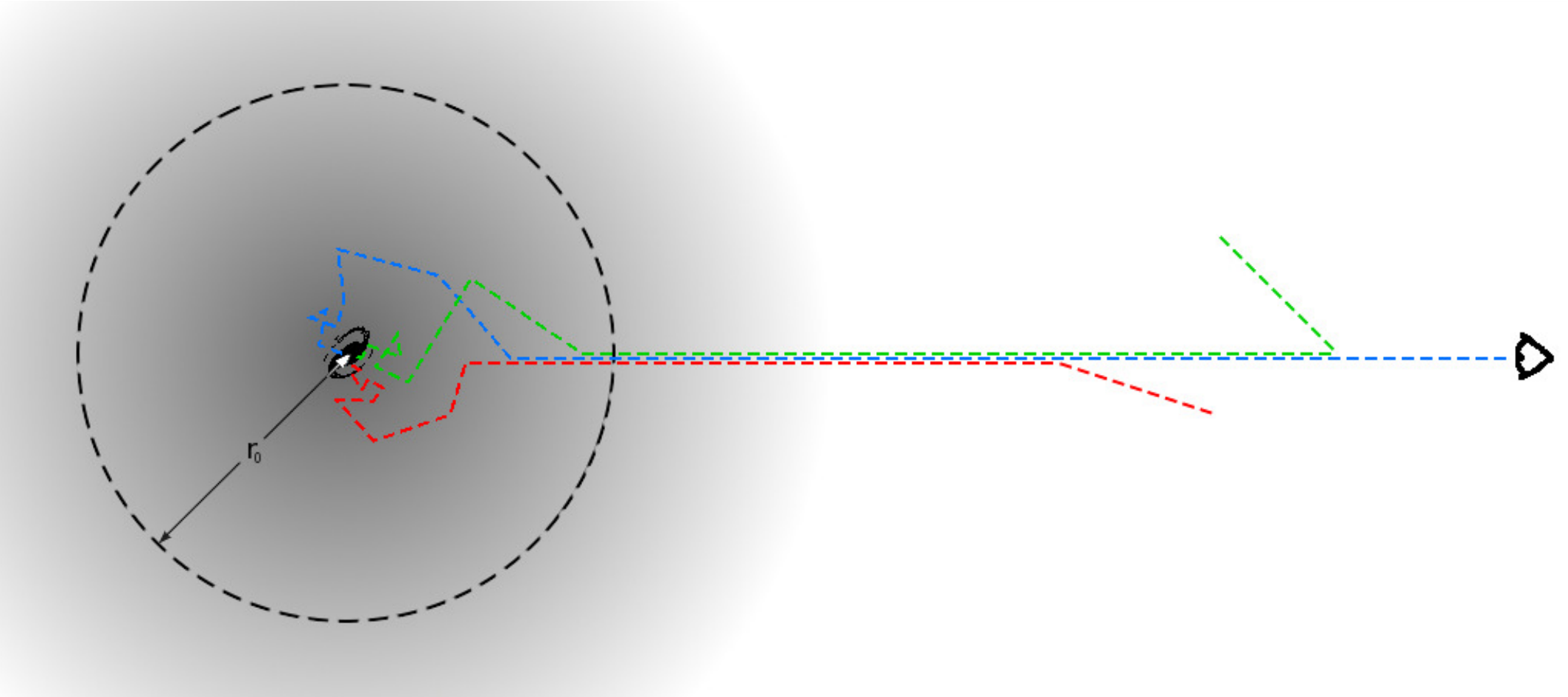}
\caption{{\cap Illustration of the difference between the galactic RT and the
               IGM RT. Close to the galaxy, photons are scattered both in and
               out of the line of sight. In the rarefied IGM, photons are
               mainly scattered out of the line of sight, obviating the need
               for a full MC RT. The exact value of $r_0$ is somewhat arbitrary,
               but is of the order of the virial radius of the galaxy.}}
\label{fig:r0}
\end{figure}

Although no formal definition of the transition from a galaxy to the IGM
exists, we have to settle on a definition of where to begin the sightlines,
i.e.~the distance $r_0$ from the center of a galaxy.
Observed Ly$\alpha$ profiles result from scattering processes in first the
galaxy and subseqently the IGM, and regardless of the chosen value of $r_0$,
for consistency galactic Ly$\alpha$ RT should be terminated at the same
value when coupling the two RT schemes.
In view of the above discussion on scattering in and out of the line of sight,
the sightlines should begin where photons are mainly scattered out of the line
of sight, and only a small fraction is scattered into the line of sight (see
\fig{r0}).

Since more neutral gas is associated with larger galaxies, as well as with
higher redshift, clearly $r_0$ depends on the galaxy and the epoch, but
measuring $r_0$ in units of the virial radius $r_{\mathrm{vir}}$
helps to compare the physical conditions around different galaxies.
In \sec{F} it is argued the a reasonable value of $r_0$ is
$1.5 r_{\mathrm{vir}}$, and showed that the final results are only mildly
sensitive to the actual chosen value of $r_0$.

\begin{figure}[!t]
\centering
\includegraphics [width=0.5\textwidth] {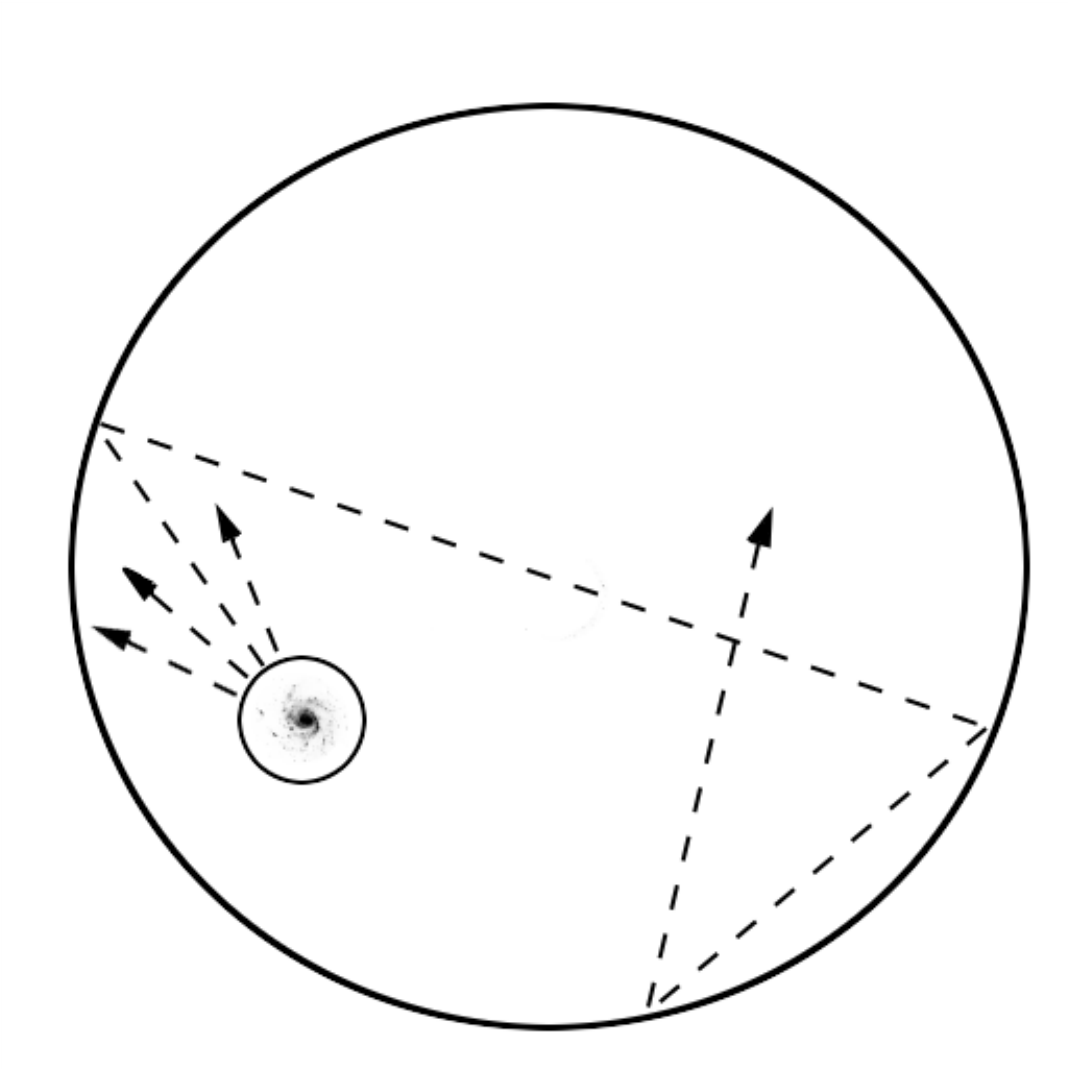}
\caption{{\cap Illustration of how the sightlines are cast through the
               cosmological volume. To sample sufficiently the full solid angle
               of $4\pi$ around the individual galaxies, $10^4$ sightlines
               are cast from each galaxy (of which four are shown here).
               Each sightline is started at a distance $r_0$ from the center of
               a galaxy and followed until the bluest wavelength of the
               emitted spectrum has been redshifted into resonance. When the
               edge of the spherical volume reached, the ray ``bounces''
               back, i.e.~continues in a random inward angle.}}
\label{fig:bounce}
\end{figure}

At a given redshift, $\Flam$ is calculated as the median of in each wavelength
bin of $10^3$
sightlines from each galaxy in the sample (using $10^4$ sightlines produces
virtually identical results). The number of galaxies amounts to
several hundreds, and increases with time.
To avoid any spuriosities at the edge of the spherical volume only a
$0.9 D_{\mathrm{box}}$ sphere is used. Due to the
limited size of the volume, in order to perform the RT until a sufficiently
short wavelength is redshifted into resonance the sightlines are allowed to
``bounce'' within the sphere, such that a ray reaching the edge of the sphere
is re-``emitted'' back in a random angle in such a way that the total volume is
equally well sampled (see \fig{bounce}).
Note however that in general the wavelength region close 
to the line that is affected by the correlation of the IGM with the source is
reached well before the the first bouncing.

The normalized spectrum is emitted at rest wavelength in the reference frame of
the
center of mass of a galaxy, which in turn may have a peculiar velocity relative
to the cell at which it is centered. This spectrum is then Lorentz transformed
between the reference frames of the cells encountered along the line of sight.
Since the expansion of space is approximately homologous, each cell can be
perceived as lying
in the center of the simulation, and hence this bouncing scheme does not
introduce any bias, apart from reusing the same volume several times for a given
sightline. However, since the sightlines scatter around stochastically and thus
pierce a given region from various directions, no periodicities arise in the
calculated spectrum.

To probe the average transmission, the sightlines are propagated until the
wavelength 1080 {\AA} has been redshifted into resonance, corresponding to
$\Delta z \simeq 0.1$. In this case the sightlines bounce roughly 30 (20) times
at $z = 2.5$ (6.5).



\partL{Scientific achievements}\label{science}

\chapter[On the consequences of resonant scattering]
        {On the consequences of resonant scattering\footnote{This chapter
        is based on \citetalias{lau09a} (\app{lau09a}).}}
\label{cha:conseq}

\init{T}{he original motivation} for the development of {\sc MoCaLaTA} was to
test whether
resonant scattering of Ly$\alpha$ photons could explain the frequent
observations that LAEs, when observed in Ly$\alpha$, seem to be more extended
on the sky compared to continuum band observations. This was mentioned in
\sec{moti}, and \fig{LEGO} showed an example of such an observation.

In most of the following sections, we will mainly focus on a particular
galaxy, namely the one poetically dubbed ``K15'', at a fiducial redshift of
 $z = 3.6$, but other galaxies, as well as other epochs, will also be
considered. The numerical and physical properties of all galaxies was showed in
\tab{num} and \tab{phy}, respectively.

The galaxy K15 is rather large, and would show up in surveys as both an LBG and
an LAE. It evolves into a Milky Way/M31-like disk galaxy at $z = 0$.
K15 consists of two small,
star-forming, disk-like structures separated by a few kpc, on one of which the
computational box is centered, and a third more extended disk, but of lower SFR,
located about 15 kpc from the center. We will refer to the two small disks as
the principal emitter, and to the more extended disk as the secondary emitter.
Additionally, the star-forming
regions are embedded in a significant amount of more diffuse, non-star-forming,
\ion{H}{i} gas in a $\sim10$--15 kpc
thick, sheet-like structure, taken to constitute the $xy$-plane.

To separate the effects of dust from the effects of pure resonant scattering,
the simulations in this chapter were performed without the effects of dust.
In \cha{dusteffect} we proceed to include dust in the calculations and discuss
the extend to which extend the conclusions of this chapter are valid.

\section{Extended surface brightness}
\label{sec:SBext}

\begin{figure}[!t]
\centering       
\includegraphics [width=0.99\textwidth] {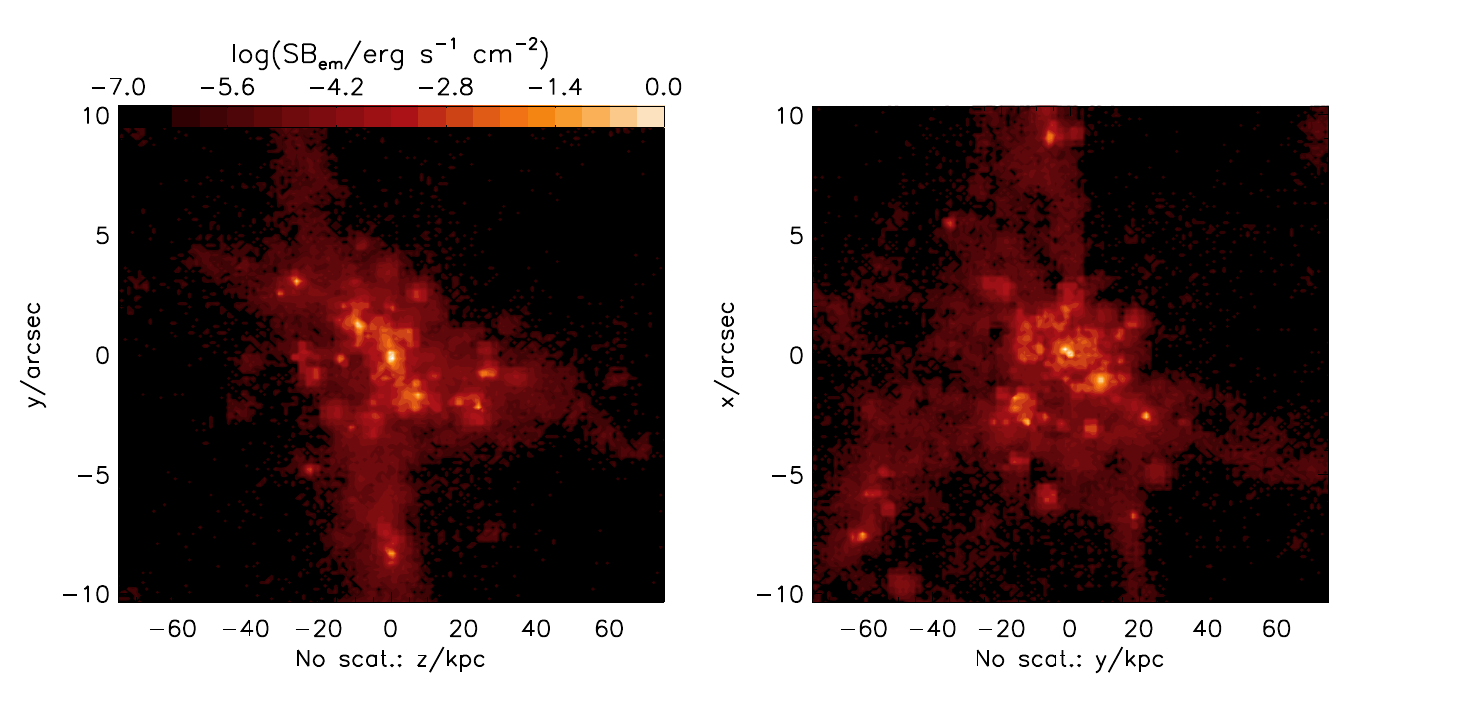}

\includegraphics [width=0.99\textwidth] {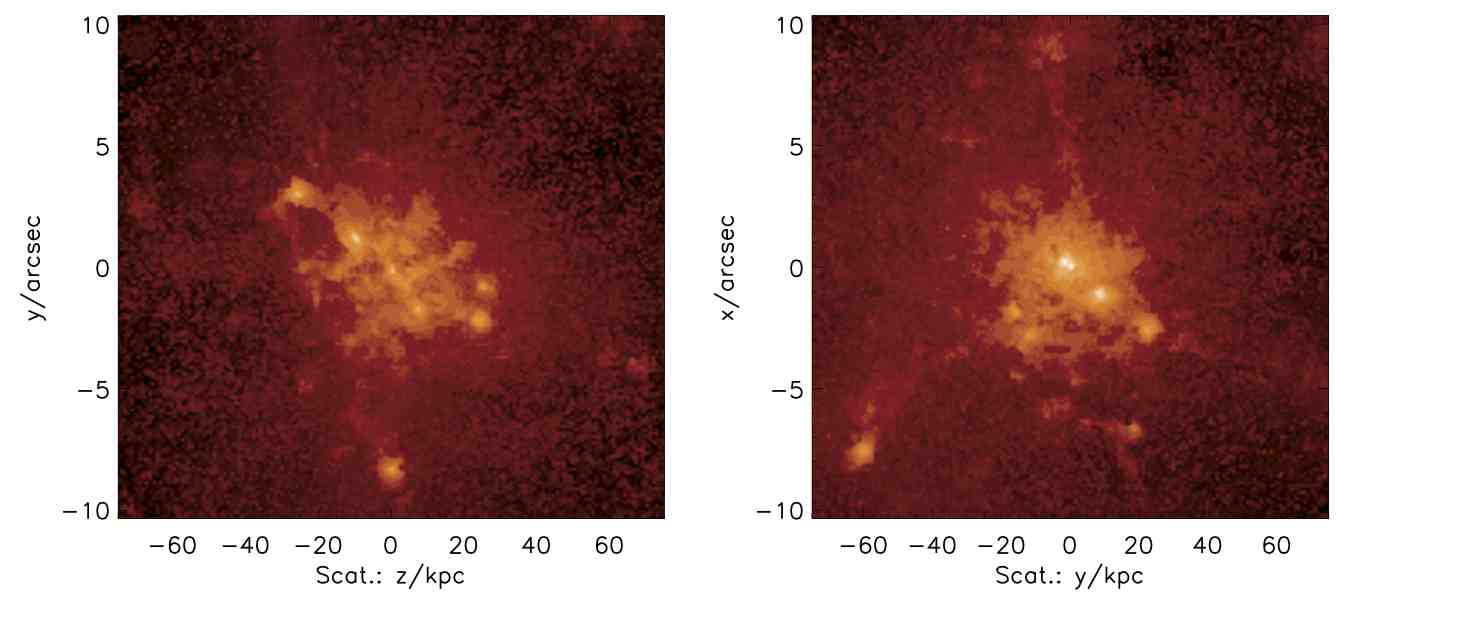}
\caption{{\small SB map of the simulated galaxy K15,
                 lying at a redshift of $z = 3.6$.
                 Left and right column show the system when viewed edge-on
                 and face-on, respectively. The top panels display the galaxy
                 as if the Ly$\alpha$ radiation was able to escape directly,
                 without scattering. The bottom panel shows the effect of the
                 scattering. For this simulation, $\sim10^7$ photons
                 was used, a good deal more than needed for acceptable
                 convergence.}}
\label{fig:SBmap}
\end{figure}

Fig.~\ref{fig:SBmap} shows how K15 would look, viewed from two different
directions --- from the negative $x$- and $z$-direction,
corresponding to an ``edge-on'' and a ``face-on'' view of the sheet-like
structure,
respectively. To emphasize the result of treating resonant scattering
properly, the galaxy is also shown as if the gas were optically thin to the
Ly$\alpha$ radiation.

The effect of the scattering is incontestable: although the original
constellation of the dominant emitters is still visible, the SB distribution is
clearly much more extended.
Certain regions can be quite bright in Ly$\alpha$ even though they have no
intrinsic emission. This astonishing result establishes the importance of
treating the scattering processes properly.

\begin{figure}[!t]
\centering
\includegraphics [width=1.0\textwidth] {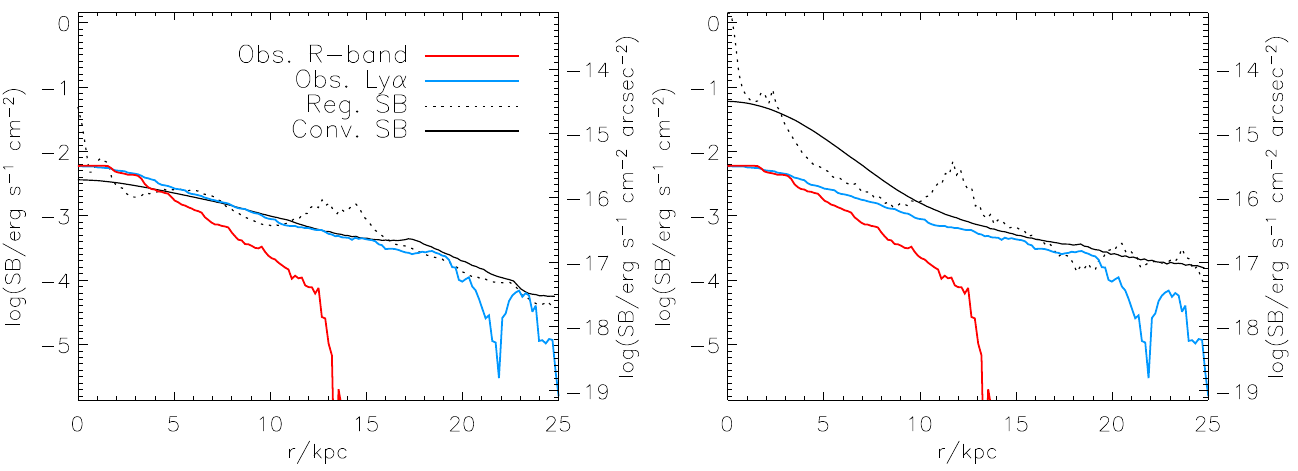}
\caption{{\small SB profiles of the simulated galaxy K15
                 when viewed edge-on (\emph{left}) and face-on (\emph{right}).
                 \emph{Dotted} lines show the true SB profiles, while
                 \emph{solid} lines show the SB profiles of the images convolved
                 with a seeing of $0\farcs8$ and omitting the luminosity of the
                 remote emitters.
                 Also shown are the SB profiles of the galaxy LEGO2138\_29
                 \citep{fyn03}
                 in Ly$\alpha$ (\emph{blue}) and in the $R$-band (\emph{red},
                 normalized to the maximum observed Ly$\alpha$ SB).
                 In particular the SB of the $yz$-plane nicely reproduces the
                 observed SB. Left axes measure the SB at the source, while
                 right axes  measure the SB observed at $z = 0$.}}
\label{fig:SBprof}
\end{figure}

\Fig{SBprof} shows the azimuthally averaged
SB profiles. To allow for a more direct comparison with observations, the
profiles are shown smoothed with a point spread function corresponding to
a seeing of $0\farcs8$ and excluding
the luminosity of the secondary emitter.

The profiles in Fig.~\ref{fig:SBprof} match the observed Ly$\alpha$ SB of the
galaxy LEGO2138\_29 \citep{fyn03} beautifully. It is worth
noting that the two
galaxies --- the simulated and the observed --- were not chosen to match each
other, although LEGO2138\_29 \emph{was} chosen due its superior extendedness in
Ly$\alpha$.


\section{Broadened spectrum}
\label{sec:spec}

\begin{figure}[!t]
\centering
\includegraphics [width=0.60\textwidth] {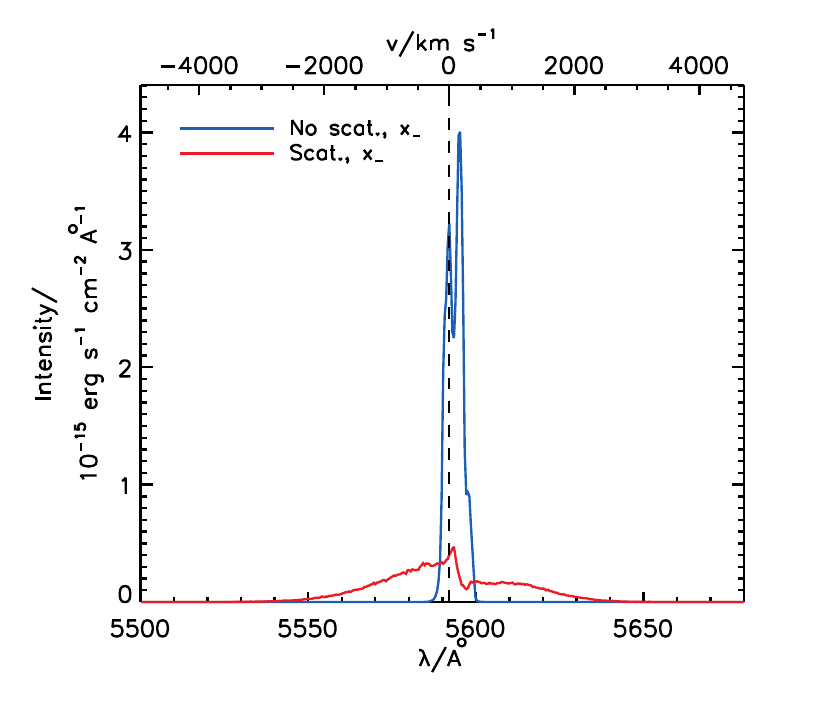}
\caption{{\cap Spectrum of the galaxy K15 as if the radiation were able to
         escape directly (\emph{blue}), and the realistic spectrum of the
         scattered photons (\emph{red}).
         Scattering broadens the spectrum by many {\AA}ngtr\"oms.
         Moreover, due to the fact that the hydrogen cross section is so large
         for photons in the line center, the spectrum is split up into two
         peaks.}}
\label{fig:K15spec}
\end{figure}

Not only the SB profile is smeared out, but also the spectrum is broadened by
resonant scattering.
\Fig{K15spec} shows the general effect of the scattering on the spectrum:
while the only broadening of the input spectrum
visible is due to the bulk motion of the gas elements emitting the photons
(both the natural and usually also the thermal broadening being much smaller),
the scattered spectrum is severely
broadened, diminished by an order of magnitude, and split up into two
peaks.

The double-peak profile seen in \fig{K15spec} is characteristic of
Ly$\alpha$ emission lines;
the high opacity for photons near the line center makes
diffusion to the either side necessary in order to escape the galaxy.
Nonetheless, unlike in previous simple models,
the intensity in the line center is not zero.
The photons that contribute to this intensity are those produced
mainly by gravitational cooling, in the outskirts of the systems.

\begin{figure}[!t]
\centering
\includegraphics [width=1.00\textwidth] {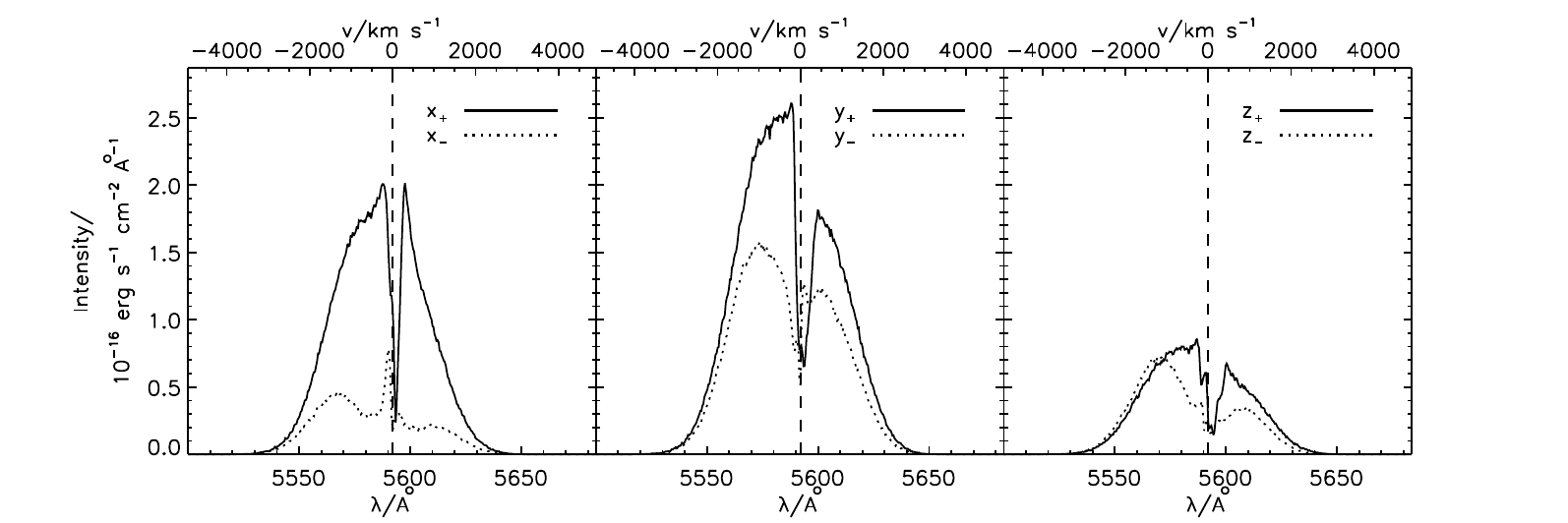}
\caption{{\cap Spectral distribution of the photons escaping the galaxy K33 in
         six
         different directions; along the positive (+) and negative ($-$) $x$-,
         $y$-, and $z$-direction. The dashed line in the middle of each plot
         indicates the line center. The lower abscissa gives the redshifted
         wavelength of the photons while on the upper abscissa, the wavelength
         distance from the line center is translated into recessional velocity.
         The resonant scattering of Ly$\alpha$ is seen to broaden the line by
         several thousands of km s$^{-1}$.}}
\label{fig:spec33}
\end{figure}

\begin{figure}[!t]
\centering
\includegraphics [width=1.00\textwidth] {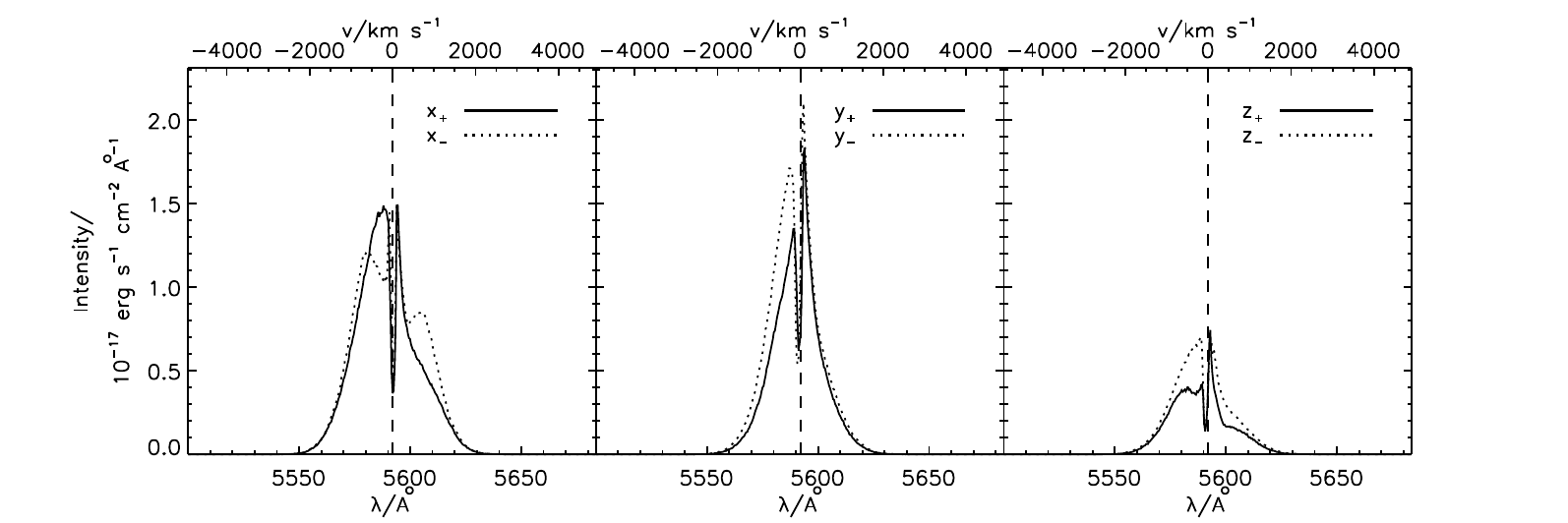}
\caption{{\cap Same as Fig.~\ref{fig:spec33}, but for the galaxy S115.}}
\label{fig:spec115}
\end{figure}

Figures \ref{fig:spec33} and \ref{fig:spec115} display the spectra emerging
from galaxies K33 and S115 at $z = 3.6$, in six different directions.
The exact shape varies quite a lot, but all spectra appear to exhibit
the double-peak profile. Moreover, they are broadened by several
1000 \kms. Although such broad LAE line profiles are indeed
observed, somewhat narrower profiles of several hundreds to
$\sim$1000 \kms are more typical.
In \cha{dusteffect} we will se how this can be explained
by the presence of dust.

Double-peaks\index{Double-peaks} have been observed on several occasions
\citep[e.g.,][]{yee91,ven05}.
\citet{tap07}, using a resolution of $R \sim 2000$, found three out of 16
LAEs at redshifts $z \sim 3$--$4$ to exhibit a double-peaked profile,
while Yamada et al. (priv.~comm.), using $R \sim 1500$ found that
26 of 94 LAEs at redshift 3.1 have double-peaked profiles.
The difference in magnitude of the two peaks can be a signature of
infalling/outflowing gas, cf.~\sec{Vbulk}.
In principle, this difference may be used as a probe of the gas dynamics, and
has indeed been used to infer the presence of galactic superwinds.
However, as shown in \cha{IGM},
the removal or diminishing of the blue peak might also be caused by
IGM resonant scattering.

Even if the double-peak survives intergalactic transmission, fairly high
resolution is required to be able to resolve them.
With a typical separation $\Delta\lambda$ of the peaks of the simulated spectra
from a few to $\simeq10$ {\AA} (at $z \sim 3.6$), the resolution must be
$R = 5600 / \Delta\lambda \simeq 500$--$2000$.

It is interesting that while the Ly$\alpha$ profile of K15 also shows
a moderate gas infall, we see some outflow signatures in the
spectrum of the lower-mass galaxy S115 from negative $x$- and $y$-directions.
Fitting a Neufeld, or Dijkstra,
profile to the observed spectra can give us an idea of
the intrinsic properties of the system. Unfortunately, due to the
degeneracy between column density and temperature, one would have to gain
knowledge either of the parameters by other means to constrain the other
(e.g.~by inferring column density from the spectrum of a coincident background
quasar, or by assuming a temperature of, say, $10^4$ K, representative of most
of the Ly$\alpha$ emitting gas).


\section{Anisotropic escape}
\label{sec:anis}

The SB profiles and the spectra presented in the previous sections reveal
another interesting property of
LAEs, namely that the radiation in general does not escape isotropically.
Qualitatively, we expect the photons to escape more easily in directions where
the column density of neutral gas is lower.
As mentioned already, K15 is embedded in a
sheet-like structure, lying at the intersection of three filaments of gas.
Since the bulk of the photons are produced in the central, star-forming
regions, the total optical depth is larger in the direction parallel
to the sheet than perpendicular to it, and hence we would expect the photons to
escape more easily in the face-on direction.
Similarly, K33 is situated in a filament of
gas, taken to lie along the $z$-axis. Here, we would expect the photons to
escape more easily in the $x$- and $y$-directions.

\begin{figure}[!t]
\centering
\includegraphics [width=1.00\textwidth] {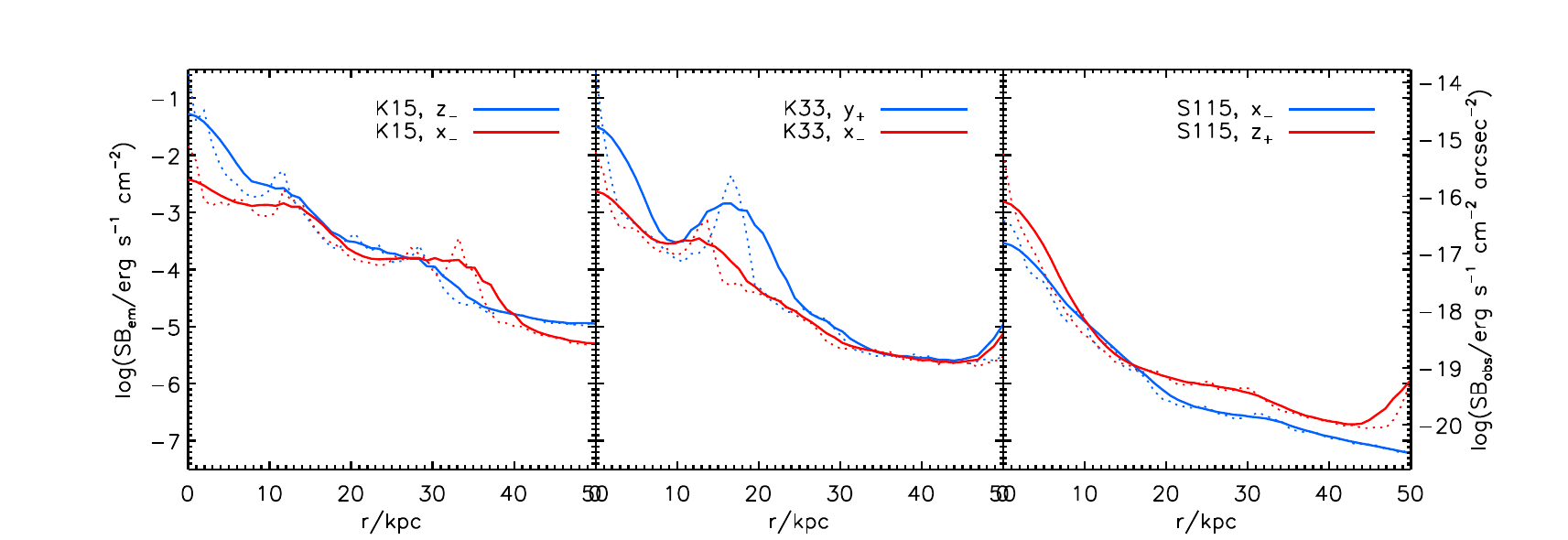}
\caption{{\cap SB profiles for the inner 50 kpc of the three galaxies K15
         (\emph{right}), K33 (\emph{middle}), and S115 (\emph{left}). For each
         galaxy, the SB is showed as observed from two different angles
         (\emph{blue} and \emph{red}, respectively). Both
         the ``true'' SB profiles (\emph{dotted})
         and the profiles of the SB convolved with a seeing of $0\farcs8$
         (\emph{solid})
         are shown. Left $y$-axis gives the SB that would be seen by an
         observer at the location of the galaxies, while right $y$-axis gives
         the SB as observed from Earth.}}
\label{fig:SBprofiles}
\end{figure}

Here this anticipation is quantified.
Averaging the SB in the azimuthal direction, the SB profiles of the three
galaxies, each as viewed from two different directions, are shown in
Fig.~\ref{fig:SBprofiles}, while Tab.~\ref{tab:SBmax} summarizes the observed
maximum surface brightnesses, SB$_{\mathrm{max}}$.
\begin{table}[!t]
\begin{center}
{\sc Maximum observed surface brightnesses\\ from different directions}
\end{center}
\centering
\begin{tabular}{lccc}
\hline
\hline
Galaxy & K15 & K33 & S115\\
\hline
log\,SB$_{\mathrm{max,}x_{+}}$ &         $8.6\times10^{-3}$  &         $6.2\times10^{-3}$  &         $1.3\times10^{-3}$  \\
log\,SB$_{\mathrm{max,}x_{-}}$ & $\mathit{3.7\times10^{-3}}$ & $\mathit{2.3\times10^{-3}}$ & $\mathbf{1.5\times10^{-3}}$ \\
log\,SB$_{\mathrm{max,}y_{+}}$ &         $9.9\times10^{-3}$  & $\mathbf{3.2\times10^{-2}}$ &         $7.6\times10^{-4}$  \\
log\,SB$_{\mathrm{max,}y_{-}}$ &         $1.2\times10^{-2}$  &         $2.1\times10^{-2}$  &         $1.2\times10^{-3}$  \\
log\,SB$_{\mathrm{max,}z_{+}}$ &         $4.9\times10^{-2}$  &         $3.4\times10^{-3}$  & $\mathit{2.9\times10^{-4}}$ \\
log\,SB$_{\mathrm{max,}z_{-}}$ & $\mathbf{5.3\times10^{-2}}$ &         $4.7\times10^{-3}$  &         $4.0\times10^{-4}$  \\
{\bf Max}/\emph{min}           &   14.3       &    13.6      &      5.3     \\
\hline
\end{tabular}
\caption{{\cap SBs are calculated from the images convolved
               with a seeing of $0\farcs8$ and are measured in \ergs \cmsq
               at the location of the galaxies.
               The maximally and the minimally observed SB$_{\mathrm{max}}$'s
               for a given galaxy are written in boldface and italics,
               respectively, and the ratios
               between these are given in the lower row.}}
\label{tab:SBmax}
\end{table}
In fact, regarding K33,
SB$_{\mathrm{max,}x_{-}}$ turns out to be smaller than both
SB$_{\mathrm{max,}z_{-}}$ and SB$_{\mathrm{max,}z_{+}}$, due to the presence in
the line of sight of hydrogen clouds with little star formation causing a
shadowing effect. As seen in Tab.~\ref{tab:SBmax}, the observed
SB of a given galaxy varies with viewing angle by approximately an
order of magnitude.

This result is intriguing in relation to the classification of galaxies.
As discussed in \cha{gals}, galaxies are commonly annotated according to the
method by which they are selected, and one
of the mysteries in the context of galaxy formation and evolution is the
connection between the different types.
As already mentioned, sufficiently high column densities of neutral hydrogen
in the line of sight toward a bright
background source give rise to broad absorption lines in their spectra, and
may be detected as the DLAs.
On the other hand, galaxies with high enough SFRs may be detected in
narrowband searches by an excess of their narrowband to continuum flux as
LAEs.

Both K15, K33, and S115 contain enough
neutral hydrogen to make them detectable as DLAs in the
spectra of hypothetical quasars.
More interestingly, the present results show that while their relatively high
SFRs may make at least K15 and K33
detectable as LAEs when viewed from a given direction, it may not be
possible to see them in Ly$\alpha$ from another direction; instead,
it may be possible to observe them as LBGs. This effect demonstrates how
galaxies selected by different means may be connected to each other.

Overlaps in the properties of LAEs and LBGs have also been inferred
observationally; \citet{gaw06b} found that more than $80$\% of a sample
of emission line-selected LAEs have the right \emph{UVB}-colors to be selected
as LBGs. The primary difference between the two populations is the
selection criteria, as only $\sim 10$\% are also brighter than the
$R_{\mathrm{AB}} < 25.5$ ``spectroscopic'' LBG magnitude cut.
Also, when correcting for dust, \citet{gro07} found comparable SFRs for the
two populations.

For high-redshift LAEs, SFRs are inferred almost exclusively from Ly$\alpha$
flux measurements, assuming isotropic luminosity. However, as is evident
from the above discussion, in general the complex morphology of a galaxy
may very well cause a preferred direction of photon escape. For
the three galaxies of the present study, the angular variation
in flux can be as high as a factor of 3.4, 6.2, and 3.3, for K15,
K33, and S115, respectively (here the flux is calculated by integrating
the SB maps over a region of radius $r = 25$ kpc, centered at $r = 0$).
Although not as pronounced as in the case of SB$_{\mathrm{max}}$,
this introduces a considerable source of uncertainty, which may propagate
into estimates of SFRs or into calculations
of the content of dust residing in galaxies.



\chapter[On the effects of dust]
        {On the effects of dust\footnote{This chapter is based on
        \citetalias{lau09b} (\app{lau09b}).}}
\label{cha:dusteffect}

\init{W}{hen dust is included} in the calculations,
the adaptive resolution of {\sc MoCaLaTA} really comes into its own.
In general, any clumpiness \index{Multi-phase medium}
of the gas elements will lower the effective optical depth. That this is true
can be seen from the following argument: taking the average over all lines of
sight, for a heterogeneous medium the effective optical depth is
\begin{equation}
\label{eq:tauhetero}
\tau_{\mathrm{het}} = -\ln \langle e^{-\tau} \rangle,
\end{equation}
while for a homogeneous medium $\tau_{\mathrm{hom}} = \langle \tau \rangle =$
constant. Since from the standard triangle inequality
\begin{equation}
\label{eq:triangle}
e^{-\langle \tau \rangle} \le \langle e^{-\tau} \rangle,
\end{equation}
it is seen that $\tau_{\mathrm{het}} \le \tau_{\mathrm{hom}}$.
This is true for any wavelength. For Ly$\alpha$, in the absence of dust,
increasing $\tau$, or decreasing clumpiness, does not increase absorption but
instead broadens the spectrum further. \emph{With} dust, however, the
situation becomes more complicated. Although the resonant scattering increases
the path length of the photons, in an inhomogeneous medium more
low-density paths exist through which the radiation can escape.

\section{The need for adaptive resolution}
\label{sec:need}

\begin{figure}[!t]
\centering
\includegraphics [width=1.10\textwidth] {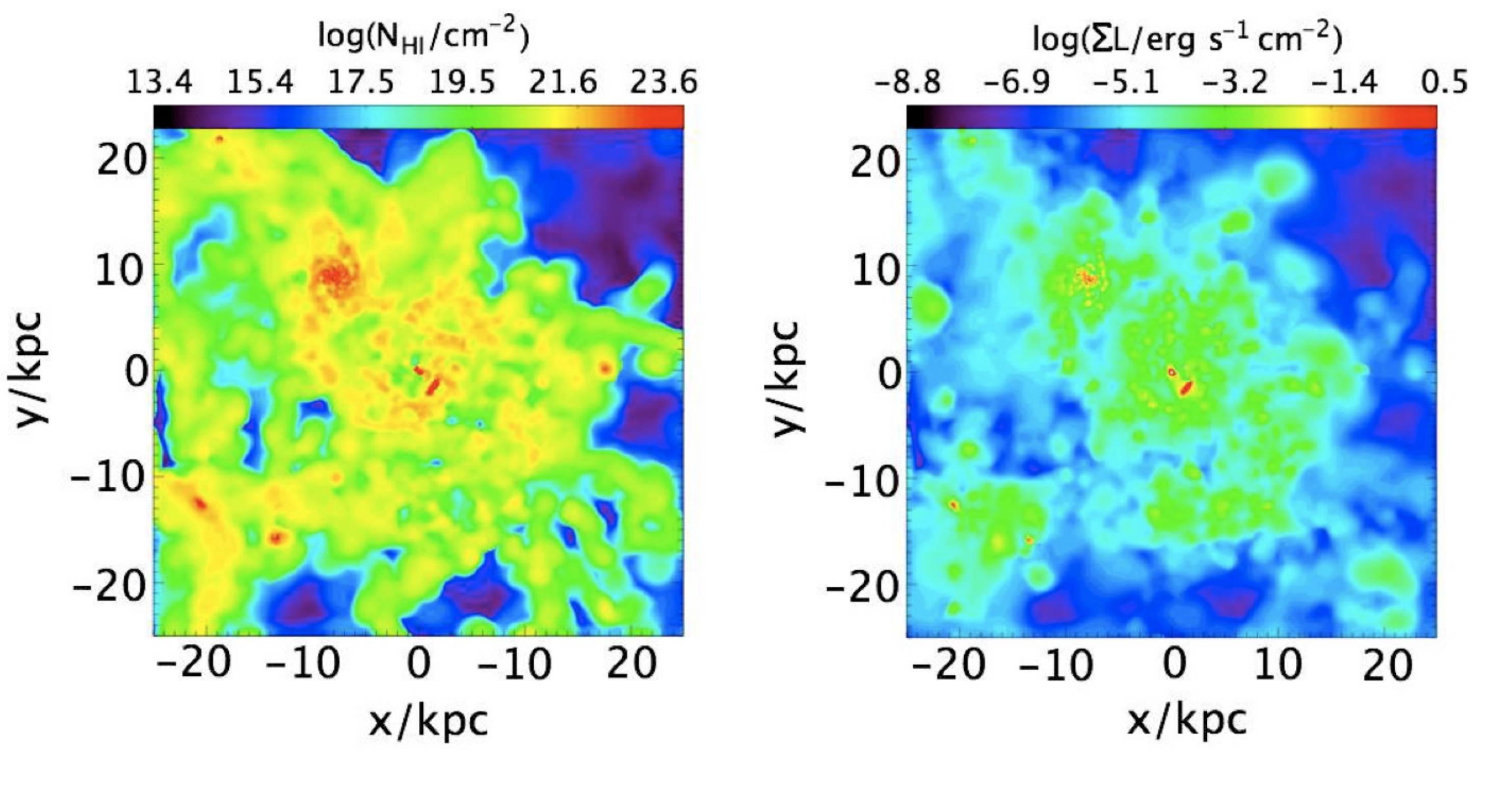}
\caption{{\cap Neutral hydrogen column density ($\Nhi$) map (\emph{left}) and
          integrated source Ly$\alpha$ emissivity
         ($\Sigma L$) map (\emph{right}) of the simulated galaxy K15.
         The vast majority of the photons are seen to be emitted in the very
         dense environments. According to the analytical solution for the
         Ly$\alpha$ escape fraction provided by
         \citet[][\eq{neufesc} in this paper]{neu90}, virtually all
         photons should be absorbed by dust, but taking into account the
         clumpiness of the ISM allows for much higher escape fractions.}}
\label{fig:NHI_L}
\end{figure}

Consequently, in simulations of Ly$\alpha$ RT with dust, high resolution is
rather crucial to accurately determine the escape of the photons.
As is evident from \fig{NHI_L},
the bulk of the photons is produced in regions of very high column densities,
exceeding $\Nhi = 10^{23}$ \cmsq.
Even for metallicities of only $1/100$ $Z_\odot$, \eq{neufesc}
implies an escape fraction of $\sim$$10^{-5}$ (for $T \sim 10^4$ K). However,
\eq{neufesc} assumes a \emph{homogeneous} medium. As was argued by
\citet{neu91} and investigated numerically by \citet{han06}, in a
multi-phase\index{Multi-phase medium}
medium, the Ly$\alpha$ photons may escape more easily. In the (academic) case
of all the dust residing in cool, dense clouds of neutral hydrogen which, in
turn, are dispersed in a hot, empty medium, the Ly$\alpha$ escape
fraction may approach unity. The reason is that the photons will scatter off of
the surface before penetrating substantially into the clouds, thus effectively
confining their journey to the dustless intercloud medium.

Although this scenario is obviously very idealized, the presence of an
inhomogeneous medium undoubtedly reduces the effective optical depth, and
has indeed been invoked to explain unusually large Ly$\alpha$ equivalent widths
\citep[e.g.][]{fin08}. Nevertheless, it is not clear to which degree the
escape fraction of Ly$\alpha$ will actually be affected.
Other scenarios have been proposed to explain the apparent paradoxal escape of
Ly$\alpha$, e.g.~galactic superwinds, as discussed in \sec{fesctheo}.
Generally, whether large or small a more or less universal escape fraction is
assumed.
This work presents the first calculations of $f_{\mathrm{esc}}$, based on fully
cosmological, numerical galaxy formation models, demonstrating that a wide
range of escape fractions are possible, and that a siginificant fraction may
escape the galaxies at
redshifts $z \sim 3$--4, even if no particularly strong winds are present.

Again, in the following we will first focus on K15, choosing the observer
placed toward negative values on the $z$-axis (the $z_-$-direction),
the direction in which most radiation escapes. In \sec{moregal} we
proceed to discuss the extend to which the results are representative for other
galaxies and directions.


\section{Where are the photons absorbed?}
\label{sec:where}

\begin{figure}[!t]
\centering
\includegraphics [width=1.10\textwidth] {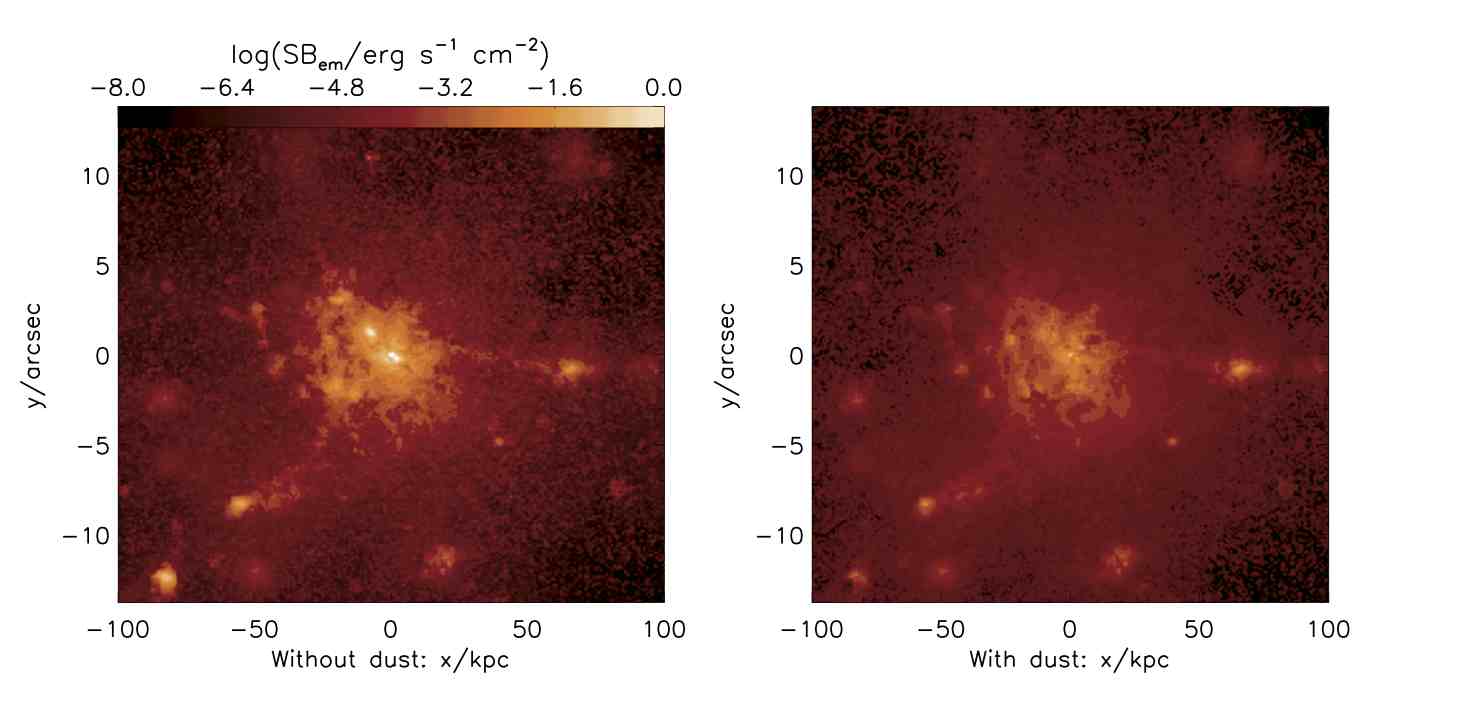}
\caption{{\cap SB maps of the galaxy K15, as viewed from the
         negative $z$-direction, without dust (\emph{left}) and with dust
         (\emph{right}). Including dust in the radiative transfer affects
         primarily the most luminous regions.}}
\label{fig:SBmapwd}
\end{figure}

\begin{figure}[!t]
\centering
\includegraphics [width=0.70\textwidth] {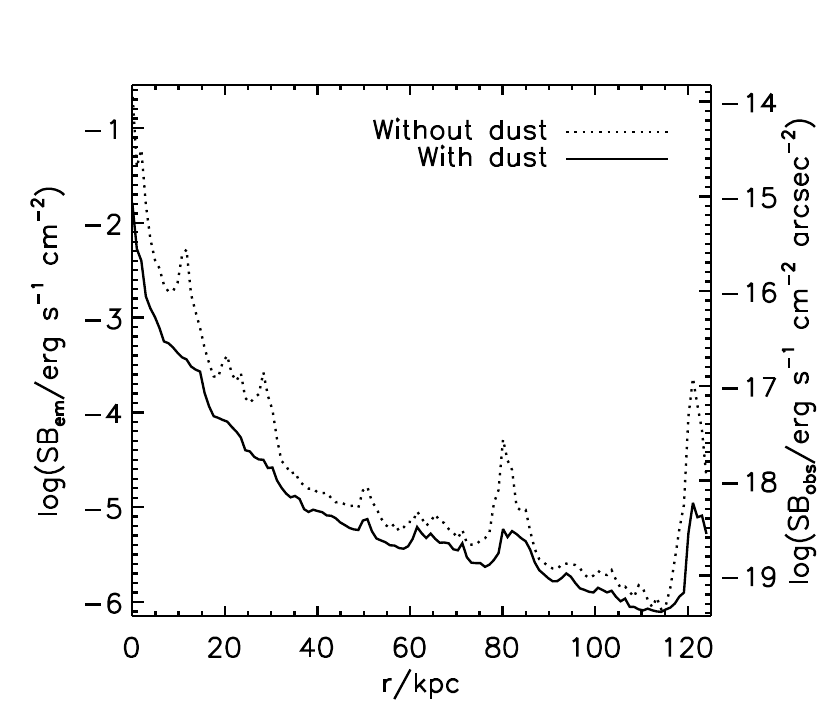}
\caption{{\cap SB profile of the galaxy K15, again without dust
         (\emph{dotted}) and with dust (\emph{solid}).
         Left ordinate axis gives the SB as measured at the source while right
         ordinate gives the values measured by an observer at a distance given
         by the luminosity distance of the galaxy.
         The decrease in SB in the less luminous regions
         is noticeable. However, this decrease is for the most part \emph{not}
         due to photons being
         absorbed in the hot and tenuous circumgalactic medium but rather
         reflects a lack of photons that in the case of no dust would have
         escaped the luminous regions and subsequently scattered on neutral
         hydrogen in the direction of the observer.}}
\label{fig:SBprofwd}
\end{figure}

\Fig{SBmapwd} shows the SB maps of the galaxy K15.
The left panel shows how the galaxy would look if the gas were dust-free,
while the right panel displays the more realistic case of a dusty medium.
Comparing the two images, it is seen that the regions that are
affected the most by dust are the most luminous regions. This is even more
evident in \fig{SBprofwd}, where the azimuthally averaged profiles of
the SB maps are shown.

The reason for this is two-fold: the most luminous regions are the regions
where the stars reside. Because dust is produced by stars, this is also where
most of the dust is. Since the stars, in turn, are born in regions of high
hydrogen density, the RT is here associated with numerous
scatterings, severely increasing the path length of the photons, and
consequently increasing further the probability of being absorbed.

\begin{figure}[!t]
\centering
\includegraphics [width=0.60\textwidth] {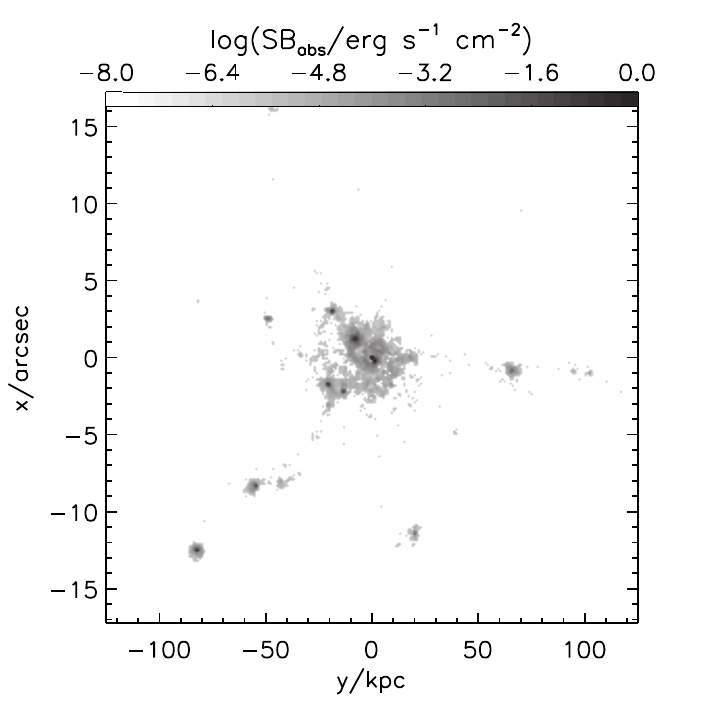}
\caption{{\cap Image of the locations of absorption of the Ly$\alpha$ radiation.
         This image corresponds roughly to the column density of dust, although
         dusty \ion{H}{ii} regions may not show up as clearly as dust-deficient
         \ion{H}{i} regions.}}
\label{fig:IR}
\end{figure}

The SB in the less luminous regions also decreases. Although dust also resides
here, having been expelled by the feedback of starbursts, only little
absorption actually takes place here because both gas and dust densities are
rather small. The decrease in luminosity is mainly
due to the photons being absorbed in the high-density regions that would
otherwise have escaped the inner regions and subsequently been scattered by the
circumambient neutral gas into the direction of the observer. This is also
discernible from \fig{IR} that displays an image of the absorbed photons.

\begin{figure}[!t]
\centering
\includegraphics [width=0.45\textwidth] {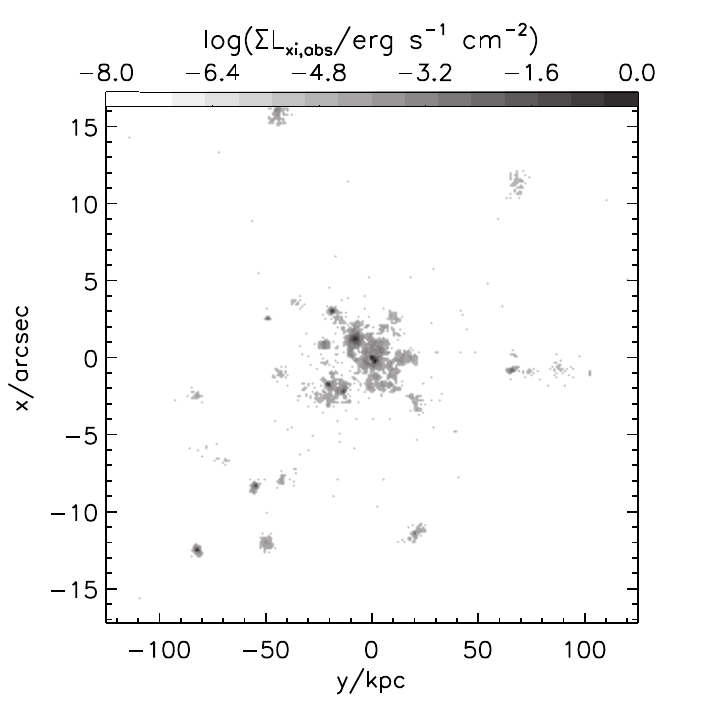}
\includegraphics [width=0.45\textwidth] {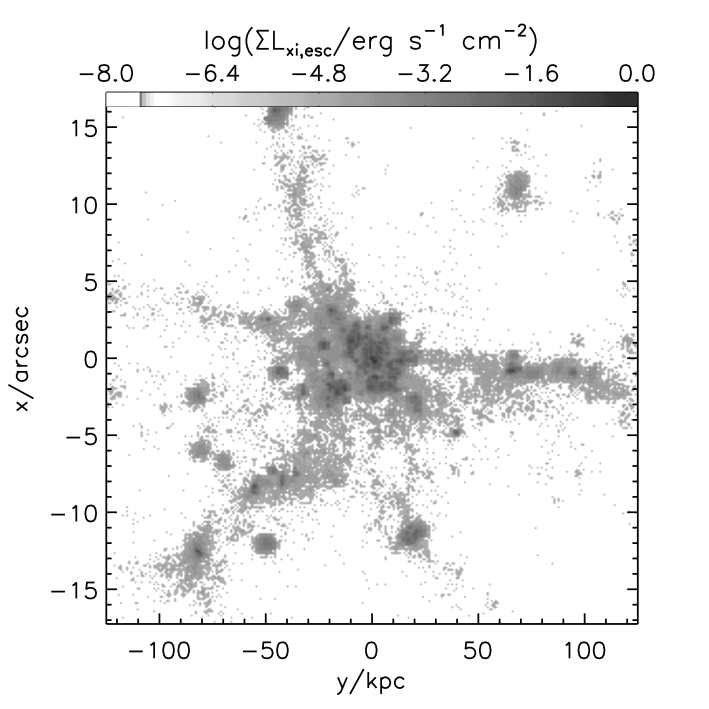}
\caption{{\cap Surface density maps of the source Ly$\alpha$ emissivity, of
         the photons that are eventually absorbed (\emph{left}), and the photons
         that eventually escape (\emph{right}). These images clearly show that
         absorbed and escaping photons do not in general probe the same
         physical domains.}}
\label{fig:Xi}
\end{figure}

\Fig{Xi} shows the \emph{source} Ly$\alpha$ emissivity of the
photons that are eventually absorbed, compared to that of the photons that
eventually escape. Here it becomes evident that virtually all the absorbed
photons are emitted from the central parts, while photons that escape are
emitted from everywhere. In particular, the radiation produced through
gravitational cooling escapes more or less freely.
%


\section{Effect on the emergent spectrum}
\label{sec:specdust}

\begin{figure}[!t]
\centering
\includegraphics [width=0.70\textwidth] {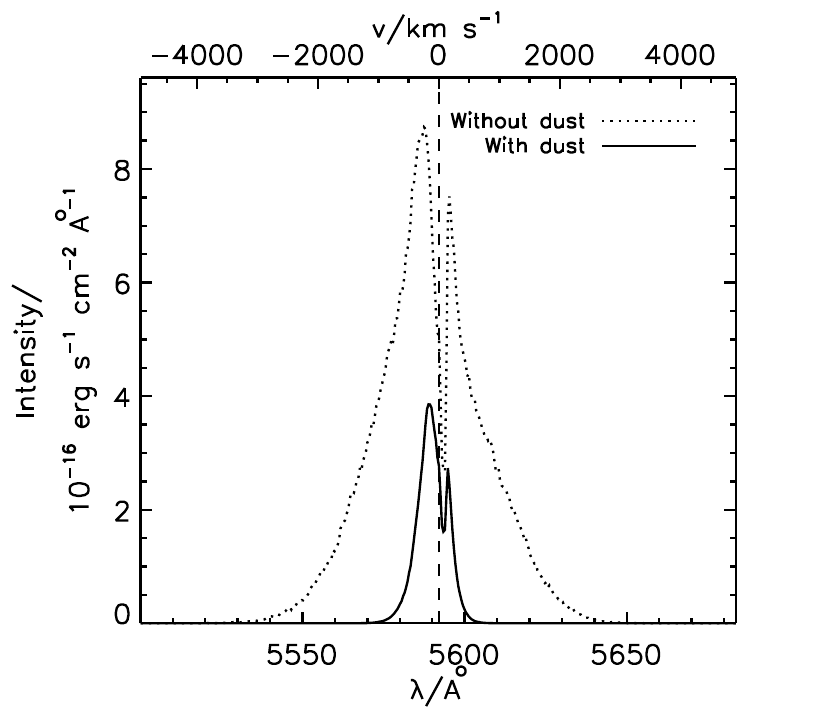}
\caption{{\cap Observed spectral distribution of the radiation escaping the
         galaxy
         K15 in the negative $z$-direction, with (\emph{solid}) and without
         (\emph{dotted}) dust. The vertical, dashed line marks the Ly$\alpha$
         line center. Although the dust is close to being gray, the
         spectrum is not affected in the same way at all wavelengths; the inner
         parts are only diminished by a factor of $\sim$2, while the wings are
         severely reduced. The reason is that the photons in the wings are the
         ones produced in the dense and dusty regions of the galaxy, so these
         photons have a higher probability of being absorbed.}}
\label{fig:specK15wd}
\end{figure}

Due to the high opacity of the gas for a Ly$\alpha$ photon at the line center,
photons generally  diffuse in frequency to either the red or the blue side in
order to escape.
Consequently, the spectrum of the radiation escaping a
dustless medium is characterized by a double-peaked profile. The broadening of
the wings is dictated by the product $a\tau_0$ (\eq{neufeld}), i.e.~low
temperatures and, in particular, high densities force the photons to diffuse
far from line center. Since such conditions are typical of the regions where
the bulk of the photons is absorbed, the emergent spectrum of a dusty medium
is severely narrowed, although the double-peaked feature persists.
Figure \ref{fig:specK15wd} displays the spatially integrated spectra of the
dustless and the dusty version of K15.

This interesting result shows that even though the dust is effectively gray
(the small wavelength dependence of \eq{sigd}) does not produce
substantially different results from using a completely wavelength
\emph{in}dependent
cross section), the emergent spectrum is affected in a highly ``non-gray''
fashion: 
whereas the escape fraction of the inner part of the spectrum is of the order
50\%, it rapidly drops when moving away from the line center.


\section{Escape fraction}
\label{sec:fescres}

In the $z_-$-direction of K15, the fraction of Ly$\alpha$ photons escaping is
0.14, when comparing to the dustless version of K15.
As mentioned earlier, the $z_-$-direction is the direction into which most
radiation escapes --- without dust, this direction is $\sim$3 times as
luminous as the $x_-$-direction, which is where the least radiation is emitted.
Including dust, because the brightest regions are affected the most, the ratio
between the luminosity in the least and the most luminous directions is
somewhat reduced, although $z_-$ is still more than twice as bright as $x_-$.

The sky-averaged escape fraction for K15 is 0.16.
Recall that in these simulations SMC dust
has been applied. However, as discussed in \sec{parstud} using LMC dust
does not alter the results significantly.

Many factors regulate the probability of escaping, and stating
$f_{\mathrm{esc}}$ for a single galaxiy is not very elucidative. In the next
section we investigate the correlation of $f_{\mathrm{esc}}$ with galaxy size.


\section{General results}
\label{sec:moregal}

The results found in the previous sections turn out to be quite illustrative
of the general outcome of Ly$\alpha$ RT in a dusty medium: photons are absorbed
primarily in the dense, luminous regions, leading to a reduced luminosity in
these parts of the galaxies and effectively smoothing out prominent features.
Furthermore, the wings of the spectrum experience a strong cut-off.

\subsection{Anisotropic escape of Ly$\alpha$}
\label{sec:aniswd}

As found in \sec{anis}, in general the radiation does \emph{not} escape
isotropically; the ratio of
luminosities observed from different directions ranges from $\sim$1.5 to
$\sim$4. Without dust, these ratios were found to be somewhat higher, up to a
factor of $\sim$6.
Although ``bright directions'' are affected more by the dust than less bright
directions, the variation in $f_{\mathrm{esc}}$ as a function of direction is
not large, and not very different from the sky-averaged $f_{\mathrm{esc}}$.


\subsection{Correlation of $f_{\mathrm{esc}}$ with galactic mass}
\label{sec:fesc_M}

\begin{figure}[!t]
\centering
\includegraphics [width=0.70\textwidth] {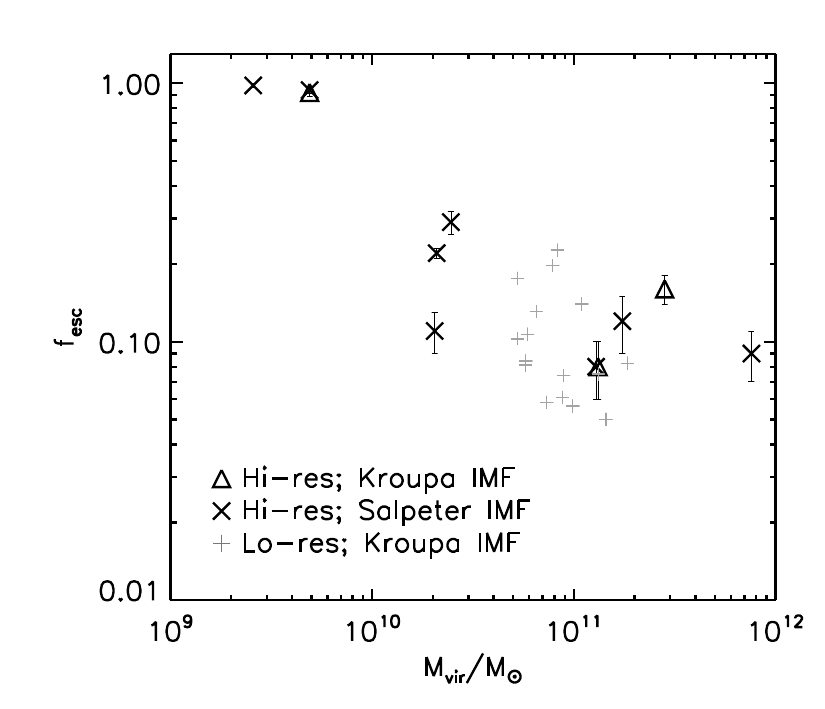}
\caption{{\cap Escape fractions $f_{\mathrm{esc}}$ as a function of galactic
         virial mass $M_{\mathrm{vir}}$.
         Errorbars denote the differences in $f_{\mathrm{esc}}$ in different
         directions.  Although the plot exhibits a large scatter,
         there is a clear trend of $f_{\mathrm{esc}}$ decreasing with
         increasing $M_{\mathrm{vir}}$.}}
\label{fig:fesc_M}
\end{figure}

\Fig{fesc_M} shows the escape fractions of the galaxies as a
function of the virial masses of the galaxies. Despite a large scatter, and
although the sample is too small to say anything definite, the figure indicates
that $f_{\mathrm{esc}}$ decreases with increasing mass of the host galaxy.
In addition to the nine high-resolution galaxies (of which two appear in
two ``versions''; with a \citeauthor{sal55} and with a \citeauthor{kro98}
IMF), 17 galaxies from the (Kroupa) cosmological simulation are shown.
The mass resolution of these galaxies is thus 8$\times$ lower. To ensure an
acceptable resolution, the galaxies were chosen according to the criterion that
the number of star particles is $\ge$1000.
Similar results are found for the escape of ionizing UV radiation
\citep{raz09}. The reason is probably a combination of two mechanisms: small
galaxies have a lower SFR and hence lower metallicity and less dust than large
galaxies.
Furthermore, due to their smaller gravitational potential the stellar feedback
will ``puff up'' small galaxies relatively more and make them less ordered,
thus reducing the column density of both dust and gas.

This effect might explain the results found by \citet{vij03}, who studied the
effects of dust in a sample of $>900$ LBGs at $2 > z > 4$:
they found that the FUV continuum (at 1600 {\AA}) dust attenuation increases
with the UV luminosity of the galaxies.

%


\subsection{Narrowing of the spectrum}
\label{sec:narrow}

\begin{figure}[!t]
\centering
\includegraphics [width=1.10\textwidth] {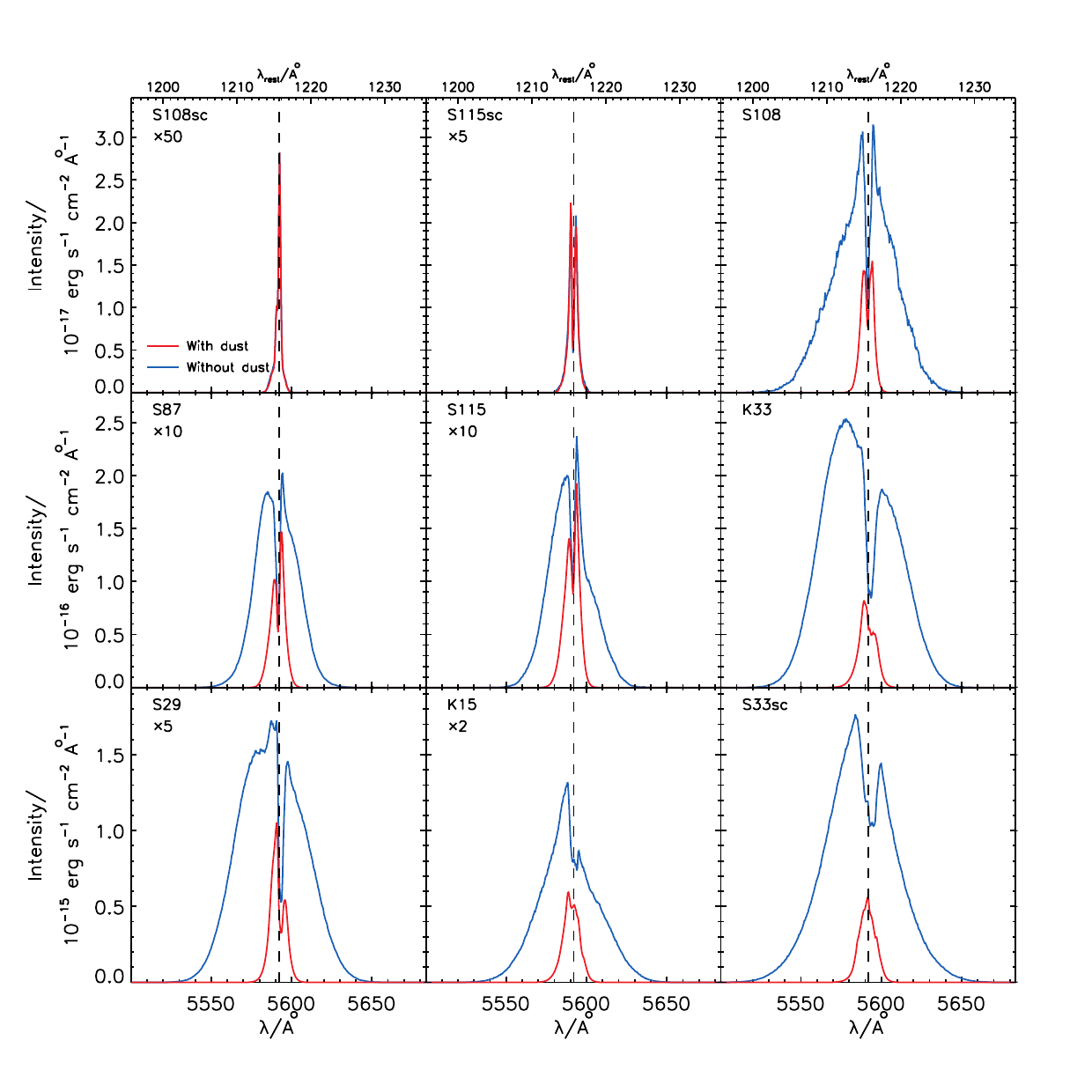}
\caption{{\cap Emergent spectra of the studied galaxies, without (\emph{blue})
         and with dust (\emph{red}), ordered after decreasing virial mass.
         In order to use the same ordinate axis for a given row, some
         intensities have been multiplied a factor indicated under the name of
         the galaxy.
         Generally, lines that are broadened by resonant scattering tend to be
         severely narrowed when including dust.}}
\label{fig:SpecPanel}
\end{figure}

Figure \ref{fig:SpecPanel} shows the emergent spectra of the studied galaxies,
arranged according to virial mass. In general, the same trend as for K15 is
seen for all galaxies: the dust primarily affects the wings of the profiles.
As discussed in \sec{specdust}, the reason is that the wings of the
Ly$\alpha$ profile are comprised by photons originating in the very dense
regions of the galaxy having to diffuse far from the line center in order to
escape, and since this is also where the most of the dust is residing, these
photons have a larger probability of being absorbed.
For the two least massive galaxies, S108sc and S115sc, the ISM is neither very
dense nor very metal-rich, meaning that photons escape rather easily.
Consequently, the line is
not particularly broadened and most of the photons escape the galaxies.

\begin{figure}[!t]
\centering
\includegraphics [width=0.70\textwidth] {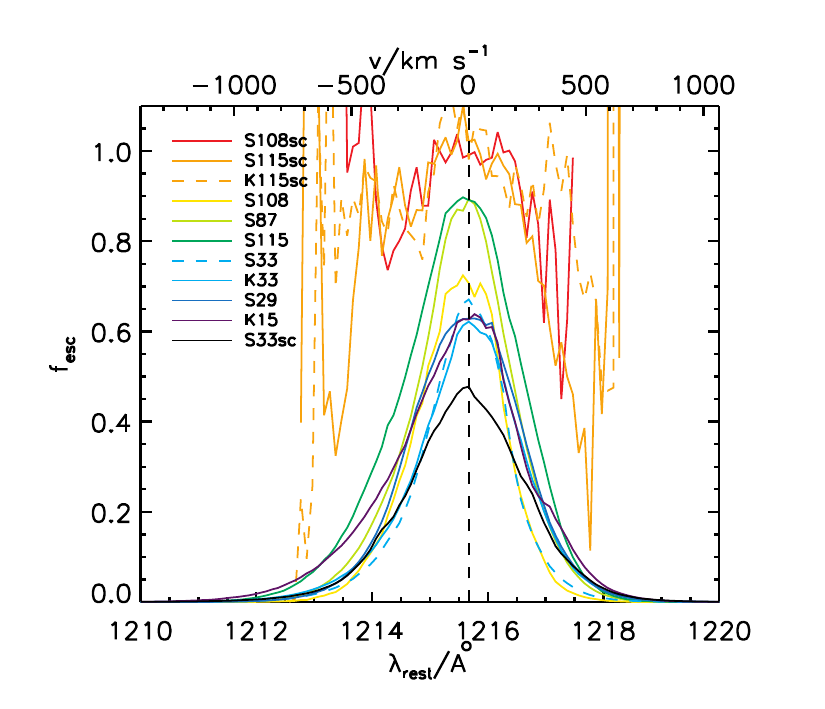}
\caption{{\cap Escape fraction $f_{\mathrm{esc}}$ as a ``function'' of
           wavelength $\lambda$ for the studied galaxies.
           The legend is ordered according the virial mass of the
           galaxies, S33sc being the most massive.
           The vertical dashed line marks the Ly$\alpha$ line center.
           Although the dust cross section is close to
           being independent of the wavelength of the light, the absorption
           profile is highly $\lambda$-dependent.}}
\label{fig:fesc_lam}
\end{figure}

An effective ``escape fraction as a function of wavelength'' can be defined as
the ratio between the calculated, realistic spectrum and the spectrum that
would be seen if there were no dust, i.e.
``$f_{\mathrm{esc}}(\lambda) =
   I_{\mathrm{dusty}} / I_{\mathrm{dustless}}$''.
\Fig{fesc_lam} shows this ratio for the nine galaxies.

Evidently, this is not a factor that can simply be multiplied on the
intrinsically emitted spectrum. Nevertheless, it illustrates which wavelengths
escape more easily. $f_{\mathrm{esc}}(\lambda)$ resembles a Gaussian; for
larger galaxies the maximum decreases, but the width does not change much.
For a homogeneous mixture of gas and (gray) dust, $f_{\mathrm{esc}}(\lambda)$
would be a flat line. Previous attempts to ascertain the impact of dust on
Ly$\alpha$ spectra predicted a constant suppression of the spectrum with
wavelength \citep[e.g.][]{ver06}.  The present results show the importance of
treating the distribution of gas and dust properly.


\subsection{Extended surface brightness profile}
\label{sec:ext}

\begin{figure}[!t]
\centering
\includegraphics [width=1.00\textwidth] {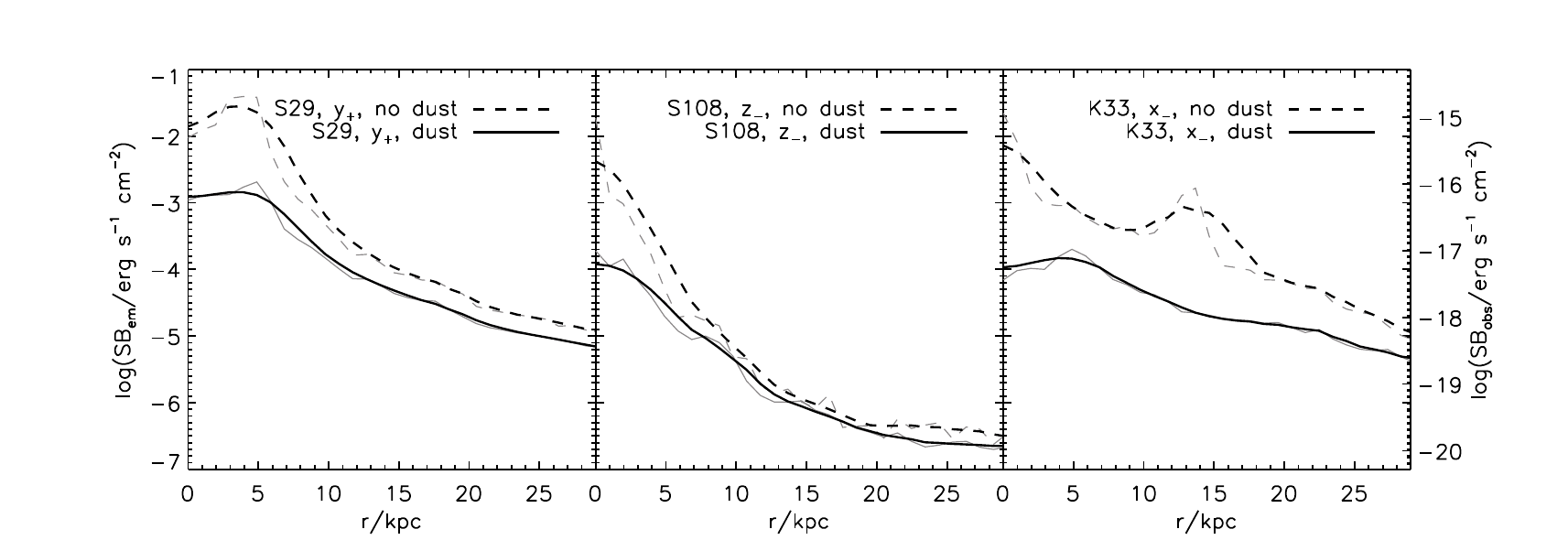}
\caption{{\cap Ly$\alpha$ SB profiles of three of the
         simulated galaxies as observed from three arbitrary directions, with
         (\emph{solid}) and without (\emph{dashed}) dust.
         While the intrinsic SB profiles are shown in \emph{gray},
         the \emph{black} lines
         show the profiles convolved with a Gaussian kernel corresponding to
         a seeing of $0\farcs5$.
         In general, the inclusion of dust tends to ``smooth out'' the
         profiles, effectively resulting in a more extended SB profile.}}
\label{fig:SBpanel}
\end{figure}

The fact that more photons are absorbed in the bright regions tend to
``smooth out'' the SB profiles of the galaxies.
As found in \sec{SBext}, even without dust, resonant scattering itself may
cause an extended Ly$\alpha$ SB profile.
Including dust merely adds to this
phenomenon, since steep parts of the profile are flattened.
\Fig{SBpanel} shows this effect for three of the galaxies.
%



\section{Temporal fluctuations}
\label{sec:t}

Many factors play a role in determining the exact value of $f_{\mathrm{esc}}$.
Although the metallicity, and hence the state of matureness of the galaxy, as
well of the size of the galaxy seem to be the most significant property
regulating $f_{\mathrm{esc}}$, less systematic factors like the specific
configuration of gas and stars are also likely to have a large influence.
The scatter in Fig.~\ref{fig:fesc_M} is probably due to this effect. To
get an idea of the fluctuations of $f_{\mathrm{esc}}$ with time, RT
calculations was run for snapshots of K15 from 100 Myr before ($z = 3.8$) to
100 Myr after ($z = 3.5$) the one already explored at $z = 3.6$.
In this relatively short time, neither $Z$ nor $M_{\mathrm{vir}}$
should not evolve much, and thus any change should be due to stochastic
scatter.

\begin{figure}[!t]
\centering
\includegraphics [width=0.70\textwidth] {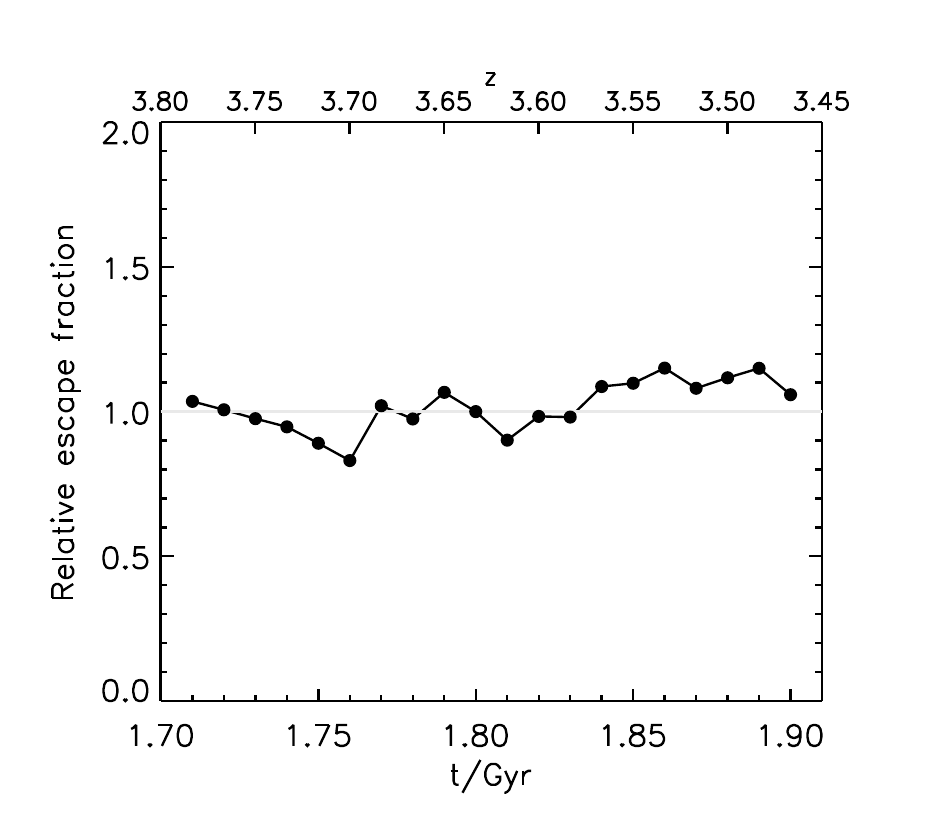}
\caption{{\cap Escape fraction $f_{\mathrm{esc}}$ from the galaxy K15 as a
         function of time $t$ over a period of 200 Myr, normalized to
         $f_{\mathrm{esc}}$ at $t = 1.8$ Gyr. The dispersion
         of $f_{\mathrm{esc}}$ over time is quite
         small, indicating that the dispersion in Fig.~\ref{fig:fesc_M} is due
         to galaxy-to-galaxy variations rather than the escape fractions of the
         individual galaxies fluctuating.}}
\label{fig:fesc_t}
\end{figure}

Figure \ref{fig:fesc_t} shows the variation of $f_{\mathrm{esc}}$ during this
time interval, demonstrating that
the temporal dispersion is of the order 10\% on a 10 Myr scale.
This suggests that the scatter seen in $f_{\mathrm{esc}}(M_{\mathrm{vir}})$
(\fig{fesc_M}) reflects galaxy-to-galaxy variations rather than
$f_{\mathrm{esc}}$ of the individual galaxies fluctuating strongly with time.

%


\section{Discussion}
\label{sec:dustdisc}

The obtained results ($f_{\mathrm{esc}}$, SB, spectra) seem to be quite
insensitive to the assumed values of various parameters characterizing the
dust, such as dust albedo, scattering asymmetry, dust
cross section, extinction curve, etc. Of the studied input parameters, the only
actual uncertainty comes from insufficient knowledge about the dust contents of
ionized gas, but this probably \emph{at most} introduces an error of
$\sim$20\%. This robustness against input parameters is convenient in the
sense that we can rely on the results, but is also a nuisance since
at least from Ly$\alpha$ observations, we should not expect to be able to
learn much about the physical properties of the dust itself.

As discussed in the introduction, previous attempts to determine Ly$\alpha$
escape fractions have been quite divergent, ranging from a few percent to
close to a hundred percent.
In this work it was shown that $f_{\mathrm{esc}}$ may indeed vary rather much
from galaxy to galaxy, and a tentative evidence for a negative correlation with
galactic mass was proposed. Obviously, many factors play a role in
regulating $f_{\mathrm{esc}}$; in particular the age of a given galaxy will
be significant, since the dust accumulates over time.

Various authors have invoked different scenarios to explain their inferred
escape fractions, e.g.~galactic outflows, ionized paths, multi-phase medium,
and viewing angle.
In this work, no evidence was found that any single of these
scenarios should be dominating entirely the magnitude of $f_{\mathrm{esc}}$.
Rather, a mixture of gas kinematics, ISM clumpiness and ionization state, as
well as viewing angle will influence the
total observed Ly$\alpha$ luminosity, and hence the deduced escape fraction,
when comparing Ly$\alpha$ luminosity to UV luminosity.
The investigated galaxies exhibit both outflows and infall, but not at
exceptionally high velocities, and artificially setting all velocities to zero
still allows plenty of radiation to escape (this was done for K15; the shape of
the spectrum is altered, but $f_{\mathrm{esc}}$ remains of the same order).
Artificially erasing the clumpiness of the ISM decreases $f_{\mathrm{esc}}$ as
expected but also in this case much radiation still escapes.



\chapter[On the impact of the intergalactic medium]
        {On the impact of the intergalactic medium\footnote{This chapter is
        based on \citetalias{lau10a} (\app{lau10a}).}}
\label{cha:IGM}

\init{T}{he results presented} in Chapters \ref{cha:conseq} and
\ref{cha:dusteffect}
ignore the effect of the IGM. At the studied redshift of $z = 3.6$, the IGM
should be almost completely neutral, and thus the simulated spectra and SB
profiles should be fairly accurate. To confirm this assertion, as well as to
study the IGM at other redshift, the IGM RT has been conducted as
described in \cha{IGMRT}.

However, this was not the only reason for this part of the project; an
additional motivation was to inquire into the possibility of an alternative
to the ``outflow scenario''\index{Outflows} as an explanation of the many
observed asymmetric Ly$\alpha$ spectra.
As shown in \sec{spec}, in general Ly$\alpha$ photons should escape a galaxy
in a double-peaked spectrum. Although this has indeed been observed,
apparently most Ly$\alpha$ profiles from
high-redshift galaxies seem to be missing the blue peak.
An immediate conclusion would be that high-redshift LAEs are in the process of
massive star formation and thus exhibit strong outflows.
In the reference frame of atoms moving away from the source, blue photons will
be shifted into resonance and be scattered, whereas red photons will be shifted
further from resonance and thus escape more easily
(cf.~the discussion in
\sec{Vbulk} and \fig{expand} on page \pageref{fig:expand}).
Indeed, this scenario
has been invoked to explain a large number of LAE spectra, most convincingly
by \citet{ver08} who --- assuming a central source and thin surrounding shell
of neutral gas, while varying its expansion velocity, temperature, and gas and
dust column density --- manage to produce nice fits to a number of observed
spectra.

Although the thin expanding shell scenario hinges on a physically plausible
mechanism, it is
obviously rather idealized. Furthermore, since most observations show only the
red peak of the profile, it seems to indicate that most (high-redshift)
galaxies exhibit outflows. However, at high redshifts many galaxies are still
forming, resulting in infall which would in
turn imply an increased \emph{blue} peak. Since this is rarely observed,
it was investigated whether absorption in the IGM could serve as a explanation.

The present study complements other recent endeavors to achieve a
comprehensive understanding of how the Ly$\alpha$ line is redistributed in
frequency and real space \emph{after} having escaped its host galaxy
\citep{ili08,bar10,fau10,zhe10a,zhe10b,zhe10c}.
Our cosmological volume is not as large as
most of these studies, and although gas dynamics are included in the
simulations, the fact that the ionizing UV RT is calculated as a post-process
rather than on the fly may mean a less accurate density field.
The reward is a highly increased resolution, allowing
us to study the circumgalactic environs in great detail. Moreover, we
inquire into the temporal evolution of the IGM.

\section{Results}
\label{sec:resigmft}

In the following sections, Model 1 is taken to be the ``benchmark'' model,
while the others are discussed in \sec{10vs6}.

\subsection{The Ly$\alpha$ transmission function}
\label{sec:F} 

\begin{figure}[!t]
\centering
\includegraphics [width=0.90\textwidth] {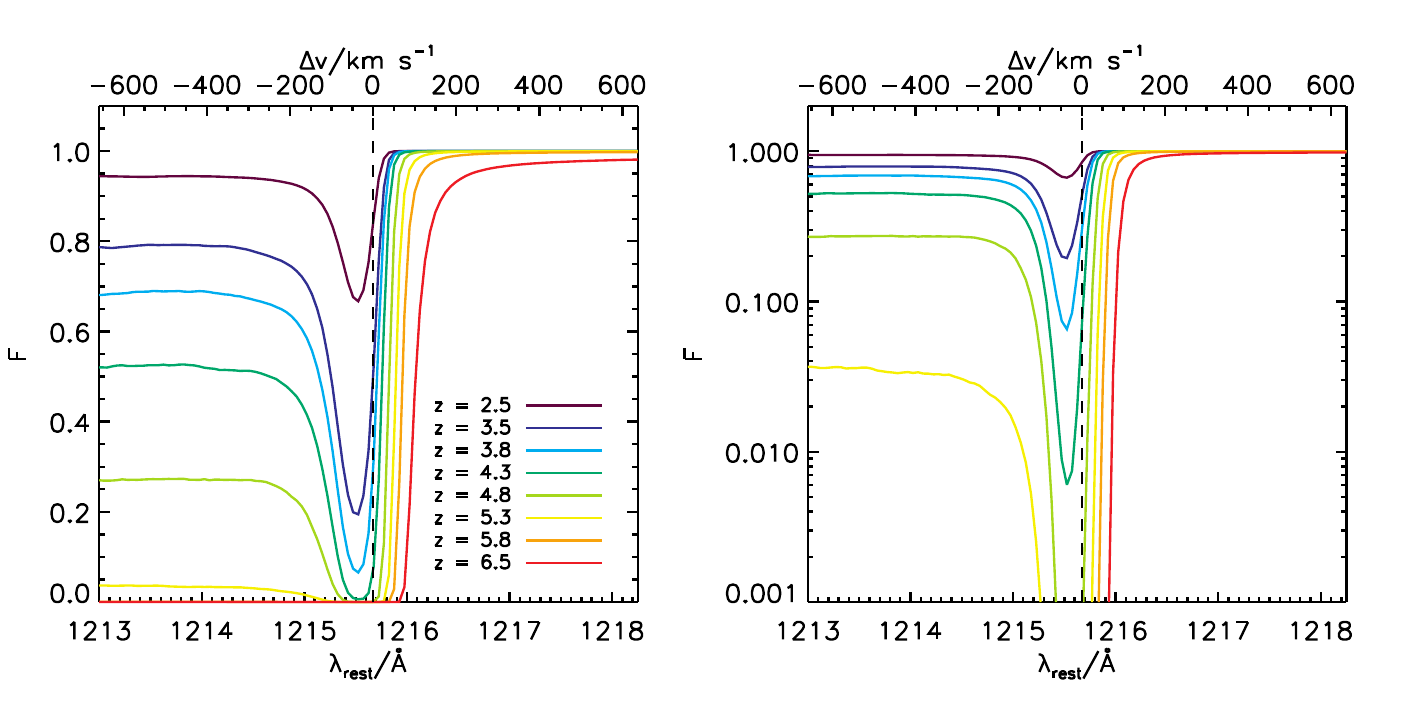}
\caption{{\cap Normalized transmission $\Flam$ at wavelengths around Ly$\alpha$,
         for different redshifts given by the color. In the \emph{left} panel,
         the vertical axis is linear, while in the \emph{right} it is
         logarithmic, emphasizing the transmission at high redshifts.}}
\label{fig:Flam_z}
\end{figure}

\begin{figure}[!t]
\centering
\includegraphics [width=0.90\textwidth] {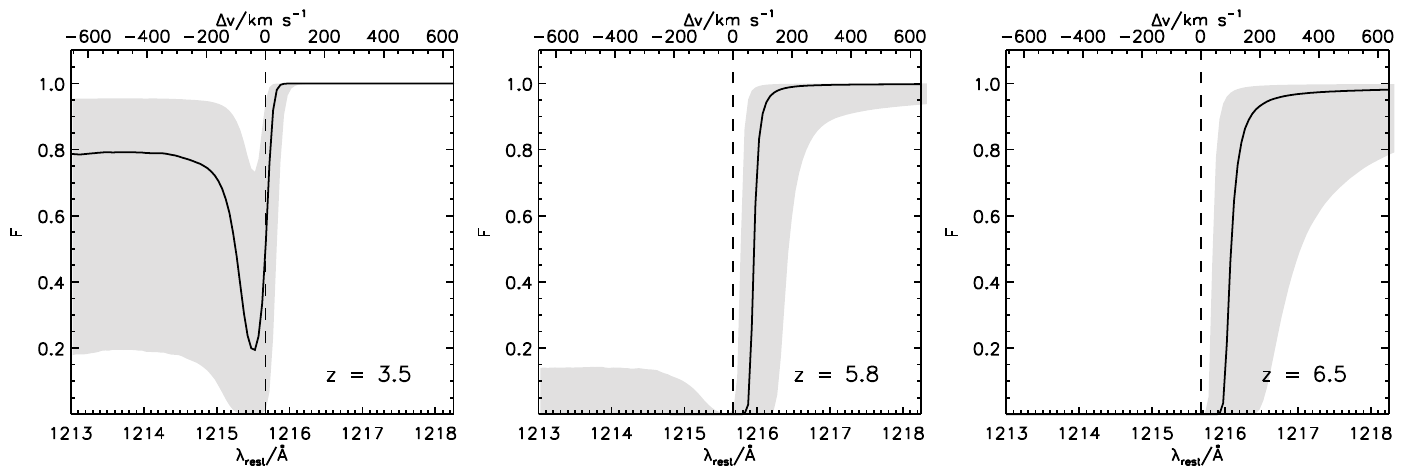}
\caption{{\cap Transmission $\F$ for $z = 3.5$ (\emph{left}), 5.8 (\emph{middle}),
         and 6.5 (\emph{right}). The shaded region indicated the
         range within which 68\% of the individual calculated transmission
         functions fall.}}
\label{fig:Fscat}
\end{figure}

\begin{figure}[!t]
\centering
\includegraphics [width=0.90\textwidth] {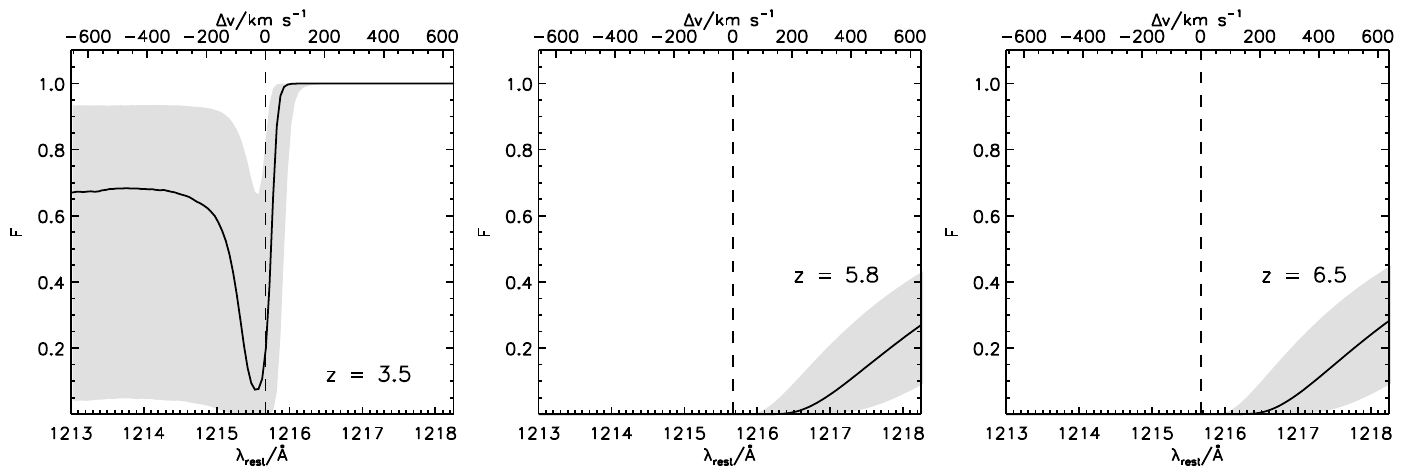}
\caption{{\cap Same as \Fig{Fscat} for Model 3, i.e. with $\zre = 6$.
         While at high
         redshifts a much more neutral IGM than in the $\zre = 10$ model causes
         a severe suppression of the Ly$\alpha$ line, by $z = 3.5$ the
         state of the IGM is not very different.}}
\label{fig:Fscat_rz6}
\end{figure}

\Fig{Flam_z} shows the calculated transmission functions $\Flam$
as a function of redshift.
Indeed, a dip just blueward of the line center is visible at all redshifts.
The results in \fig{Flam_z} were calculated as the median of sightines
emerging from all galaxies in the sample. Of course a large scatter exists,
since each sightline goes through very different regions, even if emanating
from the same galaxy.
\fig{Fscat} shows the scatter associated with $\F$ for three different
redshifts. %
The equivalent transmission functions for the $\zre = 6$ model
are shown in \Fig{Fscat_rz6}. While at high redshifts the $\zre = 6$ model
clearly results in a much more opaque universe, at lower redshifts the
transmission properties of the IGM in the different models are more similar,
although at $z = 3.5$, the $\zre = 10$ model still transmits more light.

\begin{figure}[!t]
\centering
\includegraphics [width=0.90\textwidth] {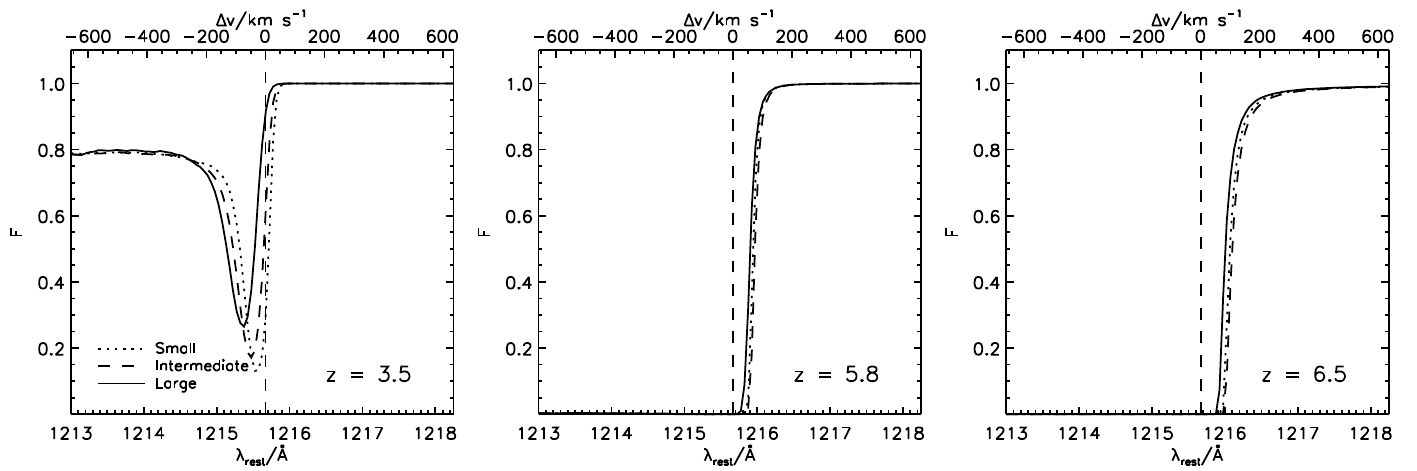}
\caption{{\cap Transmission $\F$ for $z = 3.5$ (\emph{left}),
         5.8 (\emph{middle}), and
         6.5 (\emph{right}), for three different size categories of galaxies;
         small (\emph{dotted}),
         intermediate (\emph{dashed}), and
         large (\emph{solid lines}).
         Although slightly more absorption is seen in the vicinity of smaller
         galaxies, the transmission functions are quite similar for the three
         size ranges.}}
\label{fig:LIS_z}
\end{figure}

To see the difference in transmission around
galaxies of different sizes, \fig{LIS_z}
shows transmission function for three size ranges (defined in \sec{select}).
Since the distance from the galaxies at which the sightlines start is given in
terms of virial radii, sightlines emerging from small galaxies start closer to
their source than for large galaxies, and since at lower redshifts they tend
to be clustered together in the same overdensities as large galaxies, this
results in slightly more absorption. However, the difference is not very
significant.


\subsection{Effect on the spectrum and escape fraction}
\label{sec:IGMeff}

\begin{figure}[!t]
\centering
\includegraphics [width=1.00\textwidth] {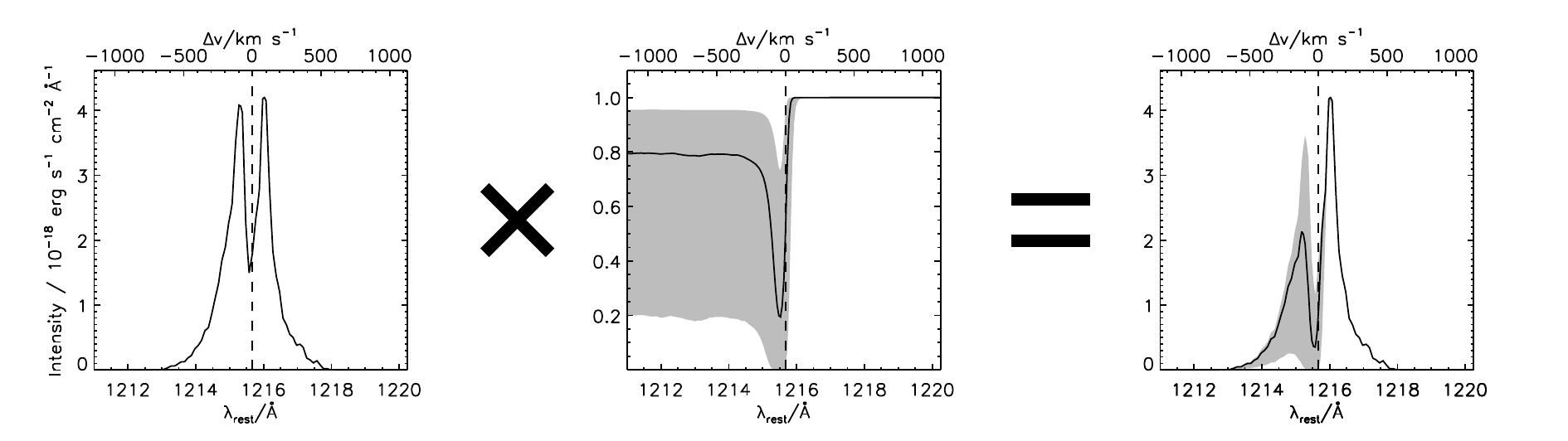}
\caption{{\cap Illustration of the effect of the IGM on the observed Ly$\alpha$
         profile emerging from a galaxy at $z \sim 3.5$. Without taking into
         account the IGM the two peaks are roughly equally high
         (\emph{left panel}). However, when the spectrum is transmitted through
         the IGM characterized by the transmission function $\Flam$
         (\emph{middle panel}), the blue peak is dimished, resulting in an
         observed spectrum with a higher red peak (\emph{right panel}).}}
\label{fig:spXF}
\end{figure}

In \fig{spXF} the ``purpose'' of the transmission function is
illustrated: the left panel shows the spectrum emerging from a galaxy of
$M_{\mathrm{vir}} = 4.9 \times 10^9 \Msun$ and Ly$\alpha$ luminosity
$L_{\mathrm{Ly}\alpha} = 4.9 \times 10^{40}$ erg s$^{-1}$ (``S115sc'').
According to the criterion stipulated in \sec{select}, its circular velocity of
42 km s$^{-1}$ characterizes it as a small galaxy. The spectrum which is
actually observed, after the light has been transferred through the IGM, is
shown in the right panel.
Although on average the effect of the IGM is not very large at this redshift,
as seen by the solid line in the right panel, due to the large dispersion in
transmission (visualized by the gray-shaded area) at least \emph{some} such
galaxies will be observed with a substantially diminished blue peak.

\begin{figure}[!t]
\centering
\includegraphics [width=1.00\textwidth] {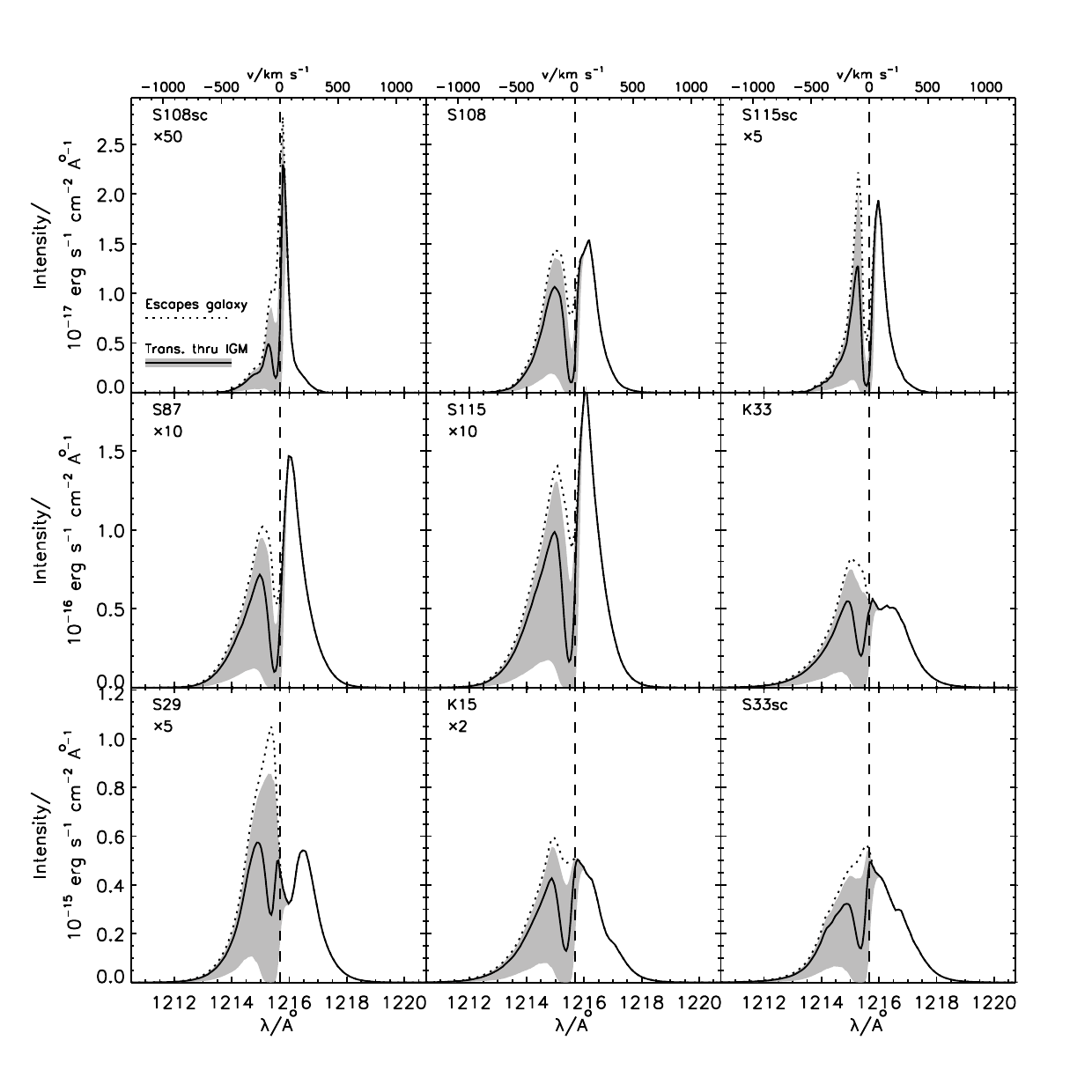}
\caption{{\cap Spectra (\emph{dotted lines}) emitted from nine different
         simulated galaxies at $\sim$3.5 --- ordered after increasing size ---
         and the corresponding spectra after being transmitted through the IGM
         (\emph{solid lines with gray regions denoting the 68\% confidence
         intervals}). The transmission functions appropriate for the given
         galaxy sizes have been used.
         In order to use the same ordinate axis for a given row, some
         intensities have been multiplied a factor indicated under the name of
         the galaxy.}}
\label{fig:spXFall}
\end{figure}

As seen in \fig{SpecPanel} on page \pageref{fig:SpecPanel}, in general the
larger a galaxy is the broader its emitted spectrum is, since
Ly$\alpha$ photons have to diffuse farther from the line center for higher
column densities of neutral gas. If dust is present, this will tend to narrow
the line.
Larger galaxies tend to have higher metallicities and hence more dust, but
the lines will still be broader than the ones of small galaxies.
The galaxy used in \fig{spXF} is quite small. In \fig{spXFall} the
impact of the IGM on the nine simulated spectra from the previous chapters is
shown.

Besides altering the shape of the emitted spectrum, the IGM also has an effect
on another quantity of much interest to observers, namely the observed fraction
$f_{\mathrm{obs}}$ of the intrinsically emitted number of Ly$\alpha$ photons.
The escape fractions $f_{\mathrm{esc}}$ shown in \fig{fesc_M} on page
\pageref{fig:fesc_M} give the fraction of the intrinsically emitted photons
that make it out of the galaxies.
As already discussed in \sec{fesctheo}, since the bulk of the
emitted Ly$\alpha$ photons is caused by young stars, the total Ly$\alpha$
luminosity of a galaxy may be used as a proxy for its SFR,
one of the main quantities characterizing galaxies.
Assuming case B
recombination, $L_{\mathrm{Ly}\alpha}$ can be converted to a total H$\alpha$
luminosity $L_{\mathrm{H}\alpha}$ (through $L_{\mathrm{H}\alpha} =
L_{\mathrm{Ly}\alpha}/8.7$), which in turn can be converted to an SFR using
the \citeauthor{ken98} relation.

The above conversion factors assumes that none of the emitted light is lost.
If dust is present in the galaxy a fraction of the emitted photons
will be absorbed
\citep[possibly making the LAE observable in FIR; see][]{day10}.
This can be corrected for if the color excess $E(B-V)$ is
measured, assuming some standard extinction curve. However, this assumes that
both the H$\alpha$ and Ly$\alpha$ radiation is simply reduced by some factor
corresponding to having traveled the same distance through the dusty medium.
But since the path of Ly$\alpha$ photons is complicated by resonant scattering,
this may be far from the truth. Comparing the Ly$\alpha$-inferred SFR with
that of H$\alpha$ (or UV continuum), the effect of scattering can be
constrained, as the quantity SFR(Ly$\alpha$)/SFR(H$\alpha$;UV) will be an
estimate of $f_{\mathrm{obs}}$.
In this way values from a few percent \citep[mostly in the nearby Universe;
e.g.][]{hay07,ate08,hay10b} to $\sim$1/3 at high redshifts \citep{gro07} have
been found, as discussed in \sec{fesctheo}.

Once the radiation enters the IGM, it is also not affected in the same way,
since the IGM is transparent to H$\alpha$, but not to Ly$\alpha$.
\citet{dij07} estimated analytically the fraction of Ly$\alpha$ photons
that are scattered out of the line of sight by the IGM (at $z \sim 4.5$--6.5)
and found a mean transmission of $f_{\mathrm{IGM}} \sim 0.1$--0.3, depending on
various assumptions (note that in \citet{dij07} this fraction is called
``$\mathcal{T}_\alpha$''). In that study, the intrinsic Ly$\alpha$ line profile
is modeled as a Gaussian, the width of which is given by the circular velocity
of the galaxies, in turn given by their mass, and assuming that no dust in the
galaxies alters the shape of the line before the light enters the IGM.

Since the transmission function is a non-trivial function of wavelength,
the exact shape and width of the Ly$\alpha$ line profile is important.
With {\sc MoCaLaTA} more realistic spectra has been
modeled, and applying the transmission function found in \sec{F},
$f_{\mathrm{IGM}}$ can be calculated for the sample of simulated galaxies.
The transmitted fraction is not a strong function of galaxy size;
on average, a fraction $f_{\mathrm{IGM}} = 0.77_{-0.34}^{+0.17}$ of the photons
escaping the galaxies is transmitted through the IGM, at the investigated
redshift of $z \sim 3.5$.
The results for the individual galaxies is given in \tab{fIGM}.
\begin{table}[!t]
\begin{center}
{\sc Transmission fractions for nine simulated galaxies at $z\sim3.5$}
\end{center}
\centering
\begin{tabular}{lcccc}
\hline
\hline
Galaxy  & $V_c$ & $f_{\mathrm{esc}}$ & $f_{\mathrm{IGM}}$              & $f_{\mathrm{obs}}$ \\
\hline
S108sc  &   17  &    $0.97\pm0.02$   & $0.66_{-0.36}^{+0.24}$          & $0.64_{-0.35}^{+0.23}$ \vspace{2mm}\\
S108    &   33  &    $0.12\pm0.02$   & $0.78_{-0.35}^{+0.16}$          & $0.09_{-0.04}^{+0.02}$ \vspace{2mm}\\
S115sc  &   42  &    $0.95\pm0.03$   & $0.76_{-0.34}^{+0.18}$          & $0.72_{-0.32}^{+0.17}$ \vspace{2mm}\\
S87     &   69  &    $0.22\pm0.01$   & $0.81_{-0.32}^{+0.15}$          & $0.18_{-0.07}^{+0.03}$ \vspace{2mm}\\
S115    &   73  &    $0.30\pm0.05$   & $0.80_{-0.34}^{+0.16}$          & $0.24_{-0.11}^{+0.06}$ \vspace{2mm}\\
K33     &  126  &    $0.08\pm0.02$   & $0.78_{-0.34}^{+0.17}$          & $0.06_{-0.03}^{+0.02}$ \vspace{2mm}\\
S29     &  137  &    $0.12\pm0.03$   & $0.75_{-0.35}^{+0.19}$          & $0.09_{-0.05}^{+0.03}$ \vspace{2mm}\\
K15     &  164  &    $0.17\pm0.02$   & $0.79_{-0.37}^{+0.17}$          & $0.13_{-0.07}^{+0.03}$ \vspace{2mm}\\
S33sc   &  228  &    $0.08\pm0.02$   & $0.80_{-0.33}^{+0.16}$          & $0.06_{-0.03}^{+0.02}$ \vspace{4mm}\\
Average &       &                    & $\mathbf{0.77_{-0.34}^{+0.17}}$ &                                    \\
\hline
\end{tabular}
\caption{{\cap Columns are, from left to right:
               galaxy name,
               circular velocity $V_c$ in km s$^{-1}$,
               fraction $f_{\mathrm{esc}}$ of emitted photons escaping the
                 galaxy (i.e.~not absorbed by dust),
               fraction $f_{\mathrm{IGM}}$ of these transmitted through the
                 IGM, and
               resulting observed fraction
                 $f_{\mathrm{obs}} = f_{\mathrm{esc}} f_{\mathrm{IGM}}$.
               Uncertainties in $f_{\mathrm{esc}}$ represent varying escape
               fractions
               in different directions, while uncertainties in $f_{\mathrm{IGM}}$
               represent variance in the IGM.
               Uncertainties in $f_{\mathrm{obs}}$ are calculated as
               $(\sigma_{\mathrm{obs}} / f_{\mathrm{obs}})^2 =
                (\sigma_{\mathrm{esc}} / f_{\mathrm{esc}})^2 +
                (\sigma_{\mathrm{IGM}} / f_{\mathrm{IGM}})^2$.}}
\label{tab:fIGM}
\end{table}

\begin{figure}[!t]
\centering
\includegraphics [width=1.00\textwidth] {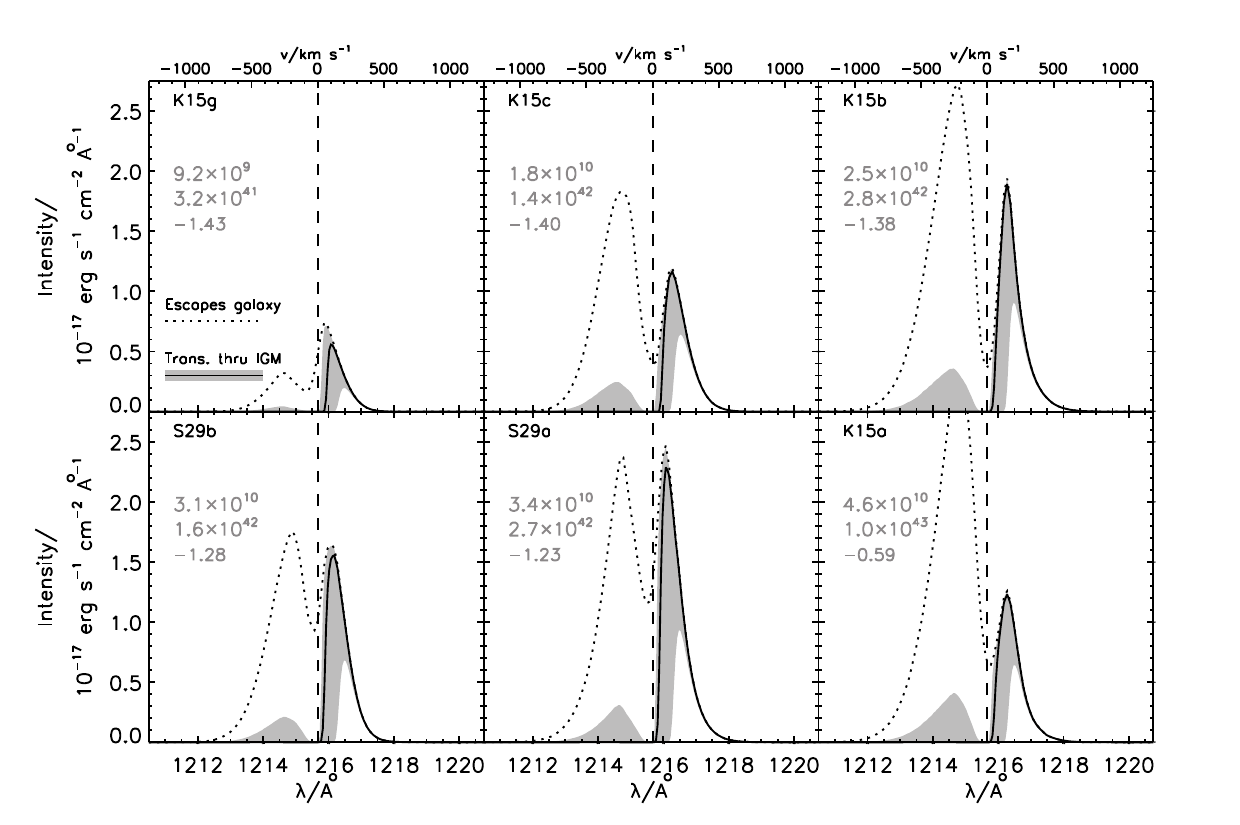}
\caption{{\cap Same as \fig{spXFall}, but for $z = 5.8$.}}
\label{fig:spXFall_z5p8}
\end{figure}

\begin{figure}[!t]
\centering
\includegraphics [width=1.00\textwidth] {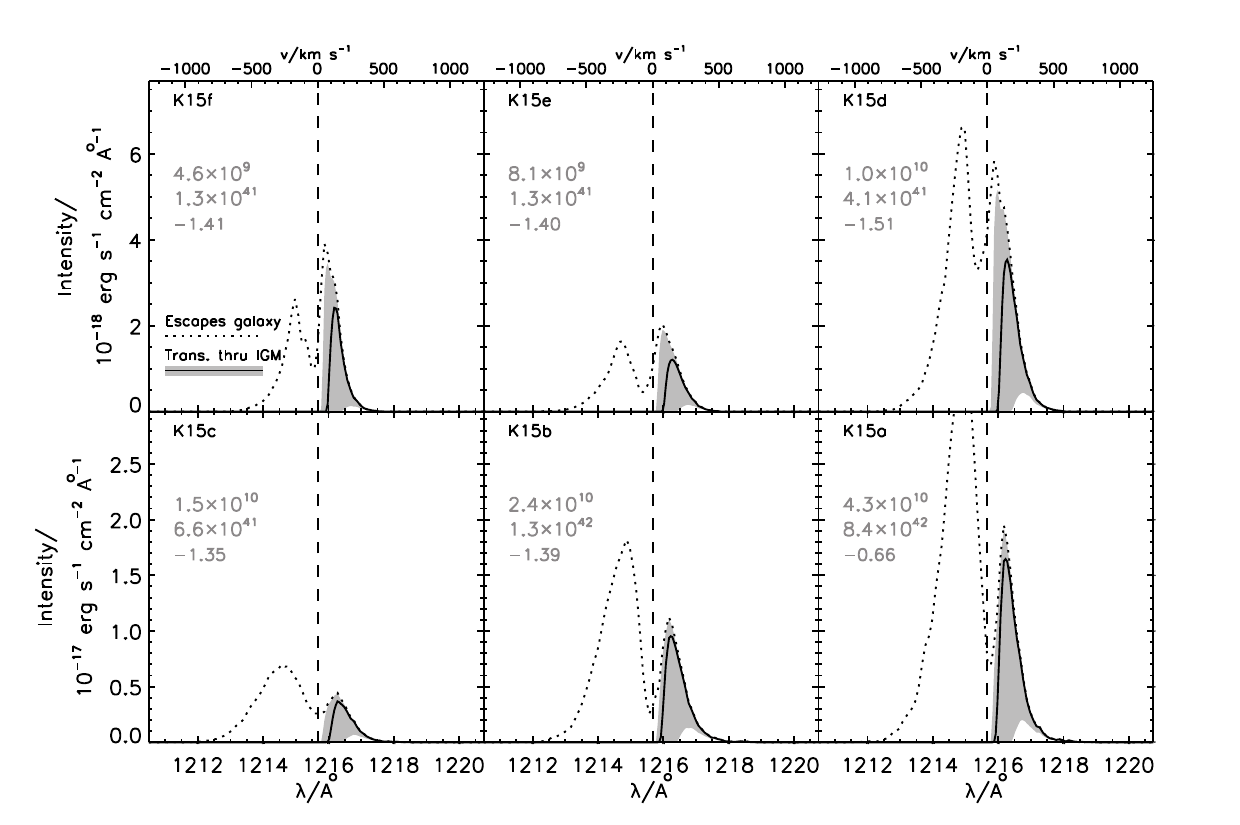}
\caption{{\cap Same as \fig{spXFall}, but for $z = 6.5$.}}
\label{fig:spXFall_z6p5}
\end{figure}

At higher redshifts the metallicity is generally lower,
leading to less dust and
hence larger values of $f_{\mathrm{esc}}$. However, the increased neutral
fraction of the IGM scatters a correspondingly higher number of photons out of
the line of sight, resulting in a smaller total observed fraction.
\Fig{spXFall_z5p8} and \Fig{spXFall_z6p5} shows the impact of the IGM on
Ly$\alpha$ profiles at $z \sim 5.8$ and $z \sim 6.5$, respectively,
and \Tab{fIGM_z5p8} and \Tab{fIGM_z6p5} summarizes the obtained fractions.
At these redshifts, for the six galaxies for which the calculations have been
carried out on average a fraction
$f_{\mathrm{IGM}}(z=5.8) = 0.26_{-0.18}^{+0.13}$ and
$f_{\mathrm{IGM}}(z=6.5) = 0.20_{-0.18}^{+0.12}$ of the photons is transmitted
through the IGM, consistent with what was obtained by \citet{dij07}.
\begin{table}[!t]
\begin{center}
{\sc Transmission fractions for six simulated galaxies at $z\sim5.8$}
\end{center}
\centering
\begin{tabular}{lcccc}
\hline
\hline
Galaxy   & $V_c$ & $f_{\mathrm{esc}}$ & $f_{\mathrm{IGM}}$              & $f_{\mathrm{obs}}$ \\
\hline
K15g     &   63  &    $0.93\pm0.02$   & $0.35_{-0.24}^{+0.16}$          & $0.33_{-0.23}^{+0.15}$ \vspace{2mm}\\
K15c     &   78  &    $0.82\pm0.02$   & $0.25_{-0.14}^{+0.11}$          & $0.21_{-0.11}^{+0.09}$ \vspace{2mm}\\
K15b     &   88  &    $0.46\pm0.02$   & $0.23_{-0.14}^{+0.12}$          & $0.11_{-0.06}^{+0.05}$ \vspace{2mm}\\
S29b     &   94  &    $0.85\pm0.01$   & $0.27_{-0.17}^{+0.14}$          & $0.23_{-0.15}^{+0.12}$ \vspace{2mm}\\
S29a     &   96  &    $0.52\pm0.08$   & $0.31_{-0.22}^{+0.14}$          & $0.16_{-0.12}^{+0.07}$ \vspace{2mm}\\
K15a     &  108  &    $0.17\pm0.05$   & $0.16_{-0.11}^{+0.11}$          & $0.03_{-0.02}^{+0.02}$ \vspace{4mm}\\
Average  &       &                    & $\mathbf{0.26_{-0.18}^{+0.13}}$ &                                    \\
\hline
\end{tabular}
\caption{{\cap Same as \tab{fIGM}, but for $z = 5.8$.}}
\label{tab:fIGM_z5p8}
\end{table}
\begin{table}[!t]
\begin{center}
{\sc Transmission fractions for six simulated galaxies at $z\sim6.5$}
\end{center}
\centering
\begin{tabular}{lcccc}
\hline
\hline
Galaxy   & $V_c$ & $f_{\mathrm{esc}}$ & $f_{\mathrm{IGM}}$              & $f_{\mathrm{obs}}$ \\
\hline
K15f     &   50  &    $0.96\pm0.03$   & $0.24_{-0.22}^{+0.18}$          & $0.23_{-0.22}^{+0.17}$ \vspace{2mm}\\
K15e     &   62  &    $0.91\pm0.04$   & $0.24_{-0.22}^{+0.14}$          & $0.22_{-0.20}^{+0.13}$ \vspace{2mm}\\
K15d     &   66  &    $0.95\pm0.02$   & $0.18_{-0.16}^{+0.13}$          & $0.17_{-0.16}^{+0.12}$ \vspace{2mm}\\
K15c     &   76  &    $0.88\pm0.02$   & $0.16_{-0.14}^{+0.07}$          & $0.14_{-0.13}^{+0.06}$ \vspace{2mm}\\
K15b     &   89  &    $0.83\pm0.03$   & $0.20_{-0.17}^{+0.06}$          & $0.17_{-0.14}^{+0.05}$ \vspace{2mm}\\
K15a     &  108  &    $0.26\pm0.09$   & $0.18_{-0.16}^{+0.06}$          & $0.05_{-0.04}^{+0.02}$ \vspace{4mm}\\
Average  &       &                    & $\mathbf{0.20_{-0.18}^{+0.12}}$ &                                    \\
\hline
\end{tabular}
\caption{{\cap Same as \tab{fIGM}, but for $z = 6.5$.}}
\label{tab:fIGM_z6p5}
\end{table}
%


\subsection{Probing the Epoch of Reionization}
\label{sec:probeEoR}

We now focus on a different topic, namely the RT in the IGM far from the
emitting galaxies.
Measuring for a sample of quasars or other bright sources the average
transmission in a wavelength interval blueward of the
Ly$\alpha$ line, the average transmission properties, and hence the physical
state, of the IGM can be ascertained.
The interval in which the transmission is calculated should be large
enough to achieve good statistics, but short enough that
the bluest wavelength does not correspond to a redshift epoch appreciably
different from the reddest.
Furthermore, in order to probe the real IGM and not the quasar's neighborhood,
the upper limit of the wavelength range should be taken to have a value
somewhat below $\lambda_0$.

\begin{figure}[!t]
\centering
\includegraphics [width=0.90\textwidth] {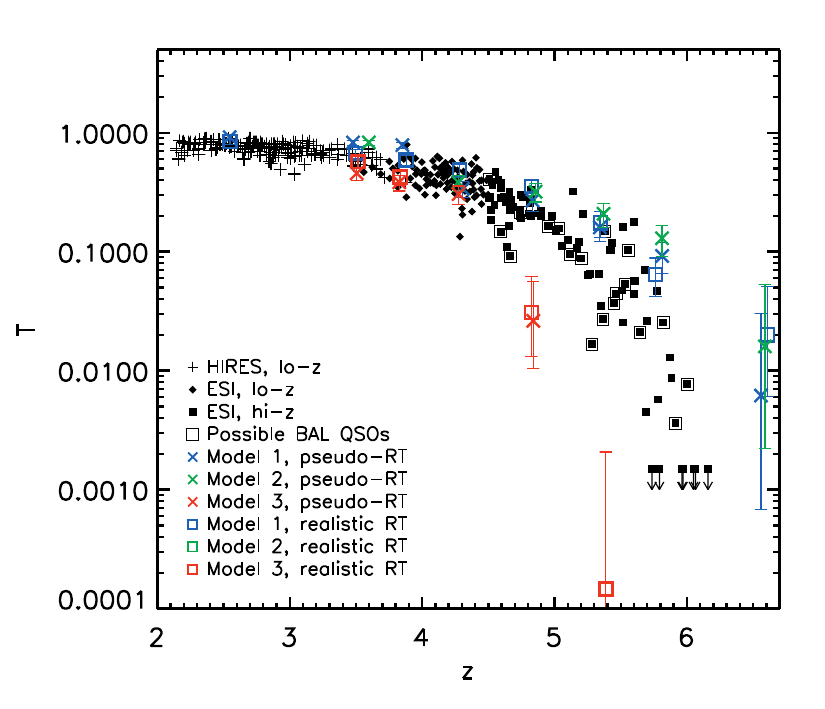}
\caption{{\cap Comparison of observations (\emph{black data points}) and
         simulations (\emph{colored data points}) of the transmitted flux
         blueward of the Ly$\alpha$ line as a function of redshift;
         The three models are represented by the colors \emph{blue},
         \emph{green}, and \emph{red} for model 1, 2, and 3, respectively.
         To highlight the significance of the improved UV RT, we show both the
         transmission in the ``original'' simulation with the ``pseudo''-RT
         (\emph{crosses}) and with the improved UV RT (\emph{squares}).
         For details on the observations see \citet{son04}, from where the data
         are kindly supplied.}}
\label{fig:Songaila}
\end{figure}

\citet{son04} measured the IGM transmission in the LAFs of a large sample of
quasars with redshifts between 2 and 6.5, in the wavelength interval
1080--1185 {\AA}. In \fig{Songaila} the simulated transmitted fractions is
compared with her sample.
As is evident from the figure, an ``early'' reionization, i.e.~with $\zre = 10$
(Model 1 and 2),
yields a slightly too transparent Universe, while a ``late'' reionization
(Model 3)
yields a too opaque Universe. Note that the log scale makes the
$\zre = 6$ data seem much farther off than the $\zre = 10$ data.



\section{Convergence test}
\label{sec:X_i}

\begin{figure}[!t]
\centering
\includegraphics [width=1.0\textwidth] {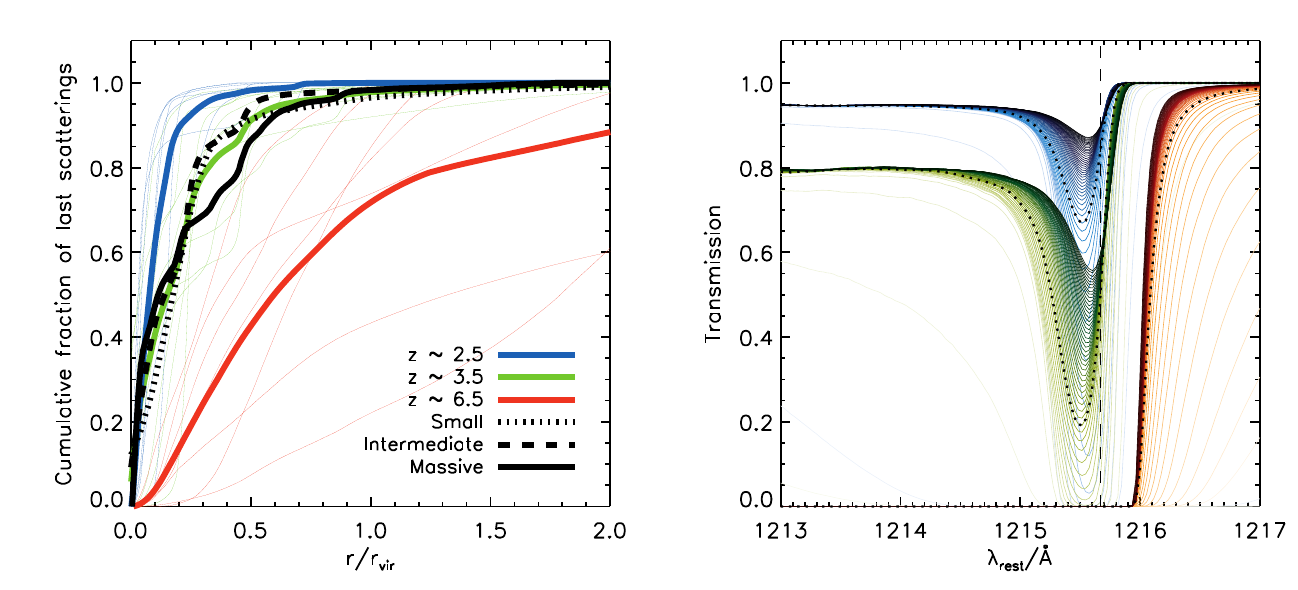}
\caption{{\cap \emph{Left:}
         Cumulative probability distribution of the distance $r$ from the
         center of a galaxy at which the last scattering takes place.
         \emph{Thin solid} lines represent individual galaxies at redshift
         2.5 (\emph{blue}), 3.5 (\emph{green}), and 6.5 (\emph{red}),
         while \emph{thick solid colored} lines are the average of these.
         Also shown, in \emph{black}, are the average of three different size
         ranges at $z = 3.5$;
         small (\emph{dotted}), intermediate (\emph{dashed}), and large
         galaxies (\emph{solid}).\\
         \emph{Right:} Resulting transmission function $\Flam$ for sightlines
         originating at various distances $r_0/r_{\mathrm{vir}}$ from galactic
         centres, increasing in steps of $0.1 r_{\mathrm{vir}}$ and ranging
         from $r_0 = 0.1 r_{\mathrm{vir}}$ to $r_0 = 5 r_{\mathrm{vir}}$.
         The three different redshifts are shown in shades of \emph{blue},
         \emph{green}, and \emph{red}, for $z = 2.5$, 3.5, and 6.5,
         respectively, and the lines go from light shades for low values of
         $r_0$ to dark shades for high values of $r_0$.
         The transmission functions corresponding to
         $r_0 = 1.5 r_{\mathrm{vir}}$ are shown in \emph{black dotted} lines.
         All results are for Model 1.}}
\label{fig:X_i}
\end{figure}

In \sec{IGMRTsims} it was stated that the sightlines should initiate at
a distance from the centers of the galaxies given by their virial radius.
\Fig{X_i} displays the cumulative probability
distributions of the distance of the last scattering from the galaxy center for
different redshifts and mass ranges, demonstrating that in most cases a photon
will have experienced its last scattering at the order of 1 $r_{\mathrm{vir}}$
from its host
galaxy. For increasing redshift, the photons tend to escape the galaxies at
larger distances due to the higher fraction of neutral hydrogen, but the change
with redshift seems quite slow. Furthermore, at a given redshift the
distinction between different galactic size ranges appear insignificant
(as long as $r_0$ is measured in term of $r_{\mathrm{vir}}$).

Also shown in \fig{X_i} are the transmission curves for sightlines
initiating at various distances $r_0$ from the centers of the galaxies.
For very small values of $r_0$, a significantly lower
fraction is transmitted due to the high density of neutral gas. Around
$r \sim r_{\mathrm{vir}}$
the change in $\Flam$ becomes slow, converging to $\Flam$ exhibiting no dip
for $r_0 \to \infty$.

In summary, scaling $r_0$ to the virial radius $r_{\mathrm{vir}}$ of the
galaxies allows us to use the same value of $r_0 = 1.5 r_{\mathrm{vir}}$ for
all sightlines.


\section{Discussion}
\label{sec:IGMdisc}

\subsection{The origin of the dip}
\label{sec:dip}

In general, the effect of the IGM --- \emph{even at relatively low redshifts}
--- is to reduce the blue peak of the Ly$\alpha$ line profile.
At $z\sim3.5$, the effect is not strong enough to fully
explain the observed asymmetry,
but at $z \gtrsim 5$, almost all radiation blueward of the line center
is lost in the IGM.

\begin{figure}[!t]
\centering
\includegraphics [width=1.00\textwidth] {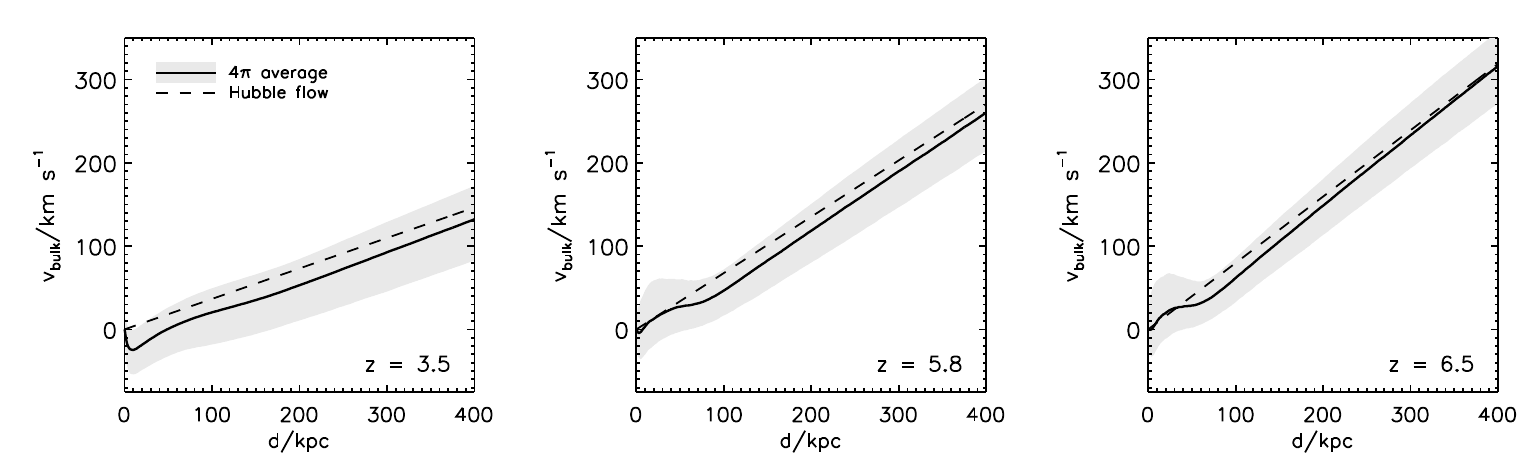}
\caption{{\cap Average recession velocity $v_{\mathrm{bulk}}$ of the IGM as a
         function of proper distance $d$ from the centers of the galaxies in
         Model 1 (\emph{solid black} lines, with \emph{gray} regions indicating
         the 68\% confidence intervals).
         At all redshift, the expansion is retarded compared to the pure Hubble
         flow (\emph{dashed} line) out to a distance of several comoving Mpc.
         At high redshifts,
         however, very close to the galaxies outflows generate higher recession
         velocities.}}
\label{fig:probeV}
\end{figure}

\begin{figure}[!t]
\centering
\includegraphics [width=1.00\textwidth] {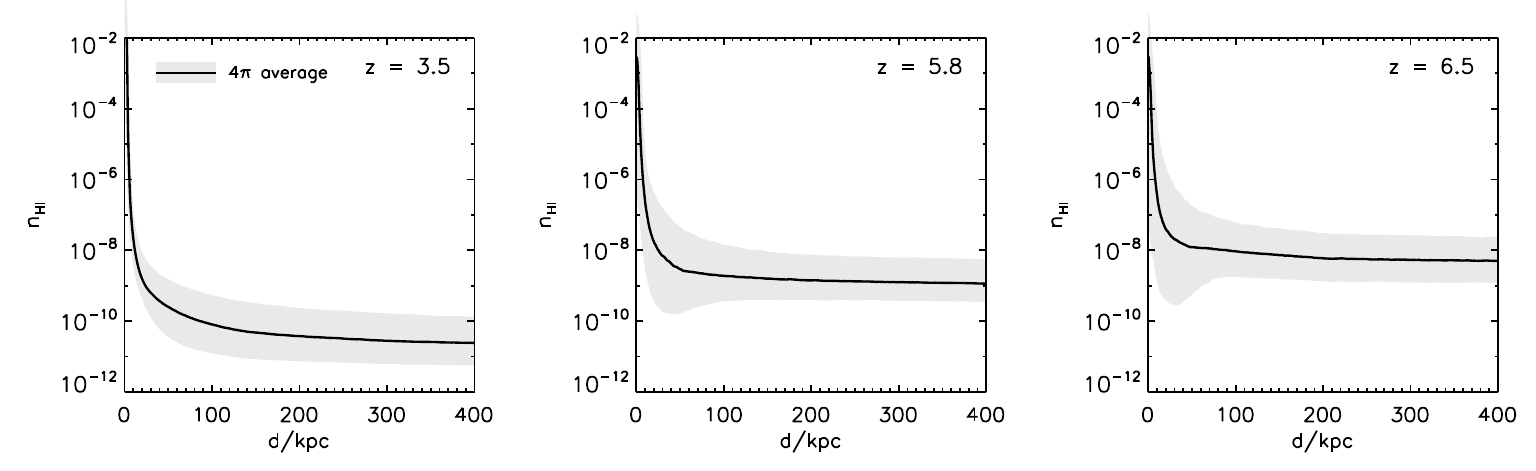}
\caption{{\cap Average density $\nhi$ of neutral hydrogen as a
         function of proper distance $d$ from the centers of the galaxies in
         Model 1 (\emph{solid black} lines, with \emph{gray} regions indicating
         the 68\% confidence intervals).
         While in general the density decreases with distance, at high
         redshifts ionizing radiation reduces $\nhi$ in the immediate
         surroundings of at least some of the galaxies, as seen by the small
         dip at $\sim$50 kpc in the lower part of the gray zone.}}
\label{fig:probenHI}
\end{figure}

Figures \ref{fig:probeV} and \ref{fig:probenHI} display for three different
epochs the velocity field and the density of neutral hydrogen, respectively,
of the IGM associated with the galaxies, taken as an average of all galaxies in
the samples, and in all directions.
At all redshifts, on average the IGM close to the galaxies recedes at a
somewhat slower rate than that given by the pure Hubble flow, slowly converging
toward the average expansion rate of the Universe.
At the higher
redshifts, the IGM in the immediate proximity of the galaxies is characterized
by higher velocities, due to outflows generated by starbursts. At $z = 3.5$,
however, this effect is overcome by the accretion of gas.

Inspecting the dips in \fig{Flam_z}, the minima are seen to be located at
roughly 50 km s$^{-1}$, almost independently of the redshift but becoming
slightly broader with increasing $z$.
At $z = 3.5$, the dip extends all the way out to 300--400 km s$^{-1}$.
This corresponds to the central absorption being caused by the IGM within
$\sim$150 kpc, and the wings of the absorption by the IGM within $\sim$1 Mpc.
As shown in \fig{probenHI}, at $z = 3.5$ the density of neutral hydrogen
decreases monotonically with distance from the source, and within $\sim$150
kpc $\nhi$ is substantially higher than the cosmic mean. Further away, the
density is close to the mean density of the Universe. However, as seen in
\fig{probeV} the recession velocity of the gas continues to lie below that of
the average, Universal expansion rate, and in fact does so until approximately
1 Mpc from the source.\footnote{Since the cosmological volume is several
Mpc across, except for the galaxies
lying close to the edge the absorption takes place before the sightlines
``bounce'', so in general there is no risk of a sightline going through the
same region of space before having escaped the zone causing the dip.}

Thus, the cause of the suppression of the blue wing of the Ly$\alpha$ line may,
at wavelengths close to the line center ($\Delta\lambda \simeq 1/2$ {\AA}),
be attributed to an increased density of neutral hydrogen close to the galaxies,
while farther away from the line center ($\Delta\lambda \lesssim$ 1--1.5 {\AA}),
to a retarded Hubble flow.

\citet{zhe10a} perform Ly$\alpha$ RT in the IGM at $z = 5.7$ through full Monte
Carlo
simulations, and make a detailed comparison with the observed line profile
obtained from simply multiplying the intrinsic line by $e^{-\tau}$, as has
been done in previous models \citep[e.g.][]{ili08}. They conclude that
neglecting scattering effects severely underestimates the transmitted fraction.
While it is certainly true that treating scattering processes as absorption
inside the galaxies is only a crude approximation, once the probability of
photons being scattered \emph{into} the line of sight becomes sufficiently
small, this approach is quite valid. In their analysis, \citet{zhe10a} start
their photons in the center of the galaxies, which are resolved only by a few
cells (their $dx$ being $\sim$28 kpc in physical coordinates and their
fiducial galaxy having $r_{\mathrm{vir}}$ = 26 kpc).
Since the side length of our smallest cells are more than 400
times smaller than the
resolution of \citet{zhe10a}, we are able to resolve the galaxies and their
surroundings in great detail, and we are hence able to determine the distance
at which the $e^{-\tau}$ model becomes realistic.
Moreover, when coupling the IGM RT with Ly$\alpha$ profiles, we use the
realistically calculated profile, whereas \citet{zhe10a} use a Gaussian set
by the galaxies' halo masses. Their line widths $\sigma_{\mathrm{init}} =
32 M_{\mathrm{10}}^{1/3}$ km s$^{-1}$,
 where $M_{\mathrm{10}}$ is the halo mass divided by
$10^{10} h^{-1} M_\odot$, thus neglect broadening by scattering.
This makes them much smaller than ours, which are typically
several hundred km s$^{-1}$.
Note, however, that our relatively small cosmological volume and the fact that
ionizing UV RT is performed as a post-process rather than on the fly may make
our density field less accurate than that of \citet{zhe10a}.

The photons that are scattered out of the line of sight are of course not lost,
but rather become part of a diffuse Ly$\alpha$ background. Since more
scatterings take place in the vicinity of galaxies, this is the reason for
the previously discussed low-surface brightness halo arounf the LAEs at
$r \sim 10$--100 kpc scales. As argued above, part of
the cause of the line suppression (mostly in the wings) is due to the IGM up
to $\sim$1 Mpc of the source. At this distance, the surface brightness is much
lower than observational thresholds, but could be detected by stacking images
of LAEs \citep{zhe10c}.

As seen in e.g.~\Fig{Fscat}, a large scatter between individual sightlines
exists, reflecting the generally quite inhomogeneous IGM. Consequently, one
cannot simply use the calculated transmission function for deconvolving
observed Ly$\alpha$ lines to obtain the intrinsic line profiles, other than in
a statistical sense. With a large sample, however, more accurate statistics on
Ly$\alpha$ profiles could be obtained.
Calculating the transmission functions as an average of all directions, as we
have done in this work, assumes that observed galaxies are randomly oriented
in space, i.e. that there is no selection effects making more or less luminous
directions pointing toward the observer. For LAEs clustered in, e.g., filaments,
the effect of the retarded Hubble flow may be enhanced perpendicular to the
filament, making the galaxies more luminous if observed along a filament than
perpendicular to it \citep[see Figure 9 in][]{zhe10b}.


\subsection{Galactic outflows}
\label{sec:outflow}\index{Outflows}

As discussed in the beginning of the chapter,
at high redshifts many galaxies are still in
the process of forming and are expected to be accreting gas. In principle, this
should result in a blueshifted Ly$\alpha$ profile, but this is rarely seen.
Evidently, IGM absorption is unable to always be the cause of this missing blue
peak.
On larger scales, mass is observed to be conveyed through large streams of gas;
the cosmic filaments. Although this has not been seen on galactic scales,
galaxy formation may be expected to occur in a similar fashion.
Indeed, numerical simulations confirm this scenario
\citep[e.g.][]{dek09,goe10}\footnote{Since the submission of this thesis,
\citet{cre10} reported on an ``inverted'' metallicity gradient
in three $z\sim3$ galaxies, which they interpret as being due to the central
gas having been diluted by the accretion of primordial gas.}.
In contrast to gas accreting through a few narrow streams, outflows are more or
less isotropic.
Even bipolar outflows have a rather large opening angle, and thus
the probability of a sightline towards a galaxy passing through outflowing gas
is larger than passing through infalling gas.
Thus, it may be expected that more observations are obtained of Ly$\alpha$
profiles lacking the blue peak than lacking the red peak.

The fact that starbursts are needed to generate large outflows also imposes a
bias on the observations;
even if outflows happen only during relatively short phases in the early life
of a galaxy (for LBGs, \citet{fer06} found that a typical starburst phase lasts
only about $30\pm5$ Myr), such galaxies are more likely to appear in surveys,
simply because they are more luminous than those with small SFRs.



\subsection{Transmitted fraction of Ly$\alpha$ photons}
\label{sec:fIGM}

Even though absorption in the IGM does not alter the line shape drastically
at an intermediate redshift of 3.5, it reduces the intensity by roughly 25\%,
as seen from \tab{fIGM}.
This fraction is not a strong function of the size of a galaxy, but since the
spectra emerging from larger galaxies tend to be broader than those of smaller
galaxies, a comparatively larger part of the small galaxy spectra will fall in
the wavelength region characterized by the dip in the transmission function.
Hence, on average the IGM will transmit a larger fraction of the radiation
escaping larger galaxies.

One-fourth is not
a lot, but since it is preferentially blue photons which are lost, the
spectrum may become rather skewed when traveling through the IGM.
The lost fraction $f_{\mathrm{IGM}}$ is in addition to what is lost internally
in the galaxies due to the presence of dust.
As mentioned in Section \ref{sec:specdust}, dust tends to make the line profile
more narrow. For galaxies with no dust, the lines can be very broad.
In this case, $f_{\mathrm{IGM}}$ will be slightly higher, since for broad
lines a relatively smaller part of the spectrum falls on the dip seen in
$\Flam$.

As expected, at higher redshifts the IGM is more opaque to Ly$\alpha$ photon; 
at $z = 5.8$ and $z = 6.5$ the intensity was found to be reduced by
approximately 75\% and 80\%, respectively.      
At $z = 6.5$ the blue wing is completely lost. In most cases this is true also
at $z = 5.8$.
However, as is seen from the gray area representing the 68\% confidence
interval, in some cases an appreciable fraction of the blue wing can make it
through the IGM.


\subsection{The significance of dust}
\label{sec:dustIGM}

The calculations performed in this work are all neglecting the effect of dust.
However, as dust is mostly (although not completely) confined to the galaxies,
especially the central parts, and since by far the greatest part of the
distance covered by a given sightline is in the hot and tenuous IGM, one may
expect this to be a fair approximation.

This anticipation was confirmed including dust in the calculations.
The factor $\nhi\sigma(\lambda)$ in \eq{tau} is then replaced by
$\nhi\sigma(\lambda) + n_{\mathrm{d}}\sigma_{\mathrm{d}}(\lambda)$.
The resulting decrease in transmission is at the $10^{-4}$ to $10^{-3}$ level.


\subsection{``Early'' vs.~``late'' reionization}
\label{sec:10vs6}

As is evident from \fig{Songaila}, although none of the models really
\emph{fit}, the early and late reionization bracket the observations.
Model 1, with $\sigma_8 = 0.74$ provides a somewhat better fit to the
observations than Model 2 with $\sigma_8 = 0.9$. Due to the more clumpy
structure of the latter, galaxies tend to form earlier, rendering the IGM more
free of gas and thus resulting in a slightly more transparent Universe.
However, this should not be taken to mean that the lower value of $\sigma_8$
is more realistic, since a higher $\sigma_8$ could be accounted for by a
(slightly) smaller $\zre$.

As discussed in \sec{CMB}, since the number density $n_e$ of
free electrons increases as the Universe gets reionized, a signature of the EoR
can be obtained by measuring the total optical depth $\tau_e$ to Thomson
scattering. In the present simulations, $\tau_e$ can be calculated by
integrating \eq{dtaue} that gives $d\tau_e/dz$.

\begin{figure}[!t]
\centering
\includegraphics [width=0.90\textwidth] {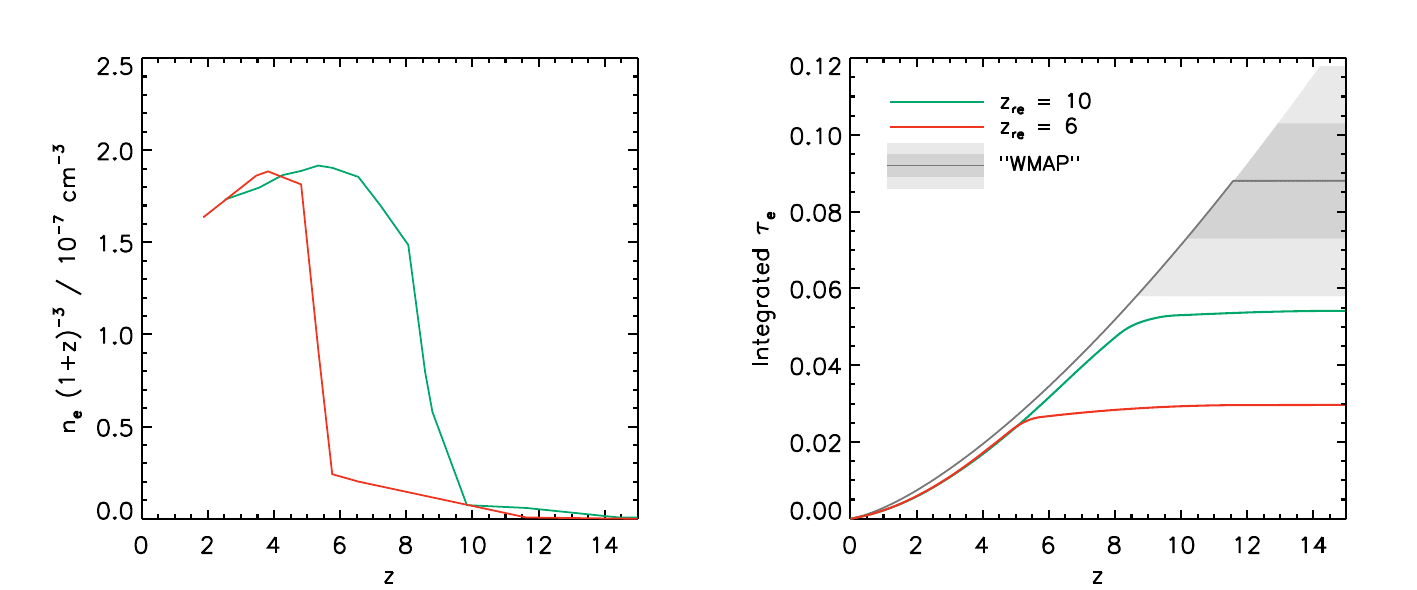}
\caption{{\cap Volume-averaged, comoving electron density $n_e/(1+z)^3$
         as a function of redshift $z$ for Model 2 (\emph{green}) and 3
         (\emph{red}) which have $\zre = 10$ and $\zre = 6$, respectively.
         Both models are characterized by a quite sharp increase in
         $n_e$ shortly after the onset of UVB. For Model 2
         the EoR is seen to take place around $z \sim 8.5$, while for
         Model the EoR lies at $z \sim 5.5$.\\
         \emph{Right:} Integrated optical depth $\tau_e$ of electrons as a
         The \emph{dark gray} line with the associated \emph{lighter gray} 68\%
         and 95\%
         confidence intervals indicate the electron density history required
         to reach the optical depth measured by WMAP, if an instant
         reionization is assumed.
         function of redshift $z$ for the two models.
         As expected, none of the models are able to reach the $0.088\pm0.015$
         inferred from the WMAP results, although the $\zre = 10$ model is
         ``only'' $\sim$2$\sigma$ away.
         Since the $n_e$ data do not extend all the way to $z = 0$, a fiducial
         value of $n_e(z=0) = 1.4\times10^{-7}$ cm$^{-1}$ has been used. The
         exact value is not very imporant, since the proper density at low
         redshift is very small.
         At low redshift, the model curves lie slightly below
         the theoretical curve. This is due to a combination of the models
         including helium
         ionization, releasing more electrons, and star formation and gas
         cool-out, removing free electrons.}}
\label{fig:WMAP}
\end{figure}

\Fig{WMAP} displays the average comoving number density $n_e$ of electrons as a
function of redshift, as well as the corresponding total optical depth
$\tau_e$ of electrons. The results for Model 2 and Model 3 are shown; those for
Model 1 lie very close to those of Model 2.
Going from larger toward smaller redshifts, both models are seen to be
characterized by a roughly constant and very low density of electrons before
reionization, then a rapid increase not \emph{at}, but shortly \emph{after}
$\zre$, and finally a slow decrease,
due to subsequent gas cool-out and resulting galaxy formation.
The rapid increase marking the EoR lasts approximately 100 Myr.
The difference in ionization history for the pseudo- and the realistic UV RT
schemes is not critical, although the creation of ionized bubbles around
stellar sources
causes a slightly earlier EoR in the realistic UV RT than the pseudo-RT is able 
to, especially in Model 3 ($\zre = 6$).

Also shown in \Fig{WMAP} is the corresponding $\tau_e$ history that would
prevail in the hypothetical case of an \emph{instant} reionization, where the
Universe is fully neutral before, and fully ionized after, some redshift
$z_{\mathrm{reion}}$ fixed to make the total $\tau_e$ match the value measured
by WMAP.
In this model, no gas is assumed to be locked up in stars, and $n_e$ is thus
given by
\begin{equation}
\label{eq:ne}
n_e(z) =  \left\{ \begin{array}{ll} 0 & \textrm{for } z > z_{\mathrm{reion}}\\
          \frac{\psi \Omega_b \rho_c (1+z)^3}{m_{\mathrm{H}}}
                                      & \textrm{for } z \le z_{\mathrm{reion}},
          \end{array} 
          \right.
\end{equation}
where $\psi = 0.76$ is the mass fraction of hydrogen, $\Omega_b = 0.046$ is the
baryonic energy density parameter \citep{jar10}, and $\rho_c$ is the critical
density of the Universe.

As seen from \fig{WMAP}, not even the early reoinization model is
able to reproduce the optical depth probed by WMAP.
However, for a model with an even earlier EoR, the transmission $\T$ of the
IGM would be even farther off the observational data, as seen in \fig{Songaila}.
In general, there seems to be a significant disagreement between the WMAP
results and the QSO results concerning the redshift of the EoR.
Many authors have tried to resolve this apparent discrepancy,
e.g.~\citet{wyi03} and \citet{cen03} who considered a \emph{double}
reionization, first at $z \sim 15$--16 and later at $z \sim 6$.

Notice nevertheless that the total $\tau_e$ of the $\zre = 10$ model is not
entirely inconsistent with the WMAP-inferred value, being roughly 2$\sigma$
away.
Furthermore, since only
the \emph{total} optical depth is measured by WMAP, the exact history of
the EoR is obviously less certain. Generally, either an instant reionization
must be assumed, or perhaps a two-step function to make the EoR slightly
extended, and possibly with an additionally
step at $z \sim 3$ to account for helium reionization.

The above results are at odds with the interpretation of the CMB EE polarization
maps as showing that the reionization of the Universe was complete at
$z = 10.5$ \citep{jar10}.
However, many factors enter the conversion of the
polarization maps into an optical depth, and
the model of this EoR may be too simplistic.
Compared to WMAP, the recently launched
Planck\footnote{http://www.rssd.esa.int/Planck/}
satellite has a much higher resolution and
sensitivity; when Planck data become available, these issues may be solved as
much tighter constraints can be put on parameters like $\tau_e$ \citep{gal10}.




\chapter[On the Ly$\alpha$ emission from damped Ly$\alpha$ absorbers]
    {On the Ly$\alpha$ emission from damped Ly$\alpha$ absorbers\footnote{This
    chapter is partly based on unpublished observations by the author and
    collaborators, partly on \citetalias{fyn10} (\app{fyn10}).}}
\label{cha:DLAs}
\index{Damped Ly$\alpha$ absorbers!In emission}

\init{A}{s already considered} in \cha{gals}, a prime goal of modern cosmology
should be to unveil the relationship between the different populations of
high-redshift galaxies. However,
whereas the observational overlap of emission selected galaxies has been
studied in reasonable detail, little is known about the connection
between absorption and emission selected galaxies.

If DLAs are really cradles of star formation, they should also emit Ly$\alpha$,
which should be observable spectroscopically, i.e.~as a small peak in the
bottom of the DLA trough. However, of the $>$1000 DLAs found to date, in
only a handful of cases associated Ly$\alpha$ has been detected (see \tab{Qs}),
\begin{table}[t]
\begin{center}
{\sc DLAs with associated Ly$\alpha$ emission}
\end{center}
\centering
\begin{tabular}{ll}
\hline
\hline
Background quasar & Reference\\
\hline
     PKS0528-250 &     \citet{mol93}      \\ 
     2233+131    &     \citet{djo96}      \\ 
     Q0151+048   &     \citet{moll98}     \\ 
     Q2059-360   &     \citet{lei99}      \\ 
{\bf Q2206-19A}  &{\bf \citet{mol02}}     \\ 
{\bf PKS0458-02} &{\bf \citet{mol04}}     \\
     HS1549+1919 &     \citet{ade06}      \\ 
{\bf Q2222-0946} &{\bf \citetalias{fyn10}}\\
\hline
\end{tabular}
\caption{{\cap List of quasars, in the spectra of which Ly$\alpha$
             emission has been detected (and published) in the trough of a
             DLA. Entries written in boldface are bona fide detections,
             i.e.~``true'' DLAs (as opposed to ``sub-DLAs'' with
             $\Nhi<10^{20.3}$ \cmsq), and at a redshift appreciably different
             from that of the quasar.}}
\label{tab:Qs}
\end{table}
even though the search for DLA emission has been going on for more than two
decades \citep[e.g.][]{smi89}, and several tens of DLAs have been showed to
have \emph{no} emission
\citep[e.g.]
[these are in addition to many null-detections that have not been published]
{cha91,low95,col02,kul06}.

Although the numerous null-detections may be largely due to the intrinsic
difficulties associated with the detection of a galaxy in the projected
proximity of the bright background source, the contrasting selection criteria
also complicates matters. While LAEs and LBGs tend to probe the bright end of
the LF, because they are flux limited, DLAs predominantly probe the faint end.
The reason is that, at least in the local Universe, the area that a galaxy
covers on the sky is proportional to its luminosity\footnote{According to the
so-called Holmberg and Bosma relations; see, e.g., \citet{wol86} and
\citet{zwa05}}. If this is true also at higher redshifts, from the slope of
the faint end of the LF it is expected that the majority of the DLAs must be
selected from the faint end \citep{fyn99,hae00,sch01,fyn08,pon08}.

\section{Q2348-011}
\label{sec:q2348}

\begin{figure}[!t]
\centering
\includegraphics [width=0.90\textwidth] {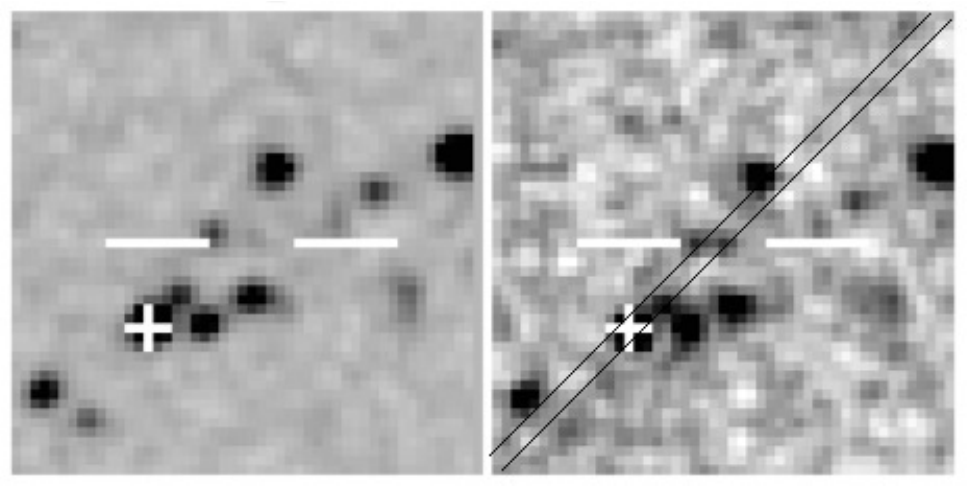}
\caption{{\cap Broadband (\emph{left}) and narrowband (\emph{right})
             images of the
             candidate (center of both images) $11''$ from the quasar Q2348-011
             (marked by the white cross). The dimension of each image is
             $40''\times40''$. North is up and east is left. The image is
             taken from \citet{man98}, and the width and position angle of
             the slit used in this study are indicated by black lines.}}
\label{fig:Q2348map}
\end{figure}

At an observing course at the Nordic Optical Telescope, Roque de los Muchachos,
La Palma, in 2007, we
--- i.e.~\href{http://www.dark-cosmology.dk/~jfynbo/}{Johan Fynbo},
\href{http://dark.nbi.ku.dk/people/christagall}{Christa Gall},
\href{http://dark.nbi.ku.dk/people/giorgosleloudas/}{Giorgos Leloudas}, and
\href{http://www.dark-cosmology.dk/~pela}{the author} --- performed
observations of the quasar Q2348-011.
This quasar, lying at a redshift of $z = 3.01$, features \emph{two} DLAs along
its line of sight, one at $z = 2.43$ and another at $2.61$ \citep{aba05}, with
evidence for an excess of UV flux associated with the former \citep{not07}.
However, as yet
searches for a galaxy responsible for the absorption have proved fruitless.
Several emitters have been identified in the immediate vicinity of the QSO.
In particular, from the excess of H$\alpha$ narrowband to broadband,
\citet{man98} find an emitter located $11''$ NW of the QSO
(\fig{Q2348map}), at the same redshift as the first absorber.

In order to search for Ly$\alpha$ emission, deep, intermediate-resolution
spectroscopy of Q2348-011 was carried out at the NOT,
with the slit of spectrograph positioned at an angle such that also
the spectrum of the emitter at $11''$ distance should be obtained, thus
enabling its redshift to be determined spectroscopically.

\subsection{Observations}
\label{sec:obsDLA}

To acquire the spectra, the ALFOSC (Andalucia Faint
Object Spectrograph and Camera) instrument together with the 2k$\times$2k
CCD on the NOT was applied.
The two DLAs lie at redshifts $z=2.43$ and $z=2.61$, and their
corresponding Ly$\alpha$ absorption troughs are thus located at wavelengths
$\lambda= 4158$ {\AA} and $\lambda = 4389$ {\AA}, respectively.
Grism \#16 of the ALFOSC was therefore used. The
wavelength coverage of this grism is in the range 3500--5060 {\AA} and its
absolute efficiency is around 50 percent at the according wavelengths.
If used with a $0\farcs5$ slit it has a resolution of $R = 2000$.
However, since high resolution was not the goal of these observations,
a slit of width $1\farcs3$ was used, resulting in a
resolution of $R\simeq770$, with no second order light in the wavelength range.

Previous searches \citep{man98} have measured the narrowband magnitudes
of the candidate objects at the H$\alpha$ line to be approximately 18.8. In
order to obtain a satisfactory signal-to-noise ratio of such a faint object,
an exposure time of at least five hours was needed. This was based on our
calculations but also on previous experience of such detections reported in the
literature \citep[e.g.,][]{mol04}. The total exposure time was divided in
shorter exposures of 1800 seconds. Since for the chosen grism the dispersion is
0.77 {\AA} per pixel, the readout noise could be reduced
by a binning of the CCD at $2\times2$ without any loss of information,
providing final pixels of size $0\farcs39$ by 1.54 {\AA}.

In order to simultaneously acquire the spectrum of the H$\alpha$ emitter,
the $1\farcs3$ long-slit was aligned at position angle PA = 45$^{\circ}$ with
respect to the E-W direction (\fig{Q2348map}).
The quasars were observed over three nights. During the first night the
conditions were not optimal, with relatively high seeing and airmass.
However, this improved radically during the last two nights (see \tab{obs}).
\begin{table}[!t]
\begin{center}
{\sc Observational data for the three nights}
\end{center}
\centering
\begin{tabular}{ccccc}
\hline
\hline
Date       & UT start  & $t_{\mathrm{exp}}$/s &   Airmass  & Seeing\\
\hline
08.18.2007 & $23^{00}$ & 3$\times$1800        & 1.48--2.05 & 2\farcs0--2\farcs4\\
08.19.2007 & $02^{00}$ & 3$\times$1800        & 1.15--1.19 & 0\farcs8--2\farcs4\\
08.21.2007 & $02^{30}$ & 4$\times$1800        & 1.15--1.20 & 0\farcs6--1\farcs0\\
\hline
\end{tabular}
\caption{{\cap }}
\label{tab:obs}
\end{table}
For the flux calibration, spectra of the standard star Feige 110 (SP2317059) was
obtained, using the same grism and slit.
This standard star was a perfect choice due its close alignment on the
sky with Q2348-011. For the wavelength calibration and flat-fielding, a helium
and a halogen lamp was used, respectively.


\subsection{Data reduction}
\label{sec:red}

The data was reduced with IRAF\footnote{IRAF is distributed by the National
Optical Astronomy Observatory (NOAO), which is operated by the Association of
Universities for Research in Astronomy, Inc.~(AURA) under cooperative agreement
with the National Science Foundation.} through a standard image reduction
procedure of overscan subtraction, bias subtraction, and flat-fielding.
The uneven illumination of the halogen lamp was corrected for, using the tasks
\verb+response+ and \verb+illumination+.

False signals in the images were removed using the task \verb+lacos_spec+
\citep{van01}
The ten 1800 sec spectra were equally weighted and averaged (although there
are alternative ways taking into account the different image variance, seeing,
etc.), to obtain one final 2D spectrum.

Subsequently, the background was removed in the following way: two stripes of
width $\sim 10$--20 pixels to the left and right, respectively, of the
spectrum and sufficiently close to
it, were used to compute an average column. This average column, representing
the average background per wavelength, was then subtracted from the original
image. Due to the bending of the skylines, this way of background subtraction
is only accurate in the center of the image, where the spectrum lies.

To obtain 1D spectra, the task \verb+apall+ was used. The 1D spectra were
wavelength calibrated by use of the task \verb+dispcor+. One wavelength
calibrated 1D spectrum were obtained per night (assuming that the wavelength
calibration did not change over the course of a single night) and the three
1D spectra are then combined with the help of  \verb+scombine+ to obtain a
final 1D spectrum (shown in \fig{1Dspec}).


\subsection{Results}
\label{sec:res}

\begin{figure}[!t]
\centering
\includegraphics [width=0.90\textwidth] {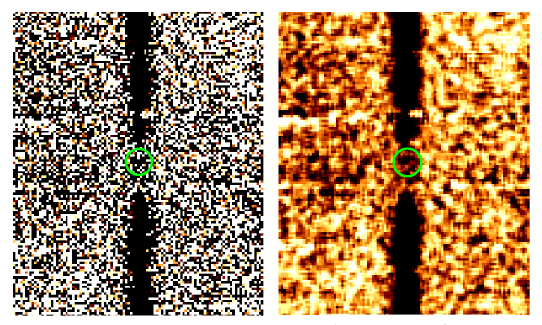}
\caption{{\cap Two-dimensional spectrum of the quasar Q2348-011,
        non-smoothed (\emph{left}) and smoothed (\emph{right}). The emission
        in the trough of the DLA is marked by a \emph{green} circle.}}
\label{fig:2Dspec}
\end{figure}

To determine the presence (and significance) of Ly$\alpha$
emission in the DLA trough, the two-dimensional,  background corrected spectrum
(\fig{2Dspec}) was used.
A quick, visual inspection is enough to reveal that even if this emission
is present, it is not as prominent as in \citet{mol04}. However, there seems
to be an excess of counts over the background.
The significance of the detection was measured in the following way:

The corresponding image statistics were computed in IRAF with the use of
\verb+imexamine+, option \verb+m+. This computes the statistics in an
aperture centered at the position of the cursor. A square box covering an
area of 25 pixels was used for this purpose.
Measuring in this way the local background values close to the trough results
in $0.61 \pm 0.37$ counts. Here the error is
the standard deviation of the mean counts in the aperture averaged over
20 local regions. In the middle of the trough, the value is $1.91$ counts,
implying that the significance of the detection is at the $3.5\sigma$ level.

\subsubsection{The effect of smoothing}
\label{sec:smooth}
 
The image in the left panel of \fig{2Dspec} is rather noisy. Smoothing the
image allows the detection to be seen more clearly.
This was done with the task \verb+boxcar+ and a
smoothing window $3 \times 3$ pixels. The resulting image can be seen in
the right image of Fig.~\ref{fig:2Dspec}. Computing the corresponding
statistics for this image,
we get $0.41 \pm 0.30$ counts for the local background and $1.69$ counts
for the middle of the trough, implying a $4.2 \sigma$ detection.


\subsubsection{Emitted flux}
\label{sec:flux}

\begin{figure}[!t]
\centering
\includegraphics [width=0.90\textwidth] {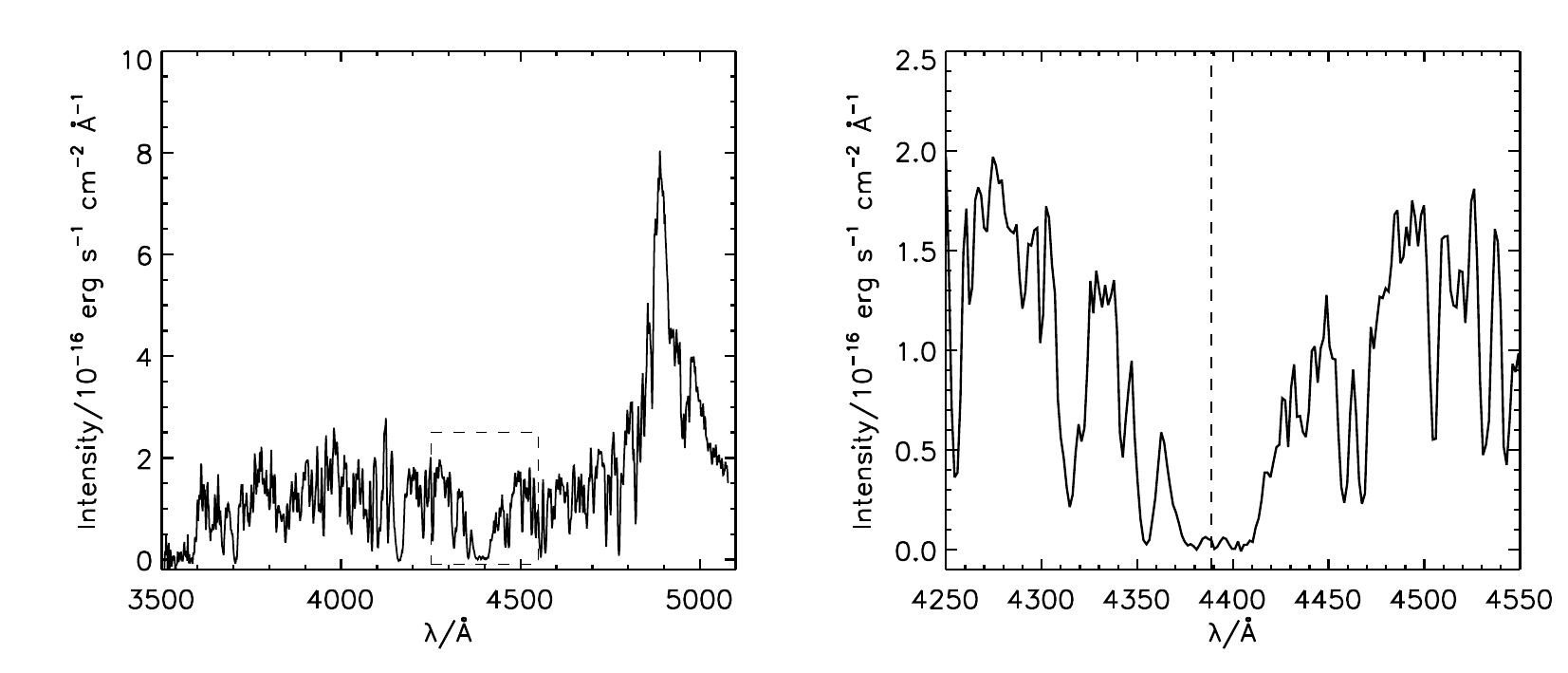}
\caption{{\cap Full, one-dimensional spectra of the quasar
        Q2348-011 (\emph{left}) and zoom-in on the region around the DLA
        situated at $z = 2.61$ (\emph{right}). The possible Ly$\alpha$
        emission is seen as two small bumps in the trough.}}
\label{fig:1Dspec}
\end{figure}

The tentative detection of the Ly$\alpha$ emission line (even at this low
significance) probably implies active star formation in the DLA. Using the tool
\verb+splot+ on the final flux calibrated 1D spectrum (\fig{1Dspec}),
a flux of $3.57\times10^{-17}$ \esca at a
wavelength of 4386 {\AA} is deduced.
There is also a second bump in the trough, with an excess over the background
of $2.77\times10^{-17}$ \esca at 4395 {\AA}.

However, flux calibration of spectra might suffer from effects like slit
losses, extraction choices, etc. For this reason, to measure the total
emitted flux in one more way, the counts level at the
bottom of the trough was compared with another well calibrated line.

For this purpose, a spectrum of the quasar obtained by the Sloan Digitized Sky
Survey (SDSS) was used. This spectrum, although it is not as deep as ours,
has the advantage that it comes from a well known pipeline with standard,
well understood and trustworthy calibrations.
Surprisingly, the NOT spectrum has more flux than the SDSS spectrum.
Since the results were very similar however, this correction was not applied.



\subsection{Follow-up observations}
\label{sec:follow}

Although 3.5$\sigma$ implies a positive detection with 99.95\% 
confidence, this is just not good enough for science! In order to secure an
unambiguous detection, 15 more hours of observing time at the NOT were applied
for, and granted (with the author as PI). Since originally the target was
observed for five hours, this quadrupled the total exposure time, thus
increasing the signal-to-noise ratio by a factor of two.

The observations were carried out during four nights, starting 9.14.2009.
Due to unfavorable weather conditions, only 11.25 hours were obtained.
The data were reduced and processed in a similar fashion as described below,
and the spectra were then stacked with the 2007-spectra.
Unfortunately, even with this total of more than 16 hours of integration, the
detection confidence was not increased, and the project was not pursued further.

Of course, the fact that the emission was only detected at a fairly low
significance does not necessarily imply that the absorber features no, or only
very little, intrinsic Ly$\alpha$ emission.
Previous identifications were all carried out on telescopes significantly
larger, so the result may simply be a consequence of the relatively small
mirror of the NOT. Moreover, as should be evident from the above numerical
results, resonant scattering effects and dust may easily suppress the
Ly$\alpha$ line below the detection limits.




\section{Q2222-0946}
\label{sec:Q2222}

As a more methodical way of probing Ly$\alpha$ emission from DLAs, one may
specifically target candidate metal-rich DLAs. In the local Universe, the
relation between the metallicity and the luminosity, as well as the metallicity
and the line profile velocity width, of galaxies are fairly
well-established.
Similar slopes of these relations are expected to exist at
high redshifts \citep{mol04,led06}. Furthermore, despite a small sample
the DLAs for which emission (in any wavelength) \emph{has} been detected tend
to have the highest metallicity among galaxies at the given redshifts
\citep{mol04,wea05}.

To this end, a survey targeting 12 high-metallicity DLAs using the newly built
echelle spectrograph
X-shooter\footnote{{\tt http://www.eso.org/sci/facilities/develop/instruments/xshooter/}},
mounted on the European Southern Observatory
(ESO) Very Large Telescope (VLT), Cerro Paranal, Chile, with Johan Fynbo as PI
(and the author as co-I).
X-shooter has the advantage that it covers the entire wavelength range from
the atmospheric cut-off at $\sim$3100 {\AA} to the $K$-band at 2.5 $\mu$
(in three different arms; the NIR, VIS, and UVB arm), thus
allowing for simultaneous identification of many important lines.
For this study, the intermediate spectral resolution is close to
optimal, since near-IR sky lines will be resolved while it is sufficiently low
not to be detector noise-limited.

The DLAs are selected from the SDSS DLA sample of \citet{not09} according to
the following criteria:
First, the rest frame EW of \ion{Si}{ii} $\lambda$1526 must be $>$1 {\AA}.
This should ensure that the metallicity is indeed high,
i.e.~$\gtrsim$0.1 $Z_\odot$ \citep[e.g.][]{pro08}.
Second, the \ion{Fe}{ii} $\lambda$2344, $\lambda$2374, $\lambda$2382 must be
well-detected. 
Finally, in order to be able to detect other emission than Ly$\alpha$ alone,
only DLAs at a redshift of $z \sim 2.4$ are included in the final sample,
since this implies that the H$\beta$, \fion{O}{ii}, \fion{O}{iii} and/or
H$\alpha$ emission lines are in the NIR transmission windows.

\begin{figure}[!t]
\centering
\includegraphics [width=0.80\textwidth] {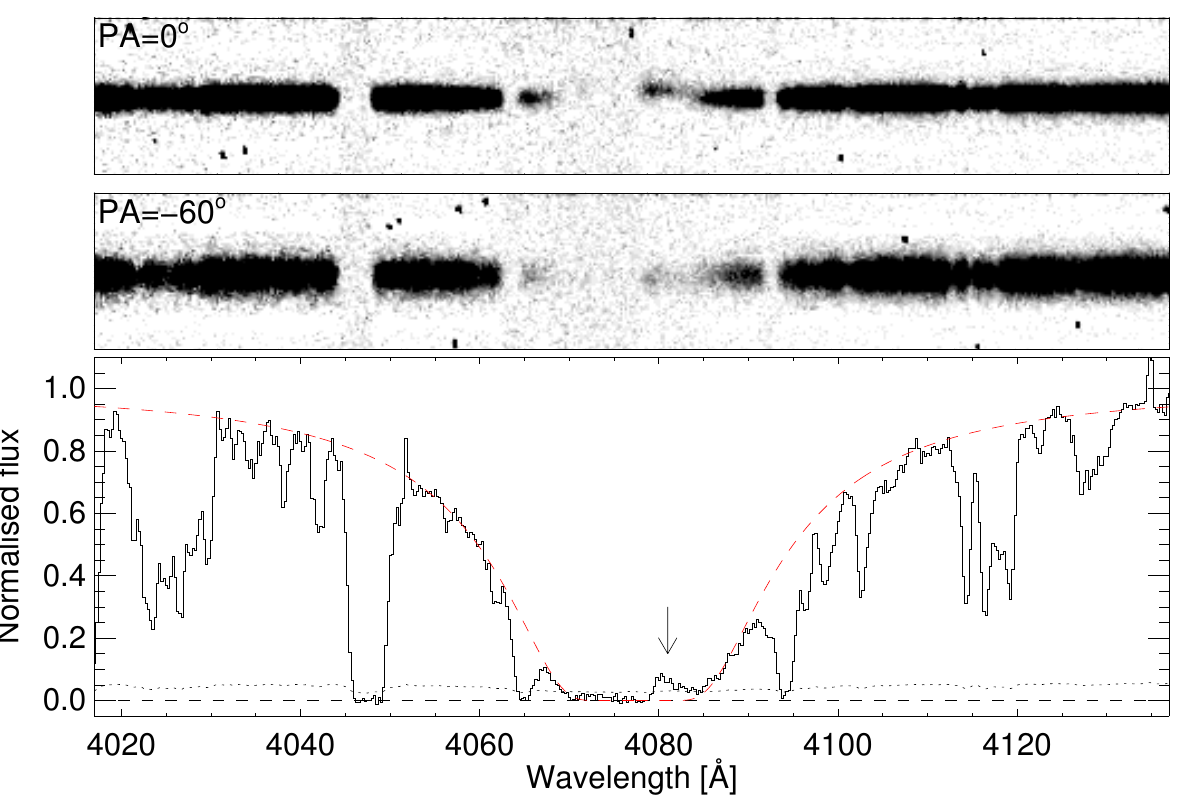}
\caption{{\cap One- and two-dimensional spectra of the DLA toward Q2222-0946.
               The two 2D spectra correspond to the PA = 0$^\circ$ with the
               best seeing (\emph{upper panel}) and the PA = $-60^\circ$
               (\emph{middle panel}).
               The 1D plot (\emph{lower panel}) is extracted from the PA =
               0$^\circ$ spectrum. In this plot the trace of the quasar spectrum
               is used, and since the Ly$\alpha$ emission line (marked by an
               arrow) is offset from the trace by 0\farcs57, the line is not
               fully recovered.
               Also seen in the plot is the noise spectrum (\emph{dotted line})
               and the best fit Voigt profile (\emph{red dashed line}).
               }}
\label{fig:Q2222}
\end{figure}

To cover the field of view in the vicinity of the quasar and allow for a
triangulation of a possible emitter, the observations are split up in three,
with position angles (PAs) 60$^\circ$, $-60^\circ$, and 0$^\circ$.
The probability of missing the galaxy counterpart is then expected to be less
than 10\%, based on the model of \citet{fyn08}.
The first target in the survey, the $z = 2.93$ quasar Q2222-0946, was observed
on 10.21.2009, and turned out to exhibit a beautiful, unambiguous Ly$\alpha$
emission line in the trough of a DLA at $z_{\mathrm{abs}} = 2.35$
(see \fig{Q2222}).
In this case, a 1\farcs3 slit was used in the UVB arm, while a 1\farcs2 slit was
used in the VIS and NIR arms, resulting in a resolution of 4700 (UVB), 6700
(VIS), and 4400 (NIR). 
The intended exposure time was 1 hr per PA, but due to an error in the
observing block
the PA = $60^\circ$ exposure was replaced by a second PA = $0^\circ$ exposure.

The spectra were processed by means of a preliminary version of the X-shooter
data reduction pipeline \citep{gol06}.

\subsection{Galaxy counterpart}
\label{sec:counter}

\subsubsection{Metallicity and dust}
\label{sec:DLAZ}

A plethora of absorption lines were detected at the redshift of the DLA,
including various ionization and excitation states of carbon, oxygen, silicon,
iron, and zinc. Since Si and Zn are thought not to deplete much to dust, these
lines can be used to estimate the metallicity.
Fitting Voigt profiles to the lines, metallicities of $[$Si/H$] = -0.51\pm0.06$
and $[$Zn/H$] = -0.46\pm0.07$ are found, indeed well above the expected
threshold of 0.1 solar.
For Ni, Fe, and Mn, metallicities lower by a factor of approximately 2, 3, and
6, respectively, are found, indicating notable dust depletion.


\subsubsection{Star formation rate}
\label{sec:DLASFR}

Thanks to the spectacular wavelength range of X-shooter, it was possible
simultaneously to obtain Ly$\alpha$, \fion{O}{iii}, and H$\alpha$ emission
lines.
This has not been seen in DLA galaxies before and is rarely seen for other
types of high-$z$ galaxies.

The total flux in the Ly$\alpha$ line is $8.9\times10^{-17}$ \ergs \cmsq,
corresponding to a total Ly$\alpha$ luminosity of
$L_{\mathrm{Ly}\alpha} = 3.8\times10^{42}$ \ergs.
Assuming standard case B recombination and the \citeauthor{ken98} relation
yields SFR = 3--4 \Mpyr.
For the H$\alpha$ line, a flux of $2.5\times10^{-17}$ \ergs \cmsq
is measured,
corresponding to a luminosity of $L_{\mathrm{H}\alpha} = 1.1\times10^{42}$
\ergs, and an SFR of 8--9 \Mpyr.

The discrepancy between the two calculated SFRs suggests a loss of Ly$\alpha$
photons, either due to dust, anisotropic emission, or IGM absorption.
According to case B recombination theory, the intrinsic ration between
Ly$\alpha$ and H$\alpha$ luminosity is 8.7 (neglecting cooling radiation).
But since in this case
$L_{\mathrm{Ly}\alpha} / L_{\mathrm{H}\alpha} \simeq 3.5$,
approximately $1 - 3.5/8.7 \simeq 60$\% of the photons seem to have been lost.

Due to an unfortunate overlap with sky lines, oxygen can only be used to place
an upper limit of the SFR of 40 \Mpyr.


\subsubsection{Morphology and kinematics}
\label{sec:DLAmorph}

Since only two PAs were obtained, an exact position is inaccessible, but in
this case an approximate impact parameter of $\sim$0\farcs8 could be estimated,
corresponding to 6.5 kpc.

The spatial profiles of both the Ly$\alpha$ and the \fion{O}{iii} line are
consistent with the spectral PSF with a seeing of 0\farcs7, meaning that the
star-forming region of the galaxy is much more compact than the neutral gas. 
Comparing to the H$\alpha$ LF for $z \sim 2$  galaxies \citep{hay10a}, the
galaxy has an H$\alpha$ luminosity of 0.1 $L^{*}$, corresponding to a dwarf
galaxy.

\begin{SCfigure}[][t]
\includegraphics[width=6.5cm,clip]{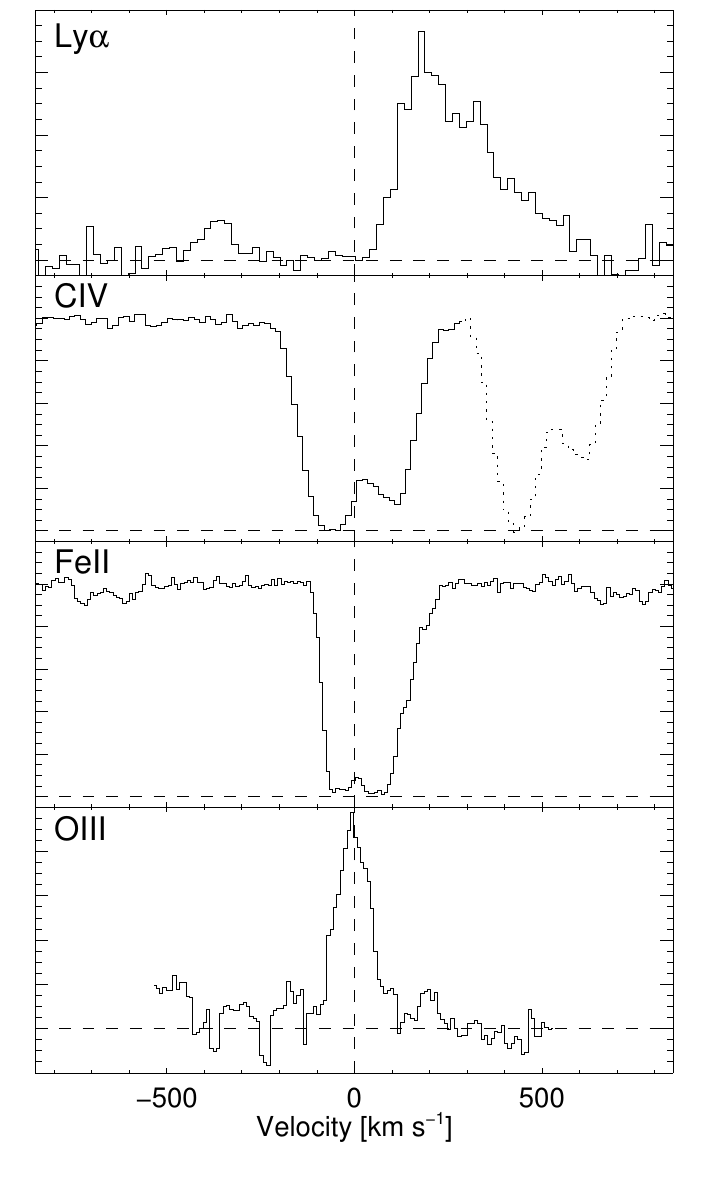}
\caption[hiort caption]{\label{fig:DLAlines} {\cap Emission and absorption
                   lines associated with the DLA
                   toward Q2222-0946.
                   \emph{Bottom} plot shows the \fion{O}{iii} emission line,
                   marking the systemic redshift of $z = 2.35406$ defining
                   the velocity $v = 0$.
                   The two \emph{middle} plots show the low- and
                   high-ionization emission lines, represented by \ion{Fe}{ii}
                   $\lambda$2600 and \ion{C}{iv} $\lambda$1548, respectively,
                   having a similar redshift as the absorption.
                   \emph{Top} plot shows the Ly$\alpha$ emission. The
                   apparent shift in velocity can be explained by resonant
                   scattering effects rather than by a different velocity
                   component. The reminiscence of the blue peak is seen at
                   $v \sim 400$ \kms.\vspace{2mm}\\
                   \ \\
                   \ \\
                   \ \\
                   \ \\
                   \ \\
                   \ \\
                   \ \\
                   \ \\
}}
\end{SCfigure}

Since \fion{O}{iii} is created mainly in \ion{H}{ii} regions and does not
resonant scatter, this line is expected to probe the systemic redshift of the
galaxy. The width of this line is 80 \kms, while the width of both low- and
high-ionization absorption lines are significantly broader and consist of two
components. This could be a signature of galactic outflows, although of course
the emission lines and the absorption lines probe two different regions.
Still, however, the absorption redshifts are very similar to the systemic, as
seen from \fig{DLAlines}.



\subsection{The Ly$\alpha$ emission line profile}
\label{sec:lyaprof}

The profile of the Ly$\alpha$ line shown in the top panel of \fig{DLAlines}
is clearly asymmetric.
Remarkably the blue edge of the profile is close to the
position of rest frame Ly$\alpha$ at the systemic redshift.
Even more striking is
the presence of a small peak on the blue side of the systemic
redshift. This can be understood as a result of resonant scattering.
As considered theoretically in \cha{ResScat} and numerically in \cha{conseq},
in general Ly$\alpha$ radiation should escape its host galaxy in with a
double-peaked profile.
Observationally, however, the blue peak is commonly found to be dimished or,
more often, entirely suppressed.
The cause of this suppression may be galactic winds, but could also be partly
due to absorption in the IGM.

Considering the relatively high metallicity inferred for the system and the
apparent depletion of refractory elements, the presence of dust is anticipated.
Due to the path length of the Ly$\alpha$ radiation being increased by resonant
scattering, dust may suppress the Ly$\alpha$ line more than other lines. Even
for ``gray'' dust, the line is not affected uniformly, the wings being
influenced more than the centre, as discussed in \sec{specdust}.

If we wish to understand the physical conditions responsible for the
formation of the observed Ly$\alpha$ line, we must take all of these
effects into account.
\citet{ver08} successfully
modeled a number of Ly$\alpha$ profiles by varying the temperature,
column density, expansion velocity and dust contents of a thin shell.
Analytical solutions exist only for homogeneous, isothermal, and
static configurations of gas, with either a central or evenly
distributed source of light.
In the case of a homogeneous gas, each of the two peaks are fairly symmetric
about their individual maxima, but taking into
account the full range of densities, and the correlation of Ly$\alpha$
emission with these densities, often results in significantly more
skewed peaks. This is the result of different parts of the spectrum
originating in physically distinct regions.
A realistic scenario of
galactic outflows probably lies somewhere in between that of the
homologously expanding sphere and that of a thin shell.
From \fig{expand}, which shows the spectra of a homologously expanding sphere
of column density $\Nhi = 10^{20.3}$ \cmsq, one sees that in the case of the
expanding sphere with a maximum velocity $v_{\mathrm{max}}$ of 20 \kms,
the red peak maximum is
approximately twice as high as the blue peak maximum, while for
$v_{\mathrm{max}} = 200$ \kms the blue peak is missing completely.
In the case of a
shell, from Fig.~14 of \citet{ver06} slightly larger expansion
velocities can occur while still allowing the blue peak to be seen.

\begin{figure}[!t]
\centering
\includegraphics [width=0.70\textwidth] {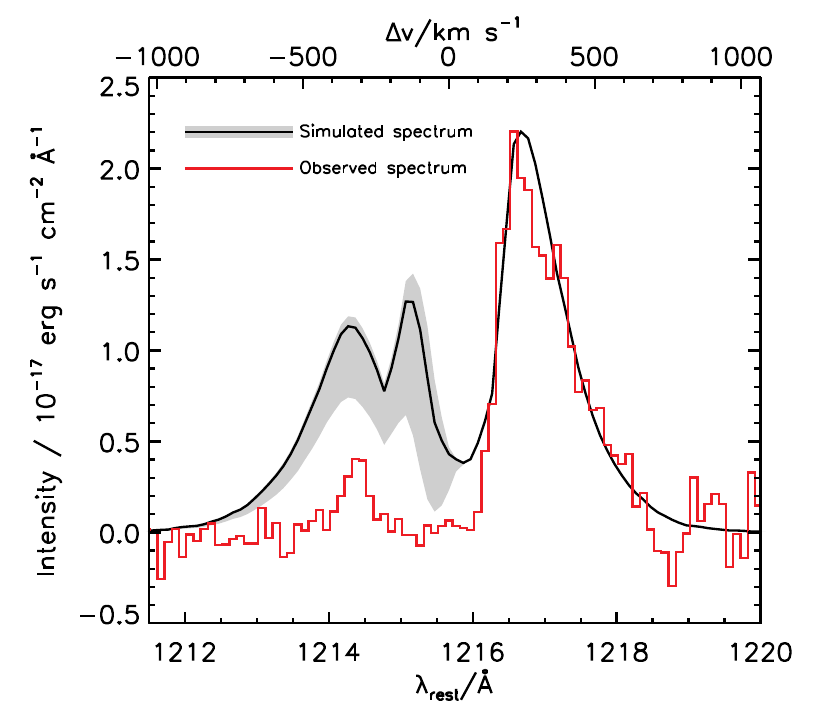}
\caption{{\cap Observed Ly$\alpha$ spectrum (\emph{red}) superimposed on a
               simulated spectrum (\emph{black}), with 68\% confidence interval
               (\emph{gray}). The latter spectrum is produced by first
               performing Ly$\alpha$ RT in a galaxy that resembles the inferred
               properties of the observed galaxy, then multiplying by the
               transmission function $\Flam$ (\sec{F}) to simulate the
               impact of the IGM, and finally normalizing such that the
               amplitude of the red peak matches that of the observed
               spectrum.}}
\label{fig:Lyafit}
\end{figure}
\Fig{Lyafit} shows the observed Ly$\alpha$ spectrum, together with a simulated
spectrum taken from the (low-resolution) cosmological simulation with
$\zre = 10, \sigma_8 = 0.74$, at a redshift of 2.5.
The galaxy used for the
Ly$\alpha$ RT has a metallicity of $[$O/H$] = -0.52$, a circular velocity of
$V_c = 80$ \kms, and a stellar mass of $1.1\times10^9$ $\Msun$,
and thus more or less resembles the above inferred properties of the observed
galaxy counterpart.

Although hardly a fit, the similarity between the two spectra is quite
remarkable, in particular the red peak. On average, the simulated galaxy is
accreting gas at $\sim$50 \kms, but in the direction in which the fitted
spectrum is ``observed'' (toward negative $x$-values), a stream of gas is
flowing out at $\sim$10--20 \kms.
If this outflow were a bit higher, the blue peak would be further reduced,
providing an even better fit.

The escape fraction of the simulated galaxy is $\sim$10\%, somewhat lower than
that inferred from comparing the H$\alpha$ and the Ly$\alpha$ luminosities.
However, $f_{\mathrm{esc}}$ might increase to some extend if the galaxy were
simulated at high resolution, matching more closely the
$f_{\mathrm{esc}} \sim 1/2$ of the observed galaxy.





\chapter{Postscript}
\label{cha:post}

\init{A}{ numerical Monte Carlo} Ly$\alpha$ radiative transfer code has been
presented
and tested against various analytical solutions. The code is capable of
propagating Ly$\alpha$ radiation on an adaptively refined mesh, with
an arbitrary
distribution of Ly$\alpha$ source emission, temperature and velocity field of
the ISM, as well as density of neutral and ionized hydrogen,
and dust.
The abundance of, and the interaction probability with, dust was
modeled by scaling known extinction curves to the
(location-specific) metallicity of the galaxies, and destruction processes were
modeled by reducing the dust density in regions where hydrogen is ionized.
Applied to simulated galaxies emerging from fully cosmological
simulations, a number of interesting characteristics of galaxies in the
high-redshift Universe has been unveiled, many of which are not possible to
ascertain in simplified model of galaxies.

To investigate the general effect of the RT in the IGM, these
calculations were performed separately. Rather than continuing the Monte Carlo
RT in the IGM, this was realized by calculating the average absortion of a
large number of sightlines starting just outside the galaxies.
The high resolution of the cosmological simulations
combined with the adaptive gridding for the RT allows us to probe the
velocity field around the galaxies in great detail.

Additionally, as part of a more comprehensive survey with the aim of building a
bridge between galaxies selected by their emission
properties and those selected by their absortion properties, observations of
DLAs were conducted, with the particular aim of finding
their galaxy counterpart. One of these observations, performed with the
X-shooter spectrograph mounted on the VLT proved successful, and the obtained
spectrum was compared to a simulated spectrum, demonstrating the potential of
the Ly$\alpha$ RT code as a tool for interpreting observations.

The main conclusions are summarized below.

\section{Summary}
\label{sec:sum}

\subsection{Extended surface brightness}
\label{sec:sumextSB}

Although the bulk of the photons are produced in the central, star-forming
regions, in general the scattering of Ly$\alpha$ photons in the neutral
hydrogen encompassing young galaxies will make a significant fraction of the
photons
escape far from the center. In contrast to continuum radiation which
escapes directly, this results in an extended SB. Comparing to
observations of a LAE reveals not only qualitative but also quantitative
agreement.

This effect is only amplified when dust is present in the galaxy.
Because the bulk of the photons are produced in the vicinity of stars, which
tend to be clustered in the central parts, in spite of scattering effects the
SB profile will often exhibit a central bump.
However, since dust is created from metals, which in turn are created by stars,
these regions are also the most dusty, and since additionally most of the
neutral gas resides here, greatly increasing the Ly$\alpha$ path length, the
probability of being absorbed in regions that would otherwise appear most
luminous a larger. Hence, the SB profiles of galaxies, when
observed in Ly$\alpha$, are ``smoothed out''. In particular the centra bump is
reduced, giving rise
to an even more ``flat'' SB profile, which can then be interpreted as an
extended SB profile when comparing to continuum bands.


\subsection{Anisotropic escape}
\label{sec:sumanis}

Ly$\alpha$ photons tend to take the shortest way out. Thus, any deviation from
spherical symmetry will make the photons escape anisotropically.
In the sample of galaxies studied in this work it was found that the maximum
observed SB on average varies by an order of magnitude, while the total
observed Ly$\alpha$ flux varies somewhat less; a factor of 3--6. If dust is
present in the galaxies, the effect is slightly reduced, but the difference is
still a factor of 2--4. This effect introduces a larger scatter in the
inferred luminosities and SFRs of LAEs than is prevalent inherently.



\subsection{Narrowing of the Ly$\alpha$ line profile by dust}
\label{sec:sumnarrow}

Although the cross section of the dust is nearly independent of the
wavelength of light across the Ly$\alpha$ line, the spectrum is affected in a
highly wavelength dependent fashion:
close to the line center, the escape fraction is of the order 50\%, while in
the wings it quickly approaches zero. Consequently, the line is severely
narrowed, although its width may still reach several hundreds and even
$\sim$1000 \kms.

The reason is that different parts of the spectrum originates in physically
distinct environments of its host galaxy.
The photons that are produced in the dense, central regions have to diffuse to
either the blue or the red side of the line center, where the cross section of
neutral hydrogen decreases rapidly.
Hence, the wings of the spectrum emanate from the star-forming 
regions, whereas the central parts of the spectrum to a high degree stem from
the regions less populated by stars, and from gravitational cooling of
infalling gas.
Since the dust originates from stars, by far most of the dust is found where
the wings are produced, implying that this is the part of the spectrum that is
affected the most by dust.


\subsection{Correlation of Ly$\alpha$ escape fraction with galaxy size}
\label{sec:sumfesc}

The escape fraction of Ly$\alpha$ seems to decrease with increasing
size of the host galaxy. This is indicated by \fig{fesc_M},
which show that $f_{\mathrm{esc}}$ is close to unity for galaxies of
$M_{\mathrm{vir}} \sim 10^9$--$10^{10}$ $M_\odot$, while it falls off to a few
\% for $M_{\mathrm{vir}} \sim 10^{11}$--$10^{12}$ $M_\odot$.
This effect could be caused partly by the fact that smaller galaxies experience
less total star formation, so that the total amount of metals, and hence dust,
is smaller than for massive galaxies. However, since the same trend is seen for
the escape of ionizing UV radiation \citep{raz09}, for which dust is less
important, the chief factor may be the feedback energy of the smaller
galaxies being able to ``puff up'' and disorder its host to a larger degree
than for massive galaxies, due to their gravitational potential being smaller.


\subsection{Absorption in the intergalactic medium}
\label{sec:sumIGMabs}

With the scheme for the IGM RT it was calculated how Ly$\alpha$ line profiles
emerging from high-redshift galaxies is reshaped by the surrounding IGM.
While earlier studies \citep{zhe10a} have shown that this approach of simply
multiplying an $e^{-\tau}$ factor on the intrinsically emitted Ly$\alpha$ line
is a poor approximation for the observed line profile and transmitted fraction,
we have argued that when the circumgalactic environs are sufficiently resolved
and combined with realistically calculated intrinsic Ly$\alpha$ lines,
this approach should be valid.
In general a larger fraction of the blue side of the line center $\lambda_0$
is lost as one moves toward higher redshifts.
Special emphasis was put on how Ly$\alpha$ line profiles
emerging from high-redshift galaxies are reshaped by the surrounding IGM.
At $z \gtrsim 5$, almost all of the light blueward of $\lambda_0$ is lost,
scattered out of the line of sight by the high neutral fraction of hydrogen.
However, even at relatively low redshift more absorption takes
place just blueward $\lambda_0$.
A transmission function $\Flam$ was calculated, giving the fraction of light
that is transmitted trough the IGM at various epochs. At all redshifts where
some of the light blueward of $\lambda_0$ \emph{is} transmitted (i.e. at
redshifts below $\sim$5.5), a significant dip with a width of the order 1 {\AA}
is seen.
The origin of this dip is a combination of an increased density of neutral
hydrogen and a retarded Hubble flow in the vicinity of the galaxies.

This extra absorption may in some cases be the reason that the blue peak of an
otherwise double-peaked Ly$\alpha$ profile is severely reduced, or lacking.
Nevertheless, it is not sufficient to be the full explanation of the numerous
observations of $z \sim 2$--4 Ly$\alpha$ profiles showing only the red peak.
The outflow scenario still seems a credible interpretation.

Combining the inferred transmission functions with simulated line profiles,
the fraction of Ly$\alpha$ photons that are transmitted through the IGM at
$z \sim 3.5$, 5.8, and 6.5 was computed, and was found to be
$f_{\mathrm{IGM}}(z=3.5) = 0.77_{-0.34}^{+0.17}$,
$f_{\mathrm{IGM}}(z=5.8) = 0.26_{-0.18}^{+0.13}$, and
$f_{\mathrm{IGM}}(z=6.5) = 0.20_{-0.18}^{+0.12}$, respectively.
This is in addition to what is lost internally in the galaxies due to dust.
The standard deviations were found to be dominated by sightline-to-sightline
variations rather than galaxy-to-galaxy variations.


\subsection{Constraints on the Epoch of Reionization}
\label{sec:sumEoR}

Considering the average fraction of light far from the line on the blue side
transmitted through the IGM, and comparing to the comprehensive set of
observations of the LAF by \citet{son04}, the EoR was constrained to have
initiated between $\zre = 10$ and $\zre = 6$, corresponding the having
ionized a significant fraction of the Universe around $z \sim 5.5$ and
$z \sim 8.5$, respectively.
Even though the ``early'' models of $\zre = 10$ produces a slightly too
transparent Universe when comparing to the LAF,
the optical depth of electrons is too low
when comparing to the observations of the CMB by the WMAP satellite, possibly
indicating a too simplistic interpretation of the CMB polarization.


\subsection{Identification of a DLA galaxy counterpart}
\label{sec:sumDLA}

Two DLAs were searched for in emission; one with the
Nordic Optical Telescope and one with the X-shooter spectrograph on the VLT.
The former resulted only in a tentative detection, but the latter turned out
to exhibit several emission lines, most notably Ly$\alpha$ emission in the
bottom of the trough. Identifying on the basis of the inferred metallicity and
velocity dispersion a similar simulated galaxy, the observed and simulated
line profiles were compared. It was concluded that it was consistent with
originating at the same redshift as the systemic (in spite of the most
prominent feature being shifted by several hundred \kms),
having a fair amount of its photons absorbed internally by dust, plus
additionally some in the IGM,
and possibly being surrounded by outflows at some tens of \kms.



\section{Outlook}
\label{sec:out}

{\sc MoCaLaTA} has proved a valuable tool, not only to predict various
observables and
physical properties of high-redshift galaxies, but also to interpret
observations by modeling real data.
Nonetheless, there is always room for improvement, as well as new fields to be
explored.
The following sections outline future prospects that are intended to be
conducted.

\subsection{Ly$\alpha$ radiative transfer in a multi-phase medium}
\label{sec:molec}
\index{Multi-phase medium}

\begin{figure}[!t]
\centering
\includegraphics [width=0.60\textwidth] {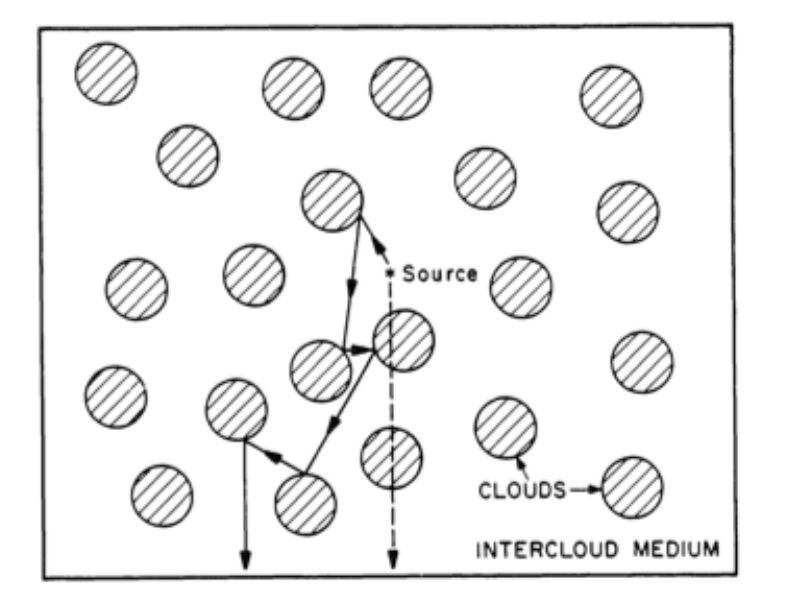}
\caption{{\cap The ``Neufeld scenario'': in a multi-phase medium, Ly$\alpha$
         radiation (\emph{solid line}) may escape more easily than continuum
         radiation (\emph{dashed line}), effectively enhancing the equivalent
         width (from \citet{neu91}).}}
\label{fig:multi}
\end{figure}

As substantiated in \sec{need}, high resolution is very important to be able to
calculate escape fractions realistically.
Although convergence tests showed us that $f_{\mathrm{esc}}$ does not change
with higher resolution, sub-grid physics could alter this picture.
The smallest scales probed by the simulation is approximately 10 pc. This
resolution is set not by {\sc MoCaLaTA}, but by the underlying cosmological
simulation.
Since cooling below $\sim$$10^4$ K is not implemented, the gas does not
condense to cold clouds. If the dust is primarily locked up in such cold clouds
of neutral hydrogen dispersed in a empty intercloud medium, the Ly$\alpha$
photon could in principle scatter off of the surface of the clouds, thus
effectively confining their journey to the dustless medium.
Continuum radiation, however, will penetrate the clouds and be subject to large
attenuation, resulting in an enhanced Ly$\alpha$ EW (see \fig{multi}).

This is the scenario described analytically and numerically by \citet{neu91}
and \citet{han06}, respectively. Numerous authors have relied on this to explain
EWs larger than what their theories could account for, in the most extreme
cases simply by tuning a ``clumpiness parameter'' to a convenient value.
Although clumpiness of the ISM undoubtedly facilitates the escape,
the ``Neufeld scenario'' \emph{is} rather simplistic, in particular because the
bulk of the photons are produced close to massive stars which predominantly are
born \emph{inside} giant molecular, and hence probably dusty, clouds.

To investigate the effect of sub-parsec resolution on $f_{\mathrm{esc}}$,
{\sc MoCaLaTA} can be applied to small-scale simulations of star-forming
regions.
To calculate EWs, an RT scheme for continuum radiation
needs to be implemented. Since the expressions for both the hydrogen and the
dust cross sections are valid far from the Ly$\alpha$ line center, this should
be rather straightforward, simply requiring knowledge about the ratio of
emitted Ly$\alpha$ and continuum photons.

Simulating simultaneously galactic scales and realistic molecular clouds is
still not feasible.
Performing the RT in a large number of different regions, however, the general
effect of the clumpiness of gas can be scrutinized. This, in turn, can be
implemented in galactic RT simulations if in each cell a sub-grid clumping
factor $C \equiv \langle \nhi^2 \rangle / \langle \nhi \rangle^2$ can be
determined, based on the physical conditions in that cell.


\subsection{Ly$\alpha$ polarization}
\label{sec:lyapol}
\index{Polarization}

When light is scattered it may be polarized, depending on the particular phase
function governing the scattering process, as well as the direction into which
it is scattered. For Ly$\alpha$ emitting galaxies, this has been investigated
numerically only in idealized cases, with spherical, homogeneous, isothermal
galaxies \citep[e.g.][]{dij08}.
Moreover, global magnetic fields has not been considered, which
may or may not have a large impact on the polarization, through the Hanle
effect, Zeeman effect, and/or Faraday rotation, as well as scattering on dust
grains oriented in some preferred direction. Magnetic fields are known to
appear on all scales, from planet formation to galaxy clusters.

With magnetohydrodynamical simulations, far more realistic
configurations of gas and magnetic fields can be modeled. Since {\sc MoCaLaTA}
follows
individual photons (or photon packets), the strategy would be to attach a
Stokes' vector to each
photon, and for each physical process determine the M\"uller matrix associated
with the interaction. The results of these simulations may provide observables
to bring new insight into the physical processes of structure formation.


\subsection{Probing the very first galaxies}
\label{sec:vista}

The Ultra-VISTA\index{Ultra-VISTA} survey mentioned in \sec{lyahist}
is a program that has been
launched recently, utilizing the ESO VISTA telescope, located close to the VLT
on a neighboring peak. In a single pointing, VISTA is capable of observing
almost 1 deg$^2$ in the NIR wavelength region.
One of the purposes of Ultra-VISTA is to secure ultra-deep imaging of LAEs at
$z = 8.8$ in the COSMOS\index{COSMOS} field \citep{sco07}, with a total
of 112 hours of integration (seven of which already have been secured at the
time of writing).
The most promising LAE candidates will probably followed up spectroscopically.
At this redshift, the Ly$\alpha$ line falls in the NIR (the central wavelength
of the filter is 1.19 $\mu$m), a
wavelength region where the sky is both transparent and relatively free of
airglow emission lines.

The deepness and the large volume surveyed will allow to establish a census of
the number and the properties of the very first (proto-)galaxies at a hitherto
unprecedented early epoch, less than 600 million years after the Big Bang.
With {\sc MoCaLaTA} it is possible not only to predict what may be observed,
but also to interpret these outstanding observations.

\begin{figure}[!t]
\centering
\includegraphics [width=1.0\textwidth] {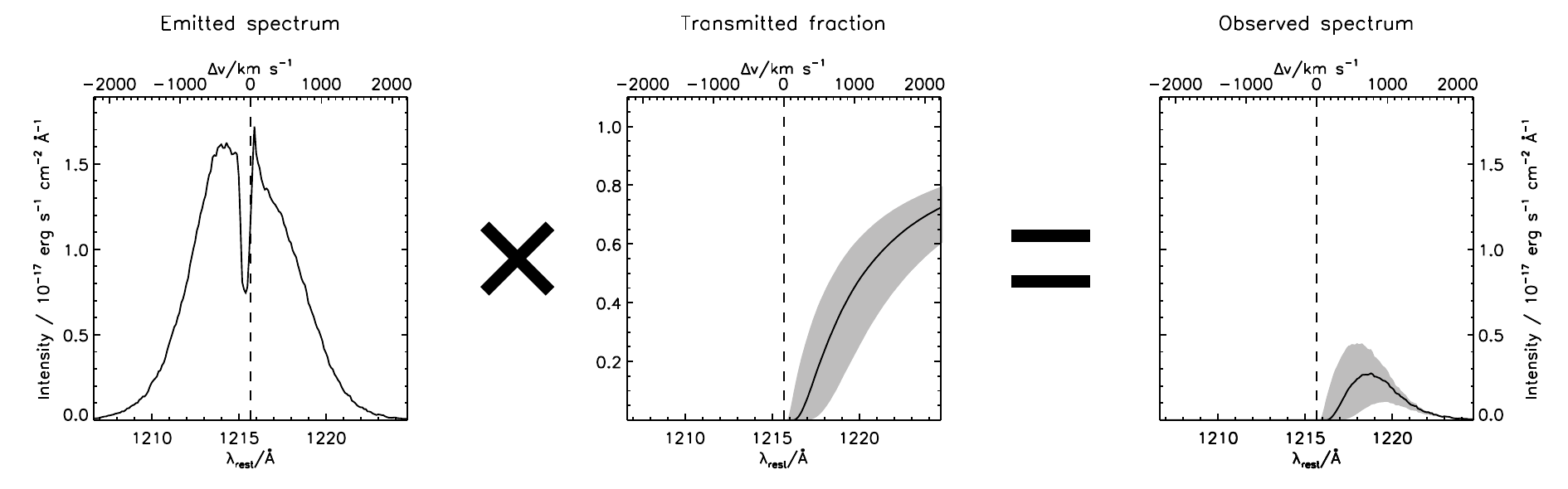}
\caption{{\cap Effect of the IGM at $z = 8.8$. The spectrum escaping a galaxy
               (\emph{right}) was taken from a dust-free version of a
               $z = 3.6$-galaxy.
               The transmission function $\Flam$ (\emph{middle}) is calculated
               at $z = 8.8$ in
               a $\zre = 10$, $\sigma_8 = 0.9$ cosmological simulation.
               The observed spectrum (\emph{left}) will appear significantly
               redshifted relative to the systemic redshift, and will be
               greatly, but not entirely suppressed.}}
\label{fig:z8p8}
\end{figure}

\Fig{z8p8} shows the tentative impact of the IGM on the observability of LAEs
at $z = 8.8$.
As is seen, the essentially completely neutral IGM hardly allows any of the
light on the blue side og the Ly$\alpha$ line center to be transmitted, and
the damping wing of the absorption even extends far into the red side.
However, since little or no dust is present at such high redshifts, the
spectrum of the light escaping the galaxies may be quite extended. As a toy
model, the transmission function $\Flam$ at $z = 8.8$ is applied to the
simulated spectrum
of a dust-free LAE at $z = 3.6$\footnote{Since unfortunately I don't have a
hi-res galaxy at $z = 8.8$ at the moment.}. In this case the fraction
transmitted through
the IGM is $f_{\mathrm{IGM}} = 0.09_{-0.06}^{+0.07}$. This is in agreement
with the analytical results of \citet{dij10}, who found that even in a
completely neutral IGM, as much as $\gtrsim$5\% of the emmitted radation can
be transmitted to the observer.

More simulations are of course needed to substantiate these preliminary
calculations, along with simulations of SBs,
but it is interesting to note that the observed profile in the
left panel of \fig{z8p8} appears redshifted relative to the systemic redshift
by $\sim$3 {\AA}, or $\sim$700 \kms, thus resulting in the probed sample lying
at a slightly lower redshift.



\ \\
\ \\
\begin{center}
--------------------------------------------------
\end{center}
\ \\
\ \\

The prospects of applying {\sc MoCaLaTA} to model individual observed
Ly$\alpha$ spectra and SB profiles are promising. In contrast to previous
attempts, with the adaptive gridding and the environment-dependent prescription
of dust {\sc MoCaLaTA} is capable of examining much more realistic scenarios.
Obviously, the simulated galaxies are limited to a rather small number
compared to model galaxies, where parameters can be tuned at will.
A different approach could be a combination, e.g.~to investigate the general
effect of outflows by emitting realistically simulated spectra from a central
source surrounded by shell of gas, where not only the column density and
expansion velocity, but also anisotropy and clumpiness is varied.

Either way, a grid of simulated spectra and SB profiles can be constructed,
providing a handy, comprehensive reference sample which, through least square
fitting, may be employed to interpret observations of Ly$\alpha$ radiation
from young, dusty galaxies.

\partL{Appendix}\label{appendix}
\appendix

\chapter{Quantum mechanical derivation of the Ly$\alpha$ cross section}
\label{app:quant}
\index{Cross section!Derivation|textbf}

In this appendix a functional form for the cross section of neutral hydrogen
is derived, i.e.~the shape of the wavelength dependent line profile and the
central wavelength. Also the oscillator strength and the lifetime of the
excited state will be derived, and the validity of various approximations is
discussed.

\section{The Hamiltonian}
\label{sec:ham}

\subsection{Total Hamiltonian}

Even in quite weak radiation fields, the density of photons is very high,
implying that we can treat the field as continuous and the interaction
between the photon and the atom as
a perturbation to the Hamiltonian $H$ of the atomic system. For an electron
with position operator $\mathbf{x}$
in a time varying electromagnetic field of scalar and vector potentials
$\Phi(\mathbf{x},t)$ and $\mathbf{A}(\mathbf{x},t)$, the
Hamiltonian is \citep[e.g.,][]{gas96}
\begin{equation}
\label{eq:ham7}
H = \frac{1}{2 m_e} \big| \mathbf{p} + \frac{e}{c} \mathbf{A} \big|^2 - e \Phi,
\end{equation}
where $\mathbf{p} = -i\hbar \mathbf{\nabla}$ is the
momentum operator and $c$
is the speed of light. Interpreting operators as matrices, the $nm$'th element
of, say, $\mathbf{x}$ is
\begin{eqnarray}
\label{eq:xnm}
\mathbf{x}_{nm} & =      & \oint \psi_n^* \mathbf{x} \psi_m dV\\
                & \equiv & \langle n | \mathbf{x} | m \rangle,
\end{eqnarray}
where $^*$ denotes the complex conjugate and the integral is over some volume
enclosing the wavefunction (in principle all space).

With a clever choice of gauge, viz. the Coulomb gauge defined by
$\mathbf{\nabla \cdot A} = 0$, \eq{ham7} can be expanded to
\begin{eqnarray}
\label{eq:ham9}
H & = & \frac{1}{2 m_e} |\mathbf{p}|^2 + e \Phi(\mathbf{x})
    +   \frac{e}{2mc} \mathbf{A \cdot p}
    +   \frac{e^2}{2mc^2} \mathbf{A \cdot A}\\
\nonumber
  & = & H_0 + H_1 + H_2,
\end{eqnarray}
where $H_0 = |\mathbf{p}|^2/2m - e \Phi(\mathbf{x})$ represents the
electrostatic field from the nucleus, and the two last terms the vacuum
radiation field. Since $H_2 \ll H_1 \ll H_0$, $H_1$ can be treated as a small
perturbation to $H_0$, while the non-linear term $H_2$ can be neglected
altogether. This
approximation is valid in the weak field case which is satisfied under the
conditions that we shall be dealing with.

Decomposing $\mathbf{A}$ into its Fourier components,
\begin{equation}
\label{eq:A10}
\mathbf{A}(\mathbf{x},t) = \sum_{\mathbf{k}} \left(
    \mathbf{a}(\mathbf{k}) e^{i (\mathbf{k\cdot x - \omega t})}
  + \mathbf{a}^*(\mathbf{k}) e^{-i (\mathbf{k\cdot x - \omega t})} \right),
\end{equation}
where $\mathbf{k}$ is the wave (or propagation)
vector and
$\omega = 2\pi\nu$ is the angular frequency.
Given the Coulomb
gauge, \eq{A10} implies that $\mathbf{k \cdot a} = 0$. Thus
$\mathbf{a}$ has two components orthogonal to $\mathbf{k}$. These are the
polarization\index{Polarization} states of the radiation field. A more complete
description is then
\begin{equation}
\label{eq:A11}
\mathbf{A}(\mathbf{x},t) = \sum_{\mathbf{k},\varepsilon}
\mathbf{\hat{e}}_{\varepsilon}(\mathbf{k}) \left(
               a_{\varepsilon}(\mathbf{k}) e^{i (\mathbf{k\cdot x - \omega t})} +
               a_{\varepsilon}^*(\mathbf{k}) e^{-i (\mathbf{k\cdot x - \omega t})}
               \right)
\end{equation}
where
$\mathbf{\hat{e}}_\varepsilon \perp \mathbf{k}$, with $\varepsilon = 1,2$.

\subsection{The Hamiltonian of the perturbation}

The Hamiltonian measures the total energy, which for a radiation field is
\citep[e.g.,][]{jac99}
\begin{equation}
\label{eq:Hrad12}
H_{\mathrm{rad}} = \frac{1}{8\pi} \int \left( |\mathbf{E}|^2
                 + |\mathbf{B}|^2 \right) dV,
\end{equation}
where $\mathbf{E}$ and $\mathbf{B}$ is the electric and magnetic field,
respectively. Using the Maxwell's equations that relate $\mathbf{E}$ and
$\mathbf{B}$ to $\mathbf{A}$,
\begin{equation}
\label{eq:max}
\mathbf{E} = - \frac{1}{c} \frac{\partial \mathbf{A}}{\partial t}, \qquad
\mathbf{B} = \nabla \times \mathbf{A},
\end{equation}
together with \eq{A11}, \eq{Hrad12} is evaluated to
\begin{equation}
\label{eq:Hrad18}
H_{\mathrm{rad}} = \sum_{\mathbf{k},\varepsilon}
    |a_{\varepsilon}(\mathbf{k})|^2 \frac{k^2 V}{2\pi}.
\end{equation}
In terms of the photon occupation number
$N_{\varepsilon}(\mathbf{k})$
(the spectrum of the radiation field),
$H_{\mathrm{rad}} = \sum_{\mathbf{k},\varepsilon} \hbar \omega
N_{\varepsilon}(\mathbf{k})$, so that comparison with \eq{Hrad18}
yields
\begin{equation}
\label{eq:ak20}
|a_{\varepsilon}(\mathbf{k})| = c \left( \frac{h N_{\varepsilon}(\mathbf{k})}
                                              {V \omega} \right)^{1/2}.
\end{equation}
From Eqs.~\ref{eq:ham9}, \ref{eq:A11}, and \ref{eq:ak20} we get the
perturbation Hamiltonian in the continuum limit
\begin{equation}
\label{eq:H1}
H_1 = \sum_{\varepsilon = 1}^2 \frac{V}{(2\pi)^3} \int \left(
      H_{\varepsilon}^{\mathrm{abs}} (\mathbf{k}) e^{-i \omega t}
    + H_{\varepsilon}^{\mathrm{em}}  (\mathbf{k}) e^{ i \omega t} \right)d^3 k,
\end{equation}
where
\begin{equation}
\label{eq:Habs}
H_{\varepsilon}^{\mathrm{abs}} (\mathbf{k}) = \frac{e}{m_e}
    \left[ \frac{h}{\omega V} N_{\varepsilon}(\mathbf{k}) \right]^{1/2}
    e^{i \mathbf{k \cdot x}} \mathbf{\hat{e}}_{\varepsilon}(\mathbf{k})
    \cdot \mathbf{p}
\end{equation}
and
\begin{equation}
\label{eq:Hem}
H_{\varepsilon}^{\mathrm{em}} (\mathbf{k}) = \frac{e}{m_e}
    \left[ \frac{h}{\omega V}
    \Big(1 + N_{\varepsilon}(\mathbf{k})\Big) \right]^{1/2}
    e^{-i \mathbf{k \cdot x}} \mathbf{\hat{e}}_{\varepsilon}(\mathbf{k})
    \cdot \mathbf{p}.
\end{equation}
The time evolution of the electron wave function is governed by the
time-dependent Schr\"odinger equation
\begin{equation}
\label{eq:tsch}
H \psi = i \hbar \frac{\partial \psi}{\partial t}.
\end{equation}
where $H = H_0 + H_1$. $H_0$ has eigenfunctions of the form
\begin{equation}
\label{eq:eig}
\psi_j (\mathbf{x},t) = \phi_j(\mathbf{x}) e^{-i E_j t / \hbar},
\end{equation}
where the energy $E_j$ of the $j$'th eigenstate $\phi_j$ is an eigenvalue of the
time-\emph{in}dependent Schr\"odinger equation $H_0 \phi_j = E_j \phi_j$.
Since for $H_0$ the set of all $\phi_j$'s forms a basis for any possible
wavefunction for stationary atomic systems, we can construct any perturbed
wavefunction from
\begin{equation}
\label{eq:psi29}
\psi (\mathbf{x},t) = \sum_j c_j(t) \phi_j(\mathbf{x}) e^{-i E_j t / \hbar},
\end{equation}
where $c_j(t)$ is the time-dependent amplitude of the $j$'th eigenstate.
With the initial conditions $\psi(\mathbf{x},t\le0) = \phi_i(\mathbf{x})$,
i.e.~$c_j(t\le0) = \delta_{ij}$, using \eq{tsch} the $c_j(t)$'s are
evaluated to first order in the perturbation $H_1$ to give
\begin{equation}
\label{eq:c34}
c_f(t) = -\frac{i}{\hbar} \int_0^t \langle f | H_1 | i \rangle
         e^{i \omega_{fi} t} dt.
\end{equation}
%


\section{Absorption}
\label{sec:abs}
\index{Absorption!Quantum mechanically}

To see how the absorption of photons happens, the transition probability for
an electron absorbing a photon is now derived. Considering only the absorption
part of \eq{H1}, integrating \eq{c34} and squaring yields
the probability of absorbing a photon of frequency $\omega = c k$:
\begin{equation}
\label{eq:cfsq}
|c_f(\mathbf{k},t)|^2 = \frac{1}{\hbar^2} |\langle f|
                        H^{\mathrm{abs}}_{\varepsilon}|i\rangle |^2
                        \frac{\sin^2[(\omega-\omega_{fi})/2]}
                             {[(\omega-\omega_{fi})/2]^2}.
\end{equation}
To get the total transition probability $P_{if}$,
\eq{cfsq} is summed
over all $\mathbf{k}$ and $\varepsilon$, so that with the relation
$d^3 k = k^2 dk\,d\Omega = (\omega^2 / c^3)d\omega\, d\Omega$,
\begin{equation}
\label{eq:Pif39}
P_{if} = \left( \frac{e}{2\pi m_e} \right)^2 \sum_{\varepsilon =1}^2 \int
         \frac{N_{\varepsilon}(\mathbf{k})}{\hbar\omega}
         \big| \langle f |
               e^{i \mathbf{k \cdot x}}
               \mathbf{\hat{e}}_{\varepsilon}(\mathbf{k}) \cdot \mathbf{p}
         |i \rangle \big|^2
         \frac{\sin^2[(\omega-\omega_{fi})/2]} {[(\omega-\omega_{fi})/2]^2}
         \frac{\omega^2}{c^3}d\omega\,d\Omega.
\end{equation}
Note that the dependence on $V$ disappeared, and that $P_{if}$ is strongly
peaked for $t \gg 2/\omega_{fi}$.

\eq{Pif39} predicts that a
transition will eventually occur. The interesting quantity is the transition
probability \emph{rate}
\begin{equation}
\label{eq:dPdt42}
\frac{d P_{if}}{d t} = \sum_{\varepsilon = 1}^2
                       \int \omega_{\varepsilon} d\Omega,
\end{equation}
where $\omega_{\varepsilon}$ is the constant probability rate per $d\Omega$ for
the radiative transition $i \to f$. Thus, we must evaluate
$\big| \langle f | e^{i \mathbf{k \cdot x}}
\mathbf{\hat{e}}_{\varepsilon}(\mathbf{k}) \cdot \mathbf{p} |i \rangle \big|^2$.
In general, this is very difficult, but expanding in a multipole expansion
greatly simplifies the problem.

Because $\mathbf{k \cdot x} \ll 1$, we can approximate $e^{i\mathbf{k\cdot x}}$
with unity. This is the so-called \emph{dipole approximation}\index{Dipole
approximation} and corresponds
to neglecting retardation across the atom. This is a very good approximation
for UV radiation, but becomes inadequate for X-ray transitions. From
Eqs.~\ref{eq:Habs} and \ref{eq:dPdt42}, and the relation
$\mathbf{p} = i m_e [H_0,\mathbf{x}]/\hbar$, we then have
\begin{equation}
\label{eq:dPdt46}
\frac{d P_{if}}{d t} = \frac{e^2}{h c^3} \sum_{\varepsilon} \oint
    \left[ N_{\varepsilon} \omega^3
    \big| \mathbf{\hat{e}}_{\varepsilon} \cdot \mathbf{x}_{fi} \big|^2
    \right]_{fi} d\Omega.
\end{equation}
\begin{figure}[!t]
\centering
\includegraphics [width=0.50\textwidth] {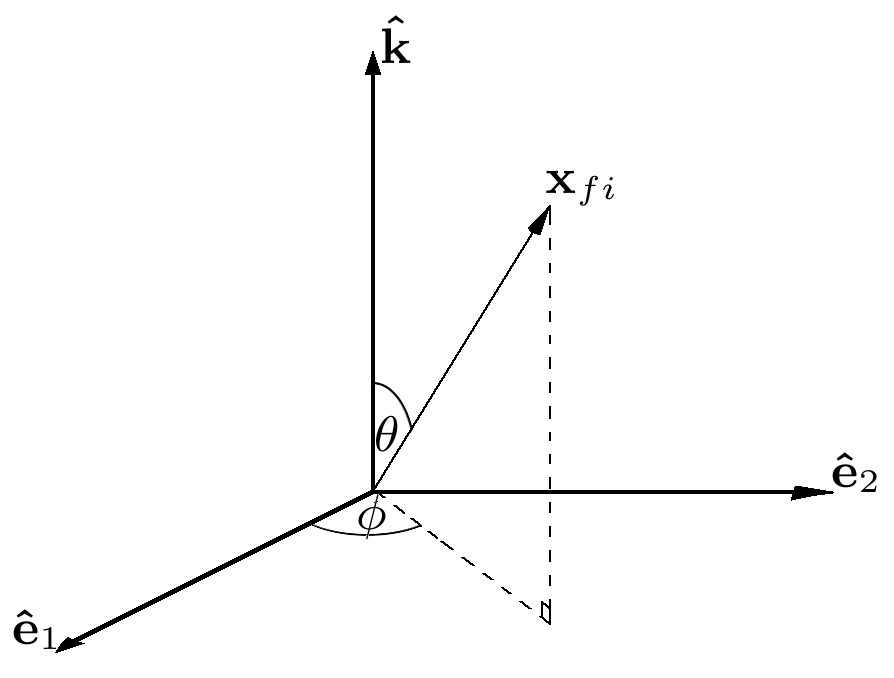}
\caption{{\small Geometrical interpretation of the matrix element
                 $\mathbf{x}_{fi}$ in spherical coordinates $r$, $\theta$ and
                 $\phi$. The vector represents the orientation of
                 the atomic charge distribution while $\mathbf{\hat{e}}_1$, 
                 $\mathbf{\hat{e}}_2$, and $\mathbf{\hat{k}}$ determine the
                 orientation of the photon field and propagation vectors,
                 respectively.}}
\label{fig:x_fi}
\end{figure}
%
From Fig.~\ref{fig:x_fi},
\begin{subequations}
\begin{eqnarray}
\label{eq:ex}
\mathbf{\hat{e}}_1 \cdot \mathbf{x}_{fi} & = &
                        |\mathbf{x}_{fi}| \sin\theta \cos\phi;\\
\mathbf{\hat{e}}_2 \cdot \mathbf{x}_{fi} & = &
                        |\mathbf{x}_{fi}| \sin\theta \sin\phi.
\end{eqnarray}
\end{subequations}
For an isotropic, unpolarized field, $N_1 = N_2 = N(\omega)/2$, so that
\eq{dPdt46} becomes
\begin{equation}
\label{eq:dPdt49}
\frac{d P_{if}}{d t} = \frac{4\pi}{3} \frac{e^2}{h c^3} N(\omega_{fi})
                       \omega_{fi}^3 |\mathbf{x}_{fi}|.
\end{equation}
Conventionally, to translate the probability of absorption into a quantity with
the dimensions of area, the bound-bound absorption cross section
$\sigma_\omega$
is defined such that\index{Cross section!Conventional definition}
\begin{equation}
\label{eq:dPdt50}
\frac{d P_{if}}{d t} = \int_0^\infty \sigma_\omega c N(\omega)
\frac{4\pi \omega^2}{(2\pi)^3 c^3} d\omega,
\end{equation}
so that, from \eq{dPdt49},
\begin{equation}
\label{eq:sigbb51}
\sigma_\omega = \frac{4\pi^2}{3} \frac{e^2}{\hbar c}
    |\mathbf{x}_{fi}|^2 \omega \delta(\omega - \omega_{fi}).
\end{equation}
Comparison with \eq{XsecClas} yields\index{Line profile!Delta
function}
\begin{equation}
\label{eq:phi53}
\phi_{if}(\nu) = \omega\delta(\omega - \omega_{fi})
               = 2\pi\nu\delta(\nu - \nu_{fi}),
\end{equation}
and\index{Oscillator strength|textbf}
\begin{equation}
\label{eq:fif54}
f_{if} = \frac{2 m_e}{3\hbar} \omega_{fi} |\mathbf{x}_{fi}|^2.
\end{equation}
%

\section{Natural broadening of the line profile}
\label{sec:nat}\index{Line profile!Natural}

A quantum mechanical interpretation of the oscillator strength has now been
derived. In the above calculation, however, the line profile was taken to be a
delta function. But since emission processes compete with absorption processes,
the absorption rate is affected by the transition rate for spontaneous
emission\index{Spontaneous emission}, i.e.~the atom decays to the ground state
again. This, in fact, is the is concept of the resonant scattering
\index{Resonant scattering} process. In principle, there is no fundamental
difference between ``absorbing and re-emitting'' and ``scattering'' a photon,
as long as the decay is to the same state as the initial state, since the
energy of the photon is the same before and after the event in both cases (in
the reference frame of the atom).

Before evaluating the oscillator strength, we will derive the proper line
profile. Let\footnote{This is the so-called \emph{Einstein A coefficient};
sometimes written $A_{fi}$.}
$\Gamma \equiv dP_{fi}/dt$. Then the transition rate is
\begin{eqnarray}
\label{eq:cdot55}
\nonumber
\frac{d}{dt} \Big[ \textrm{occupation of level $f$} \Big] & = & 
           - \Big[ \textrm{probability of decay}    \Big] \times
             \Big[ \textrm{occupation of level $f$} \Big] \\
\nonumber
\frac{d}{dt} |c_f|^2 & = & - \Gamma |c_f|^2 \\
\dot{c}_f\Big|_{\mathrm{spon.em.}} & = & - \frac{1}{2} \Gamma c_f,
\end{eqnarray}
where dot denotes differentiation with respect to time. For spontaneous
emission $N(\omega) = 1$, so with $|\mathbf{x}_{fi}| = |\mathbf{x}_{if}|$,
adding the above result to the expression for $\dot{c}_f(t)$ yields
\begin{equation}
\label{eq:sig60}
\sigma_\omega = \frac{4\pi^2}{3} \frac{e^2}{\hbar c}
    |\mathbf{x}_{fi}|^2 \omega_{fi} \mathcal{L}(\omega),
\end{equation}
where\index{Lorentzian}
\begin{equation}
\label{eq:L61}
\mathcal{L}(\omega) = \frac{1}{\pi} \frac{\Gamma/2}
                      {(\omega - \omega_{fi})^2 + (\Gamma/2)^2}
\end{equation}
is the Lorentzian profile associated with the natural line profile. This
profile is characterized by a sharp peak of width $\Gamma$ centered at
$\omega_{fi}$, and broad wings.


\section{Transition rate}
\label{sec:tran}

To calculate the transition probability for the downward transition
$(n',\ell',m') \to (n,\ell,m)$, where $n$, $\ell$ and $m$ are the
principal, orbital, and
magnetic quantum numbers, respectively, we must evaluate the matrix
element $\langle i | \mathbf{x} | f \rangle$. Since the wavefunction for the
hydrogen atom is rotationally invariant,
it can be separated as
\begin{equation}
\label{eq:f62}
| f \rangle = | n \ell m \rangle = R_{n\ell}(r) Y_\ell^m(\theta,\phi)
\end{equation}
where $R_{n\ell}(r)$ and $Y_\ell^m(\theta,\phi)$ are the radial wavefunction
and spherical harmonic, respectively \citep[see, e.g.,][]{sak94}. Writing
$\mathbf{x}$ as
\begin{equation}
\label{eq:x65}
\mathbf{x} = \frac{1}{2} \big[ (x+iy)(\mathbf{\hat{e}}_y - i\mathbf{\hat{e}}_y)
                             + (x-iy)(\mathbf{\hat{e}}_y + i\mathbf{\hat{e}}_y)
                         \big] + z\mathbf{\hat{e}}_z,
\end{equation}
we see that $\mathbf{\hat{e}}_{\varepsilon} \cdot \mathbf{x} = z$ represents
interaction with a wave polarized\index{Polarization} in the
$z$-direction, and hence traveling in the $xy$-plane, while
$\mathbf{\hat{e}}_{\varepsilon} \cdot \mathbf{x} \propto x \pm iy$ represents
a wave traveling in the $z$-direction and having left or right circular
polarization. These directions of polarization can be expressed in terms of
spherical harmonics as
\begin{eqnarray}
\label{eq:pol66}
z        & = & r\cos\theta = \left( \frac{4\pi}{3} \right)^{1/2} r Y_1^0;\\
\label{eq:pol67}
x \pm iy & = & r\sin\theta e^{\pm i\phi}
           = \left( \frac{8\pi}{3} \right)^{1/2} r Y_1^{\pm 1}.
\end{eqnarray}
For the Ly$\alpha$ decay $|21m'\rangle \to |100\rangle$, there are three
possible values of
$m'$, and $Y$ must be evaluated separately in each case. However, after a bit
of algebra, it is found from Eqs.~\ref{eq:f62}--\ref{eq:pol67} for all $m'$
that
\begin{equation}
\label{eq:x75}
\big| \langle 21m'| \mathbf{x} | 100 \rangle \big|^2
     = \frac{2^{15}}{3^{10}} a_0^2,
\end{equation}
where $a_0 = \hbar^2/m_e e^2$ is the Bohr radius. The transition rate is then
evaluated to
\begin{equation}
\label{eq:rate}
\Gamma = 6.25\times10^{8} \textrm{ s}^{-1}.
\end{equation}
This implies that the half-life\index{Half-life!Natural} $t_{1/2}$ of
the excited state is
\begin{equation}
\label{eq:t12app}
t_{1/2} = \frac{1}{\Gamma} = 1.60\times10^{-9} \textrm{ s}.
\end{equation}

By a coordinate transformation $\mathbf{x} \to -\mathbf{x}$, it can be
shown that the scalar product
$\mathbf{\hat{e}}_\varepsilon \cdot \mathbf{x}_{fi}$, and
hence the transition probability, is  non-vanishing only if $\ell + \ell'$
is odd, i.e.~if the initial and final state have opposite parity. For this
reason, the electron can be excited from the ground state $| 100 \rangle$ (the
``$1S$'' state) to $| 21m' \rangle$ (the three ``$2P$'' states), but not to
$| 200 \rangle$ (the ``$2S$'' state), since $\ell_{1S} = \ell_{2S}$. 
Expanding the term $e^{i\mathbf{k \cdot x}}$ without making the dipole
approximation yields a non-zero probability of this ``forbidden'' transition.
\index{Destruction of photons} The lifetime of the $2S$ is of the order of a
second, i.e.~$\sim8$ orders of
magnitude larger than that of the $2P$ states, and in fact the most
probable decay is through an intermediate state of opposite parity, with the
emission of two photons as result. Obviously, this would lead to the destruction
of the Ly$\alpha$ photon. However, since the probability of being excited to
the $2S$ state is then $\sim8$ orders of magnitude smaller than to a $2P$
state, this effect can be safely neglected.

For the physical conditions governing the situations that we shall be dealing
with, $t_{1/2}$ can effectively be regarded as instantaneous.
Even in the very dense regions,
the density $\nhi$ of neutral
hydrogen rarely exceeds $\sim10^4$ cm$^{-3}$. In these regions, the
temperature is $\sim10^4$ K, corresponding to a typical velocity of the order
of 10 km s$^{-1}$. With a collisional cross section
\index{Cross section!Collisional}\index{Collisions}
of the order of $n^2 a_0^2$ for hydrogen atoms,
in one second the hydrogen atom sweeps out a volume of $\sim10^{-11}$ cm$^3$,
so the probability of colliding with another atom is at most
$10^{-6}$ s$^{-1}$.  However,
for high densities and temperatures, collisions may perturb the atom before
de-excitation, in which case the above calculations are no longer valid.

The excited atom can also collide with free electrons and protons. For thermal
protons and electrons at $T = 10^4$ K, carrying out calculations similar to the
above one finds that collisional cross sections $q_p$ and $q_e$ for the
transitions
\begin{equation}
\label{eq:2P2S}
\textrm{H}(2P) + p,e \longrightarrow \textrm{H}(2S) + p,e
\end{equation}
are $q_p = 4.74\times10^{-4}$ cm$^3$ s$^{-1}$ and $q_e = 5.70\times10^{-5}$
cm$^3$ s$^{-1}$, respectively \citep{ost89}. For proton densities lower than
$\sim10^4$ cm$^{-3}$, the $2P \to 2S$ transition can be relatively important,
but for higher densities, the reverse transition $2S \to 2P$ cancels out the
destruction effect \citep{ost89}. The ratio between the transition rates gives
the probability $P_{\mathrm{destr.}}$ of photons being destroyed by two-photon
processes\index{Two-photon process}\index{Destruction of photons}:
\begin{eqnarray}
\label{eq:Pdestr}
P_{\mathrm{destr.}} & = & \frac{P_{\textrm{{\tiny $2P\!\!\to\!\!2S$}}}}
                               {P_{\textrm{{\tiny $2P\!\!\to\!\!1S$}}}}\\
                    & = & \frac{q_p n_p + q_e n_e}{\Gamma}\\
               & \simeq & 8.5\times10^{-13} n_p,
\end{eqnarray}
where the density $n_p$ of protons is assumed to be approximately equal to the
density $n_e$ of electrons. Taking into account the temperature dependence of
$q_p$ and $q_e$ introduces a factor of $(T/10^4\textrm{ K})^{0.17}$
\citep{neu90}. Obviously, collisional transitions can be safely neglected.

The energy $E_1$ of the ground state is
\begin{equation}
\label{eq:E1}
E_1 = - \frac{e^2}{2a_0} = -13.6 \textrm{ eV},
\end{equation}
so the energy of the $2P$ state is $E_2 = E_1 / n^2 = -3.4$ eV. The angular
frequency of the Ly$\alpha$ transition is then
\begin{equation}
\label{eq:om21}
\omega_{21} = \frac{E_2 - E_1}{\hbar} = 1.55 \times 10^{16} \textrm{ s}^{-1},
\end{equation}
so that the true \emph{line center} frequency $\nu_0 \equiv \omega_{fi}/2\pi$
is
\begin{equation}
\label{eq:nu0}
\boxed{
\nu_0 = 2.466\times10^{15} \textrm{ s}^{-1}.
}
\end{equation}

We are now ready to evaluate the oscillator strength and hence the absorption
cross section. Due to the three-fold degeneracy of the $2P$, the effective
oscillator strength of the upward transition is three times that of the
downward transition. Thus, from Eqs.~\ref{eq:fif54} and \ref{eq:om21}
\index{Oscillator strength}
\begin{equation}
\label{eq:f12}
\boxed{
f_{12} = 0.4162.
}
\end{equation}

By the uncertainty principle, the finite lifetime of the excited state
tranlates into an uncertainty $\Delta E_2 = \hbar/t_{1/2}$ in the energy of the
state. Expressed in terms of frequency this yields an uncertainty
$\Delta\nu_L = \Delta E_2 / h$, or\index{Natural line width}
\index{Line width!Natural}
\begin{equation}
\label{eq:DnuLapp}
\boxed{
\Delta\nu_L = 9.936\times10^7 \textrm{ s}^{-1}.
}
\end{equation}
The absorption cross section $\sigma_\nu$ for the Ly$\alpha$ transition is then
\index{Cross section!Absorption|textbf}
\begin{equation}
\label{eq:signu}
\boxed{
\sigma_\nu = f_{12} \frac{\pi e^2}{m_e c}
                   \frac{\Delta\nu_L/2\pi}{(\nu-\nu_0)^2 + (\Delta\nu_L/2)^2}.
}
\end{equation}
%



\chapter{Paper I}
\label{app:lau09a}

\begin{center}
The following page displays the abstract from the paper\\
\ \\
\textbf{\LARGE{Lyman $\alpha$ Radiative Transfer in Cosmological Simulations
using Adaptive Mesh Refinement}}\\
\ \\
by\\
\ \\
{\Large Peter Laursen, Alexei O. Razoumov, \& Jesper Sommer-Larsen}\\
\ \\
published in The Astrophysical Journal, vol.~696, pp.~853--869, 2009 May 1
\citep{lau09a}.\\
\ \\
\ \\
\underline{\ \ \ \ \ \ \ \ \ \ \ \ \ \ \ \ \ \ \ \ \ \ \ \ \ \ \ \ \ \ \ \ \ \ \ \ \ \ \ \ \ \ \ \ \ \ }
\end{center}
\ \\
\ \\
A version of this thesis with the full papers in appendices can be downloaded
from the URL
\href{http://www.dark-cosmology.dk/~pela/Phd/PhDthesis_full.pdf}
     {http://www.dark-cosmology.dk/\~{}pela/Phd/PhDthesis.pdf}.\vspace{1mm}
\ \\
Note that in the text following immediately after Eq.~17 in the published
paper a typographical error has sneaked in, misplacing
inequality symbols ``$\le$'' and ``$>$'' by
``$\ge$'' and ``$<$'', respectively. An erratum has been made \citep{lau09c}.

\newpage
\ \\
\newpage

 \begin{figure}
 \centering
 \hspace*{-20mm}
  \includegraphics [width=1.50\textwidth] {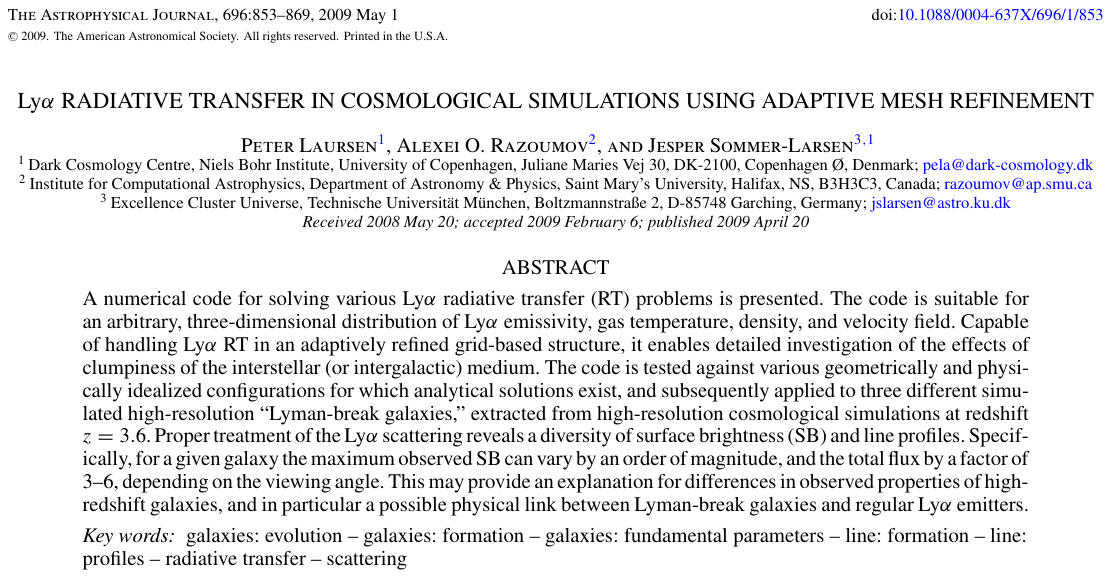}
 \end{figure}
 %


\chapter{Paper II}
\label{app:lau09b}

\ \\
\ \\
\begin{center}
The following page displays the abstract from the paper\vspace{1mm}\\
\ \\
\textbf{\LARGE{Lyman $\alpha$ Radiative Transfer with Dust:\\
       Escape Fractions from Simulated High-Redshift Galaxies}}\\
\ \\
by\\
\ \\
{\Large Peter Laursen, Jesper Sommer-Larsen, \& Anja C. Andersen}\\
\ \\
published in The Astrophysical Journal, vol.~704, pp.~1640--1656, 2009
October 20 \citep{lau09b}.\\
\ \\
\ \\
\underline{\ \ \ \ \ \ \ \ \ \ \ \ \ \ \ \ \ \ \ \ \ \ \ \ \ \ \ \ \ \ \ \ \ \ \ \ \ \ \ \ \ \ \ \ \ \ }
\end{center}

\newpage
\ \\
\newpage

 \begin{figure}
 \centering
 \hspace*{-20mm}
  \includegraphics [width=1.50\textwidth] {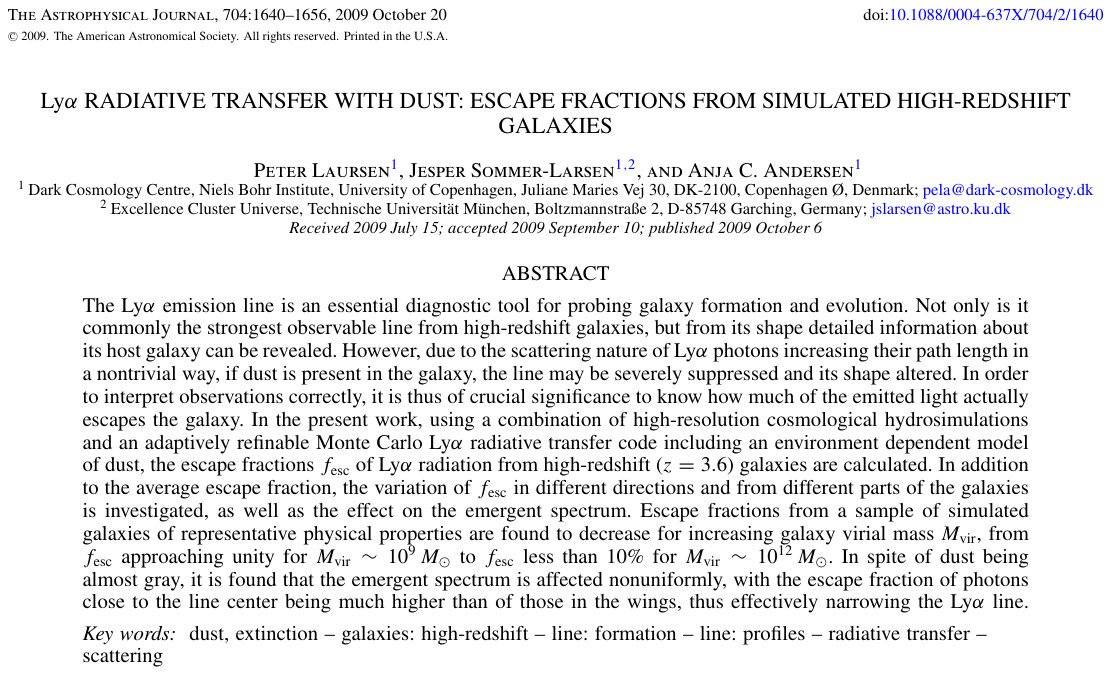}
 \end{figure}
 %


\chapter{Paper III}
\label{app:lau10a}

\ \\
\ \\
\begin{center}
The following page displays the abstract from the paper\vspace{1mm}\\
\ \\
\textbf{\LARGE{Intergalactic Transmission and its Impact on the Ly$\alpha$
Line}}\\
\ \\
by\\
\ \\
{\Large Peter Laursen, Jesper Sommer-Larsen, \& Alexei O. Razoumov}\\
\ \\
accepted for publication in The Astrophysical Journal in January 2011
\citep{lau10a}.\\
\ \\
\ \\
\underline{\ \ \ \ \ \ \ \ \ \ \ \ \ \ \ \ \ \ \ \ \ \ \ \ \ \ \ \ \ \ \ \ \ \ \ \ \ \ \ \ \ \ \ \ \ \ }
\end{center}

\newpage
\ \\
\newpage

 \begin{figure}
 \centering
 \hspace*{-20mm}
  \includegraphics [width=1.50\textwidth] {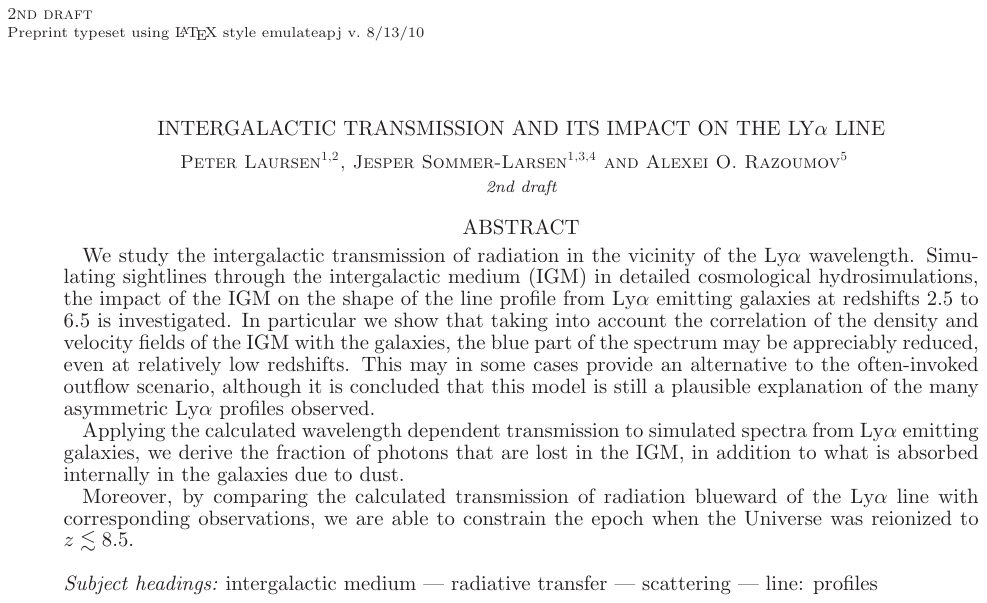}
 \end{figure}
 %


\chapter{Paper IV}
\label{app:fyn10}

\ \\
\ \\
\begin{center}
The following page displays the abstract from the paper\vspace{1mm}\\
\ \\
\textbf{\LARGE{Galaxy Counterparts of metal-rich Damped Lyman-$\alpha$
Absorbers -- I:\\
The case of the $z=2.35$ DLA towards\vspace{2mm}\\ Q2222-0946}}\\
\ \\
by\\
\ \\
{\Large J. P. U. Fynbo,
P. Laursen,
C. Ledoux,
P. M{\o}ller,
P. Goldoni,
B. Gullberg,
L. Kaper,
J. R. Maund,
P. Noterdaeme,
G. \"Ostlin,
M. L. Strandet,
S. Toft,
P. M. Vreeswijk, \&
T. Zafar
}\\
\ \\
published in Monthly Notices of the Royal Astronomical Society
vol.~408, pp.~2128--2136, November 2010 \citep{fyn10}.\\\
\ \\
\ \\
\underline{\ \ \ \ \ \ \ \ \ \ \ \ \ \ \ \ \ \ \ \ \ \ \ \ \ \ \ \ \ \ \ \ \ \ \ \ \ \ \ \ \ \ \ \ \ \ }
\end{center}

\newpage
\ \\
\newpage

 \begin{figure}
 \centering
 \hspace*{-20mm}
  \includegraphics [width=1.50\textwidth] {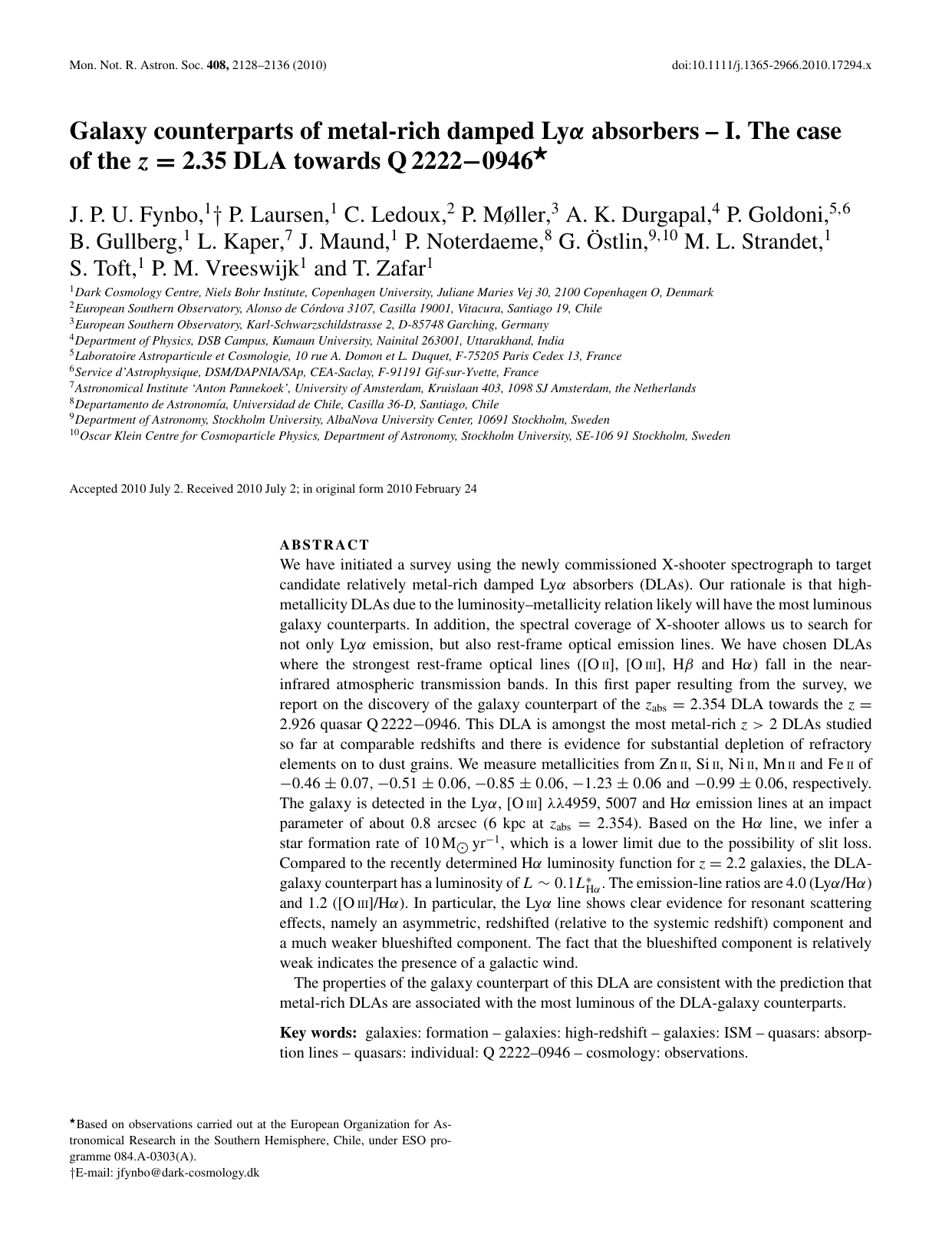}
 \end{figure}
 %



{}

\newpage
\ \\
\newpage

\label{ind}
\printindex

\end{document}